\documentclass[twocolumn,astrosymb]{aastex701}

\usepackage{CJK,amssymb}

\usepackage{graphicx}
\usepackage{caption}
\usepackage{subcaption}
\usepackage{rotating}

\accepted{for publication in ApJ}
\begin{document}
\begin{CJK*}{UTF8}{gbsn}

\title{Time Evolution of Optical Darkness in GRB Afterglow: The Case of GRB~240825A}
\shorttitle{Optical Darkness in GRB~240825A afterglow}
\shortauthors{Li et al.}

\correspondingauthor{Jirong Mao}
\email{jirongmao@mail.ynao.ac.cn}

\author[orcid=0000-0002-4205-0933, sname='R.-Z.', gname='Li']{Rui-Zhi Li}
  \affiliation{Yunnan Observatories,
    Chinese Academy of Sciences, 
    Kunming 650216, People's Republic of China}
  \affiliation{University of Chinese Academy of Sciences,
    Beijing 101408, People's Republic of China}
  \affiliation{Center for Astronomical Mega-Science, 
    Chinese Academy of Sciences,
    Beijing 100012, People's Republic of China}
  \email[show]{liruizhi@ynao.ac.cn}

\author[orcid=0000-0002-7077-7195, sname='J.', gname='Mao']{Jirong Mao}
  \affiliation{Yunnan Observatories,
    Chinese Academy of Sciences, 
    Kunming 650216, People's Republic of China}
  \affiliation{Center for Astronomical Mega-Science, 
    Chinese Academy of Sciences,
    Beijing 100012, People's Republic of China}
  \affiliation{Key Laboratory for the Structure and Evolution of Celestial Objects, 
    Chinese Academy of Sciences, 
    Kunming 650216, People's Republic of China}
  \email[]{jirongmao@mail.ynao.ac.cn}

\author[orcid=0000-0001-6374-8313, sname='Y.-P.', gname='Yang']{Yuan-Pei Yang}
  \affiliation{South-Western Institute for Astronomy Research, 
    Yunnan University, 
    Kunming 650504, 
    People's Republic of China}
 \email[]{ypyang@ynu.edu.cn}

\author[orcid=0000-0001-9342-1485, sname='B.-T.', gname='Wang']{Bo-Ting Wang}
  \affiliation{Yunnan Observatories,
    Chinese Academy of Sciences, 
    Kunming 650216, People's Republic of China}
  \affiliation{University of Chinese Academy of Sciences,
    Beijing 101408, People's Republic of China}
  \affiliation{Center for Astronomical Mega-Science, 
    Chinese Academy of Sciences,
    Beijing 100012, People's Republic of China}
  \email[]{wangbaiting@ynao.ac.cn}
    
\author[orcid=0009-0004-9835-353X, sname='F.-F.', gname='Song']{Fei-Fan Song}
  \affiliation{Yunnan Observatories,
    Chinese Academy of Sciences, 
    Kunming 650216, People's Republic of China}
  \affiliation{University of Chinese Academy of Sciences,
    Beijing 101408, People's Republic of China}
  \affiliation{Center for Astronomical Mega-Science, 
    Chinese Academy of Sciences,
    Beijing 100012, People's Republic of China}
  \email[]{songfeifan@ynao.ac.cn}
    
\author[orcid=0000-0003-1991-776X, sname='Y.-X.', gname='Xin']{Yu-Xin Xin}
  \affiliation{Yunnan Observatories,
    Chinese Academy of Sciences, 
    Kunming 650216, People's Republic of China}
  \affiliation{Center for Astronomical Mega-Science, 
    Chinese Academy of Sciences,
    Beijing 100012, People's Republic of China}
  \affiliation{Key Laboratory for the Structure and Evolution of Celestial Objects, 
    Chinese Academy of Sciences, 
    Kunming 650216, People's Republic of China}
  \email[]{xyx@ynao.ac.cn}
    
\author[sname='J.-M.', gname='Bai']{Jin-Ming Bai}
  \affiliation{Yunnan Observatories,
    Chinese Academy of Sciences, 
    Kunming 650216, People's Republic of China}
  \affiliation{Center for Astronomical Mega-Science, 
    Chinese Academy of Sciences,
    Beijing 100012, People's Republic of China}
  \affiliation{Key Laboratory for the Structure and Evolution of Celestial Objects, 
    Chinese Academy of Sciences, 
    Kunming 650216, People's Republic of China}
  \email[]{baijinming@ynao.ac.cn}

%% Use the \collaboration command to identify collaborations. This command
%% takes an optional argument that is either a number or the word "all"
%% which tells the compiler how many of the authors above the command to
%% show. For example "\collaboration[all]{(DELVE Collaboration)}" wil include
%% all the authors above this command.
%%
%% Mark off the abstract in the ``abstract'' environment. 
\begin{abstract}
    % Context. 
    % Optically dark long-duration gamma-ray bursts (GRBs) are believed to occur in star-forming regions. 
    Long-duration gamma-ray bursts (GRBs) are believed to occur in star-forming regions.
    % Aims. 
    % The multiwavelength follow-up observations of the early afterglow of GRB 240825A provided insights into the evolution of the afterglow, the interaction between the relativistic jet and the circumburst medium, and the physical properties of its host galaxy. 
    The multiwavelength follow-up observations of the early afterglow of GRB 240825A provided insights into the evolution of the optical-to-X-ray spectral feature of the afterglow.
    % Methods. 
    % We comprehensively investigate the evolution of X-ray spectral properties through time-resolved spectral analysis and fit the X-ray-to-optical SEDs of afterglow in different time intervals to derive the extinction curves of the host. The optical darkness ($\beta_\mathrm{OX}$) is calculated in different optical bands to reveal the physical properties of the circumburst medium. 
    % Results. 
    % We find that the initial X-ray column density ($N_\mathrm{H,X}\sim 1.7 \times 10^{22} \mathrm{\,cm^{-2}}$) rapidly decreased by about a factor of two at $\sim 150\mathrm{\,s}$ post-trigger, then gradually increased to $\sim 1.2 \times 10^{22} \mathrm{\,cm^{-2}}$ and stabilized at later times. The $\beta_\mathrm{OX}$ exhibits a trend of decreasing and then increasing, reaching its minimum value at $\sim3000\mathrm{\,s}$ post-trigger. 
    We comprehensively investigate the evolution of X-ray spectral properties through time-resolved spectral analysis and calculate optical darkness ($\beta_\mathrm{OX}$) to reveal the physical properties of the afterglow. 
    The X-ray-to-optical SEDs of afterglow in different time intervals are fitted to derive the extinction curves. 
    % We find that an likely thermal component in the early X-ray emission of the afterglow, which may explain the underlying spectral curvature at $\sim10\mathrm{\,keV}$. 
    The $\beta_\mathrm{OX}$ exhibits a trend of decreasing and then increasing, reaching its minimum value at $\sim1000\mathrm{\,s}$ post-trigger. However, at 11 hours post-trigger, $\beta_\mathrm{OX}$ does not meet the criteria for an optically dark burst.
    The extinction curves in different time intervals indicate that GRB 240825A occurred in a dust-obscured environment.
    % Conclusions.
    % These results might be interpreted that the dense medium near the progenitor is rapidly photoionized by intense radiation during the early stages of the burst. Subsequent recombination processes or the photo-destruction of dust contribute to the extinction of short-wavelength emission. Finally, the afterglow emission propagates through a more diffuse medium, leading to the stabilization of X-ray absorption and the transition from an optically dark to an optically bright state.
    % \textbf{Considering the special physical properties of the burst, we propose that the thermal component may originate from the shock breakout of a supernova. This provides key insights into the underlying mechanisms of the GRB possibly associated with a supernova.}
\end{abstract}

%% Keywords should appear after the \end{abstract} command. 
%% The AAS Journals now uses Unified Astronomy Thesaurus (UAT) concepts:
%% https://astrothesaurus.org/concept-select/
%% You will be asked to selected these concepts during the submission process
%% but this old "keyword" functionality is maintained in case authors want
%% to include these concepts in their preprints.
%%
%% You can use the \uat command to link your UAT concepts back its source.

%% From the front matter, we move on to the body of the paper.
%% Sections are demarcated by \section and \subsection, respectively.
%% Observe the use of the LaTeX \label
%% command after the \subsection to give a symbolic KEY to the
%% subsection for cross-referencing in a \ref command.
%% You can use LaTeX's \ref and \label commands to keep track of
%% cross-references to sections, equations, tables, and figures.
%% That way, if you change the order of any elements, LaTeX will
%% automatically renumber them.

% Main body with filler text
\section{Introduction} \label{sec:intro}
Long-duration gamma-ray bursts (LGRBs) are among the most luminous astronomical transient events. They are believed to originate from the core collapse of massive stellar progenitors \citep{2003ApJ...591..288H}, typically situated in dense gas and/or dust regions within star-forming regions where the lifecycle of stars unfolds—from birth to death \citep{1998ApJ...494L..45P,2002ApJ...565..174R,2007ApJ...654L..17C,2010MNRAS.402.2429C,2012MNRAS.421.1697C}. The multi-wavelength afterglow originates from the interaction between the relativistic jet and the circumburst medium, typically described as synchrotron radiation within the framework of the standard afterglow model \citep{1998ApJ...497L..17S,10.1016/S0370-1573(98)00127-6,2015PhR...561....1K}. According to the afterglow model, the high-energy spectrum in the case of slow cooling is predicted to follow a power-law distribution of $\nu^{-p/2}$, where $p$ is the energy index of accelerated electrons. When the cooling frequency falls between the optical and X-ray bands, the spectral shape gradually shifts from $\nu^{-(p-1)/2}$ to $\nu^{-p/2}$. Therefore, the intrinsic spectral energy distribution (SED) of a GRB afterglow, assuming no attenuation by dust or other intervening systems, can be described by a broken power-law model. The flux density $F_\nu$ as a function of frequency $\nu$ is given by \citep[e.g.,][]{1998ApJ...497L..17S,2004ApJ...617L..21J,2009ApJ...699.1087V,2024A&A...690A.373R}:
\begin{equation}\label{eq:no_attenuation}
  F_\nu =
  \left\{
      \begin{array}{ll}
          F_0 \cdot \nu^{-\beta_\mathrm{O}}, & \nu \leq \nu_\mathrm{break}, \\[3pt]
          F_0 \cdot \nu_\mathrm{break}^{\Delta\beta} \cdot \nu^{-\beta_\mathrm{X}}, & \nu > \nu_\mathrm{break},
      \end{array}
  \right.
  \end{equation} where $F_0$  is the normalization constant, $\beta_\mathrm{O}$  and $\beta_\mathrm{X}$ are the spectral indices in the optical and X-ray regimes, respectively, and $\nu_\mathrm{break}$ is the break frequency at which the transition between the two power-law segments occurs. The parameter  $\Delta\beta = \beta_\mathrm{X} - \beta_\mathrm{O}=\frac{p}{2}-\frac{p-1}{2}=0.5$ accounts for the change in the spectral slope across the break frequency. This form captures the characteristic double power-law behavior of GRB afterglows, as expected from synchrotron radiation within the framework of the fireball model. For frequencies below $\nu_\mathrm{break}$, the emission is dominated by optical photons, while above $\nu_\mathrm{break}$, the X-ray regime governs the observed spectrum. The break frequency is often associated with the cooling frequency or the synchrotron peak frequency, depending on the physical conditions of the emitting region. For the typical case of $p \gtrsim 2$, a lower limit of 0.5 for the broad-band synchrotron spectral index is obtained by such electron distribution. In particular, an optical-to-X-ray spectral index $\beta_\mathrm{OX} \geq 0.5$ is expected when $\nu_\mathrm{break}$ lies between the optical and X-ray bands. However, the hints from the theoretical analysis in the framework of the fireball model are not always suitable for explaining all the complicated cases of the optical-to-X-ray spectral indices in GRB multiwavelength observations. Instead, from the observational point of view, we may consider examining the optical-to-X-ray spectral index from multiwavelength data by a direct way.

Optically dark bursts, a distinct subclass of GRBs, are identified using two primary classification methods, both serving as rapid diagnostic tools for distinguishing these bursts from the general GRB population \citep{2004ApJ...617L..21J,2009ApJ...699.1087V}. Although the classification criteria differ in methodology, both methods share a common underlying principle: optically dark GRBs are identified as those with optical flux levels falling below predictions made by the X-ray spectral index, as outlined in the standard afterglow model. \citet{2004ApJ...617L..21J} classified a GRB as `dark' based on the optical-to-X-ray spectral index, denoted as $\beta_\mathrm{OX}$. They proposed that bursts with $\beta_\mathrm{OX} < 0.5$ could be considered optically dark after analyzing a sample of 52 GRBs, thereby establishing this threshold as a key criterion for classification. By contrast, \citet{2009ApJ...699.1087V} leveraged X-ray flux and spectral data from the Neil Gehrels Swift Observatory \citep[hereafter Swift;][]{2004ApJ...611.1005G}, enabling a more nuanced diagnostic method that accounts for the possibility of $p < 2$. In this method, a GRB is categorized as `dark' if $\beta_\mathrm{OX}$ is shallower than the X-ray spectral index $\beta_\mathrm{X}$ reduced by 0.5, i.e., $\beta_\mathrm{OX} < \beta_\mathrm{X} - 0.5$. In this paper, a burst is classified as optically dark if the criteria of both $\beta_\mathrm{OX} < \beta_\mathrm{X} - 0.5$ and $\beta_\mathrm{OX} < 0.5$ are satisfied.

Some reasons for the existence of optically dark GRBs have been presented by \citet{2011A&A...526A..30G} and \citet{2012MNRAS.421.1265M}. Below, several key factors are summarized.

\begin{itemize}
    \item \textit{Observational limitations and delays.} GRBs may appear optically dark due to the timing and constraints of observations. Delays in follow-up observations can result in missing the rapidly fading optical afterglow, particularly for bursts with faint optical emission \citep{2011A&A...526A..30G}. Additionally, adverse observing conditions, such as high sky brightness, poor weather, or the unavailability of optical telescopes, can impact optical detection \citep{2011A&A...526A..30G, 2012MNRAS.421.1697C,2022JApA...43...11G}. The optical darkness of a GRB is initially evaluated at a fiducial time, commonly defined as the 11 hours post-trigger. This choice of a fiducial time ensures that any eventual central engine activity has ceased and that afterglow follow-up observations could have been carried out. This implies that any delays and weather conditions mentioned above may not change the classification. %(in the worst conditions, a classification cannot be made)
    % The optical flux should then be evaluated in the R/r band to weaken the contribution of intrinsic absorption.
    Advances in arcsecond localization of GRB afterglows using the X-Ray Telescope \citep[XRT;][]{2004SPIE.5165..201B} onboard the Swift Observatory, combined with rapid-response ground-based follow-up observations, have significantly enhanced the acquisition of high-quality 0.3-10\,keV X-ray spectra and complementary ultraviolet/optical/infrared data for GRB afterglows. These photometric and spectroscopic datasets facilitate the measurement of redshifts and the equivalent widths of absorption lines from the circumburst medium. With the advent of rapid multiwavelength follow-up observations of GRB afterglows, the nature of optically dark bursts can be unveiled, shedding light on their physical origins and properties. 
    \item \textit{Intrinsic faintness of the optical afterglow.} Many GRBs do not produce bright optical afterglows. Some bursts exhibit intrinsically faint afterglows across all wavelengths, particularly in the optical range. 
    This faintness may arise from physical properties such as weak synchrotron radiation due to low jet energy, a low-density circumburst medium, and variations in magnetic field strength or electron energy distribution \citep{2013ApJ...774..132W}. %\textbf{Additionally, steep decay rates of an afterglow are often linked to viewing geometry or jet structure, such as observations made at the edge of a narrow jet \citep{2006PhDT.........2N,2024arXiv241109609C}.} 
    However, we note that optical darkness is commonly quantified using the flux ratio between the X-ray afterglow and optical afterglow. In principle, a GRB with a relatively bright X-ray afterglow compared to a relatively faint optical afterglow may be identified as a dark GRB.
    % rapidly fading optical afterglows can fall below the detection threshold of most telescopes before observations begin, resulting in a dark classification \citep{2019MNRAS.484.5245H}. 
    \item \textit{High redshift and cosmological distance.} GRBs may occur at high redshifts, where cosmic expansion shifts their optical afterglows into the infrared, rendering them undetectable by traditional optical instruments \citep{2012MNRAS.421.1265M}. In extreme cases, very high redshifts can shift the afterglow entirely outside standard ground-based observational windows \citep[Lyman-$\alpha$ absorption occurring at $\lambda_\mathrm{obs} < 1216 ( 1 + z ) $\AA;][]{2009Natur.461.1254T,2011ApJ...736....7C}, necessitating infrared or space-based observations for detection.
    \item \textit{Dense circumburst medium.} A common explanation for optical dark is the presence of a dust-rich environment within the host galaxies of GRBs \citep{2012MNRAS.421.1265M}. The medium in dense molecular clouds or star-forming regions, where LGRBs often occur, can absorb or scatter optical light, thereby impeding the detection of afterglows in optical wavelengths \citep{2009RAA.....9.1103Z}. In contrast, X-rays/$\gamma$-rays can penetrate such dense environments, enabling detection at higher energies. The absorbing column density typically includes contributions from the circumburst medium of GRB host galaxy, the intergalactic medium (IGM), and the Galactic interstellar medium.
\end{itemize}

In this study, we analyze multi-wavelength observational data to characterize the evolution of X-ray spectral properties and optical darkness for GRB~240825A and to infer the physical properties of the circumburst medium environment, especially the extinction curve responsible for the observed attenuation. These results shed light on the physical processes influencing the afterglow evolution and provide valuable constraints on the properties of the circumburst medium. The paper is organized as follows. The observational properties of GRB 240825A are described in Section~\ref{sec:GRB240825A}. The multiwavelength observations and data reduction are described in Section~\ref{sec:Observation_data}. The X-ray spectral analysis, the calculation of optical darkness, and the extinction curve of the circumburst medium are presented in Section~\ref{sec:Results_and_Discussions}, along with discussions of the possible physical processes driving the evolution of X-ray spectral properties, optical darkness, and the extinction curve. A summary of our findings is provided in Section~\ref{sec:Summery}.

Throughout the paper, we assign the Swift Burst Alert Telescope \citep[BAT;][]{2005SSRv..120..143B} trigger time of GRB 240825A as $T_{0}$ \citep{2024GCN.37274....1G}. All times presented in the results and figures are calculated relative to $T_{0}$, unless stated otherwise. %All errors are reported at the $1\sigma$ confidence level unless stated otherwise.

\section{Observational Properties of Long GRB 240825A}\label{sec:GRB240825A}
At 15:52:59 UT on 2024 August 25 (referred to as $T_{0}$), the Swift/BAT triggered and located GRB 240825A \citep{2024GCN.37274....1G,2024GCN.37355....1M}. The XRT promptly detected X-ray afterglow emission \citep{2024GCN.37274....1G,2024GCN.37294....1G}, while the Ultra-Violet/Optical Telescope \citep[UVOT;][]{2005SSRv..120...95R} captured early afterglow emission in the $White$ and $UVW1$ filters \citep{2024GCN.37274....1G,2024GCN.37296....1K}. The UVOT detection of a bright optical counterpart of the early afterglow ruled out both the high-redshift and the intrinsically optical-faint scenarios for optically dark bursts.

GRB 240825A was rapidly followed up by many optical facilities after the trigger, 
including Nanshan/HMT \citep{2024GCN.37275....1J}, 
Skynet \citep{2024GCN.37276....1D}, 
AKO \citep{2024GCN.37277....1O,2024GCN.37299....1O}, 
Mephisto \citep{2024GCN.37278....1Z}, 
Global MASTER-Net \citep{2024GCN.37279....1L,2024GCN.37283....1L}, 
GMG \citep{2024GCN.37280....1L,2024GCN.37306....1W}, 
LCO \citep{2024GCN.37287....1I}, 
%Fermi-LAT \citep{2024GCN.37288....1D}, 
MASTER \citep{2024GCN.37289....1L}, 
Montarrenti \citep{2024GCN.37291....1L,2024GCN.37400....1L}, 
SVOM/C-GFT \citep{2024GCN.37292....1S,2024GCN.37373....1S}, 
VLT/X-shooter \citep{2024GCN.37293....1M}, 
REM \citep{2024GCN.37295....1B}, 
%AstroSat CZTI \citep{2024GCN.37298....1J}, 
MISTRAL/T193 OHP \citep{2024GCN.37300....1L},
%Konus-Wind \citep{2024GCN.37302....1F},
PRIME \citep{2024GCN.37303....1G},
KAIT \citep{2024GCN.37304....1Z},
NUTTelA-TAO / BSTI \citep{2024GCN.37307....1M},
TNG \citep{2024GCN.37310....1M},
SAO RAS \citep{2024GCN.37313....1M,2024GCN.37336....1M},
%ALMA \citep{2024GCN.37314....1L},
%GECAM \citep{2024GCN.37315....1W},
%VLA \citep{2024GCN.37322....1P},
%IceCube \citep{2024GCN.37326....1I},
iTelescope \citep{2024GCN.37335....1G},
SVOM/VT \citep{2024GCN.37338....1S},
%MeerKAT \citep{2024GCN.37353....1P},
SOAR \citep{2024GCN.37361....1F},
TSHAO \citep{2024GCN.37367....1B},
%MeerLICHT \citep{2024GCN.37372....1D},
%ATCA \citep{2024GCN.37388....1G},
7DT \citep{2024GCN.37454....1P},
%INTEGRAL SPI ACS \citep{2024GCN.37537....1P}, 
and LBT \citep{2024GCN.37638....1M}. The Swift-BAT light curve displayed a very bright fast rise and exponential decay-like pulse, with the burst characterized by a duration ($T_{90}$) of $57.20 \pm 8.57$ seconds in the 15-350\,keV energy band, identifying it as an LGRB \citep{2024GCN.37355....1M}. The optical counterpart of the afterglow was precisely localized at equatorial coordinates (J2000): $\mathrm{R.A. = 22^{h}58^{m}17^{s}.27, Dec. = +01^\circ01^\prime36^{\prime\prime}.7}$, with a positional uncertainty of $0^{\prime\prime}.1$ in each coordinate \citep{2024GCN.37275....1J,2024GCN.37276....1D,2024GCN.37277....1O,2024GCN.37280....1L}. This position is consistent with the Swift/UVOT position %of $\mathrm{R.A. = 22^{h}58^{m}17^{s}.26, Dec. = +01^\circ01^\prime36^{\prime\prime}.7}$ 
\citep{2024GCN.37274....1G,2024GCN.37290....1E}. In the direction of the optical afterglow, the Galactic foreground extinction corresponds to a reddening of $E_{B-V} = 0.0533\mathrm{\,mag}$, obtained from the Galactic Dust Reddening and Extinction Service\footnote{\url{https://irsa.ipac.caltech.edu/applications/DUST/}} \citep{2011ApJ...737..103S}. The Galactic hydrogen column density, $N^\mathrm{Gal}_\mathrm{H} = 5.28\times10^{20}\mathrm{\,cm^{-2}}$, was retrieved using the \texttt{NHtot} tool\footnote{\url{https://www.swift.ac.uk/analysis/nhtot/}}, which accounts for contributions from both atomic and molecular hydrogen \citep{2013MNRAS.431..394W}.

The spectroscopic observations determined the redshift of GRB~240825A to be $z = 0.659$ \citep[][]{2024GCN.37293....1M, 2024GCN.37310....1M}.
% , \textbf{further ruling out a high-redshift origin for this burst.}
Adopting the Lambda Cold Dark Matter ($\Lambda\mathrm{CDM}$) cosmological model with a Hubble constant $H_{0} = 67.3\mathrm{\,km\,s^{-1}\,Mpc^{-1}}$, baryon and cold dark matter density $\Omega_\mathrm{m} = 0.315$, and dark energy density $\Omega_{\Lambda} = 0.685$ \citep{2014A&A...571A..16P}, the Konus-Wind team reported the isotropic energy release as 
$E_\mathrm{iso} = (2.00 \pm 0.96) \times 10^{53}\mathrm{\,erg}$ and the peak luminosity as 
$L_\mathrm{iso,p} = (3.04 \pm 0.14) \times 10^{53}\mathrm{\,erg\,s^{-1}}$. Both values are given at the 68\% confidence level and are estimated over the rest-frame bolometric energy band from 1\,keV to $(1+z)\cdot10$\,MeV \citep{2017ApJ...850..161T,2024GCN.37302....1F}.

The simultaneous multiband photometric observations from the Multi-channel Photometric Survey Telescope \citep[Mephisto;][]{2020SPIE11445E..7MY} have unveiled critical insights into the optical afterglow of GRB~240825A. The optical afterglow can be described by a power-law dependence of flux density on time ($t$) and frequency ($\nu$), $F_\nu \propto t^{-\alpha_\mathrm{obs,O}} \nu^{-\beta_\mathrm{obs,O}}$, with an observed temporal decay index of  $\alpha_\mathrm{obs,O}=1.340\pm0.002$ and an observed spectral index of $\beta_\mathrm{obs,O}=2.477\pm0.006$ \citep{2025ApJ...979...38C}. Firstly, such a steep spectral index contrasts with the typical value of $\beta_\mathrm{obs} \sim 1$ observed in most GRBs \citep[e.g.,][]{2010ApJ...720.1513K,2011ApJ...734...96K}. For instance, in the golden sample of 48 GRBs compiled by \citet{2010ApJ...720.1513K}, only two events (GRB 070802 and GRB 080310) exhibit $\beta_\mathrm{obs} \gtrsim 2$, while most of the spectral indices are around $\beta_\mathrm{obs} \sim 1$ with uncertainties of 0.01-0.1. Secondly, the X-ray spectrum generally does not exhibit strong spectral evolution at late times relative to the trigger \citep{2007ApJ...663..407B,2016ApJ...831..111M}. The late-time X-ray spectrum measured in photon counting (PC) mode of Swift-XRT provides a relatively reliable basis for calculating the robust X-ray spectral index. The X-ray spectral index was derived using the relation $\beta_\mathrm{X} = \Gamma - 1 = 0.79 \pm 0.05$, where $\Gamma$ represents the photon index of the time-averaged X-ray spectrum measured in PC mode. The corresponding X-ray spectral fitting results are given in Table~\ref{tab:time-sliced}. It is worth noting that the optical spectral energy distribution, with a spectral index of $\beta_\mathrm{obs,O} \simeq 2.48$ measured from the early optical afterglow, is steeper than what is predicted by the X-ray spectral index within the framework of the standard afterglow model. The anomalous optical spectral index of GRB 240825A may indicate substantial absorption of the afterglow by a dense circumburst medium, leading to a strong extinction. This motivates a further investigation into the complex optical-to-X-ray spectral features of GRB 240825A. Here, we note that data around 11 hr after GRB trigger is usually applied to identify a dark GRB, as optical spectral index of optical afterglow varies with time.

\section{Observations and Data Reduction} \label{sec:Observation_data}

\subsection{Swift data}\label{sec:swift_data}

The Swift Observatory autonomously reoriented its narrow-field instrument, XRT, in response to the trigger of GRB~240825A by the BAT. The $0.3-10$\,keV X-ray data were obtained by the XRT during two distinct time intervals: from 86 to 1297 seconds post-trigger in windowed timing (WT) mode, and from 4773 seconds to $\sim36.2$ days post-trigger in PC mode. The XRT count-rate light curve was extracted from the UK Swift Science Data Center\footnote{\url{https://www.swift.ac.uk/xrt_curves/01250617/}} (UKSSDC).

For the time-resolved XRT spectral analysis, the count-rate light curve was segmented into intervals of relatively constant count rates using the Bayesian blocks algorithm \citep{1998ApJ...504..405S,2013ApJ...764..167S}, as shown in the upper panel of Figure~\ref{fig:properties}. Based on these intervals, we used the time-sliced spectra builder\footnote{\url{https://www.swift.ac.uk/xrt_spectra/addspec.php?targ=01250617}} to extract the time-resolved spectra \citep{2007A&A...469..379E,2009MNRAS.397.1177E}. The XRT spectra were processed through an automated pipeline that accounts for all instrumental features\footnote{\url{https://www.swift.ac.uk/xrt_spectra/01250617/}} \citep[][]{2007A&A...469..379E,2009MNRAS.397.1177E}. Each spectrum was automatically fitted with an absorbed power-law (PL) model (i.e., \texttt{TBabs*zTBabs*powerlaw} in \texttt{XSPEC}) using the Efficient Library for Spectral Analysis in High-Energy Astrophysics \citep[ELISA;][]{ELISA}. The model accounts for X-ray absorption contributions from both the Milky Way and the intrinsic circumburst medium of the GRB host galaxy, while neglecting contributions from the IGM. The solar abundances are adopted from \citet{2000ApJ...542..914W}, and the cross-sections are taken from \citet{1995A&AS..109..125V} and \citet{1996ApJ...465..487V}. The Galactic absorption, modeled with \texttt{TBabs} component, was fixed at the value of $N^\mathrm{Gal}_\mathrm{H}$. The intrinsic absorption, modeled as $N_\mathrm{H,X}$ with \texttt{zTBabs} component, was estimated at the redshift of the GRB host galaxy. The photon index ($\Gamma$) and normalization, described by the \texttt{powerlaw} component, were treated as free parameters.

The results of the time-resolved analysis for the X-ray afterglow are summarized in Table~\ref{tab:time-sliced} of Appendix~\ref{sec:spectral_analysis} and ploted in the lower panel of Figure~\ref{fig:properties}. The 0.3-10\,keV flux light curve was converted into a flux density light curve using the power-law spectral indices of the time-resolved XRT spectra, calculated at the logarithmic median of the 0.3-10\,keV range. The derived 1.732\,keV flux density light curve is shown in the sub-figures (a) and (c) of Figure~\ref{fig:lightcurve}.

The post-slew BAT data cover the time interval $T_0+[71,832]\mathrm{\,s}$,
while the XRT observations began at $T_0+87\mathrm{\,s}$. Therefore, the overlapping time interval between the BAT and XRT observations is $T_0 + [87, 832]\mathrm{\,s}$. The BAT and XRT data for this time interval were processed using HEASoft (v6.35.2) with the latest calibration files. For the overlapping time interval between the BAT and XRT observations, we first performed spectral fitting using a single absorbed power-law model for the XRT data and a single unabsorbed power-law model for the BAT data (see Table~\ref{tab:XRT+BAT}). We then performed a time-averaged joint spectral analysis using the BAT and XRT data within the overlapping time interval. Initially, an absorbed smoothly broken power-law (SBPL) model was used for the joint spectral analysis. The SBPL model typically is described by the low-energy power law index $\Gamma_1$, the high-energy power law index $\Gamma_2$ and the break energy $E_\mathrm{b}$. All parameters in an absorbed SBPL model were allowed to vary freely. We found that the addition of a thermal blackbody (BB) component with a temperature of $kT$ (i.e., an absorbed SBPL+BB model) may improve the goodness of fit (see Table~\ref{tab:XRT+BAT} of Appendix~\ref{sec:spectral_analysis}). %A more detailed time-resolved spectral analysis of the BAT data is hindered by the limited photon counts. 
Considering the curvature feature presented in the early X-ray spectrum, the 10\,keV flux density light curve was extracted from the burst analyser page of GRB 240825A\footnote{\url{https://www.swift.ac.uk/burst_analyser/01250617/}} and is displayed in the sub-figures (b) and (c) of Figure~\ref{fig:lightcurve}.

\subsection{Optical Photometric Data} \label{sec:photometric_data}

% The ground-based facility of the Space-based multi-band astronomical Variable Objects Monitor \citep[SVOM;][]{2023hxga.book..149W}, the Chinese Ground Follow-up Telescope (C-GFT; C. Wu et al., in preparation), automatically conducted photometric observations of GRB~240825A in the $gri$ bands about 65 seconds after the Swift trigger.

\subsubsection{Mephisto Data}
The early simultaneous multiband photometric observations in the wide $uvgriz$ bands were conducted by Mephisto, starting 128 seconds after the trigger and continuing until the fourth day, during which period the afterglow rapidly faded, %and 
then the apparent magnitude approached that of the host galaxy. The optical photometric data with dense sampling and high signal-to-noise ratios (SNRs) were uniformly obtained using a standardized image processing pipeline developed specifically for Mephisto (Y. Fang et al. 2025, in preparation). The magnitudes for GRB~240825A obtained with Mephisto, which have already been corrected for the Galactic extinction, are provided by \citet{2025ApJ...979...38C}.

\subsubsection{GMG Data}
The photometric observations were %also
performed by the Gao-Mei-Gu (GMG) 2.4-meter telescope at Lijiang Observatory \citep[IAU code: 044;][]{2019RAA....19..149W,2020RAA....20..149X}, part of the Yunnan Astronomical Observatories, Chinese Academy of Sciences. These observations began about 1.83 hours after the trigger and continued until the fourth night post-trigger \citep{2024GCN.37280....1L,2024GCN.37306....1W}. The refined results of the GMG photometric observations are presented in Appendix~\ref{sec:opt_data}.

\subsubsection{Optical Data Extracted from GCN}
The remaining optical photometric data provided in the General Coordinates Network (GCN) Circulars\footnote{\url{https://gcn.nasa.gov/circulars?query=240825A}} were manually parsed and compiled as supplementary data. For identical measurements, data from peer-reviewed publications were preferred over those from GCN. The photometric results of follow-up observations reported in the GCN Circulars are also summarized in Table~\ref{tab:gcn-data} of the Appendix. All magnitudes listed in Table~\ref{tab:gcn-data} have not been corrected for the Galactic foreground and host-galaxy dust extinction.

% https://swift.gsfc.nasa.gov/uvot_tdrss/1250617/index.html
% https://www.mpe.mpg.de/~jcg/grb240825A.html

%\section{Data Analysis and Discussions} 
\section{Results and Discussions}\label{sec:Results_and_Discussions}

\begin{figure}[htbp]
  \centering
  \includegraphics[width=\linewidth]{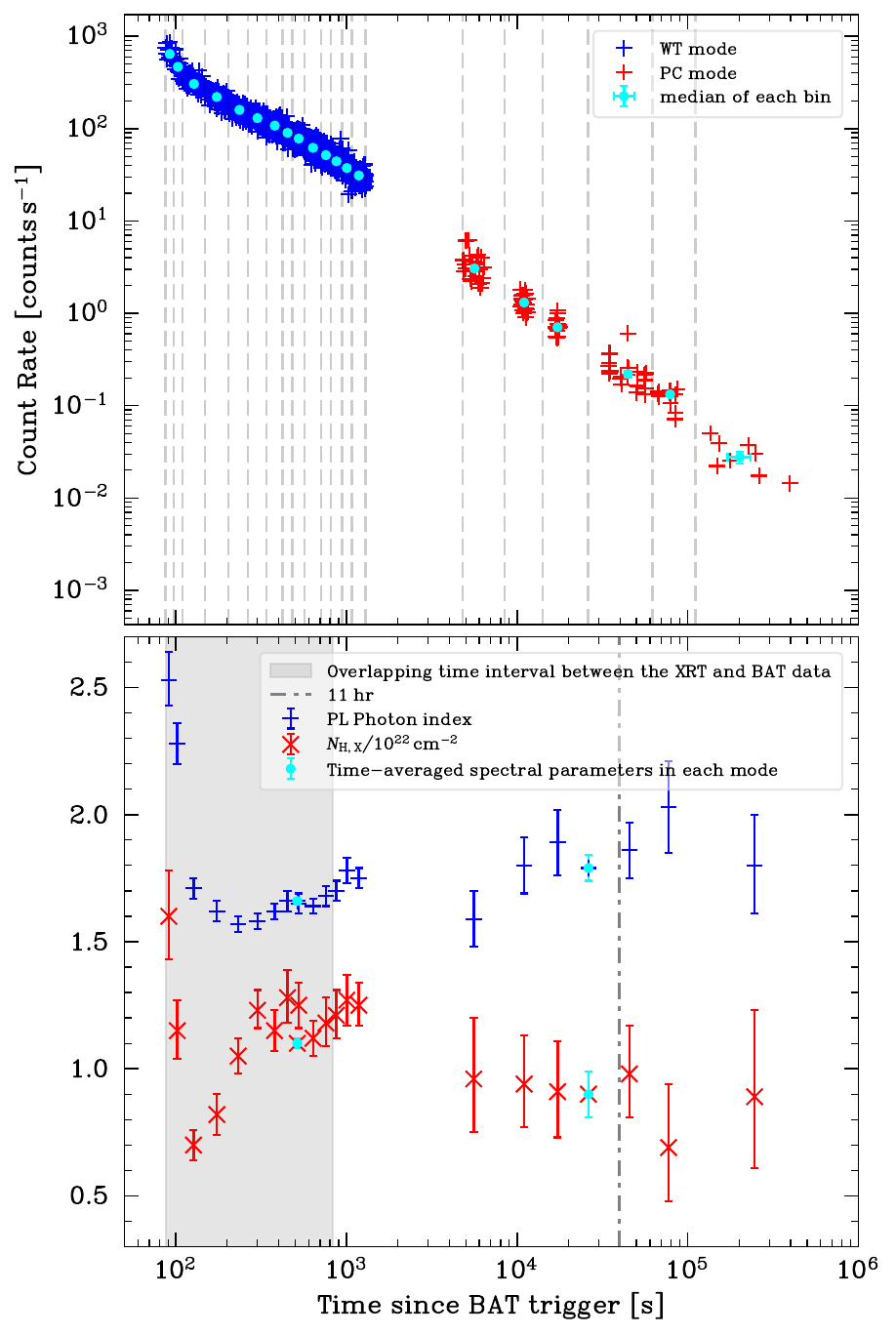}
  \caption{
  \textit{Upper panel:} Time-sliced count-rate light curve used for time-resolved spectral analysis. The method for dividing the time intervals is described in Section~\ref{sec:swift_data}. 
  \textit{Lower panel:} Evolution of time-resolved spectral properties, including the PL photon index ($\Gamma$) and the intrinsic X-ray absorbing column density ($N_\mathrm{H,X}$). 
  % \textbf{The first break time in the 10\,keV flux density light curve is indicated by a vertical dotted line (see Appendix~\ref{sec:Temporal_Decay_Indices} for details). Note that the PL photon indices and the corresponding intrinsic column densities before the first break time of the 10\,keV flux density light curve should be interpreted with caution due to possible spectral curvature, and thus are plotted as lighter-colored data points.} 
  The overlapping time interval between the XRT and BAT data is indicated by the vertical gray region. Note that the PL photon indices and the corresponding intrinsic column densities during this interval should be interpreted with caution due to possible spectral curvature.
  The 11 hr post-trigger time is marked by the vertical dashed-dotted line.
  }
  \label{fig:properties}
\end{figure}

\subsection{Evolution of X-Ray Absorption and Photon Index}\label{sec:Evolution_of_X-Ray_Absorption_and_Photon_Index}

The potential degeneracy between the intrinsic X-ray absorption ($N_\mathrm{H,X}$ evaluated at the GRB redshift of $z = 0.659$) and the photon index ($\Gamma$) has proven to be complex \citep[e.g.,][]{2021NatCo..12.4040R,2023ApJ...944...73V}. Earlier studies, such as those by \citet{2007ApJ...663..407B}, emphasized the risk of misinterpreting intrinsic spectral evolution as a decline in $N_\mathrm{H,X}$. In this work, a correlation between X-ray absorption and the photon index is demonstrated in Figure~\ref{fig:properties}. This correlation can likely arise because a soft spectrum with substantial absorption may produce a similar observed flux at low energies as a hard spectrum with weaker absorption. Detailed temporal analysis, however, provides additional clarity. The variations in $N_\mathrm{H,X}$ have been observed on finer timescales, suggesting that the evolution is genuinely linked to physical processes, and not solely due to degeneracy \citep[see Section~4 of][]{2023ApJ...944...73V}.

The X-ray spectrum during the overlapping time interval between the XRT and BAT data ($T_0 + [87,832]\mathrm{\,s}$) consists of a SBPL component and an additional BB component, as shown in Appendix~\ref{sec:spectral_analysis}.
% The PL spectral parameters may not accurately reflect the intrinsic spectral properties before the first break time of the 10\,keV flux density light curve (within $T_0 + [87,272]\mathrm{\,s}$), owing to the presence of spectral curvature in the early X-ray emission. Therefore, the PL spectral parameters prior to this epoch are not discussed, but are only presented in Figure~\ref{fig:properties}. After the first break time of the 10\,keV flux density (within $T_0 + [272,1325]\mathrm{\,s}$), the intrinsic column density becomes roughly stable at $N_\mathrm{H,X}\sim1.23\times 10^{22}\mathrm{\,cm^{-2}}$.
Therefore, the PL spectral parameters may not accurately reflect the intrinsic spectral properties during this time interval, owing to the presence of spectral curvature in the early X-ray emission. It is not appropriate to calculate the X-ray flux density using a simple PL model. The flux density light curve at 10 keV, constructed from both XRT and BAT data, was then utilized. The details are provided in Section~\ref{sec:evolution_of_optical_darkness} to further illustrate the time evolution of the optical darkness.   
%the PL spectral parameters during this interval are not discussed, but are only presented in Figure~\ref{fig:properties}.

In the late stages ($t \gtrsim 5000\mathrm{\,s}$), $N_\mathrm{H,X}$ stabilizes, as indicated by measurements in PC mode. The lack of further decline suggests that the dominant absorbing material is located at larger distances from the GRB and is almost unaffected by the ionizing radiation from the burst. Such distant absorbers, likely associated with the host galaxy, exhibit little variability and contribute a nearly constant absorption component \citep{2021A&A...649A.135C}. We noted that the X-ray absorption during this period, $N_\mathrm{H,X}=(9.0\pm0.9) \times 10^{21}\mathrm{\,cm^{-2}}$, listed as spectrum No.22 in Table~\ref{tab:time-sliced}, is notably greater than the total column density of hydrogen nuclei in molecular clouds identified along the spiral arms in the solar neighborhood \citep[see Figure~8 of][]{2024AJ....168..223L}. This value of the intrinsic X-ray absorbing column density is also consistent with observational statistics of the GRB samples \citep{2010MNRAS.402.2429C,2012MNRAS.421.1697C} and theoretical simulations \citep{2002ApJ...565..174R,2004A&A...415..171V}.

\begin{figure*}[htbp]
    \centering
    \gridline{\fig{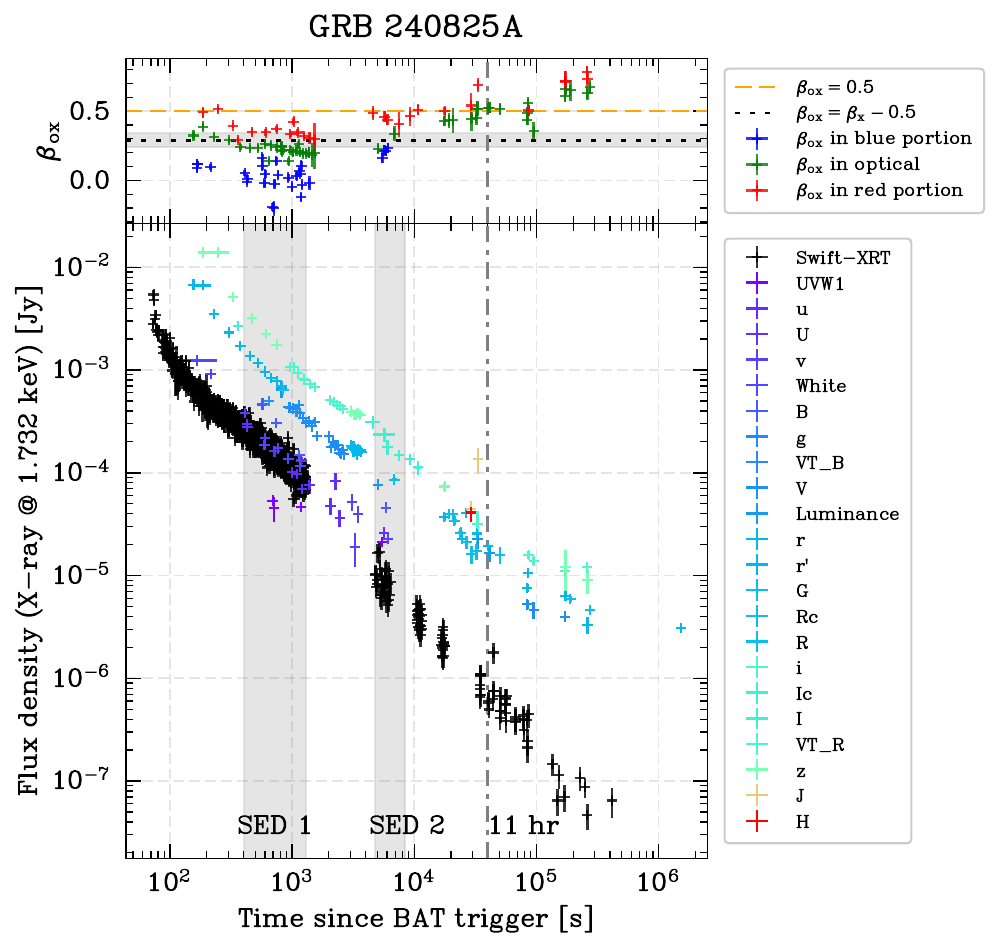}{0.5\linewidth}{(a)}
            \fig{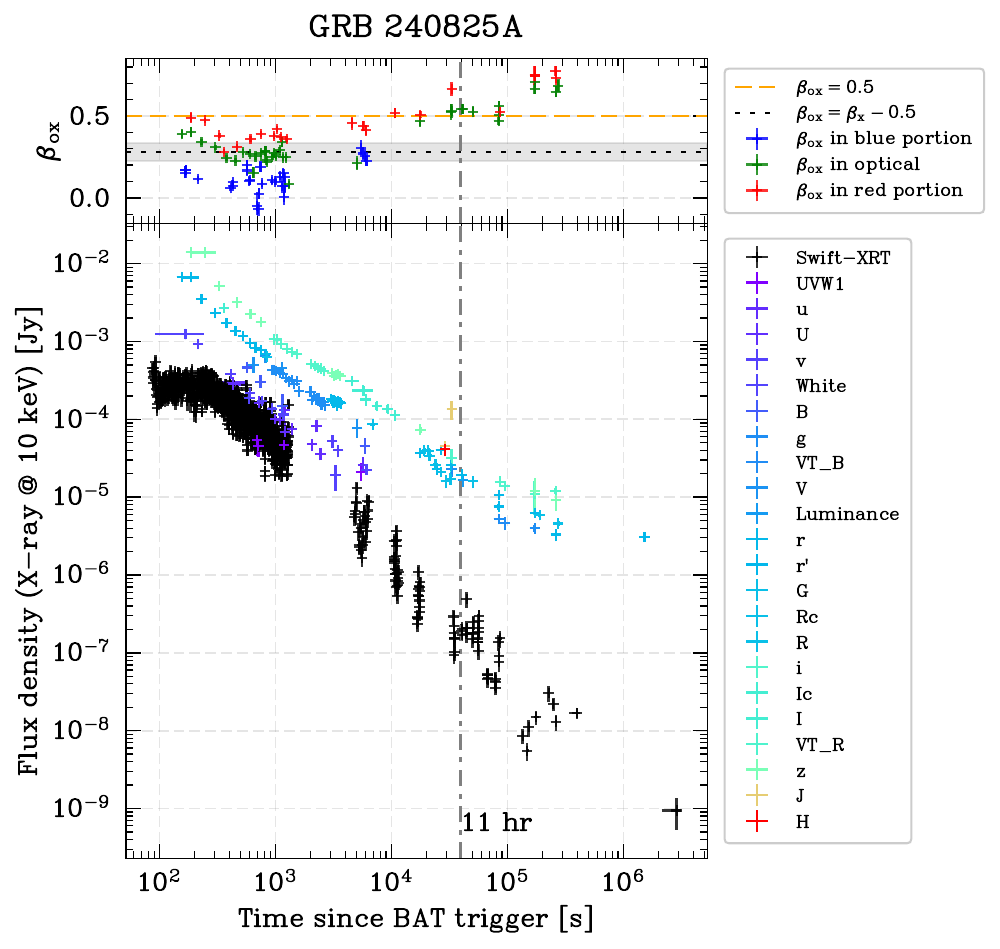}{0.5\linewidth}{(b)}}
    \gridline{\fig{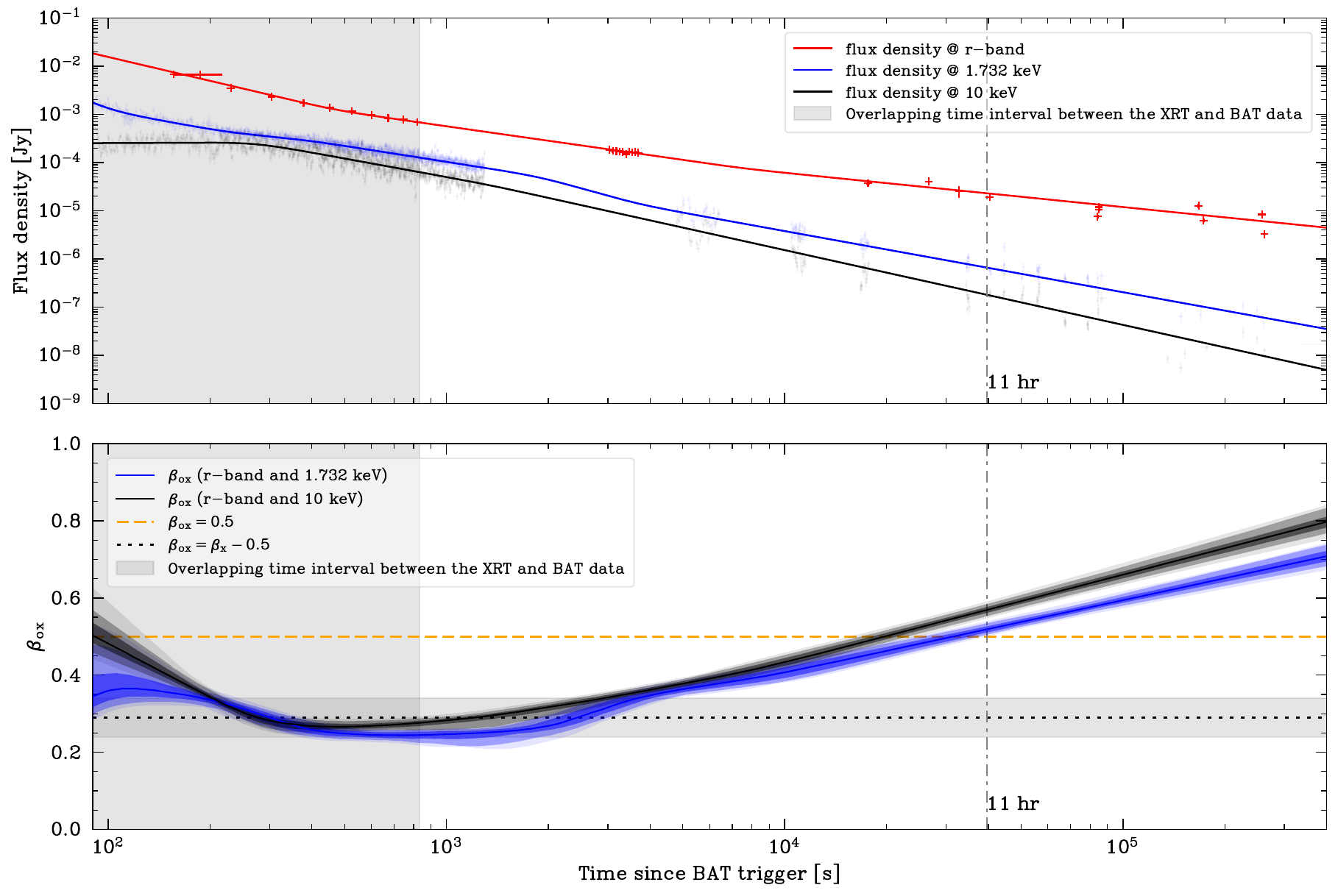}{0.8\linewidth}{(c)}}
    \caption{Evolution of optical darkness ($\beta_\mathrm{OX}$) and multiwavelength light curves for the GRB~240825A afterglow.
    \textit{Sub-figure (a)}: \textit{Upper panel:} Evolution of optical darkness ($\beta_\mathrm{OX}$) for the GRB~240825A afterglow. 
    The orange-dashed line indicates the threshold ($\beta_\mathrm{OX}=0.5$) defined by \citet{2004ApJ...617L..21J}. 
    The black-dotted line and gray shaded region represent the threshold ($\beta_\mathrm{OX} = \beta_\mathrm{X} - 0.5$， where the X-ray spectral index is calculated from the time-averaged X-ray spectrum measured in PC mode) outlined by \citet{2009ApJ...699.1087V}. 
    % Data points falling below the respective threshold lines are classified as optically dark according to the corresponding method.  
    \textit{Lower panel:} Multiwavelength light curve of GRB~240825A afterglow. The X-ray flux density light curve is calculated at 1.732\,keV. 
    % The effective wavelengths of different filters are mapped with rainbow colors. 
    % Two vertical gray shadings indicate the two time intervals of flux density used for SED fittings, as presented in Section~\ref{sec:host_extinction_curve}. 
    % The format of this figure follows a structure similar to that of Figure~5 in \citet{2023MNRAS.523..775G}.
    \textit{Sub-figure (b)}: This figure is analogous to sub-figure (a), but the X-ray flux density light curve is calculated at 10\,keV.
    \textit{Sub-figure (c)}: \textit{Upper panel:} Flux density light curves in the $r$-band, at 1.732\,keV and at 10\,keV. 
    % The fitting procedures of the light curves is presented in Appendix~\ref{sec:Temporal_Decay_Indices}. 
    % The first break time of the 10\,keV flux density light curve is marked with a vertical dotted line.
    \textit{Lower panel:} Temporal evolution of $\beta_\mathrm{OX}$ derived from fitting the $r$-band, 1.732\,keV, and 10\,keV flux density light curves. 
    % The solid lines and shaded regions represent the median values of $\beta_\mathrm{OX}$ calculated at two different frequencies and their corresponding $1\sigma$, $2\sigma$, and $3\sigma$ credible intervals, respectively. 
    % The 11 hr post-trigger time is denoted by the vertical dashed-dotted line in all sub-figures. 
    % The first break time of the 10\,keV flux density light curve is also marked with a vertical dotted line for comparison in the sub-figures (b) and (c).
    % The overlapping time interval between the XRT and BAT data is marked with a vertical gray region in the sub-figures (c).
    }
    \label{fig:lightcurve}
\end{figure*}

% \begin{figure*}[htbp]
%   \centering
%   \includegraphics[width=\linewidth]{calculated_betaOX}
%   \caption{\textit{Upper panel:} Flux density light curves at $r$-band, 1.732\,keV and 10\,keV. The fitting procedures of the light curves is presented in Appendix~\ref{sec:Temporal_Decay_Indices}. 
%     % The first break time of the 10\,keV flux density light curve is marked with a vertical dotted line.
%     \textit{Lower panel:} Temporal evolution of $\beta_\mathrm{OX}$ derived from fitting the $r$-band, 1.732\,keV, and 10\,keV flux density light curves. The solid lines and shaded regions represent the median values of $\beta_\mathrm{OX}$ calculated at two different frequencies and their corresponding $1\sigma$, $2\sigma$, and $3\sigma$ credible intervals, respectively. 
%     % The 11 hr post-trigger time is denoted by the vertical dashed-dotted line in all sub-figures. 
%     % The first break time of the 10\,keV flux density light curve is also marked with a vertical dotted line for comparison in the sub-figures (b) and (c).
%     The overlapping time interval between the XRT and BAT data is marked with a vertical gray region in the figure.}
%   \label{fig:calculated_betaOX}
% \end{figure*}

% \subsection{Optical Data Analysis}\label{sec:optical_data_analysis}
\subsection{Evolution of Optical Darkness}\label{sec:evolution_of_optical_darkness}

\citet{2012MNRAS.421.1265M} suggested the evolution of the GRB optical darkness feature. Previous studies have reported optical darkness measurements at varying post-burst times, including 600 s, 11 h \citep{2012MNRAS.421.1265M}, 5.5 h \citep{2015MNRAS.449.2919L}, and 1000 s \citep{2009ApJ...693.1484C}. 
In particular, \citet{2009ApJS..185..526F} selected observations taken a few hours after the burst whenever possible, to avoid contamination from the early phases of the canonical X-ray light curve. In addition, for low-redshift bursts, late-time optical afterglow flux is affected by the contribution from the host galaxy \citep{2009ApJS..185..526F}. The X-ray flux density is usually estimated at 1\,keV \citep{2015MNRAS.449.2919L}, 1.732\,keV \citep{2023MNRAS.523..775G} or 3\,keV \citep{2004ApJ...617L..21J}. The optical flux density is generally estimated in the $R/r$ band by taking into account the observational efficiency and aiming to weaken the contribution of intrinsic absorption. Thanks to the rapid response and early-time dense sampling with high SNRs from Mephisto, the GMG telescope, and other ground-based facilities, we can directly calculate the optical darkness ($\beta_\mathrm{OX}$) using the muilt-band optical data and X-ray data of GRB 240825A.

We followed the methodology outlined in \citet{2023MNRAS.523..775G} to investigate the evolution of the optical darkness of the GRB~240825A afterglow. The empirical extinction data presented in \citet{2011ApJ...737..103S} was interpolated to calculate the ratio of the Galactic extinction to reddening $R_\lambda$. All optical photometric data described in Section~\ref{sec:photometric_data} were converted to flux density at the effective wavelengths of the respective filters using the relationship between the magnitude system and flux density, with corrections for Galactic extinction $A^\mathrm{Gal}_\lambda=R_\lambda\cdot E_{B-V}$. Accounting for the quasi-simultaneous matching of X-ray and optical flux densities, as well as additional uncertainties arising from time-dependent afterglow decay \citep[see Section~2.4 of][]{2023MNRAS.523..775G}, the power-law slope between the optical and the X-ray flux densities was derived for each temporally matched pair, as follows:
\begin{equation}
    \beta_\mathrm{OX}=-\frac{\log (F_{\nu,\mathrm{x}}/F_{\nu,\mathrm{o}})}{\log (\nu_\mathrm{x}/\nu_\mathrm{o})}\,,
\end{equation} where the flux densities in the X-ray and optical bands are denoted as $F_{\nu,\mathrm{x}}$ and $F_{\nu,\mathrm{o}}$, respectively, and the corresponding frequencies as $\nu_\mathrm{x}$ and $\nu_\mathrm{o}$. For GRB 240825A, the early X-ray spectral curvature was identified (see Appendix~\ref{sec:spectral_analysis}). The 10 keV flux density light curve, constructed from both BAT and XRT data, was used (see Figure~\ref{fig:lightcurve}). Consequently, we presented the X-ray flux density light curves at both 1.732 keV and 10 keV to compare the values of $\beta_\mathrm{OX}$. %derived from each curve.
The evolution of optical darkness ($\beta_\mathrm{OX}$) and the multiwavelength light curves of the GRB~240825A afterglow are presented in the sub-figures (a) and (b) of Figure~\ref{fig:lightcurve}. Due to the lack of temporally matched pairs of optical and X-ray flux densities, $\beta_\mathrm{OX}$ cannot be estimated during the time interval $1500 \lesssim t \lesssim 5000\mathrm{\,s}$ (see sub-figures (a) and (b) of Figure~\ref{fig:lightcurve}). The $\beta_\mathrm{OX}$ points are color-coded according to the effective wavelengths of the filters for reference: blue for blue portion ($\lambda<4500$\AA), green for optical ($\lambda $ within $4500-7000$\AA), and red for red portion ($\lambda>7000$\AA). As a matter of fact, the blue portion of $\beta_\mathrm{OX}$ is consistently lower than the optical and red portions. This can be easily understood as the extinction of afterglow emission by dust in the circumburst medium, with smaller grains primarily contributing to the extinction of shorter-wavelength emission.

Based on the fitted flux density light curves in the $r$ band, at 1.732\,keV, and at 10\,keV (see Appendix~\ref{sec:Temporal_Decay_Indices} for details), we analytically calculated $\beta_\mathrm{OX}$ to evaluate the impact of early X-ray spectral curvature on optical darkness. However, the trend of $\beta_\mathrm{OX}$ derived from the $r$-band and 1.732\,keV is roughly consistent with that derived from the $r$-band and 10\,keV after $\sim 1000\mathrm{\,s}$ post-trigger.
% the first break time of the 10\,keV flux density light curve ($t>$, see Appendix~\ref{sec:Temporal_Decay_Indices} for details). 
% Therefore, the main results and conclusions regarding optical darkness are not significantly affected (see sub-figure (c) of Figures~\ref{fig:lightcurve}). 
Despite the varying quality of the optical photometric data, the trend of optical darkness at first slightly decreasing then increasing over time is clearly evident, with the minimum value likely occurring at $t\sim 1000\mathrm{\,s}$, as shown in the sub-figure (c) of Figures~\ref{fig:lightcurve}.

The evolutionary process can be divided into two phases: during \textit{the initial decreasing phase} ($t \lesssim 1000 \mathrm{\,s}$), all $\beta_\mathrm{OX}$ values fall below the threshold proposed by \citet{2004ApJ...617L..21J}, i.e., $\beta_\mathrm{OX} < 0.5$. 
% At the beginning of the phase ($t \sim 150\,\mathrm{s}$), only the blue portion of the $\beta_\mathrm{OX}$ values satisfies the criterion $\beta_\mathrm{OX} < \beta_\mathrm{X} - 0.5$, as defined by \citet{2009ApJ...699.1087V}.
At the end of the phase ($t \sim 1000\,\mathrm{s}$), all $\beta_\mathrm{OX}$ values fall below both threshold lines within their error margins (i.e., $\beta_\mathrm{OX} < 0.5$ and $\beta_\mathrm{OX} < \beta_\mathrm{X} - 0.5$), meeting the criteria for optical dark as defined by \citet{2004ApJ...617L..21J} and \citet{2009ApJ...699.1087V}. During \textit{the subsequent increasing phase} ($t \gtrsim 1000 \mathrm{\,s}$), $\beta_\mathrm{OX}$ derived from the fitted flux density light curves provide an estimated trend of $\beta_\mathrm{OX}$ (see sub-figures (c) of Figure~\ref{fig:lightcurve}).
% \textbf{The PL photon index may exhibit an underlying spectral softening evolution (within $T_0 + [272,1325]\mathrm{\,s}$; see Figure~\ref{fig:properties}). The temporal behavior of the PL spectral parameters can be interpreted that the thermal emission initially peaking in the X-ray band followed by a gradual cooling of the blackbody component, which causes the thermal emission to shift into the optical band and contribute to the observed optical flux \citep{2012IAUS..279..375O}. This process provides a natural explanation for the observed decline and subsequent rise in $\beta_\mathrm{OX}$.} % Alternatively, extended periods of shallow decay may increase the X-ray flux, leading to a decrease in $\beta_\mathrm{OX}$ \citep{2008ApJ...686.1209M,2009ApJ...693.1484C}. 
Both X-ray and optical afterglow fluxes generally decay with time as power laws. %, but the contribution from the host galaxy to the optical flux at late times may lead to a rise in $\beta_\mathrm{OX}$.
At the typical time of $11\mathrm{\,h}$ post-trigger ($t \sim 4 \times 10^{4}\mathrm{\,s}$), the calculated $\beta_\mathrm{OX}$ exceeds the thresholds $\beta_\mathrm{OX}=\beta_\mathrm{X}-0.5$ and $\beta_\mathrm{OX}=0.5$, indicating that the conditions for an optically dark burst are not satisfied (see Figure~\ref{fig:lightcurve}).
The above results suggest that dust, being more concentrated in the high-density core regions, is more susceptible to destruction, while the gas farther from the GRB tends to be more diffuse. This may explain the increase in $\beta_\mathrm{OX}$ due to the lack of dense dust during this phase, while $N_\mathrm{H,X}$ remains nearly constant. 

\begin{figure}[htbp]
  \centering
  \includegraphics[width=\linewidth]{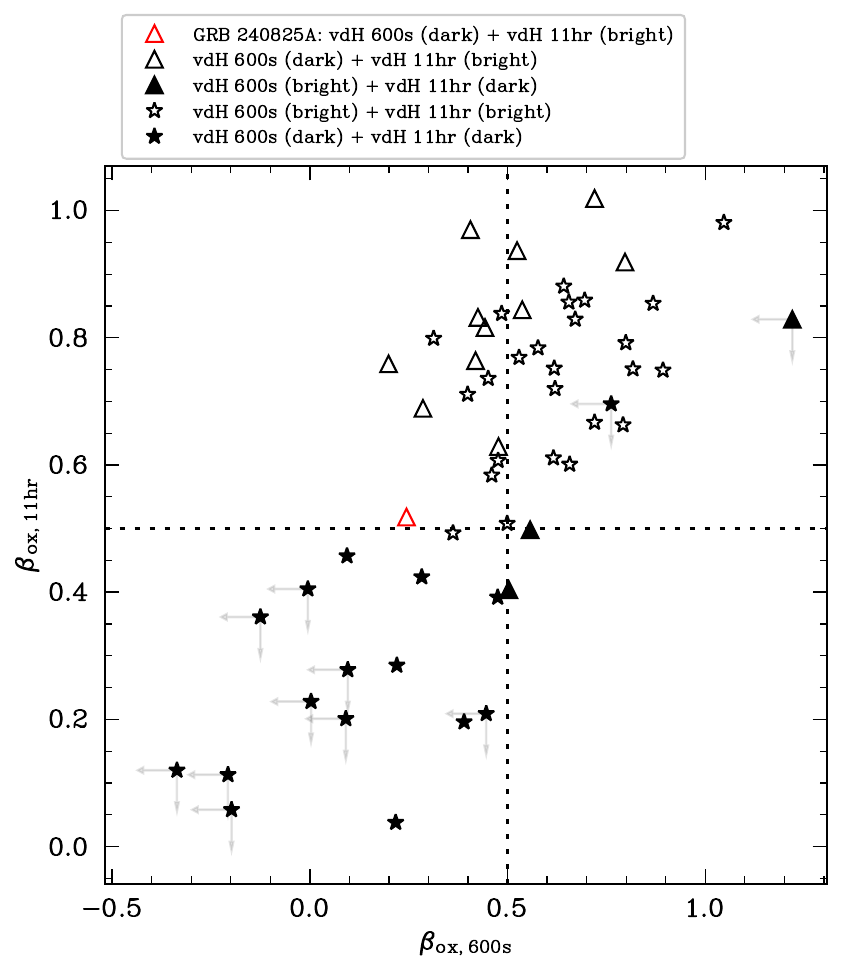}
  \caption{Darkness evolution from 600 s to 11 hr of the events in a sample from \citet{2012MNRAS.421.1265M}. For GRB 240825A, the derived optical darkness at 600 s and 11 hr post-trigger are $\beta_\mathrm{ox, 600 s}=0.244^{+0.006}_{-0.007}$ and $\beta_\mathrm{ox, 11 hr}=0.518^{+0.008}_{-0.007}$, respectively.}
  \label{fig:Melandri12_BetaOX_600s_vs_11hr}
\end{figure}

The dark burst population studied with complete samples of GRBs suggested that many bursts exhibit an evolution in their optical darkness \citep{2012MNRAS.421.1265M}. Based on the evolution of the optical darkness of GRB 240825A, this burst can be classified in the category of events that are optically dark at early times and bright at late times, which is typically evidence of a dusty circumburst environment. In the broader context of dark burst population studies, we plotted a diagram of darkness evolution in Figure~\ref{fig:Melandri12_BetaOX_600s_vs_11hr}. GRB 240825A is classified as dark at 600 s and bright at 11 hr, and therefore falls in the upper left-hand quadrant of the \citet{2004ApJ...617L..21J} classification diagram, marked with a red open triangle according to the \citet{2009ApJ...699.1087V} scheme. Some bursts located in the upper left-hand quadrant and represented by open triangles are classified in the same category as GRB 240825A, namely GRBs 050318, 050416A, 060614, 080605, 090715B, 110205A, and 110503A.

\subsection{Color Index of Optical Afterglow}
The optical spectral index, $\beta_\mathrm{obs,O} \simeq 2.48$, derived from the early-time multiband photometric data in the $uvgriz$ bands observed by Mephisto \citep{2025ApJ...979...38C}, is significantly higher than the typical values of $\beta \sim 1$ commonly seen in most GRBs, such as those in the `golden sample' of \citet{2010ApJ...720.1513K}. This highlights the peculiarity of GRB~240825A. This unusually soft optical SED is likely linked to significant extinction from the circumburst medium, as the blue portion of the optical afterglow appears partially absorbed, resulting in an additional reddening. Over time as the afterglow decays, the optical SED becomes progressively harder, accompanied by a decrease in the inferred color excess, $\Delta E_{B-V} \sim 0.26$\,mag within 100-3000\,s \citep{2025ApJ...979...38C}. According to the discussion presented in Section~4 of \citet{2025ApJ...979...38C}, this hardening is unlikely to result from the passage of a synchrotron characteristic frequency. We note that these results indicate an interaction between the jet and the dust/gas in the immediate vicinity of the burst \citep{2002ApJ...569..780D,2013ApJ...779...66S,2014MNRAS.440.1810M,2020ApJ...895...16H}.

\begin{figure}[htbp]
  \centering
  \includegraphics[width=\linewidth]{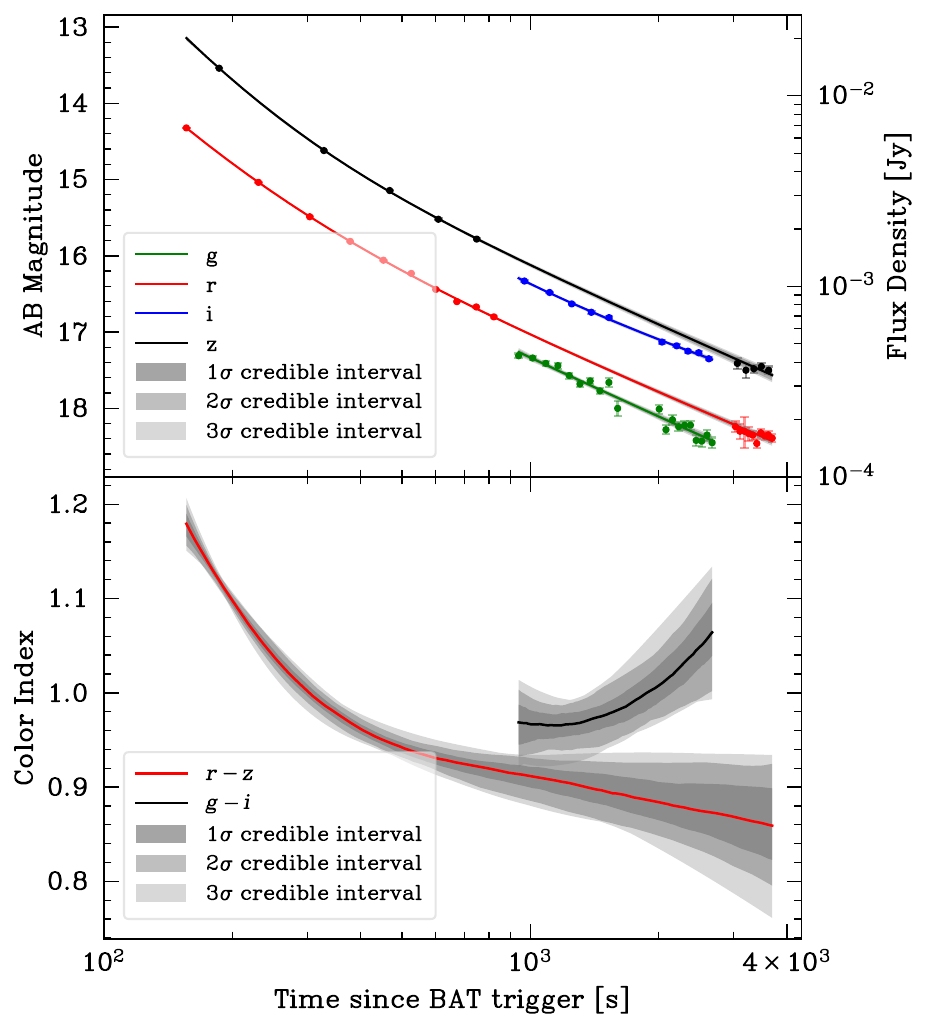}
  \caption{
  \textit{Upper panel:} Early light curve of the optical afterglow from Mephisto. The best-fitting light curves in different bands are shown as colored lines.
  \textit{Lower panel:} Color index evolution of early optical afterglow. The color index evolution is exhibited by the continuous lines derived from the differences between the best-fitting light curves of the corresponding bands. The solid lines and varying gray shadings represent the posterior median values of the color indices and their corresponding credible intervals, respectively. The optical photometric results in the $uv$ bands from Mephisto are not shown in this figure.
  Considering the influence of the host galaxy on the color index evolution of the afterglow, only the color index of the afterglow for $t \lesssim 4000\mathrm{\,s}$ is presented here.}
  \label{fig:color}
\end{figure}

We also investigated the color index evolution of early optical afterglow, a uniform optical spectral index is not adopted over the entire time period in this work. The light curve of early optical afterglow, observed by Mephisto in each $griz$ band, is independently fitted by  Equation~\ref{eq:temporal_decay}.
% \begin{equation}\label{eq:temporal_decay}
% F_{\nu,\mathrm{o}}^\mathrm{obs}=10^c\cdot\left[\left(\frac{t}{t_b}\right)^{-\alpha_1n}+\left(\frac{t}{t_b}\right)^{-\alpha_2n}\right]^{\frac{1}{n}}\,,
% \end{equation} where the normalization is represented by the constant $c$, the break time is indicated by $t_b$, and $\alpha_{1}$ and $\alpha_{2}$ are two temporal decay indices before and after the break time. The smoothness parameter of the break is denoted by $n$. The parameter space is explored via Bayesian inference. 
The best-fitting light curves in Mephisto-$griz$ bands and the color index evolution of early optical afterglow are shown in Figure~\ref{fig:color}. The color index of the optical afterglow ($r-z$) decreased from approximately 1.18 to 0.86 within $\sim150-4000\mathrm{\,s}$ post-trigger.

\subsection{Extinction Curve of Circumburst Medium}\label{sec:host_extinction_curve}
A significant fraction of optically dark GRBs is hosted by dusty, metal-rich galaxies \citep{2010ApJ...719..378H, 2015ApJ...806..250H, 2018ApJ...863...95H}. 
From a theoretical point of view, GRB host galaxies have been investigated in the framework of galaxy formation and evolution \citep{2008MNRAS.386..608L, 2017ApJ...835...37L}. In particular, the absorption features of GRB afterglow, incorporated with GRB host galaxy properties, were theoretically presented by \citet{2010ApJ...717..140M}. In this paper, for GRB 240825A, the $\beta_\mathrm{OX}$ values satisfy the criterion $\beta_\mathrm{OX} < 0.5$, during the first 11 hours after the burst. During this period, the process of dust destruction likely converts larger grains into smaller ones. These smaller grains, which exhibit stronger extinction of shorter-wavelength emission, may alter the extinction curve in the direction of the optical afterglow \citep{2020ApJ...895...16H}. The extinction curves toward the host galaxies of GRBs can generally be described by a Small Magellanic Cloud-like (SMC-like) shape with a steep far-UV rise, indicating an abundance of small grains in the circumburst medium \citep{2017A&A...601A..83H}. Additionally, the UV bump feature at $\sim2175\mathrm{\,\AA}$ in the extinction curves, which produces a broad absorption feature in the spectrum, is weaker than that in the Milky Way \citep[e.g.,][]{2011A&A...532A.143Z,2012ApJ...753...82Z,2014MNRAS.440.1810M,2015A&A...579A..74J,2018A&A...617A.141C,2018MNRAS.479.1542Z,2018MNRAS.480..108Z,2018ApJ...860L..21Z,2019MNRAS.486.2063H,2024A&A...690A.373R}.
Considering the early optical photometric data in this work, a complex extinction model may not be necessary. As a result, the observed attenuation in the UV-to-optical afterglow can be modeled using a simple extinction curve given by \citep{2024A&A...690A.373R}:
\begin{equation}\label{eq:extinction_curve}
    A^\mathrm{host}_\lambda=A^\mathrm{host}_V\left[\left(\frac{5500\mathrm{\AA}}{\lambda}\right)^\gamma+D(\lambda_0,\Delta\lambda,E_\mathrm{bump};\lambda)\right]\,,
\end{equation} where $\lambda$ is the wavelength in the rest-frame of the GRB. The amounts of extinction in the direction of the optical afterglow at $\lambda$ and in the $V$-band are represented by $A^\mathrm{host}_\lambda$ and $A^\mathrm{host}_V$, respectively. The slope of the extinction curve is denoted by $\gamma$. The UV bump feature is described by the Lorentzian-like Drude profile $D(\lambda_0, \Delta\lambda, E_\mathrm{bump}; \lambda)$, as follows \citep{2007ApJ...663..320F}:
\begin{equation}\label{eq:Drude_profile}
    D(\lambda_0,\Delta\lambda,E_\mathrm{bump};\lambda)=\frac{E_\mathrm{bump}(\lambda\Delta\lambda)^2}{(\lambda^2-{\lambda_0}^2)^2+(\lambda\Delta\lambda)^2}\,,
\end{equation} where the central wavelength $\lambda_0$ and the width $\Delta\lambda$ of the bump are fixed at values of $2175\mathrm{\,\AA}$ and $470\mathrm{\,\AA}$, respectively \citep{2024A&A...690A.373R}. The maximum amplitude of the bump in excess of the UV linear extinction is given by $E_\mathrm{bump}$. 

The afterglow is significantly influenced by the absorption effects of the circumburst medium in GRB host galaxy, which attenuate the observed light and modify the spectral properties of the afterglow. The intrinsic SED of GRB afterglow is attenuated by the following components: \textit{the intrinsic absorption} at the redshift of GRB host galaxy (the X-ray absorption column density $N_\mathrm{H,X}$ and the extinction curve $A^\mathrm{host}_\lambda$), as well as \textit{the Galactic components} (the X-ray absorption column density $N^\mathrm{Gal}_\mathrm{H}$ and the extinction $A^\mathrm{Gal}_\lambda$). The contribution from the IGM is neglected due to the closeness of the burst \citep[redshift $z=0.659$;][]{2024GCN.37293....1M, 2024GCN.37310....1M}. Based on Equations~\ref{eq:no_attenuation},~\ref{eq:extinction_curve},~and~\ref{eq:Drude_profile}, the attenuated flux can be summarized as \citep[e.g.,][]{2011A&A...532A.143Z,2015A&A...579A..74J,2018A&A...617A.141C,2024A&A...690A.373R}:
\begin{equation}
    F_\nu^\mathrm{obs} =  F_\nu\cdot10^{-0.4A^\mathrm{host}_\lambda}\cdot \mathrm{e}^{-N_\mathrm{H,X}\sigma(\nu)}\,,
\end{equation} where $\sigma(\nu)$ is the cross-section from \citet{1995A&AS..109..125V} and \citet{1996ApJ...465..487V}.

\begin{figure*}[htbp]
    \centering
    \gridline{\fig{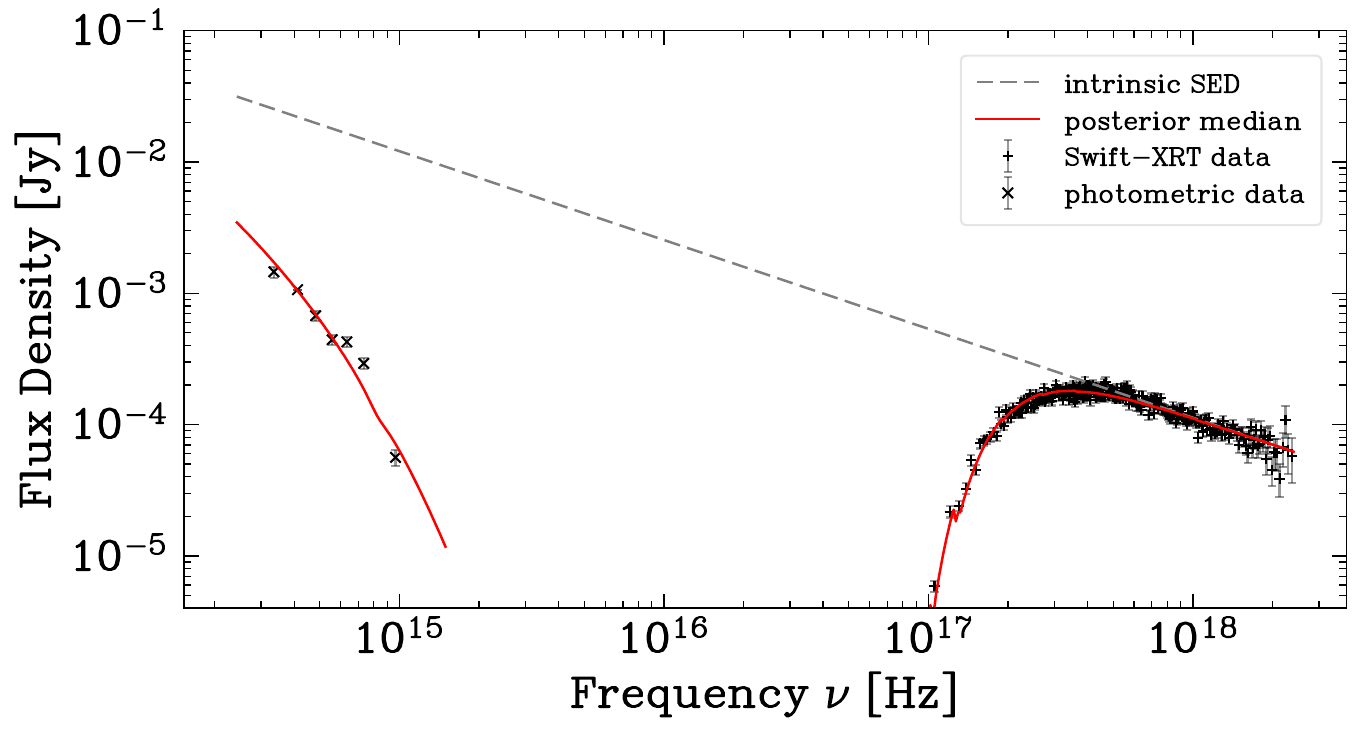}{0.5\linewidth}{(a) SED fitting for the afterglow in the time interval of $407-1290\mathrm{\,s}$}
                \fig{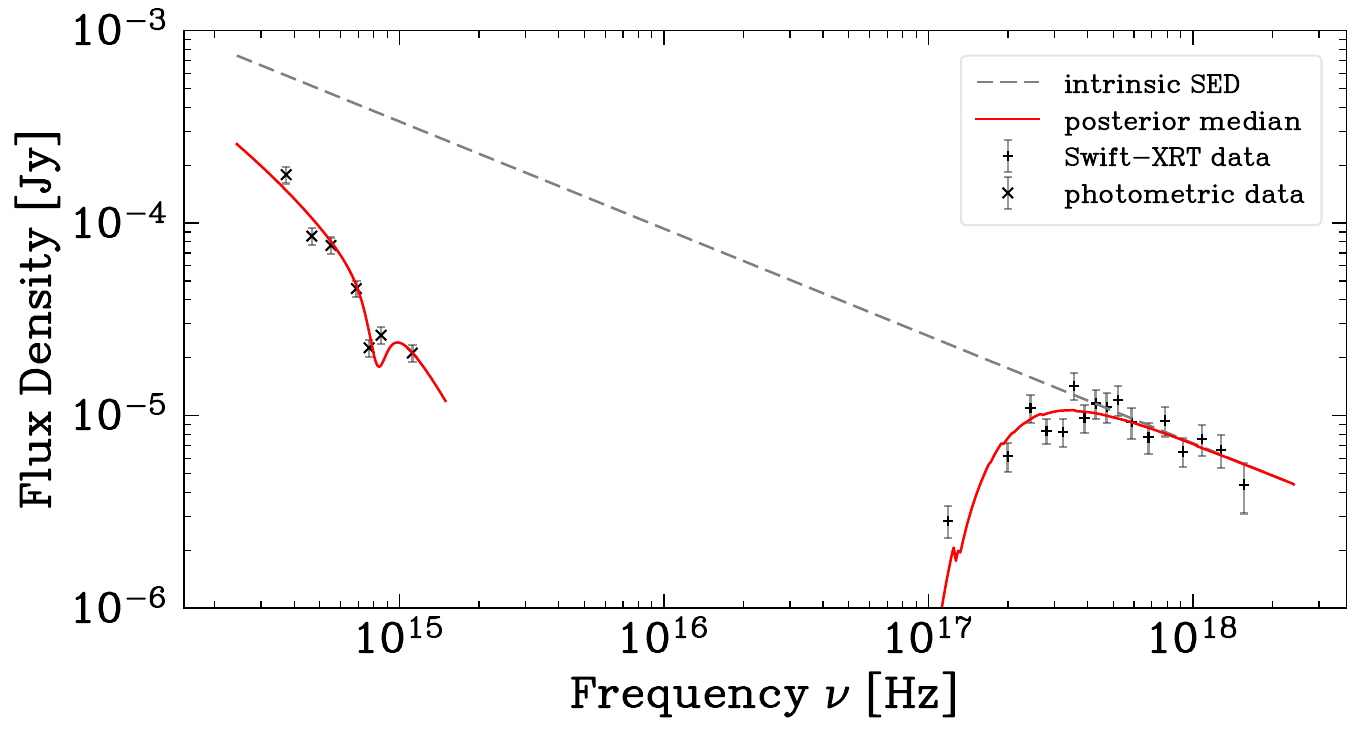}{0.5\linewidth}{(b) SED fitting for the afterglow in the time interval of $4797-8430\mathrm{\,s}$}}
    \caption{Time-sliced SED fits for the afterglow of GRB 240825A in the observer frame. 
    % These SEDs are shifted to the source frame using a redshift of $z = 0.659$. 
    The dashed line represents the intrinsic SED of afterglow. The red lines denote the posterior median. The Swift-XRT data and the optical photometric data are marked with `+' and `$\times$' symbols, respectively.
    \textit{Panel (a)}: The SED is extracted from the time interval $407-1290\mathrm{\,s}$. The fitting results are as follows: 
    the slope of the extinction curve is $\gamma = 0.61_{-0.02}^{+0.02}$, 
    the maximum amplitude of the bump is $E_\mathrm{bump} = (5.11_{-3.69}^{+6.68})\times10^{-2}$, 
    the host extinction is $A^\mathrm{host}_V = 2.87_{-0.12}^{+0.12}\mathrm{\,mag}$, 
    the intrinsic X-ray absorption column density is $N_\mathrm{H,X} = (1.19 \pm 0.03)\times10^{22}\mathrm{\,cm^{-2}}$, and 
    the intrinsic spectral index is $\beta_\mathrm{X} = 0.68 \pm 0.01$. 
    \textit{Panel (b)}: Same format as panel (a), but the SED is extracted from the time interval $4797 - 8430 \, \mathrm{s}$, with 
    $\gamma = 0.59_{-0.20}^{+0.26}$, 
    $E_\mathrm{bump} = 0.67_{-0.29}^{+0.54}$, 
    $A^\mathrm{host}_V = 1.37_{-0.55}^{+0.81}\mathrm{\,mag}$, 
    $N_\mathrm{H,X} = (0.92_{-0.17}^{+0.21})\times10^{22}\mathrm{\,cm^{-2}}$, and 
    $\beta_\mathrm{X} = 0.56_{-0.07}^{+0.10}$.
    }
    \label{fig:SED}
\end{figure*}

\begin{figure}[htbp]
  \centering
  \includegraphics[width=\linewidth]{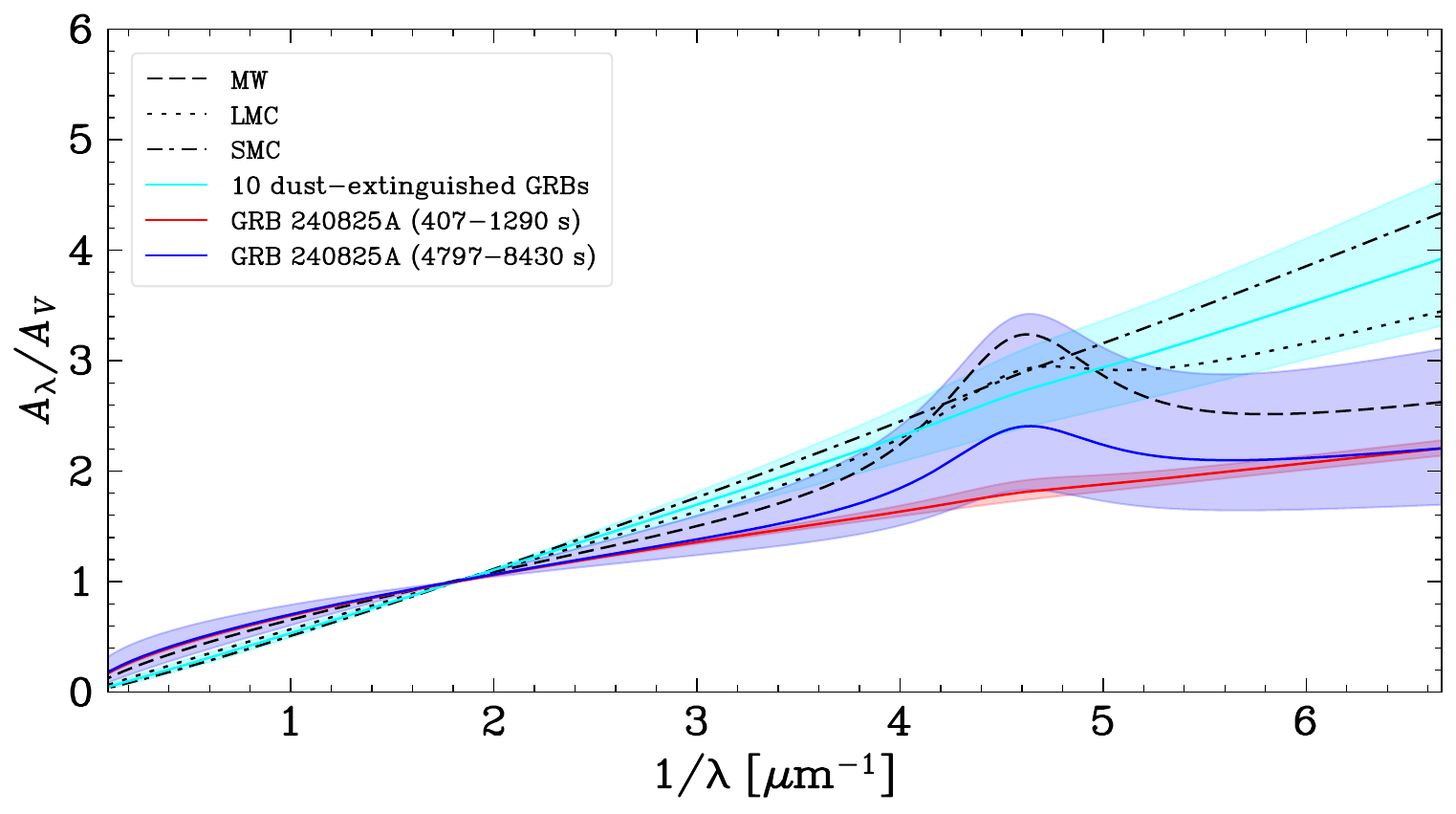}
  \caption{Comparison of extinction curves toward different targets. The mean extinction curves of the MW, LMC, SMC, and the 10 dust-extinguished GRBs are represented by the dashed, dotted, dashed-dotted lines, and the sky-blue line with shading, respectively \citep{2024A&A...690A.373R}. 
  % The steep extinction curve toward the host of GRB 140506A is also depicted by the cyan line for comparison \citep{2017A&A...601A..83H}.
  The fitting results of the extinction curves for two time intervals in this work are illustrated by the red and blue lines with shading, respectively.}
  \label{fig:extinction_curves}
\end{figure}

We further investigated the evolution of the extinction curve based on the extinction curve model, optical photometric data and X-ray data. Two time-sliced spectra were used, corresponding to the time intervals $T_0 + [407, 1288]\mathrm{\,s}$ (SED 1) and $T_0 + [4797, 8430]\mathrm{\,s}$ (SED 2), respectively. These two time intervals are marked by the vertical gray shadings in the sub-figure (a) of Figure~\ref{fig:lightcurve}. We note that SED 1 may be partially affected by the underlying spectral curvature, since the time interval of SED 1 partially overlaps with that of BAT data coverage. The flux densities in different photometric bands were also extracted within these intervals to ensure simultaneity. The posterior distributions of the model parameters were estimated using the Python package ELISA for Bayesian inference \citep{ELISA}. The fitting strategy for the X-ray model is described in Section~\ref{sec:swift_data}. For the intrinsic SED of the afterglow, we tested both the power-law model and the broken power-law model with a fixed $\Delta\beta=0.5$. The resulting extinction curves showed small variations within the error range. Therefore, we retained only the fitting results from the simple power-law model. The X-ray-to-optical afterglow SED fitting results for the two time intervals are presented in Figure~\ref{fig:SED}. The resulting extinction curves are shown in Figure~\ref{fig:extinction_curves}.

The UV bump feature described by $E_\mathrm{bump}$ is poorly constrained due to the redshift of the host and the limited coverage of photometric bands (see Figures~\ref{fig:SED} and~\ref{fig:extinction_curves}). Thanks to the photometric measurements of the blue portion of the early optical afterglow, the slope of the extinction curves is well constrained. The extinction curves derived from the two time intervals exhibit minor differences within the error range, suggesting a similar size distribution of grains in the two time intervals. 
% The derived slopes of these extinction curves are comparable to the steep slope observed in the extinction curve of GRB 140506A in the blue portion \citep{2017A&A...601A..83H}.
In contrast, the derived slopes display noticeable differences compared to the mean extinction curves of the Milky Way (MW), the Large Magellanic Cloud (LMC), the SMC, and the 10 dust-extinguished GRBs \citep{2024A&A...690A.373R}. The mean extinction of the host decreased significantly from $A^\mathrm{host}_V = 2.87_{-0.12}^{+0.12}\mathrm{\,mag}$ to $A^\mathrm{host}_V = 1.37_{-0.55}^{+0.81}\mathrm{\,mag}$ between the two time intervals, providing evidence of dust destruction in the circumburst medium \citep{2013ApJ...779...66S,2014MNRAS.440.1810M,2023MNRAS.520.6104K,2025ApJ...979...38C}. The color-magnitude diagram of galaxies and the SED fitting for the host galaxy consistently indicate that the host galaxy of GRB~240825A has a stellar mass of $\log(M_{*}/M_\odot) \sim 10$ \citep{2025ApJ...979...38C}. We suggest that GRB 240825A occurred in a dust-obscured environment, within a medium-mass host galaxy.

\subsection{Thermal Component in X-ray Emission}
We found that the addition of a BB component may improve the goodness of fit in the time-averaged joint XRT and BAT spectral analysis.
The total luminosity of the BB component is estimated from the spectral normalization of the BB component and the luminosity distance of the burst, yielding $L_\mathrm{BB} \simeq 7.81\times10^{48}\mathrm{\,erg\,s^{-1}}$. 
The corresponding BB radius is estimated to be $R_\mathrm{BB} \simeq 4.51\times10^{10}\mathrm{\,cm}$. The BB component contributes approximately 17.74\% to the total observed flux in the observer-frame 0.3-150\,keV, i.e., $F_\mathrm{BB}/F_\mathrm{X} \simeq 17.74\%$.
The finding of the BB component might explain the spectral curvature at $\sim10\mathrm{\,keV}$, but other explanations for this curvature can not be ruled out. 
% Although the X-ray flux density evaluated at 1.732 keV—based on a single power-law model within the overlapping time interval—may be influenced by the thermal component, the effect is negligible.

% \textbf{
% To fit the time-resolved XRT data with the SBPL model, a relatively reasonable approach is to fix the high-energy power-law index $\Gamma_2$ at 2.3 (as derived from the joint fitting of BAT and XRT data with the SBPL+BB model), although $\Gamma_1$ and $N_\mathrm{H,X}$ are poorly constrained. The intrinsic column density derived from a single SBPL model can be treated as a rough upper limit, if additionally incorporating a BB component into the fit tends to reduce the inferred intrinsic column density.
% }

% In the prompt emission of GRB~240825A, a two-hump spectrum was observed \citep{2025arXiv250103082Z}, along with the evolution of an early thermal component \citep{2025ApJ...985L..30W}. Additionally, a quasi-periodic oscillation at 6.37 Hz was identified in the photospheric emission of GRB~240825A \citep{2025arXiv250716538L}.
The X-ray emission of GRB~240825A observed by Swift/XRT significantly exceeds the model predictions for both the forward and reverse shocks, as shown in Figure~7 of \citet{2025arXiv250702806W}. The thermal X-ray emission has been reported in previous studies of Swift GRBs \citep [e.g.,][]{2011MNRAS.416.2078P,2012MNRAS.427.2950S,2012MNRAS.427.2965S}. We found a thermal component in the early X-ray emission of GRB 240825A, providing an important clue regarding the underlying physical mechanisms of the burst.

\section{Summary} \label{sec:Summery}
GRB~240825A, recently triggered by Swift \citep{2024GCN.37274....1G} and Fermi \citep{2024GCN.37273....1F}, was rapidly followed up across multi-wavelengths. The soft optical spectral index of the early optical afterglow \citep[$\beta_\mathrm{obs,O} \simeq 2.48$;][]{2025ApJ...979...38C} indicates that GRB~240825A is most likely located in the dense circumburst medium. In this paper, we present the X-ray and optical observations to comprehensively investigate the evolution of X-ray spectral properties and the optical darkness of GRB~240825A.

% \textbf{The X-ray spectral analysis reveals that an likely thermal component in the early X-ray emission, which may explain the the early-time X-ray spectral curvature at $\sim10\mathrm{\,keV}$.}
% \textbf{Given the particular physical properties of the burst, we suggest that the thermal component may be related to the shock breakout of a supernova.}
% \textbf{The finding of the thermal component can enrichs the physical picture of the burst possibly associated with a supernova.}

The temporal evolution of $\beta_\mathrm{OX}$ shows an initial decline followed by a subsequent rise, reaching its minimum at $\sim1000$\,s. 
% This temporal behavior may be attributed to the gradual cooling of an additional thermal component. 
However, GRB 240825A is not classified as an optically dark burst according to the criteria $\beta_\mathrm{OX} < \beta_\mathrm{X} - 0.5$ and $\beta_\mathrm{OX} < 0.5$ at 11 hr post-trigger.

We also obtained the characteristics of the circumburst medium environment, particularly the extinction curve that governs the observed attenuation. The extinction curves, derived by fitting the X-ray-to-optical SEDs of the afterglow in different time intervals, indicate that GRB 240825A occurred in a dust-obscured environment. Additionally, the evidence of dust destruction in the circumburst medium was observed.

\begin{acknowledgments}
    We thank the anonymous referee for instructive comments that improved the current work significantly.
    This work is supported by National Key R\&D Program of China (2023YFE0101200), Natural Science Foundation of China 12393813, %CSST grant CMS-CSST-2021-A06, 
    and the Yunnan Revitalization Talent Support Program (YunLing Scholar Project). Y.P.Y. is supported by the National Key Research and Development Program of China (2024YFA1611603), the National Natural Science Foundation of China grant No.12473047 and the National SKA Program of China (2022SKA0130100). We acknowledge the valuable discussions with Drs. Xiao-Hong Zhao, Guo-Bao Zhang, Hong-Tao Liu, and Ny Avo Rakotondrainibe.
    This research has made use of NASA's Astrophysics Data System, as well as the following GitHub repositories: \href{https://github.com/yymao/adstex}{\texttt{yymao/adstex}}, \href{https://github.com/cgobat/dark-GRBs}{\texttt{cgobat/dark-GRBs}}, \href{https://github.com/cgobat/asymmetric_uncertainty}{\texttt{cgobat/asymmetric\_uncertainty}}, \href{https://github.com/karpov-sv/stdpipe}{\texttt{karpov-sv/stdpipe}} and \href{https://github.com/wcxve/elisa}{\texttt{wcxve/elisa}}. We also extend our gratitude to the software developers for providing their code as free software. Additionally, this research utilized data provided by the UKSSDC at the University of Leicester.
\end{acknowledgments}

\software{Astropy \citep{2013A&A...558A..33A,2018AJ....156..123A,2022ApJ...935..167A},
          PyMC \citep{pymc2023}, 
          Pandas \citep{2023zndo..10107975T},
          Matplotlib \citep{2007CSE.....9...90H},
          smplotlib \citep{2023zndo...8126529L},
          NumPy \citep{2020Natur.585..357H},
          asymmetric\_uncertainty \citep{2022ascl.soft08005G},
          STDPipe \citep{2021ascl.soft12006K,Karpov_2025},
          ELISA \citep{ELISA}
          }

%% Appendix material should be preceded with a single \appendix command.
%% There should be a \section command for each appendix. Mark appendix
%% subsections with the same markup you use in the main body of the paper.
%%
%% Each Appendix (indicated with \section) will be lettered A, B, C, etc.
%% The equation counter will reset when it encounters the \appendix
%% command and will number appendix equations (A1), (A2), etc. The
%% Figure and Table counter will not reset.
\clearpage
\appendix
\restartappendixnumbering
\section{Spectral Analysis Results of the X-ray Emission of GRB 240825A}\label{sec:spectral_analysis}
% For the time-resolved X-ray spectral analysis, the count-rate light curve was segmented into intervals of relatively constant count rates, determined using the Bayesian blocks algorithm \citep{1998ApJ...504..405S,2013ApJ...764..167S}, as implemented in the Python package \texttt{Astropy} \citep[v6.1.5;][]{astropy_collaboration_2024_14053341}. Based on these intervals, we used the time-sliced spectra builder\footnote{\url{https://www.swift.ac.uk/xrt_spectra/addspec.php?targ=01250617}} to extract the time-resolved spectra \citep{2007A&A...469..379E,2009MNRAS.397.1177E}. Each spectrum was automatically fitted with an absorbed power-law model (\texttt{TBabs*zTBabs*powerlaw} in \texttt{XSPEC}) using the Efficient Library for Spectral Analysis in High-Energy Astrophysics (ELISA\footnote{\url{https://github.com/wcxve/elisa}}). The model accounts for X-ray absorption contributions from both the Milky Way and the intrinsic circumburst medium of the GRB host galaxy, while neglecting contributions from the IGM. The solar abundances are adopted from \citet{2000ApJ...542..914W}, and the cross-sections are taken from \citet{1995A&AS..109..125V} and \citet{1996ApJ...465..487V}. The Galactic absorption, modeled with \texttt{TBabs} component, was fixed at the value of $N^\mathrm{Gal}_\mathrm{H}$. The intrinsic absorption, modeled as $N_\mathrm{H,X}$ with \texttt{zTBabs} component, was estimated at the redshift of the GRB host galaxy. The photon index ($\Gamma$) and normalization, described by the \texttt{powerlaw} component, were treated as free parameters. The results of the time-resolved analysis for the X-ray afterglow are summarized in Table~\ref{tab:time-sliced}. 

\begin{figure*}[ht!]
    \centering
    \begin{subfigure}[t]{0.49\textwidth}
        \centering
        \includegraphics[width=\linewidth]{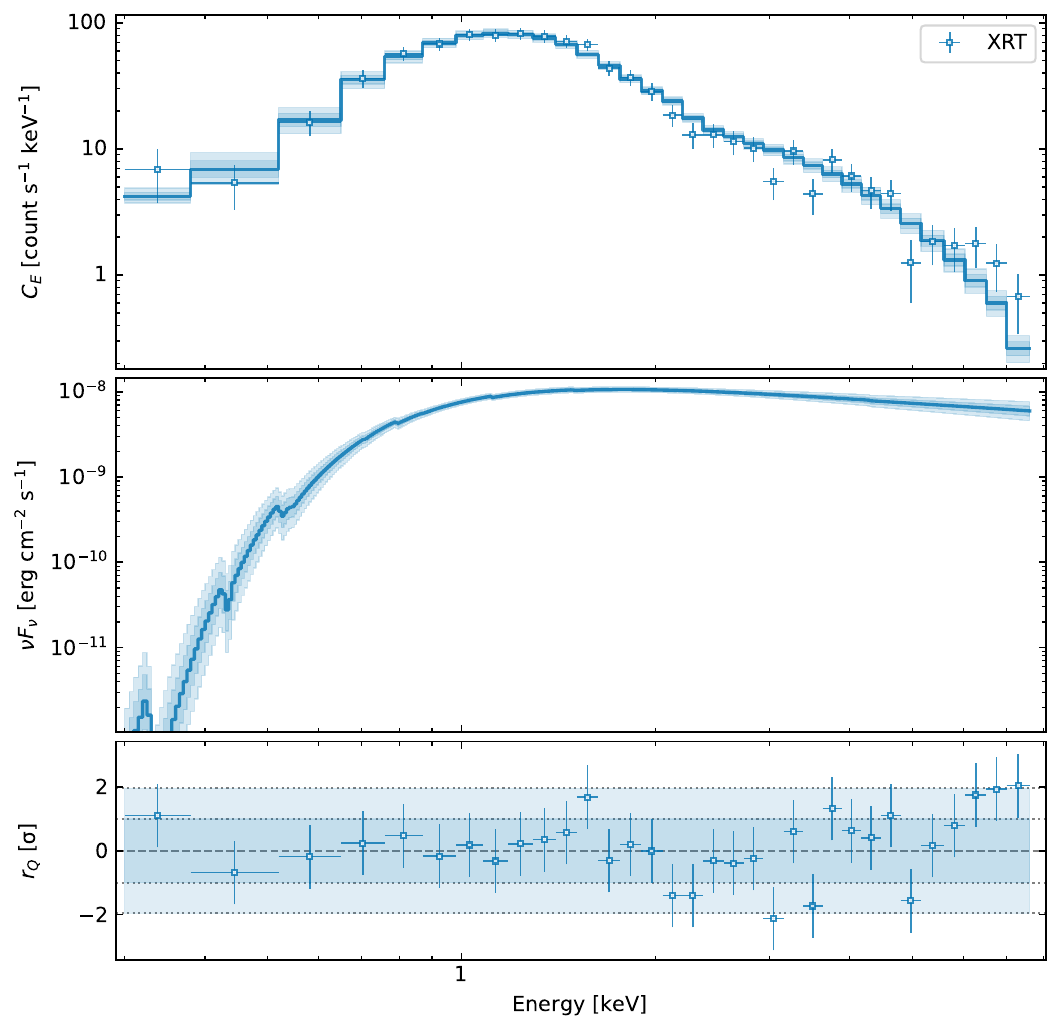}
        \caption{XRT spectrum, PL model, and residuals.}
    \end{subfigure}
    \hfill
    \begin{subfigure}[t]{0.49\textwidth}
        \centering
        \includegraphics[width=\linewidth]{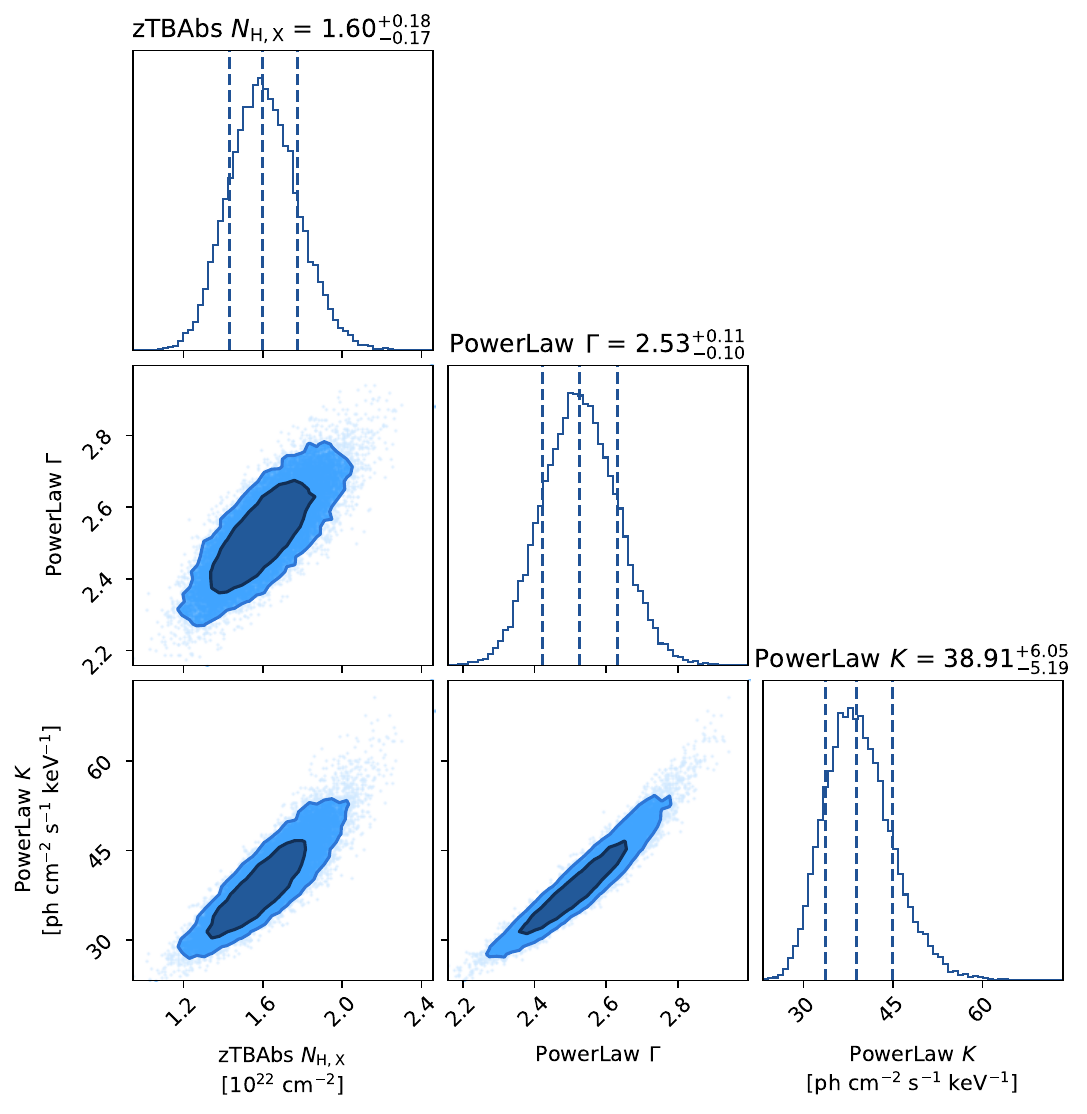}
        \caption{Posterior distributions of model parameters.}
    \end{subfigure}
    \vspace{1em}
    \begin{subfigure}[t]{0.49\textwidth}
        \centering
        \includegraphics[width=\linewidth]{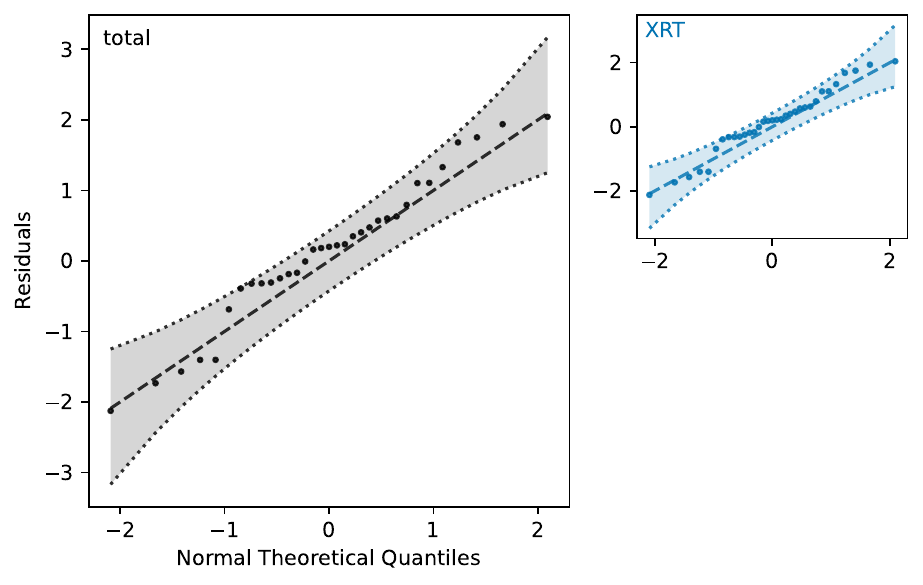}
        \caption{Q-Q plots of residuals.}
    \end{subfigure}
    \hfill
    \begin{subfigure}[t]{0.49\textwidth}
        \centering
        \includegraphics[width=\linewidth]{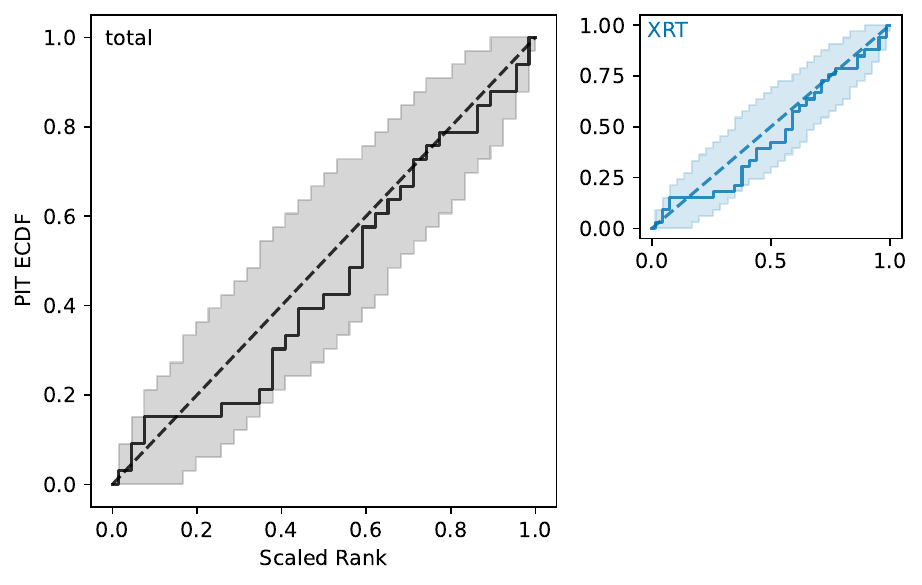}
        \caption{PIT-ECDF diagnostic.}
    \end{subfigure}
    \caption{XRT spectrum extracted from 87-97 s post-trigger, fitted with an absorbed PL model and shown with posterior diagnostics. 
    Panel (a): Spectral fit to the data with PL model and residuals.
    Panel (b): Corner plot of the posterior distributions of the model parameters.
    Panel (c): Quantile-Quantile (Q-Q) plot of residuals.
    Panel (d) Posterior predictive probability integral transformation-empirical cumulative distribution function (PIT-ECDF) check.}
    \label{fig:PowerLawSpec1}
\end{figure*}

\begin{figure*}[ht!]
    \centering
    \begin{subfigure}[t]{0.49\textwidth}
        \centering
        \includegraphics[width=\linewidth]{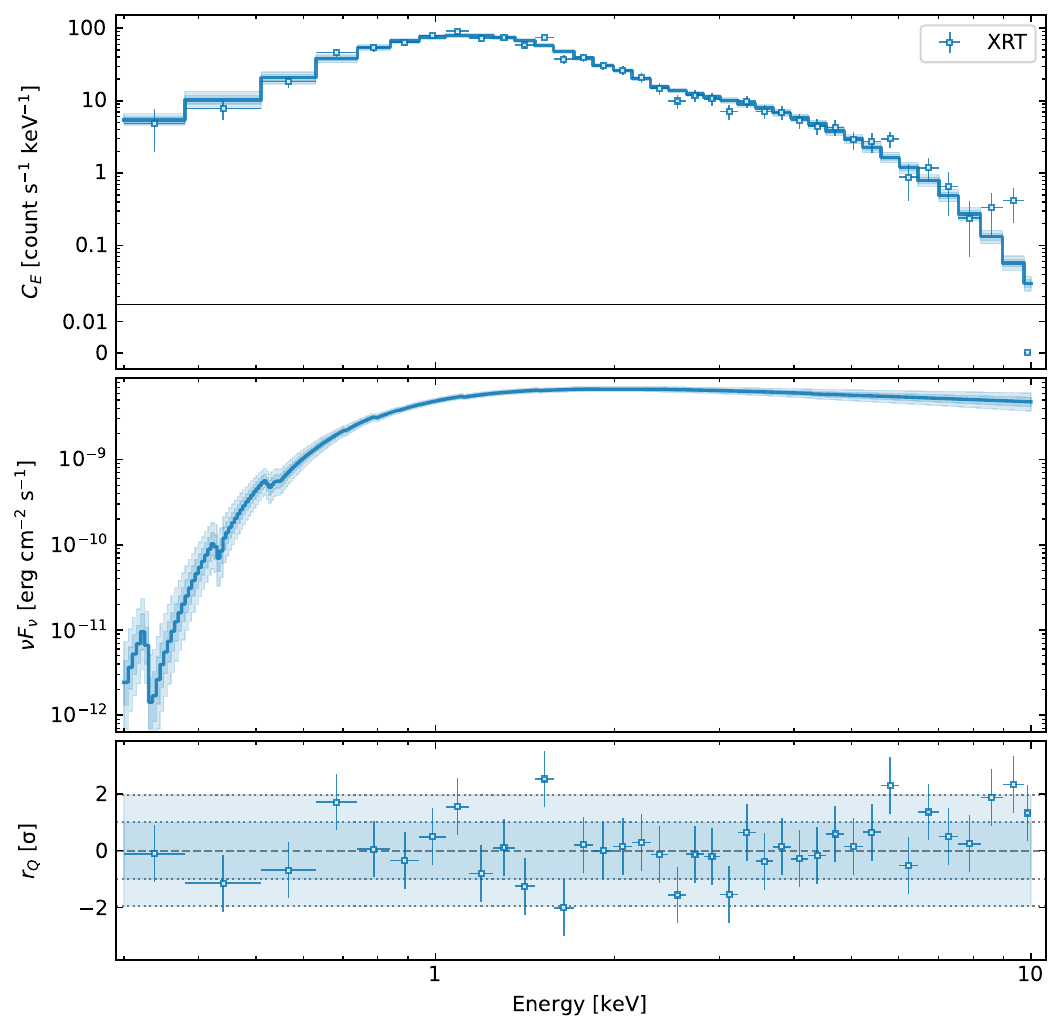}
        \caption{XRT spectrum, PL model, and residuals.}
    \end{subfigure}
    \hfill
    \begin{subfigure}[t]{0.49\textwidth}
        \centering
        \includegraphics[width=\linewidth]{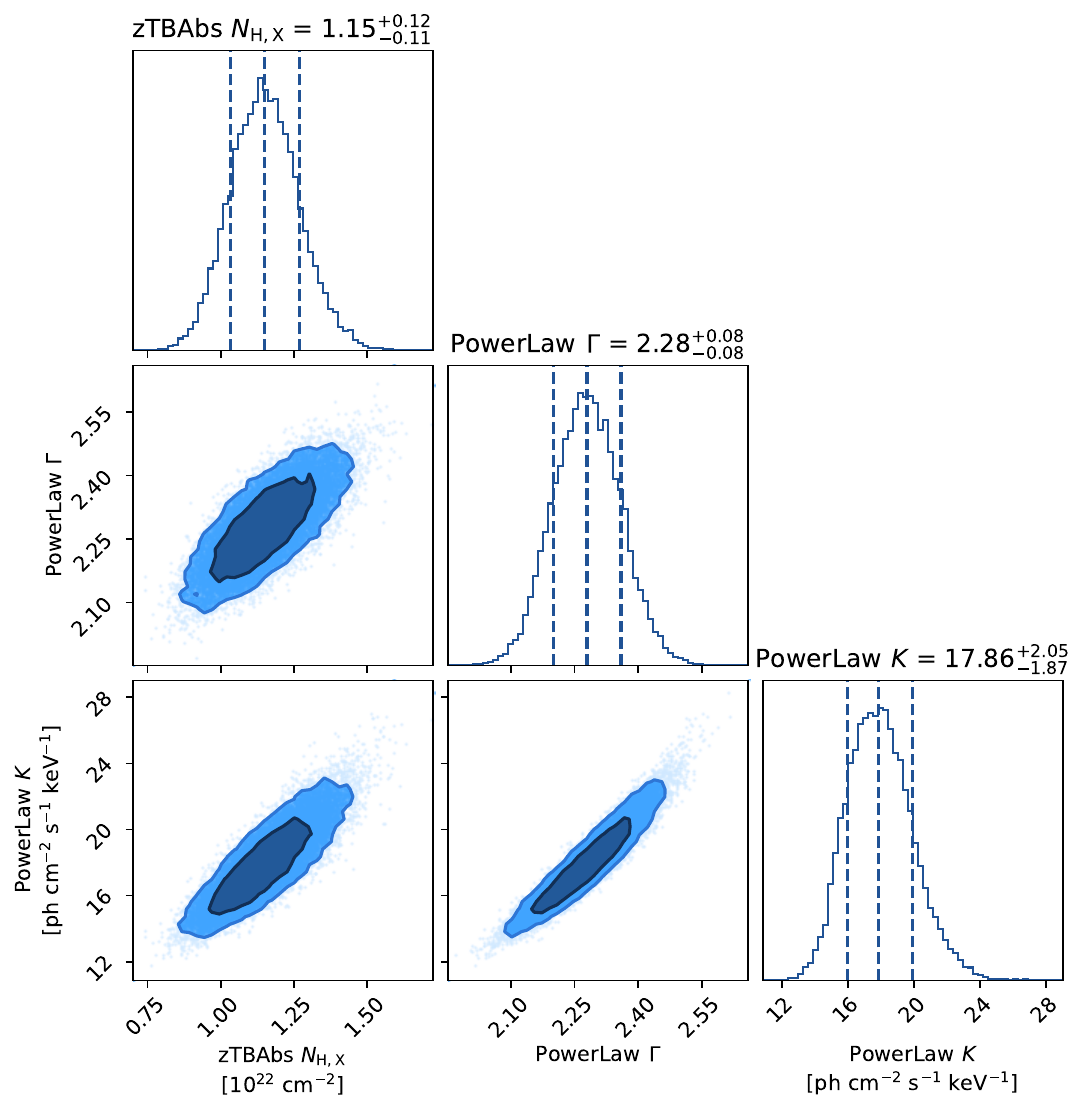}
        \caption{Posterior distributions of model parameters.}
    \end{subfigure}
    \vspace{1em}
    \begin{subfigure}[t]{0.49\textwidth}
        \centering
        \includegraphics[width=\linewidth]{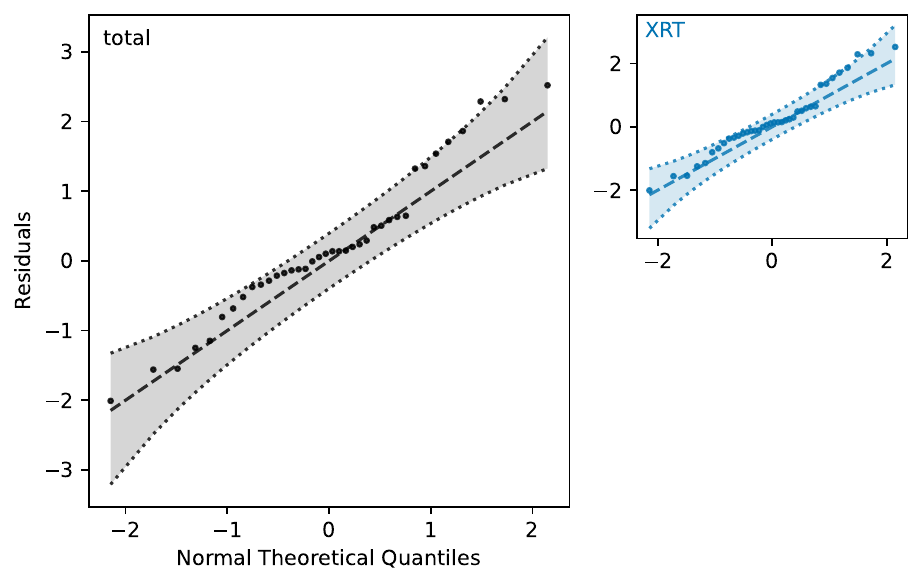}
        \caption{Q-Q plots of residuals.}
    \end{subfigure}
    \hfill
    \begin{subfigure}[t]{0.49\textwidth}
        \centering
        \includegraphics[width=\linewidth]{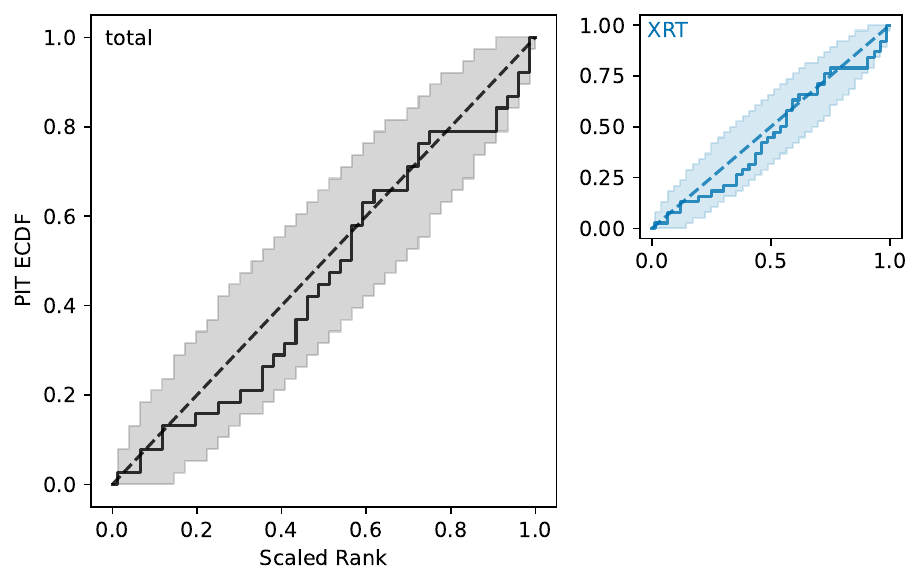}
        \caption{PIT-ECDF diagnostic.}
    \end{subfigure}
    \caption{XRT spectrum extracted from 97-109 s post-trigger, fitted with an absorbed PL model and shown with posterior diagnostics. This figure follows the format of Figure~\ref{fig:PowerLawSpec1}.}
    \label{fig:PowerLawSpec2}
\end{figure*}

\begin{figure*}[ht!]
    \centering
    \begin{subfigure}[t]{0.49\textwidth}
        \centering
        \includegraphics[width=\linewidth]{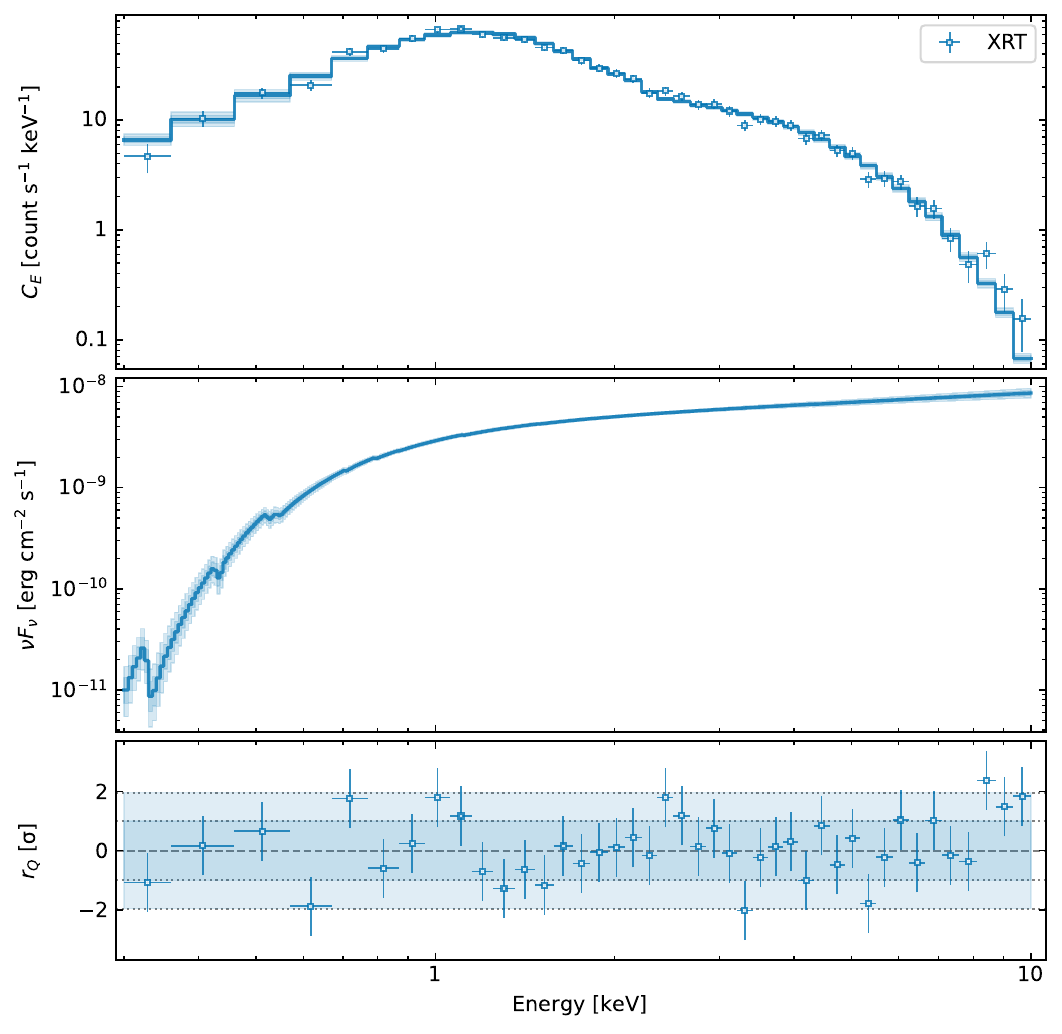}
        \caption{XRT spectrum, PL model, and residuals.}
    \end{subfigure}
    \hfill
    \begin{subfigure}[t]{0.49\textwidth}
        \centering
        \includegraphics[width=\linewidth]{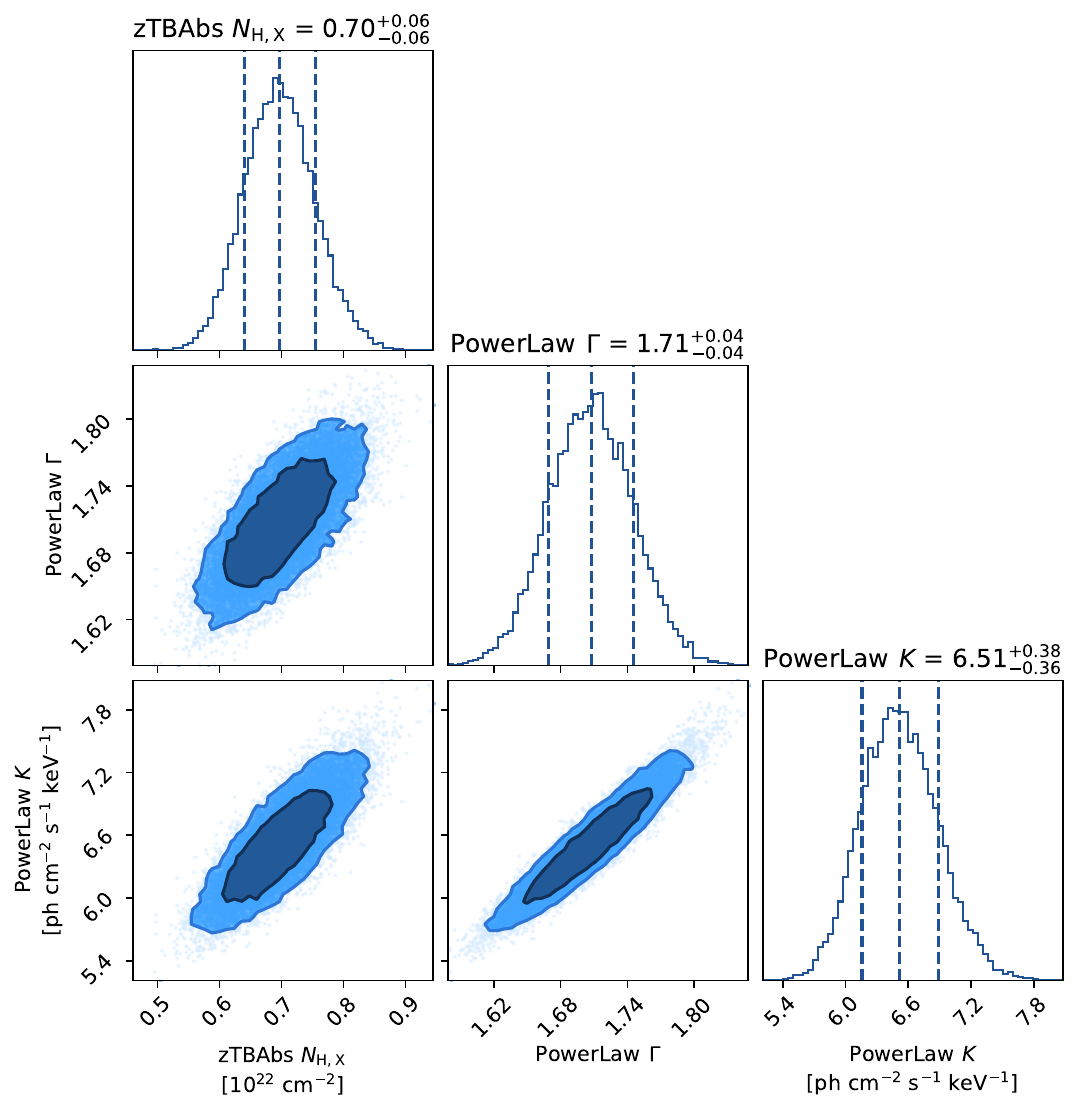}
        \caption{Posterior distributions of model parameters.}
    \end{subfigure}
    \vspace{1em}
    \begin{subfigure}[t]{0.49\textwidth}
        \centering
        \includegraphics[width=\linewidth]{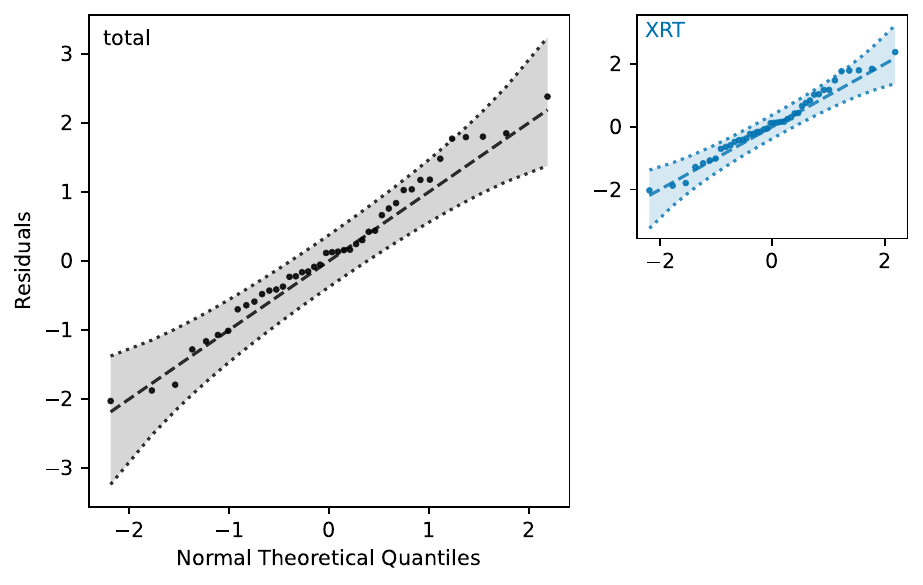}
        \caption{Q-Q plots of residuals.}
    \end{subfigure}
    \hfill
    \begin{subfigure}[t]{0.49\textwidth}
        \centering
        \includegraphics[width=\linewidth]{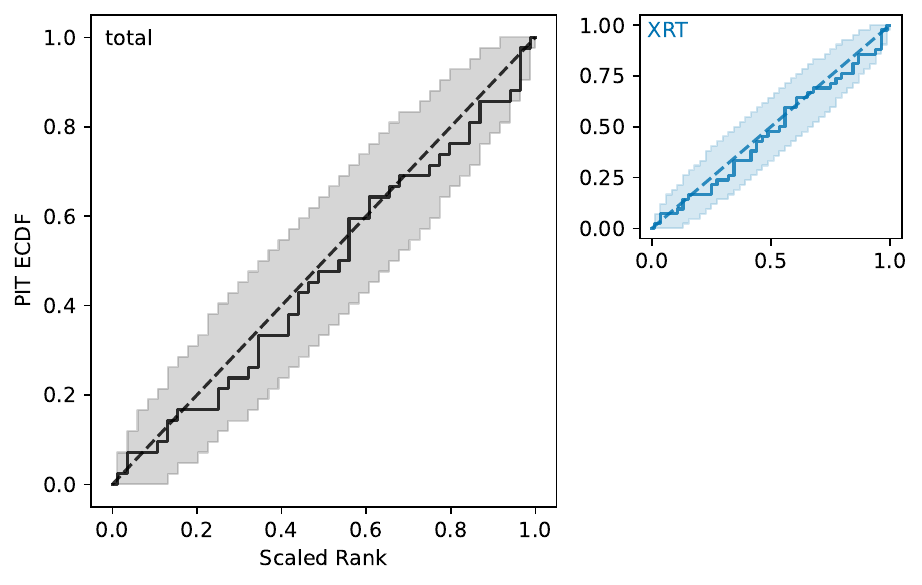}
        \caption{PIT-ECDF diagnostic.}
    \end{subfigure}
    \caption{XRT spectrum extracted from 109-149 s post-trigger, fitted with an absorbed PL model and shown with posterior diagnostics. This figure follows the format of Figure~\ref{fig:PowerLawSpec1}.}
    \label{fig:PowerLawSpec3}
\end{figure*}

\begin{figure*}[ht!]
    \centering
    \begin{subfigure}[t]{0.49\textwidth}
        \centering
        \includegraphics[width=\linewidth]{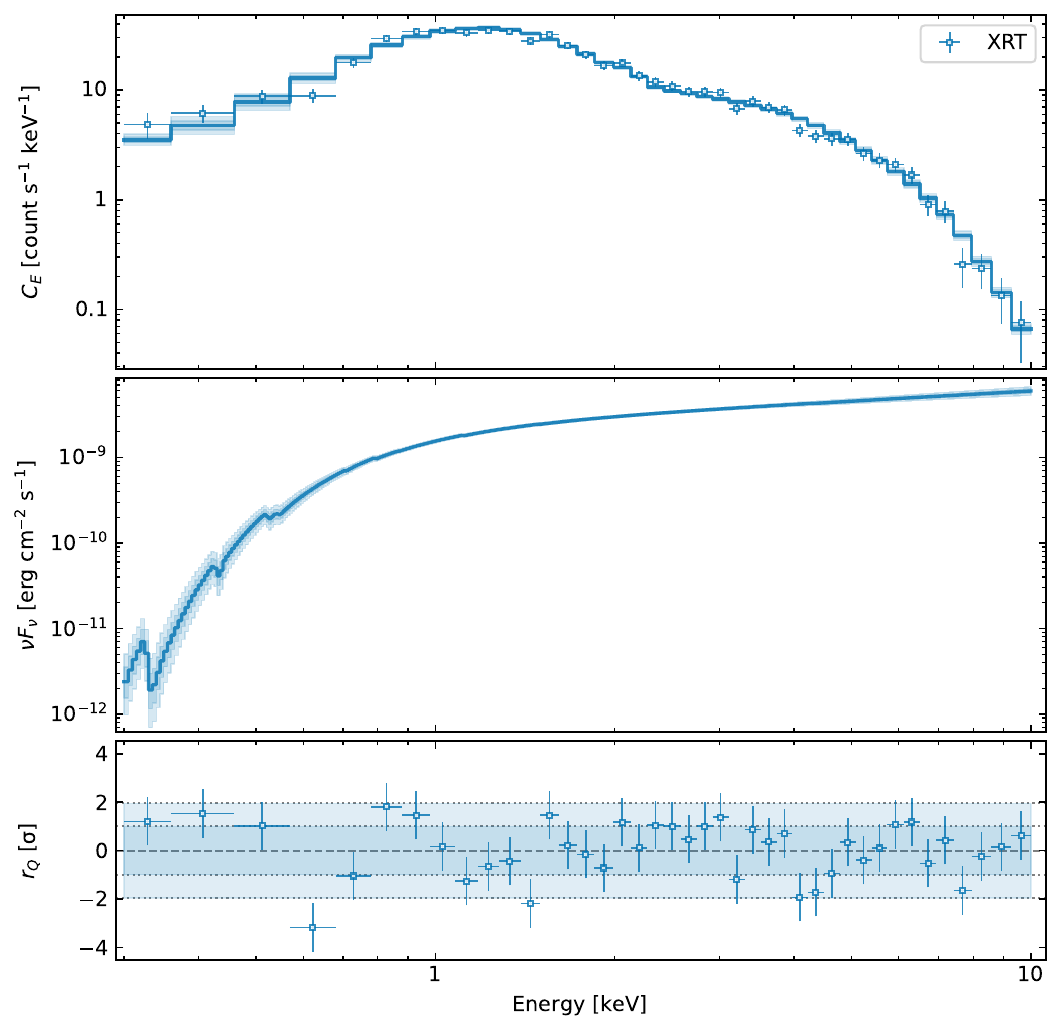}
        \caption{XRT spectrum, PL model, and residuals.}
    \end{subfigure}
    \hfill
    \begin{subfigure}[t]{0.49\textwidth}
        \centering
        \includegraphics[width=\linewidth]{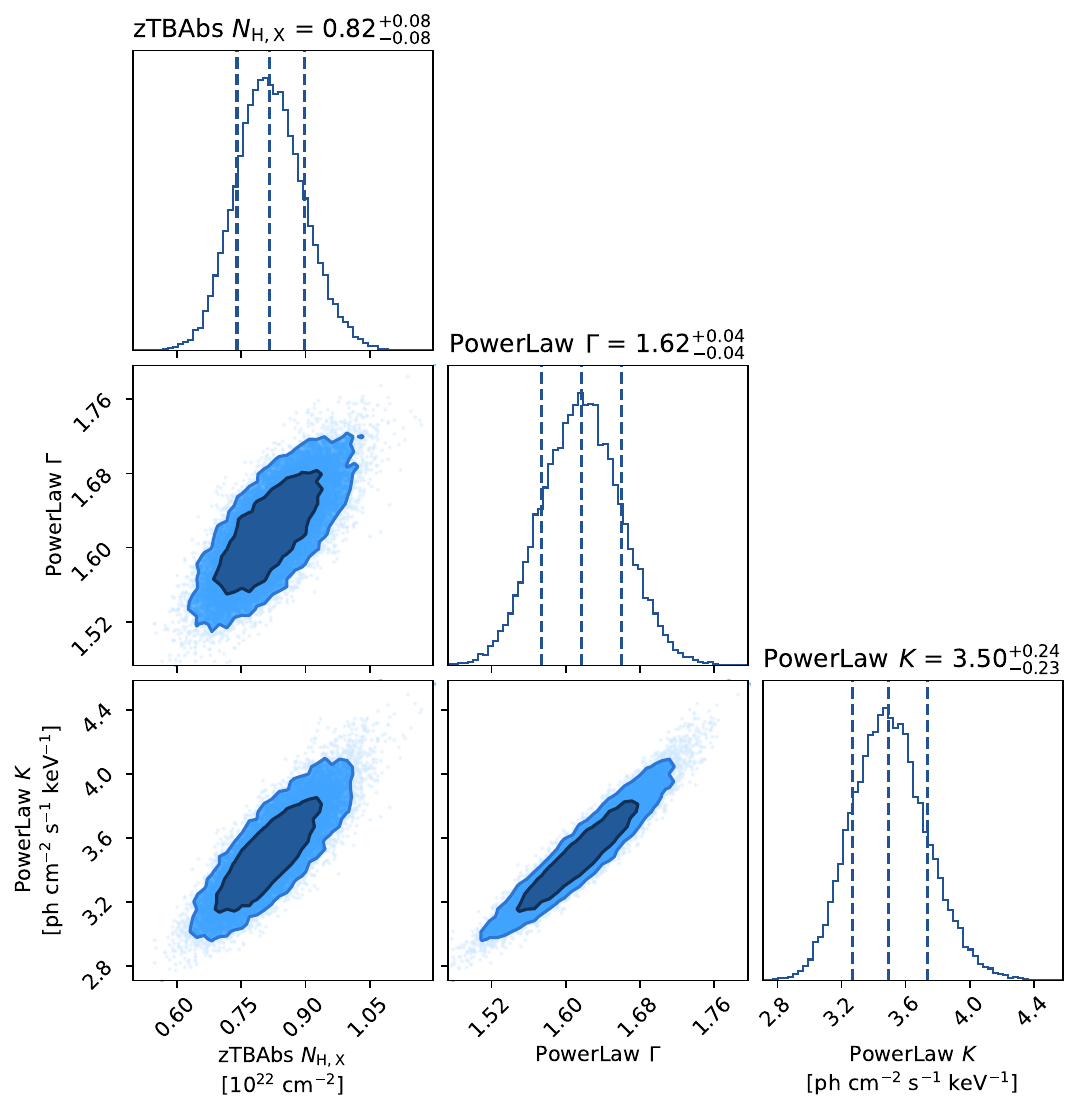}
        \caption{Posterior distributions of model parameters.}
    \end{subfigure}
    \vspace{1em}
    \begin{subfigure}[t]{0.49\textwidth}
        \centering
        \includegraphics[width=\linewidth]{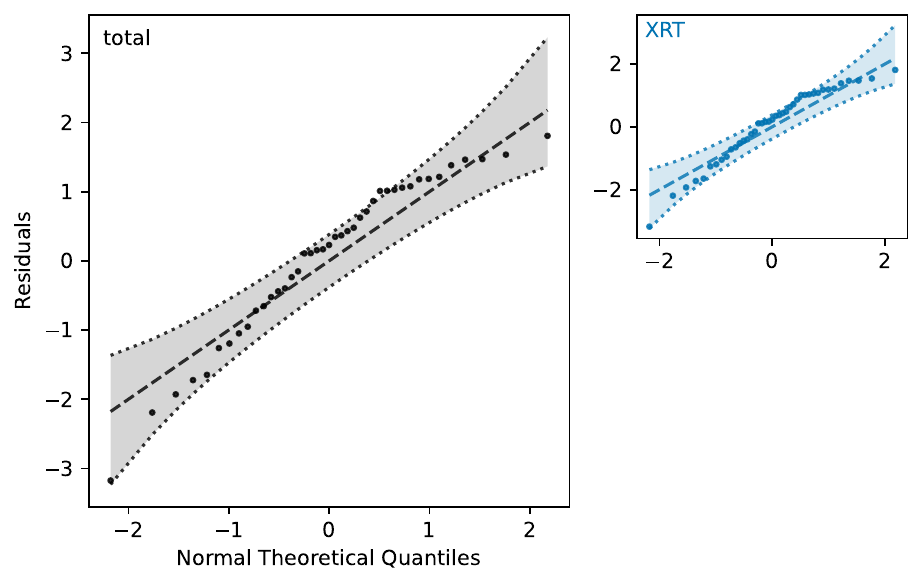}
        \caption{Q-Q plots of residuals.}
    \end{subfigure}
    \hfill
    \begin{subfigure}[t]{0.49\textwidth}
        \centering
        \includegraphics[width=\linewidth]{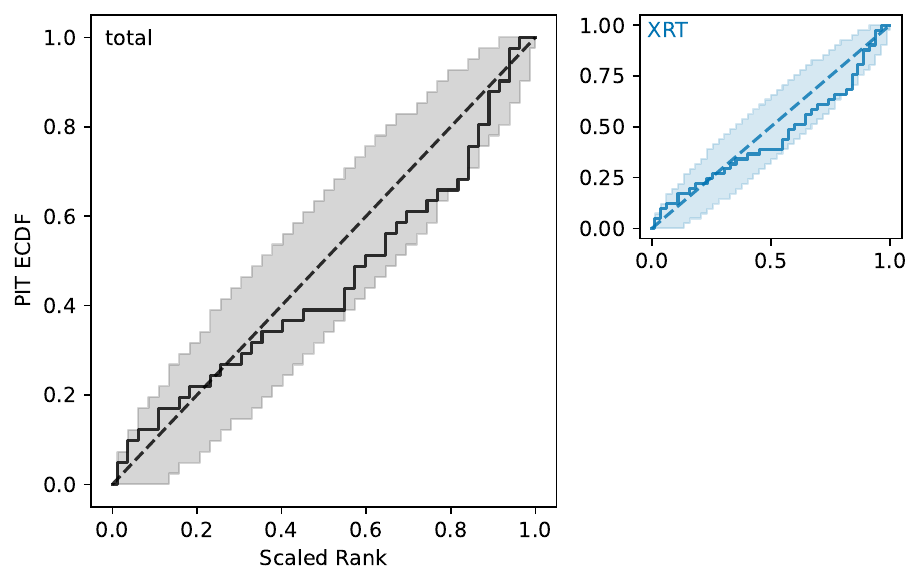}
        \caption{PIT-ECDF diagnostic.}
    \end{subfigure}
    \caption{XRT spectrum extracted from 149-203 s post-trigger, fitted with an absorbed PL model and shown with posterior diagnostics. This figure follows the format of Figure~\ref{fig:PowerLawSpec1}.}
    \label{fig:PowerLawSpec4}
\end{figure*}

\begin{figure*}[ht!]
    \centering
    \begin{subfigure}[t]{0.49\textwidth}
        \centering
        \includegraphics[width=\linewidth]{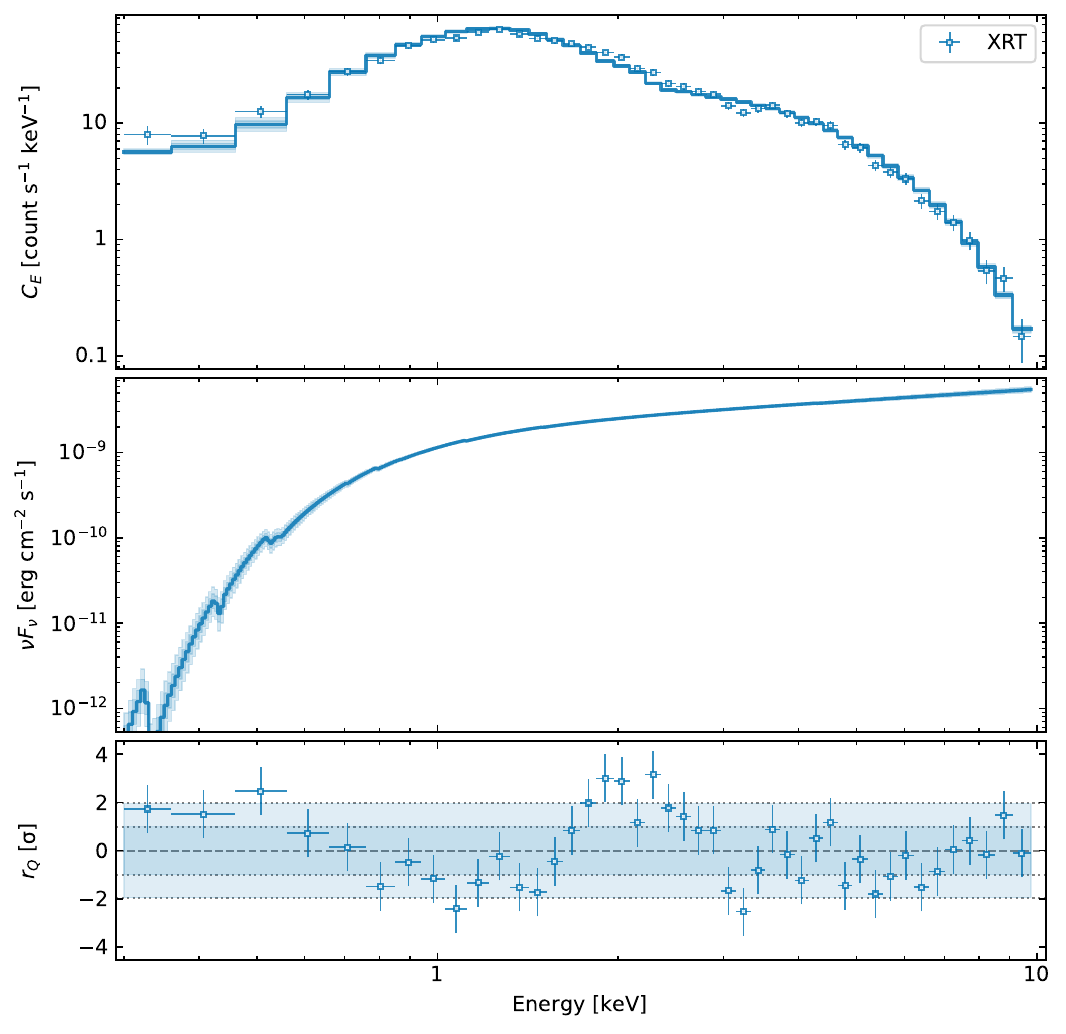}
        \caption{XRT spectrum, PL model, and residuals.}
    \end{subfigure}
    \hfill
    \begin{subfigure}[t]{0.49\textwidth}
        \centering
        \includegraphics[width=\linewidth]{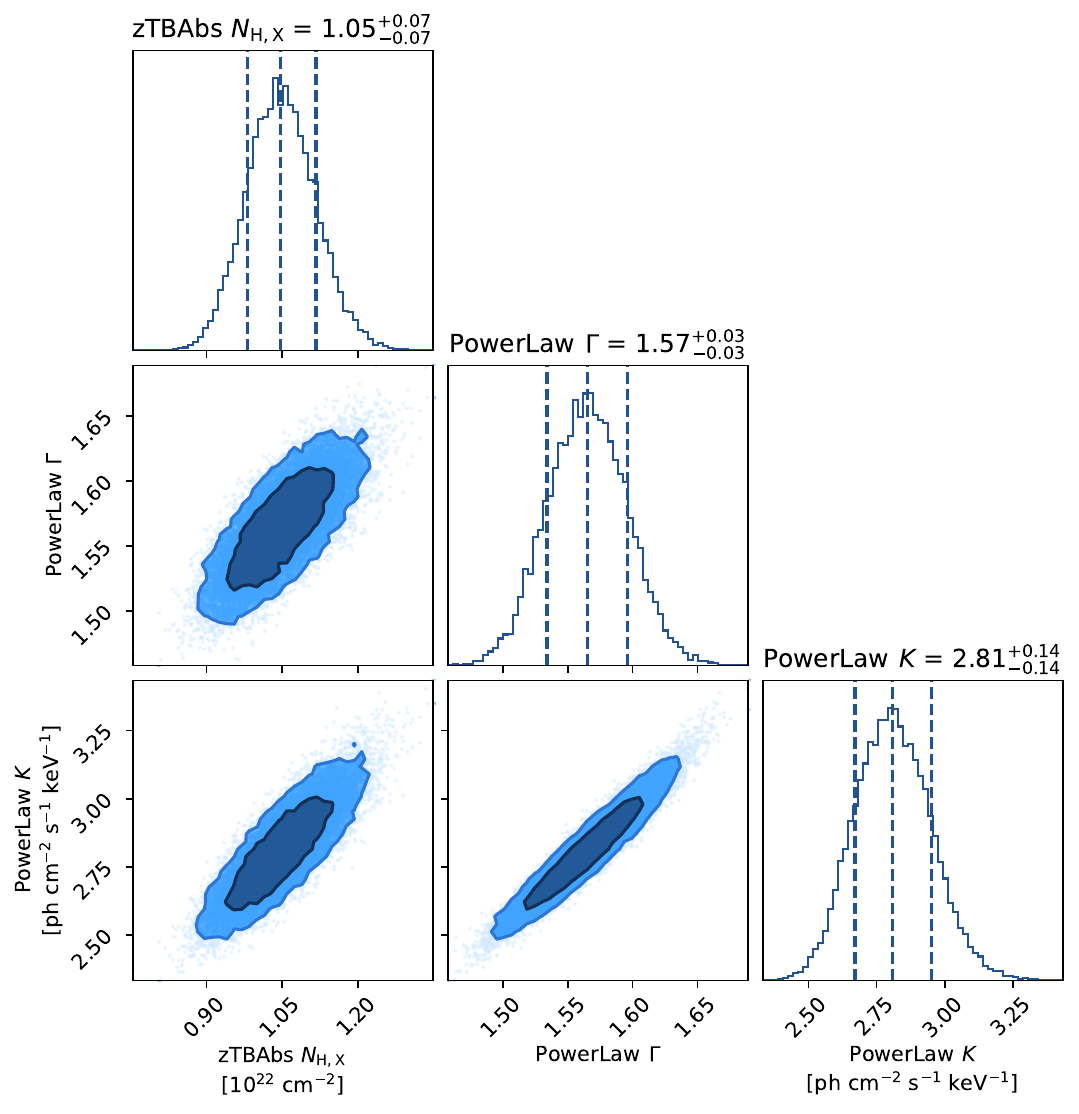}
        \caption{Posterior distributions of model parameters.}
    \end{subfigure}
    \vspace{1em}
    \begin{subfigure}[t]{0.49\textwidth}
        \centering
        \includegraphics[width=\linewidth]{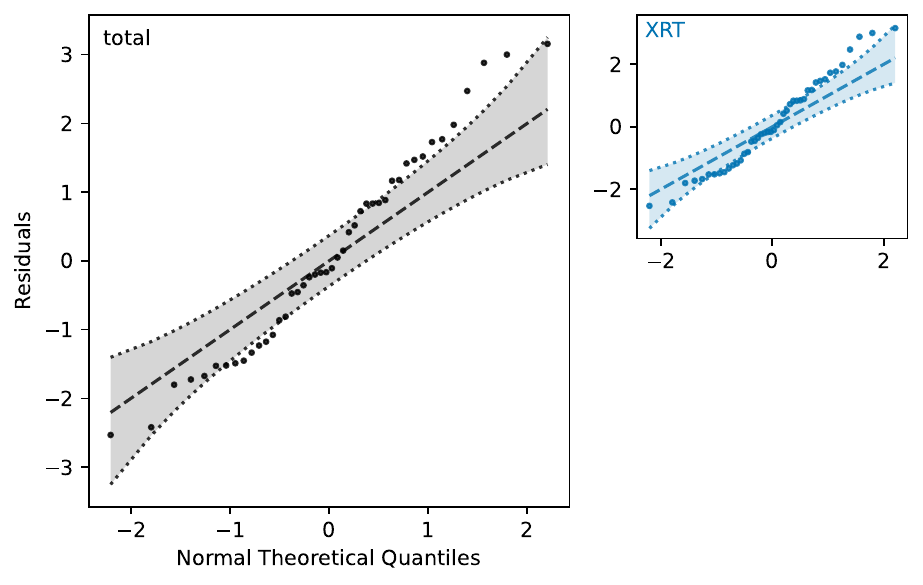}
        \caption{Q-Q plots of residuals.}
    \end{subfigure}
    \hfill
    \begin{subfigure}[t]{0.49\textwidth}
        \centering
        \includegraphics[width=\linewidth]{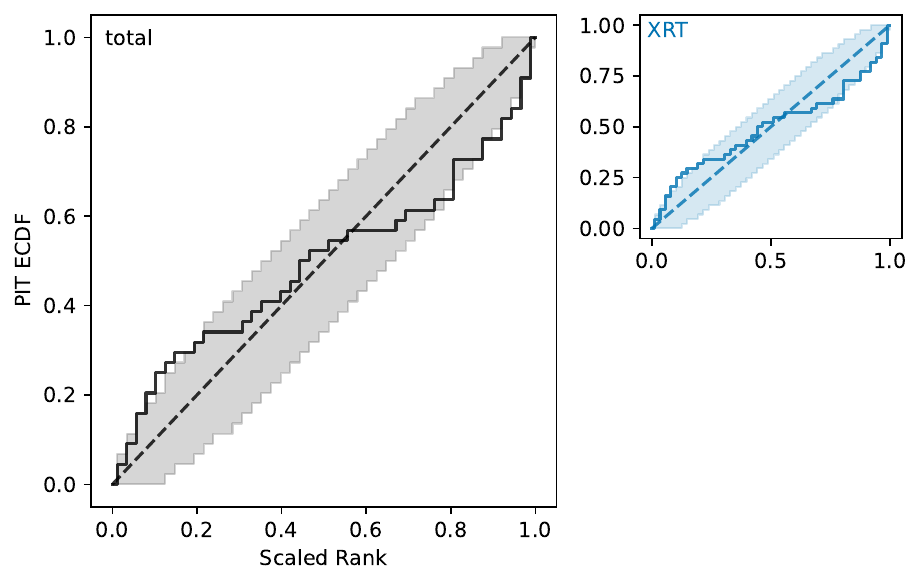}
        \caption{PIT-ECDF diagnostic.}
    \end{subfigure}
    \caption{XRT spectrum extracted from 203-265 s post-trigger, fitted with an absorbed PL model and shown with posterior diagnostics. This figure follows the format of Figure~\ref{fig:PowerLawSpec1}.}
    \label{fig:PowerLawSpec5}
\end{figure*}

\begin{figure*}[ht!]
    \centering
    \begin{subfigure}[t]{0.49\textwidth}
        \centering
        \includegraphics[width=\linewidth]{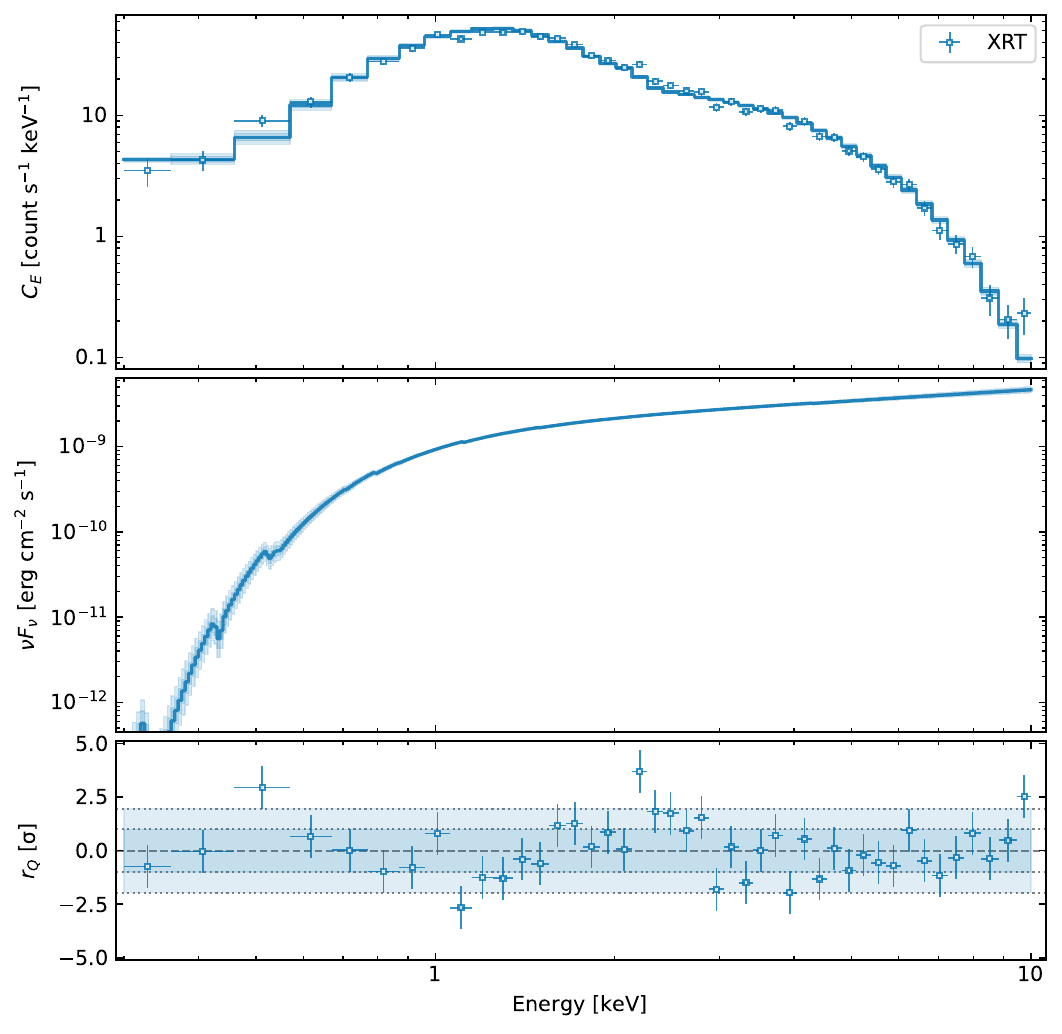}
        \caption{XRT spectrum, PL model, and residuals.}
    \end{subfigure}
    \hfill
    \begin{subfigure}[t]{0.49\textwidth}
        \centering
        \includegraphics[width=\linewidth]{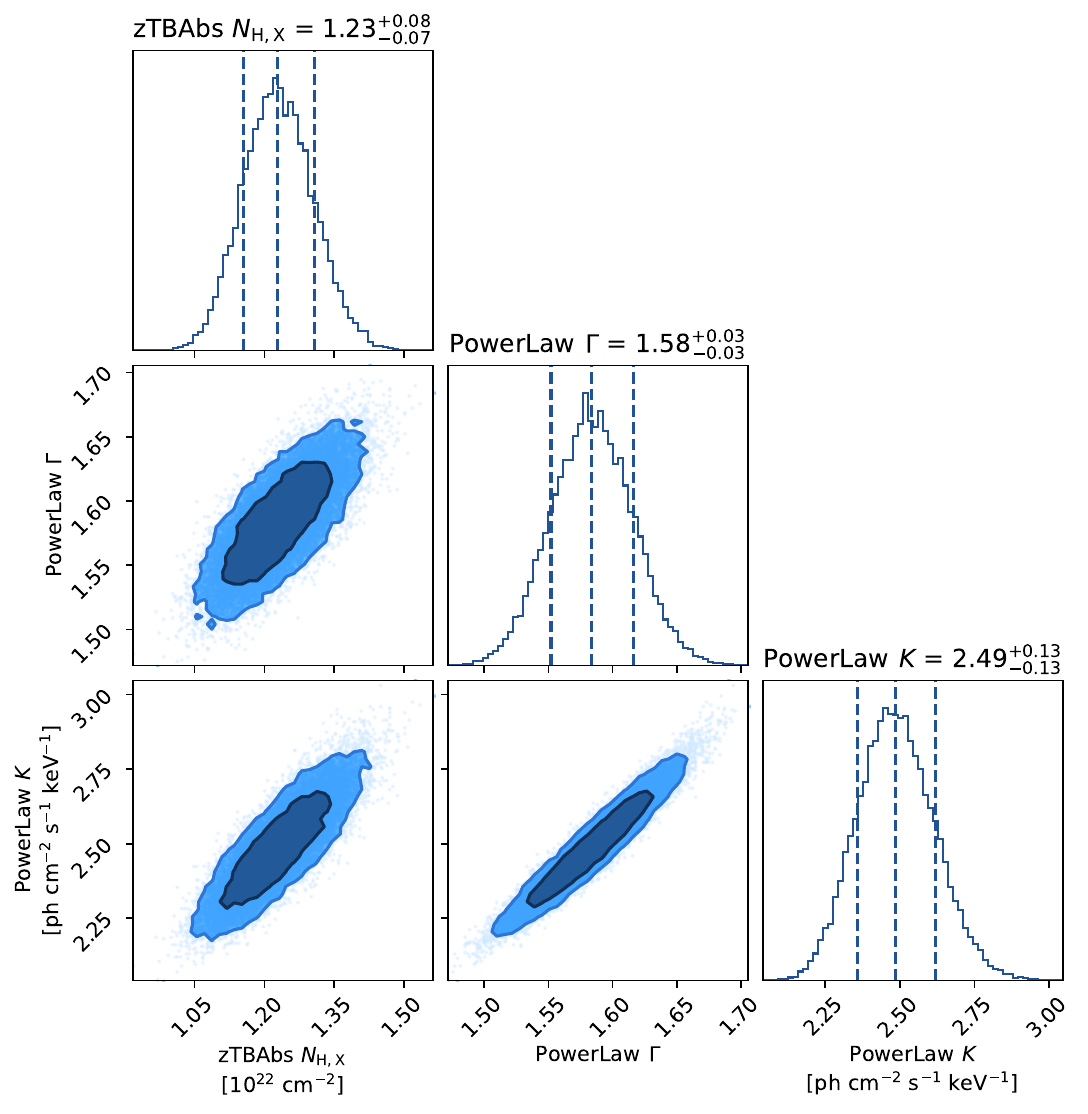}
        \caption{Posterior distributions of model parameters.}
    \end{subfigure}
    \vspace{1em}
    \begin{subfigure}[t]{0.49\textwidth}
        \centering
        \includegraphics[width=\linewidth]{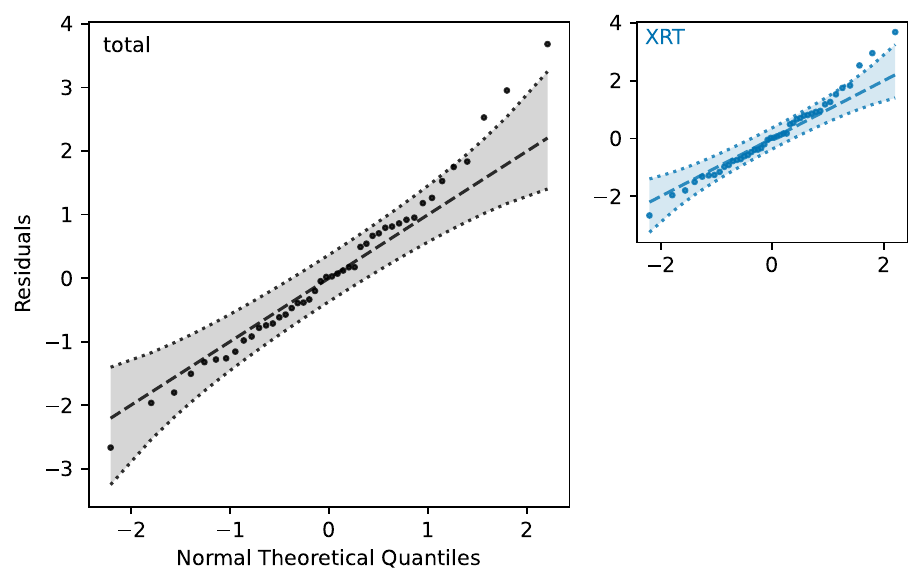}
        \caption{Q-Q plots of residuals.}
    \end{subfigure}
    \hfill
    \begin{subfigure}[t]{0.49\textwidth}
        \centering
        \includegraphics[width=\linewidth]{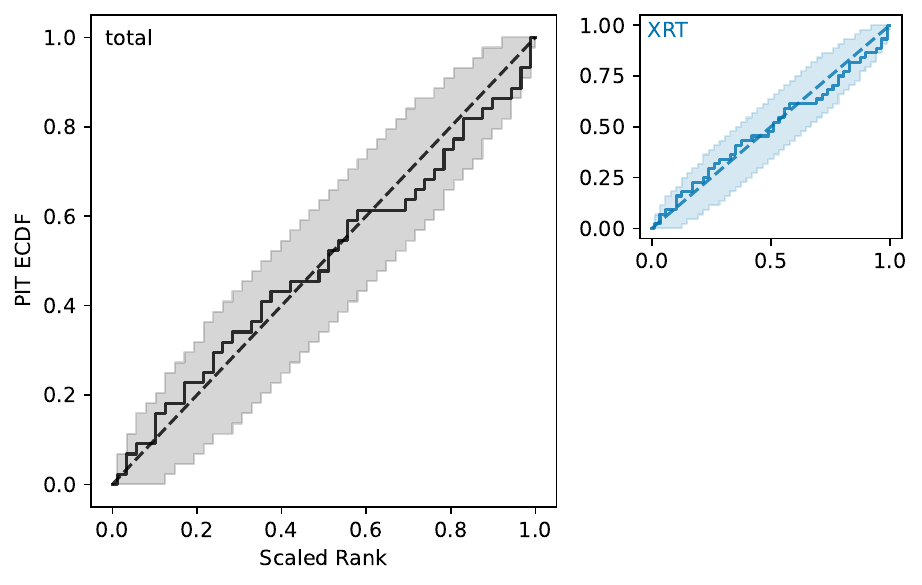}
        \caption{PIT-ECDF diagnostic.}
    \end{subfigure}
    \caption{XRT spectrum extracted from 265-340 s post-trigger, fitted with an absorbed PL model and shown with posterior diagnostics. This figure follows the format of Figure~\ref{fig:PowerLawSpec1}.}
    \label{fig:PowerLawSpec6}
\end{figure*}

\begin{figure*}[ht!]
    \centering
    \begin{subfigure}[t]{0.49\textwidth}
        \centering
        \includegraphics[width=\linewidth]{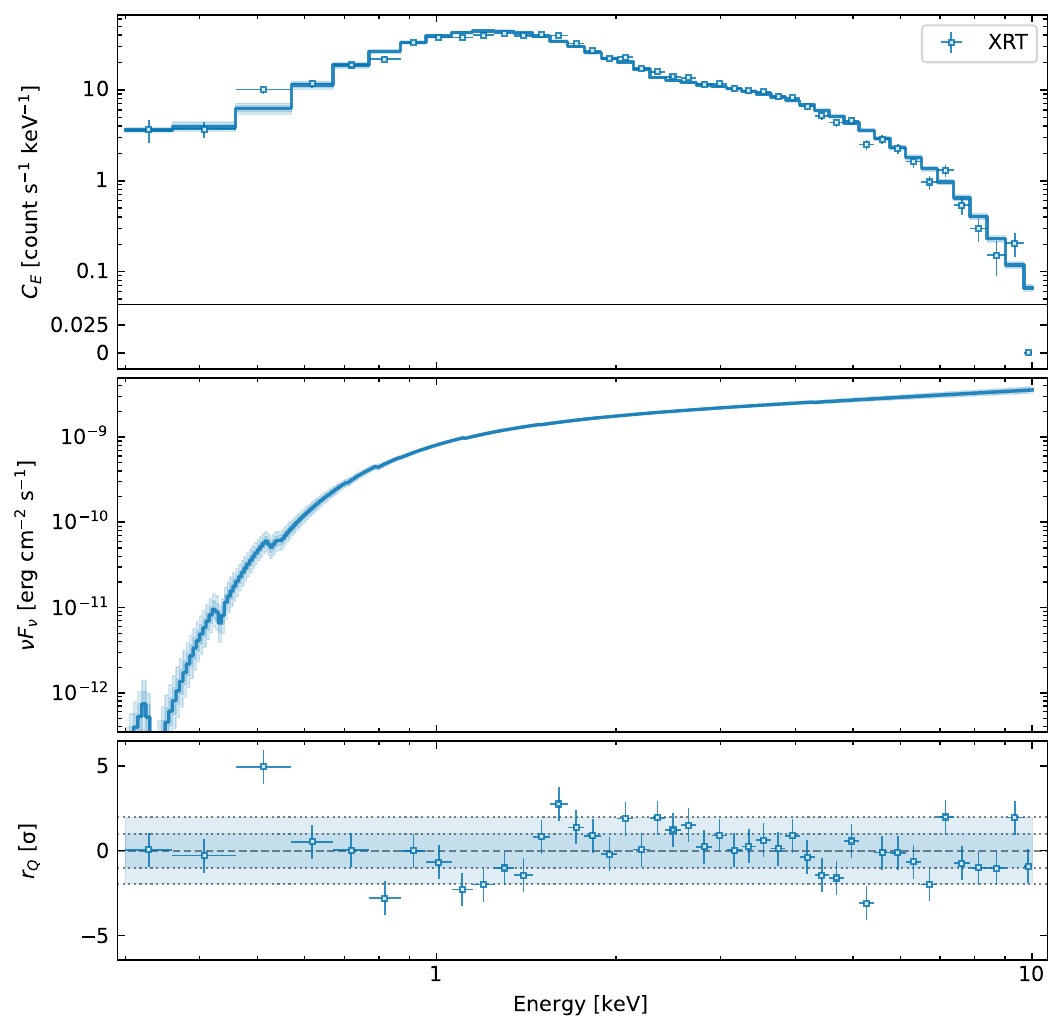}
        \caption{XRT spectrum, PL model, and residuals.}
    \end{subfigure}
    \hfill
    \begin{subfigure}[t]{0.49\textwidth}
        \centering
        \includegraphics[width=\linewidth]{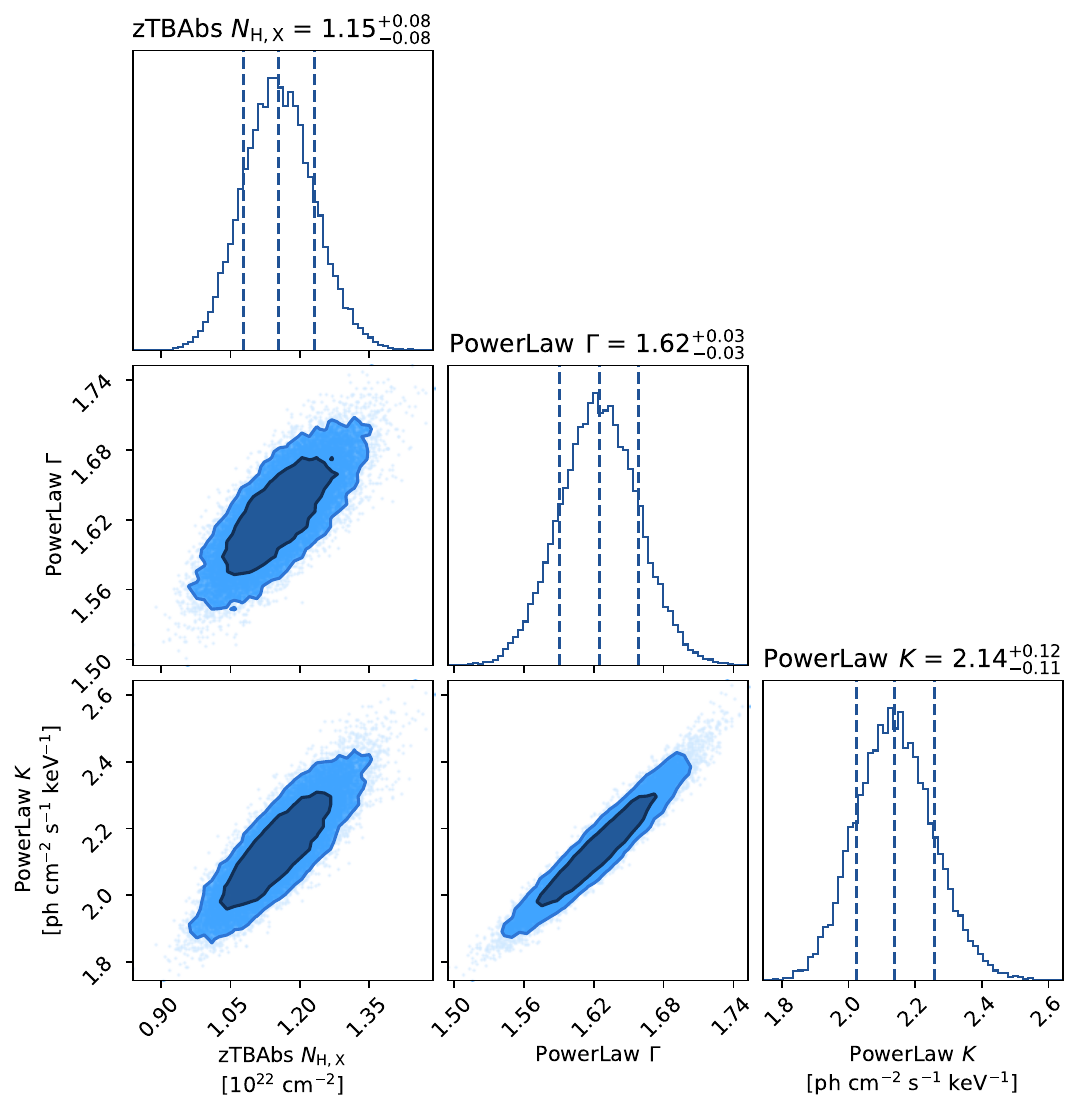}
        \caption{Posterior distributions of model parameters.}
    \end{subfigure}
    \vspace{1em}
    \begin{subfigure}[t]{0.49\textwidth}
        \centering
        \includegraphics[width=\linewidth]{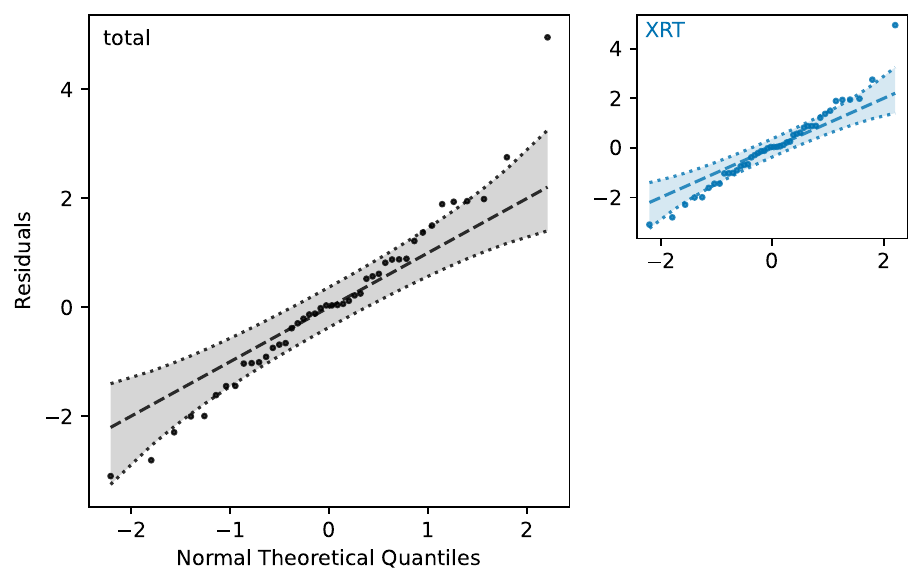}
        \caption{Q-Q plots of residuals.}
    \end{subfigure}
    \hfill
    \begin{subfigure}[t]{0.49\textwidth}
        \centering
        \includegraphics[width=\linewidth]{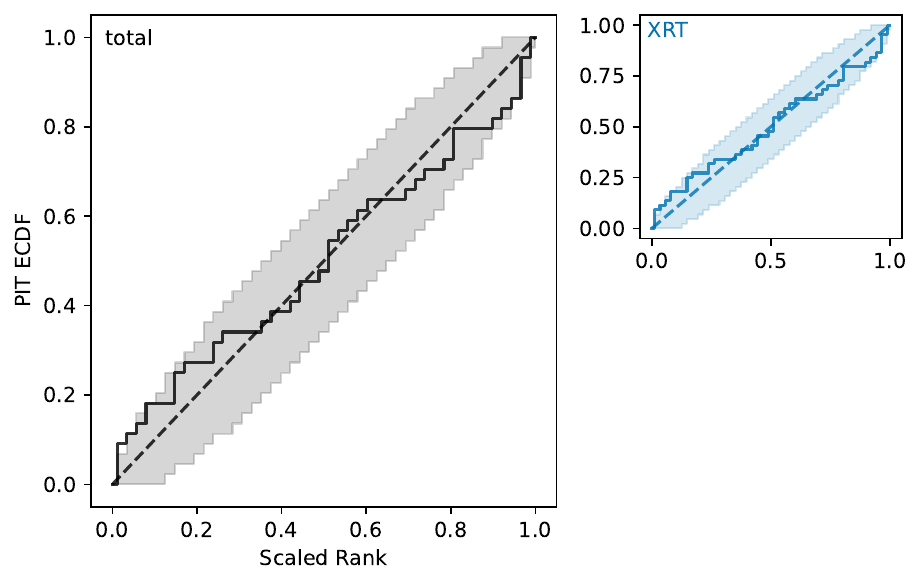}
        \caption{PIT-ECDF diagnostic.}
    \end{subfigure}
    \caption{XRT spectrum extracted from 340-421 s post-trigger, fitted with an absorbed PL model and shown with posterior diagnostics. This figure follows the format of Figure~\ref{fig:PowerLawSpec1}.}
    \label{fig:PowerLawSpec7}
\end{figure*}

\begin{figure*}[ht!]
    \centering
    \begin{subfigure}[t]{0.49\textwidth}
        \centering
        \includegraphics[width=\linewidth]{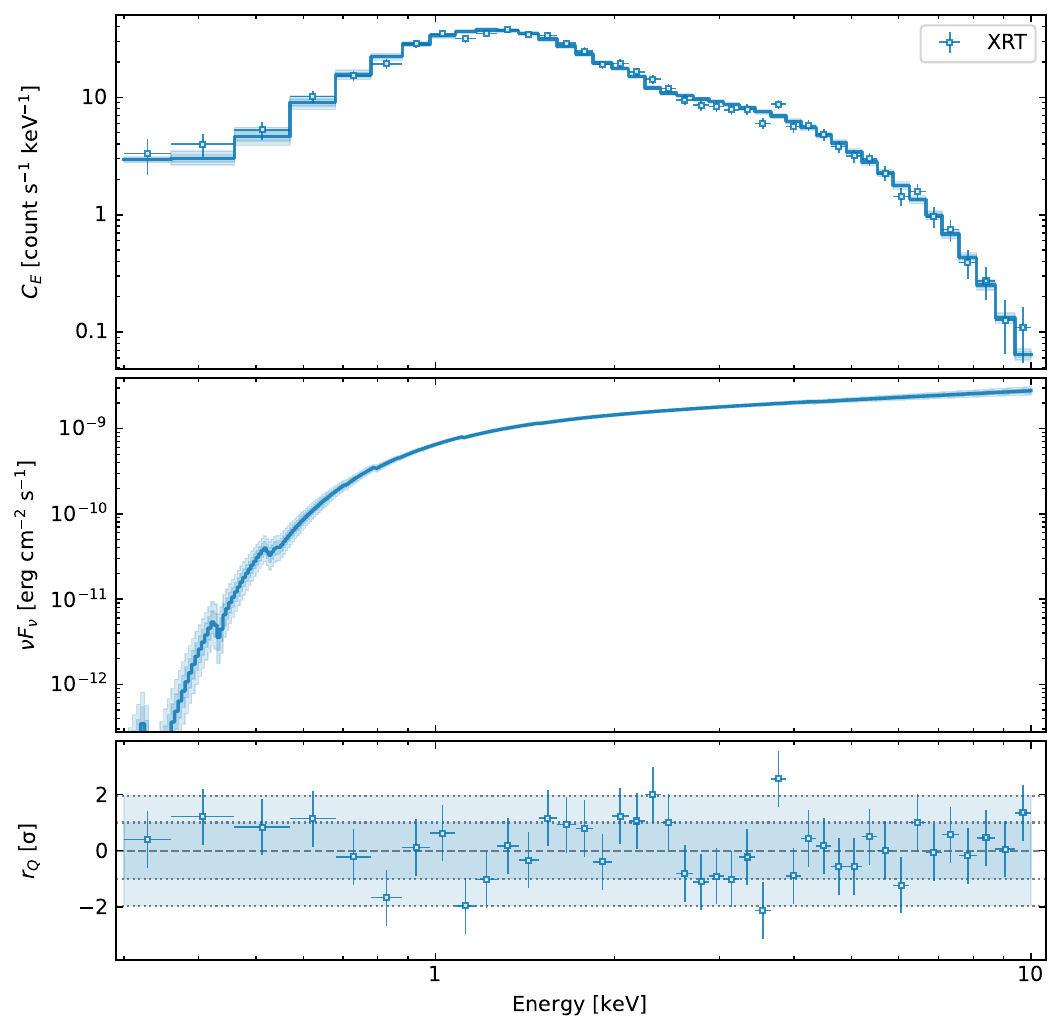}
        \caption{XRT spectrum, PL model, and residuals.}
    \end{subfigure}
    \hfill
    \begin{subfigure}[t]{0.49\textwidth}
        \centering
        \includegraphics[width=\linewidth]{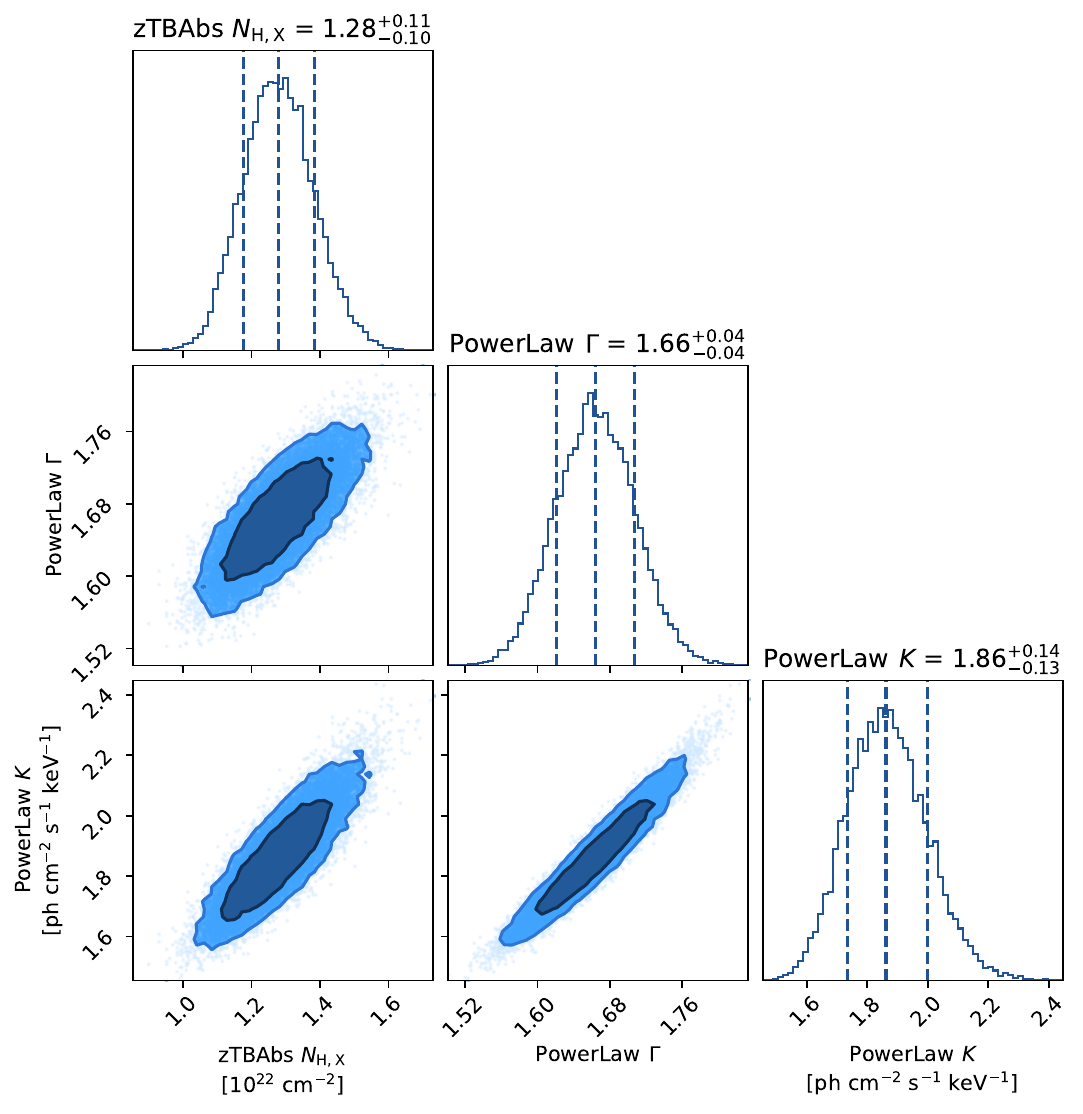}
        \caption{Posterior distributions of model parameters.}
    \end{subfigure}
    \vspace{1em}
    \begin{subfigure}[t]{0.49\textwidth}
        \centering
        \includegraphics[width=\linewidth]{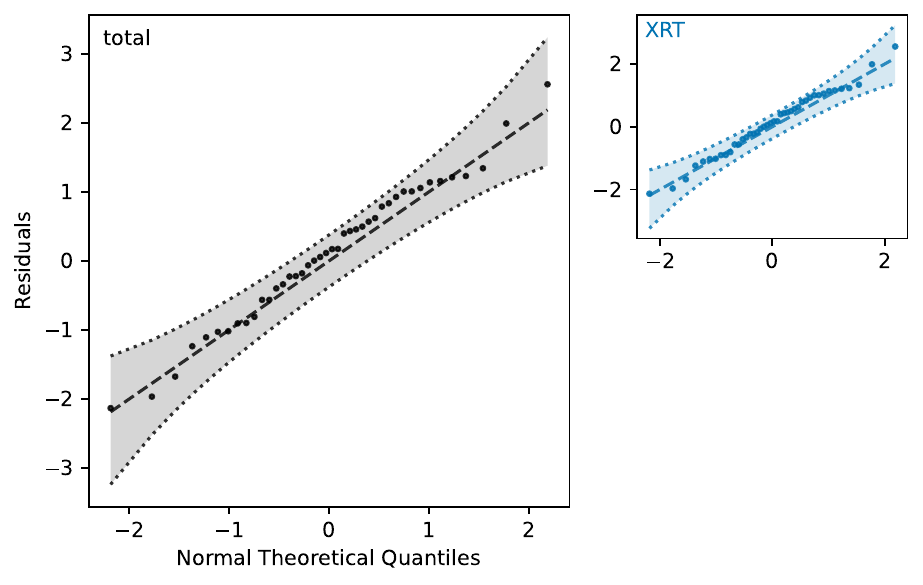}
        \caption{Q-Q plots of residuals.}
    \end{subfigure}
    \hfill
    \begin{subfigure}[t]{0.49\textwidth}
        \centering
        \includegraphics[width=\linewidth]{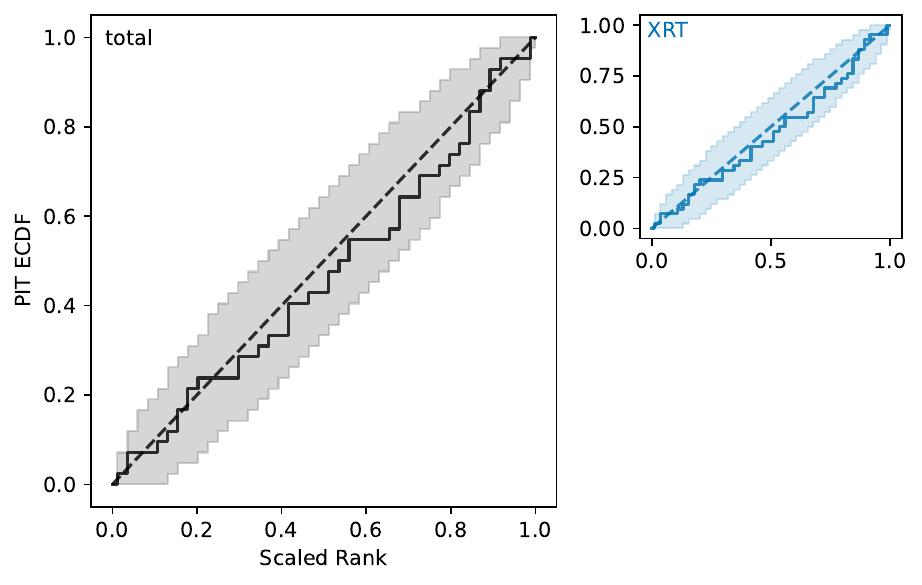}
        \caption{PIT-ECDF diagnostic.}
    \end{subfigure}
    \caption{XRT spectrum extracted from 421-482 s post-trigger, fitted with an absorbed PL model and shown with posterior diagnostics. This figure follows the format of Figure~\ref{fig:PowerLawSpec1}.}
    \label{fig:PowerLawSpec8}
\end{figure*}

\begin{figure*}[ht!]
    \centering
    \begin{subfigure}[t]{0.49\textwidth}
        \centering
        \includegraphics[width=\linewidth]{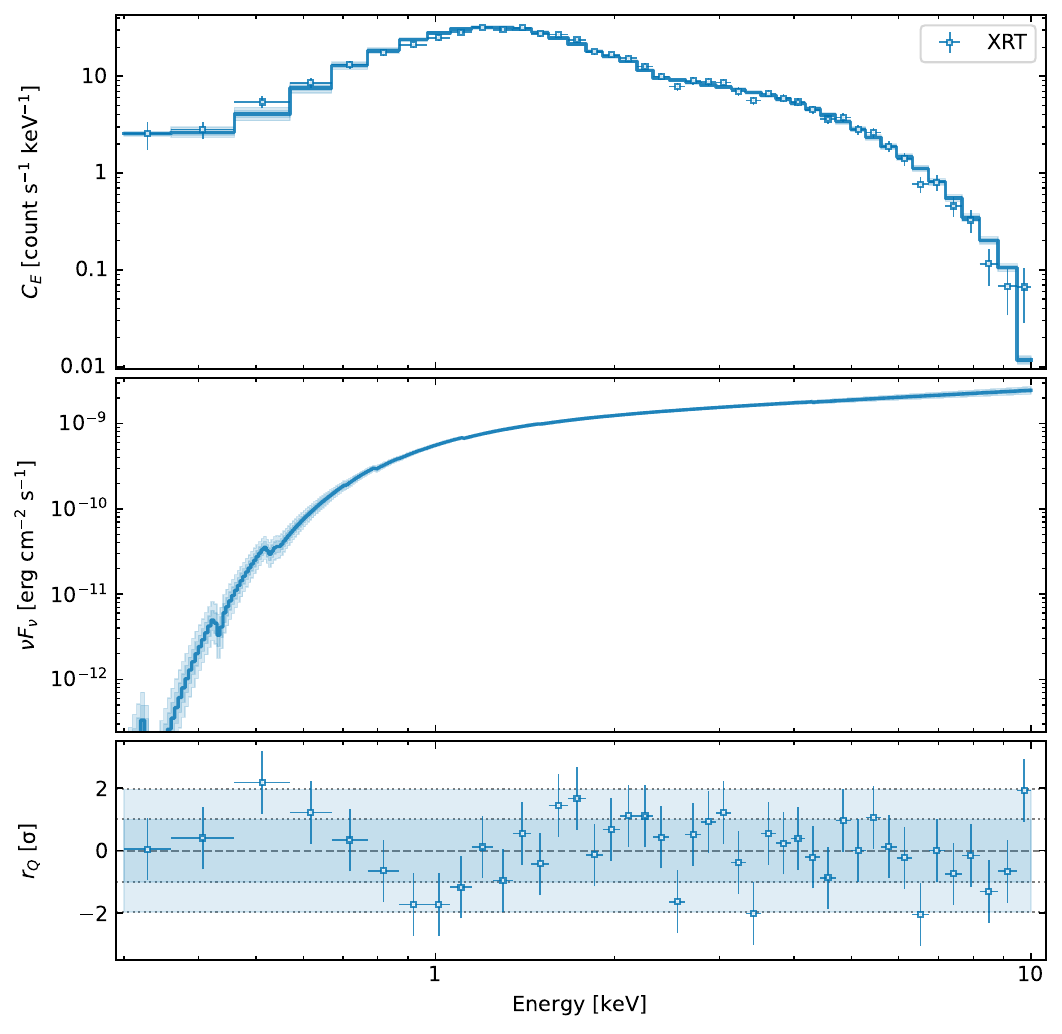}
        \caption{XRT spectrum, PL model, and residuals.}
    \end{subfigure}
    \hfill
    \begin{subfigure}[t]{0.49\textwidth}
        \centering
        \includegraphics[width=\linewidth]{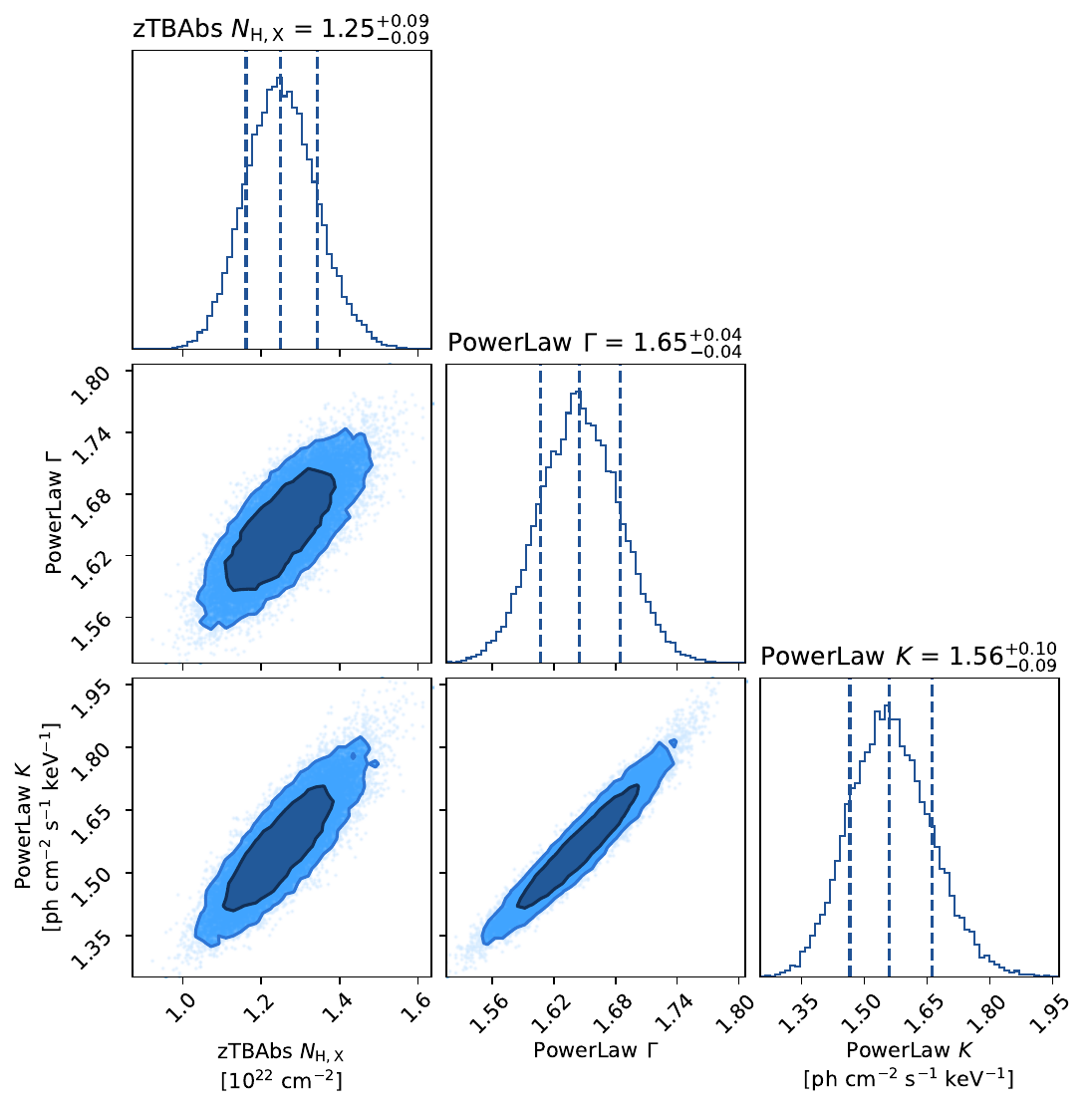}
        \caption{Posterior distributions of model parameters.}
    \end{subfigure}
    \vspace{1em}
    \begin{subfigure}[t]{0.49\textwidth}
        \centering
        \includegraphics[width=\linewidth]{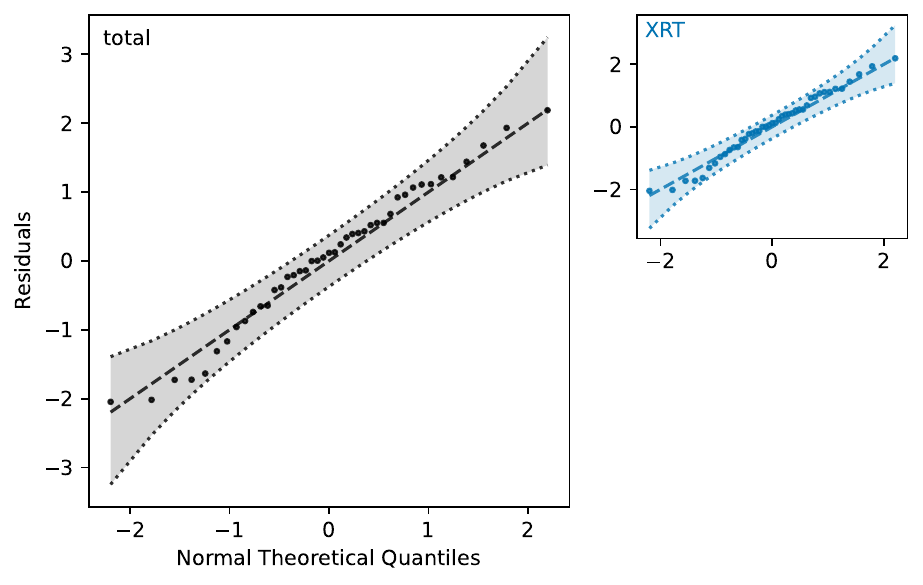}
        \caption{Q-Q plots of residuals.}
    \end{subfigure}
    \hfill
    \begin{subfigure}[t]{0.49\textwidth}
        \centering
        \includegraphics[width=\linewidth]{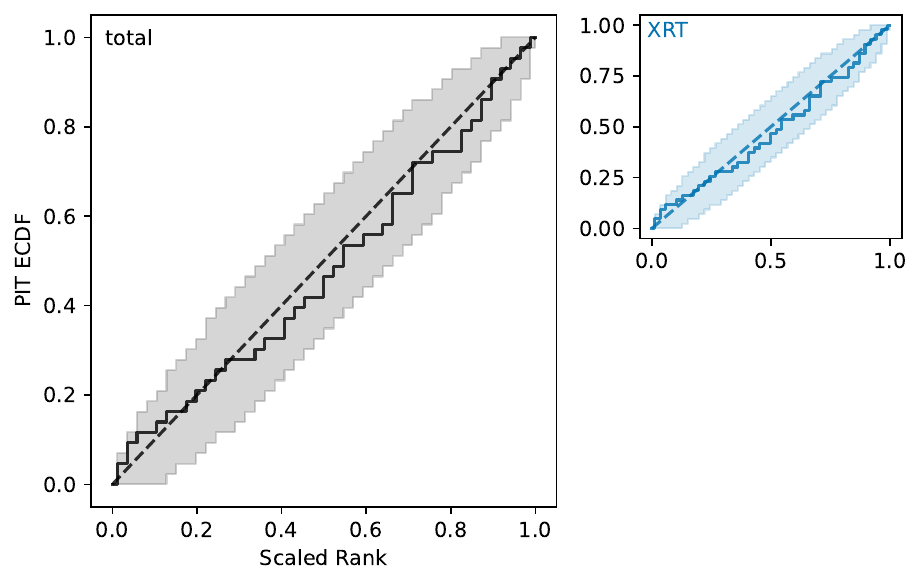}
        \caption{PIT-ECDF diagnostic.}
    \end{subfigure}
    \caption{XRT spectrum extracted from 482-569 s post-trigger, fitted with an absorbed PL model and shown with posterior diagnostics. This figure follows the format of Figure~\ref{fig:PowerLawSpec1}.}
    \label{fig:PowerLawSpec9}
\end{figure*}

\begin{figure*}[ht!]
    \centering
    \begin{subfigure}[t]{0.49\textwidth}
        \centering
        \includegraphics[width=\linewidth]{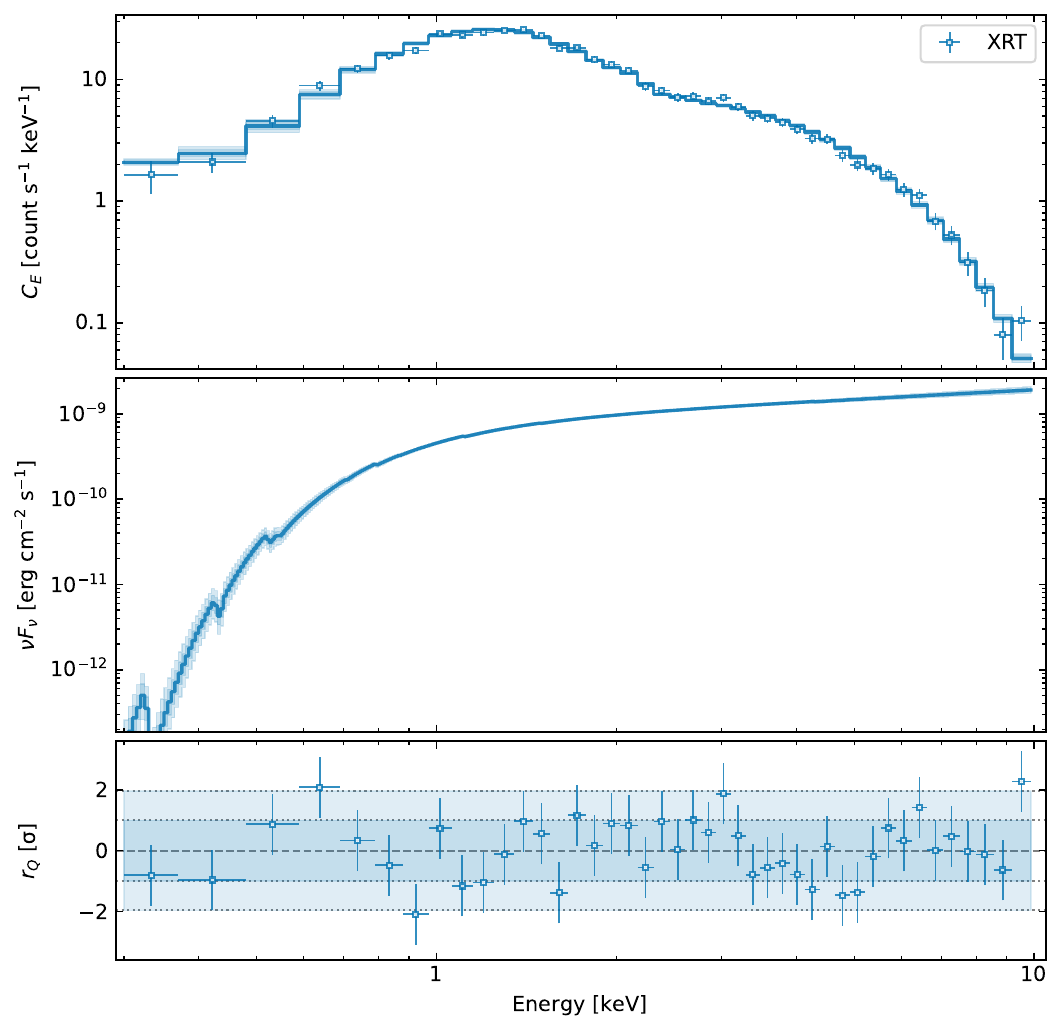}
        \caption{XRT spectrum, PL model, and residuals.}
    \end{subfigure}
    \hfill
    \begin{subfigure}[t]{0.49\textwidth}
        \centering
        \includegraphics[width=\linewidth]{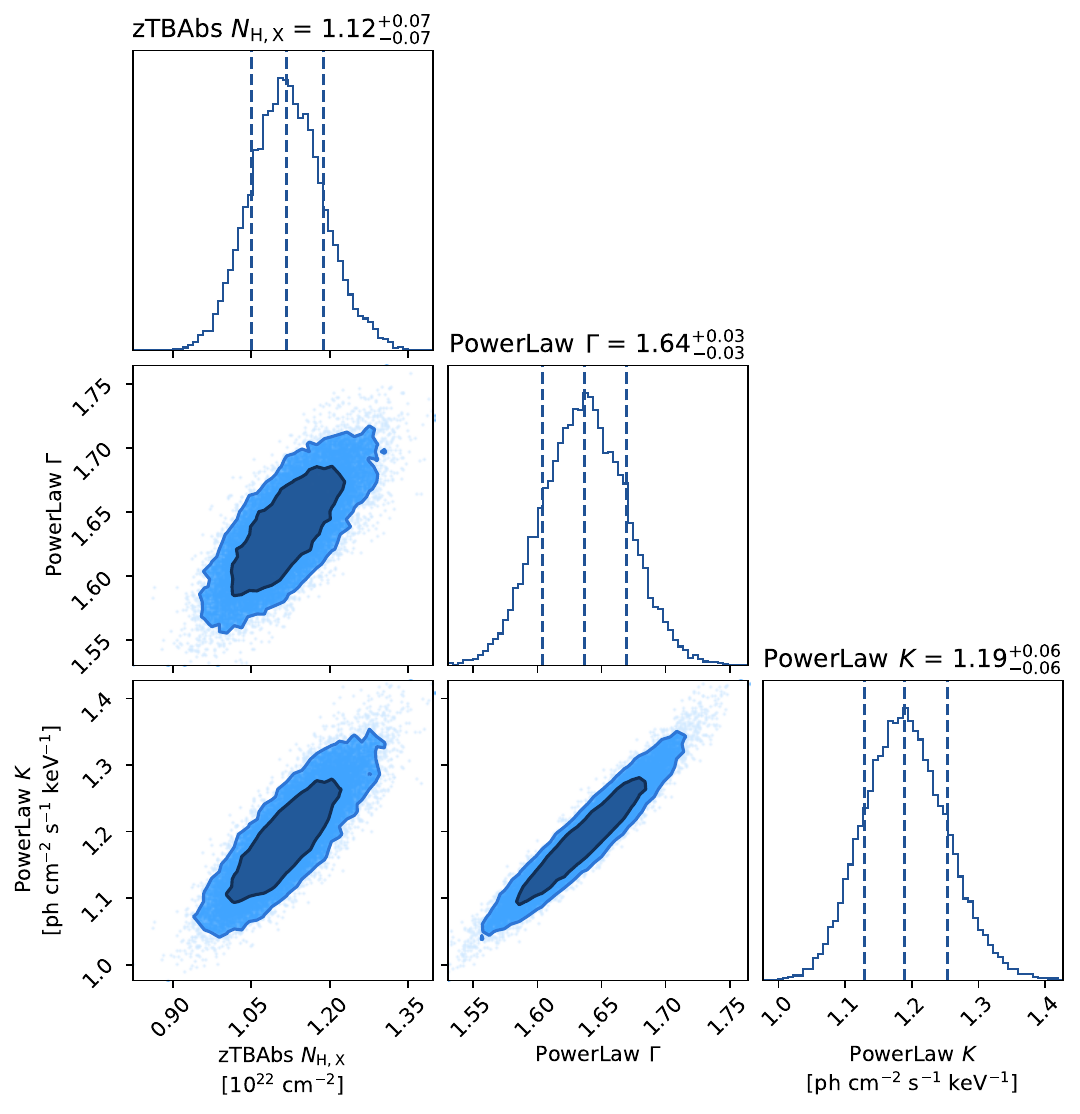}
        \caption{Posterior distributions of model parameters.}
    \end{subfigure}
    \vspace{1em}
    \begin{subfigure}[t]{0.49\textwidth}
        \centering
        \includegraphics[width=\linewidth]{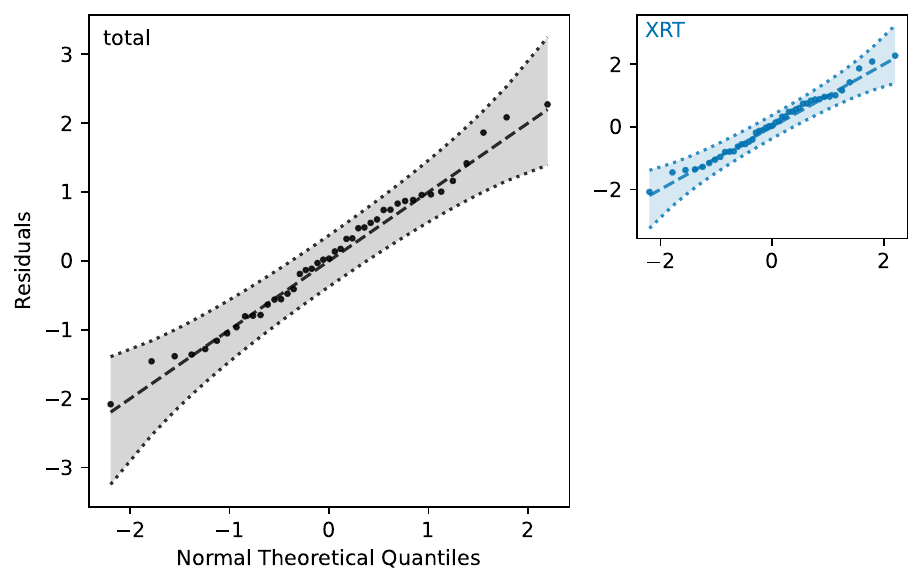}
        \caption{Q-Q plots of residuals.}
    \end{subfigure}
    \hfill
    \begin{subfigure}[t]{0.49\textwidth}
        \centering
        \includegraphics[width=\linewidth]{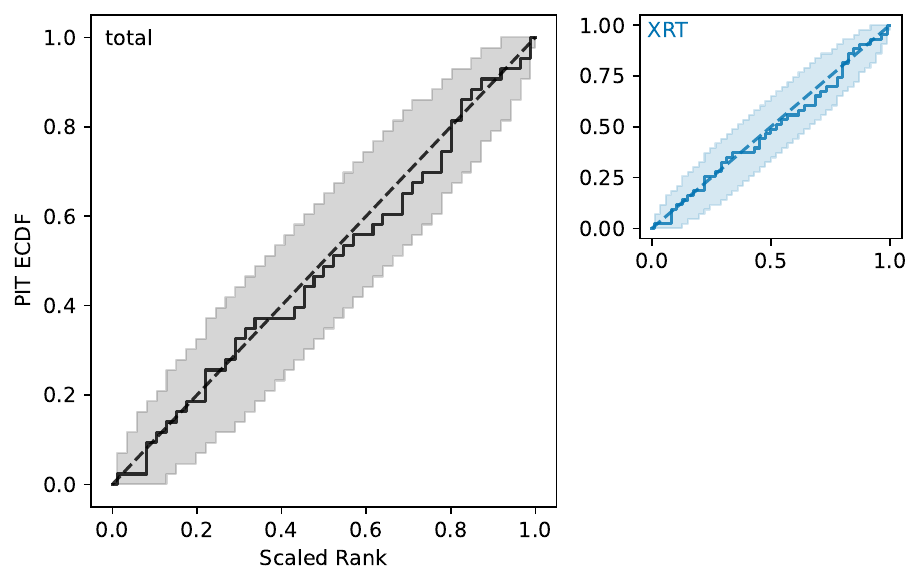}
        \caption{PIT-ECDF diagnostic.}
    \end{subfigure}
    \caption{XRT spectrum extracted from 569-709 s post-trigger, fitted with an absorbed PL model and shown with posterior diagnostics. This figure follows the format of Figure~\ref{fig:PowerLawSpec1}.}
    \label{fig:PowerLawSpec10}
\end{figure*}

\begin{figure*}[ht!]
    \centering
    \begin{subfigure}[t]{0.49\textwidth}
        \centering
        \includegraphics[width=\linewidth]{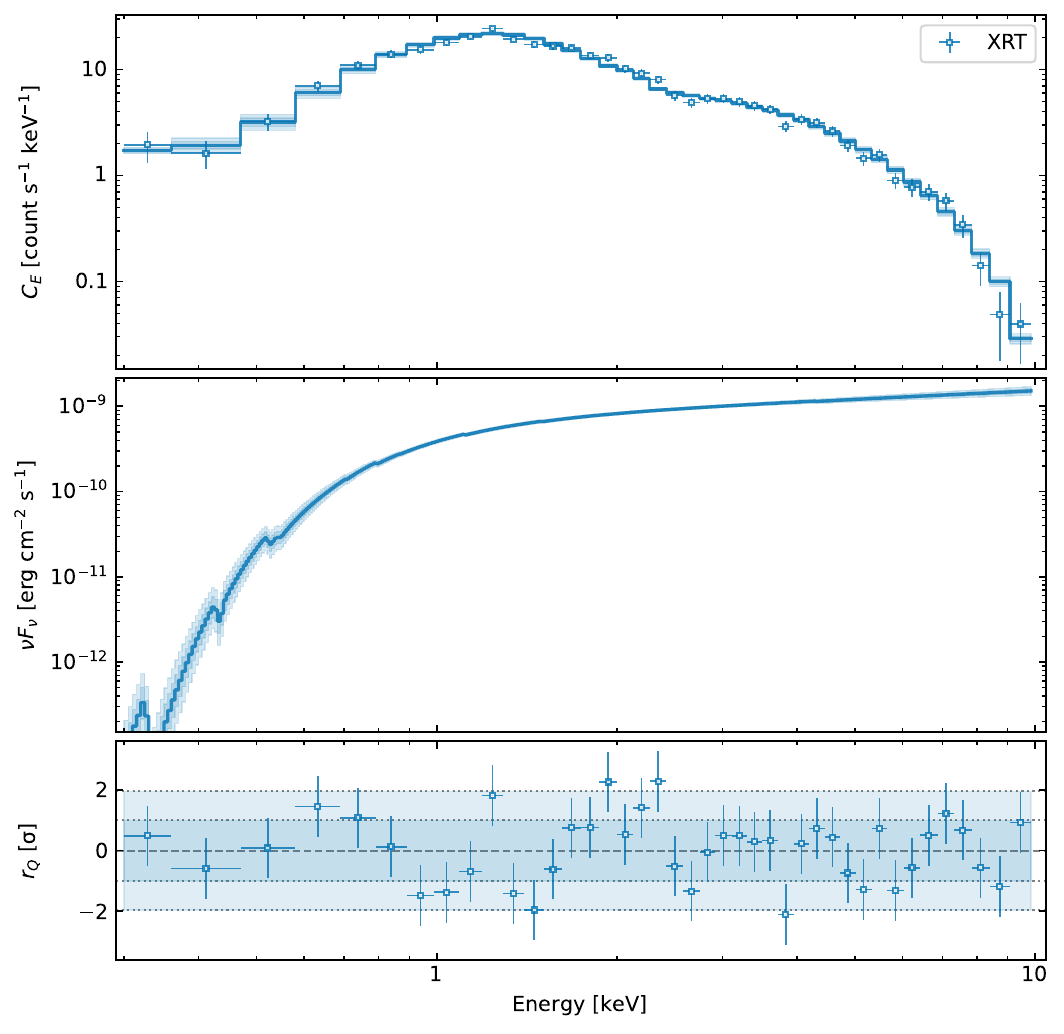}
        \caption{XRT spectrum, PL model, and residuals.}
    \end{subfigure}
    \hfill
    \begin{subfigure}[t]{0.49\textwidth}
        \centering
        \includegraphics[width=\linewidth]{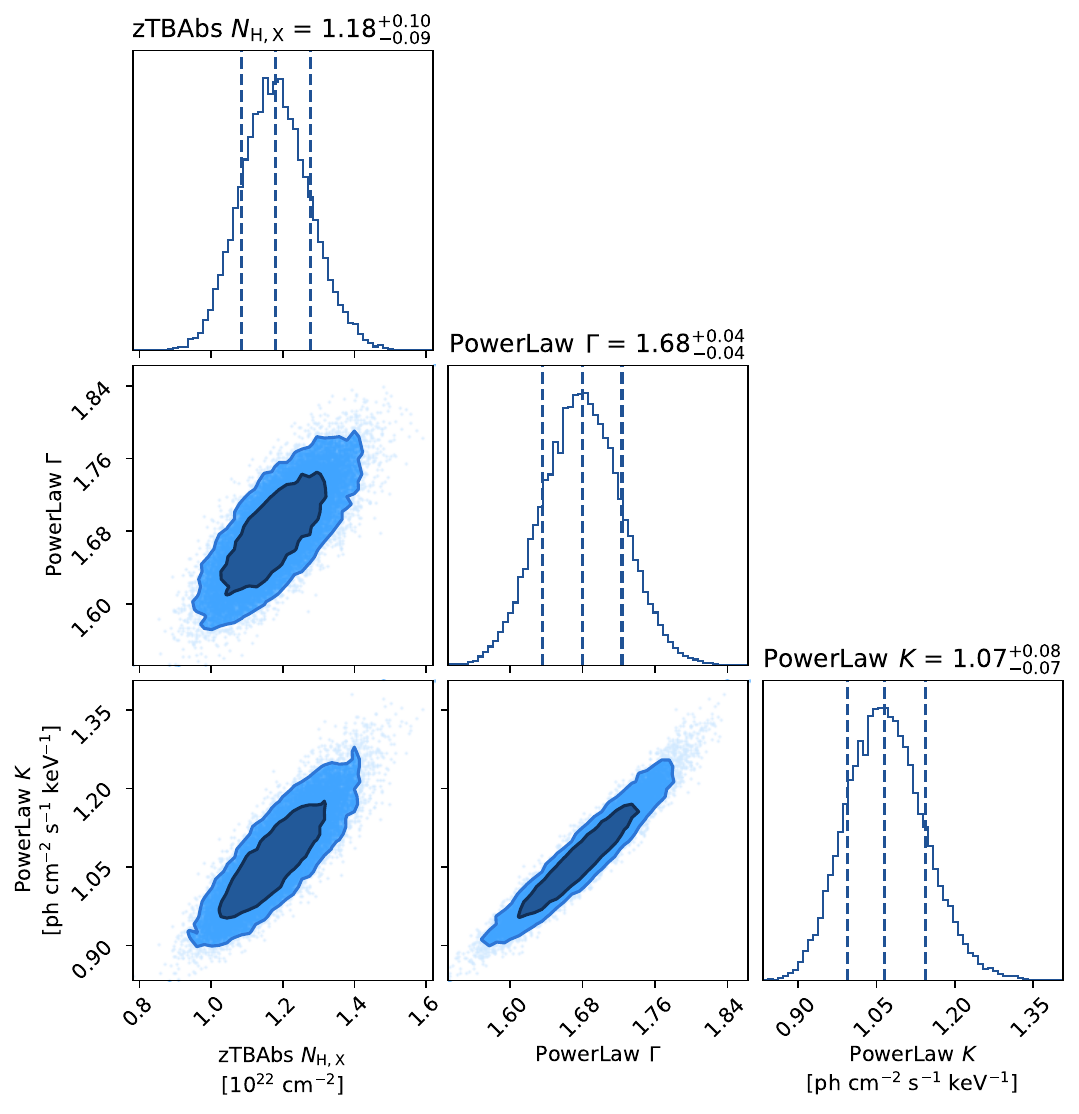}
        \caption{Posterior distributions of model parameters.}
    \end{subfigure}
    \vspace{1em}
    \begin{subfigure}[t]{0.49\textwidth}
        \centering
        \includegraphics[width=\linewidth]{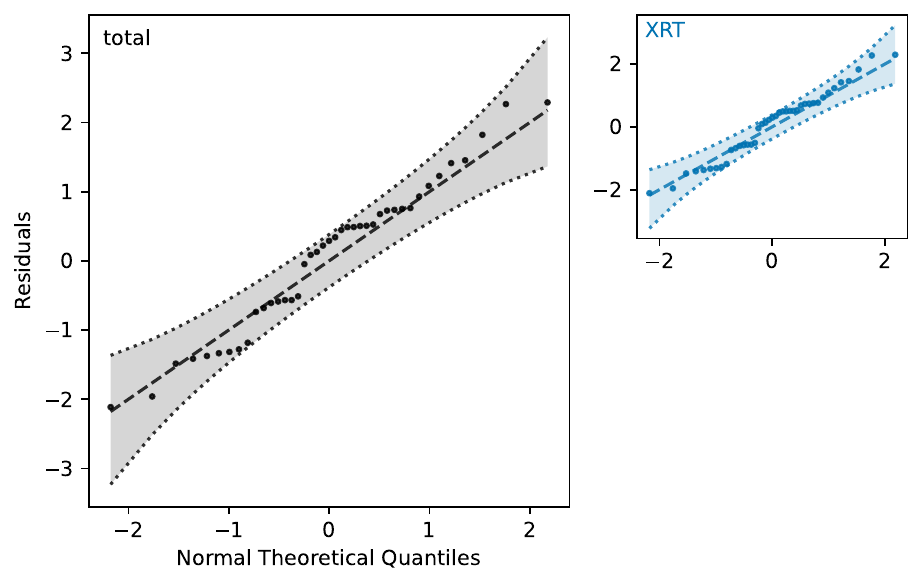}
        \caption{Q-Q plots of residuals.}
    \end{subfigure}
    \hfill
    \begin{subfigure}[t]{0.49\textwidth}
        \centering
        \includegraphics[width=\linewidth]{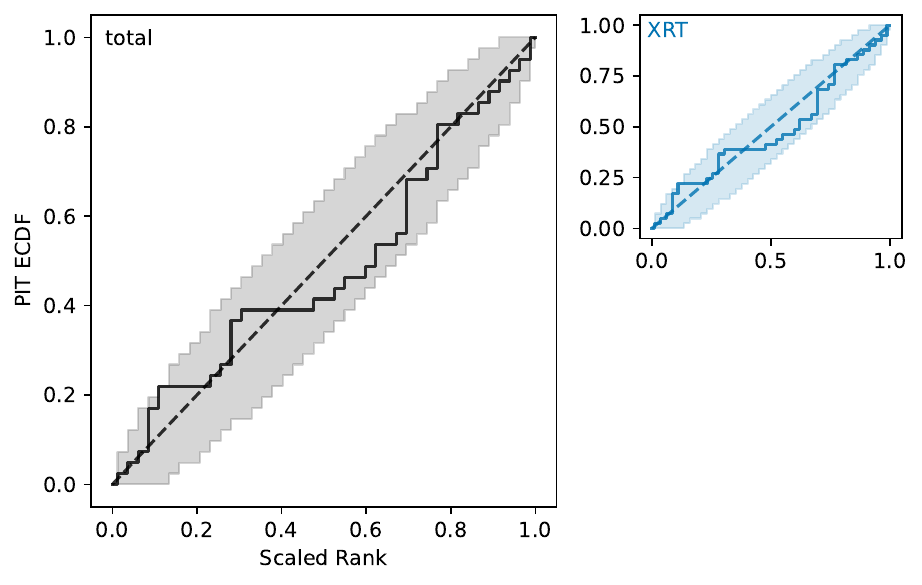}
        \caption{PIT-ECDF diagnostic.}
    \end{subfigure}
    \caption{XRT spectrum extracted from 709-808 s post-trigger, fitted with an absorbed PL model and shown with posterior diagnostics. This figure follows the format of Figure~\ref{fig:PowerLawSpec1}.}
    \label{fig:PowerLawSpec11}
\end{figure*}

\begin{figure*}[ht!]
    \centering
    \begin{subfigure}[t]{0.49\textwidth}
        \centering
        \includegraphics[width=\linewidth]{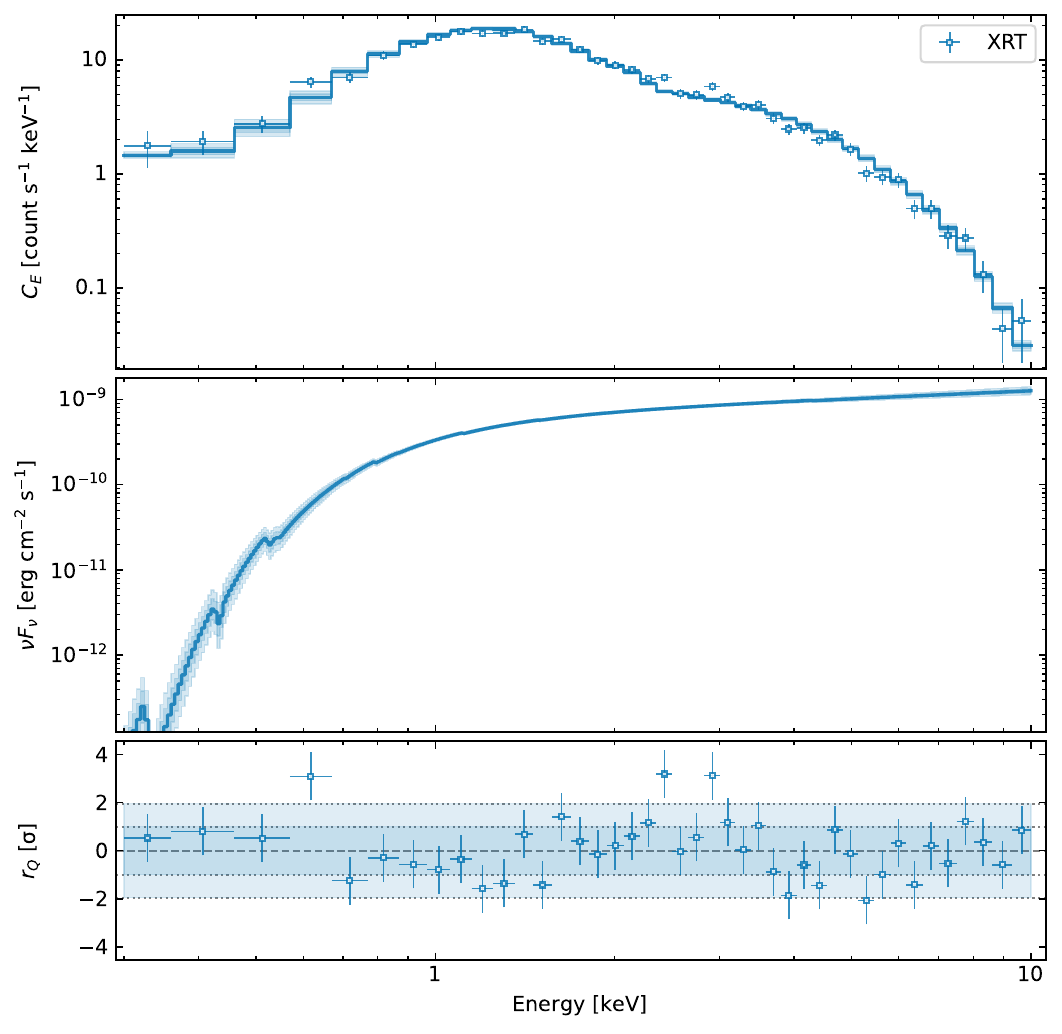}
        \caption{XRT spectrum, PL model, and residuals.}
    \end{subfigure}
    \hfill
    \begin{subfigure}[t]{0.49\textwidth}
        \centering
        \includegraphics[width=\linewidth]{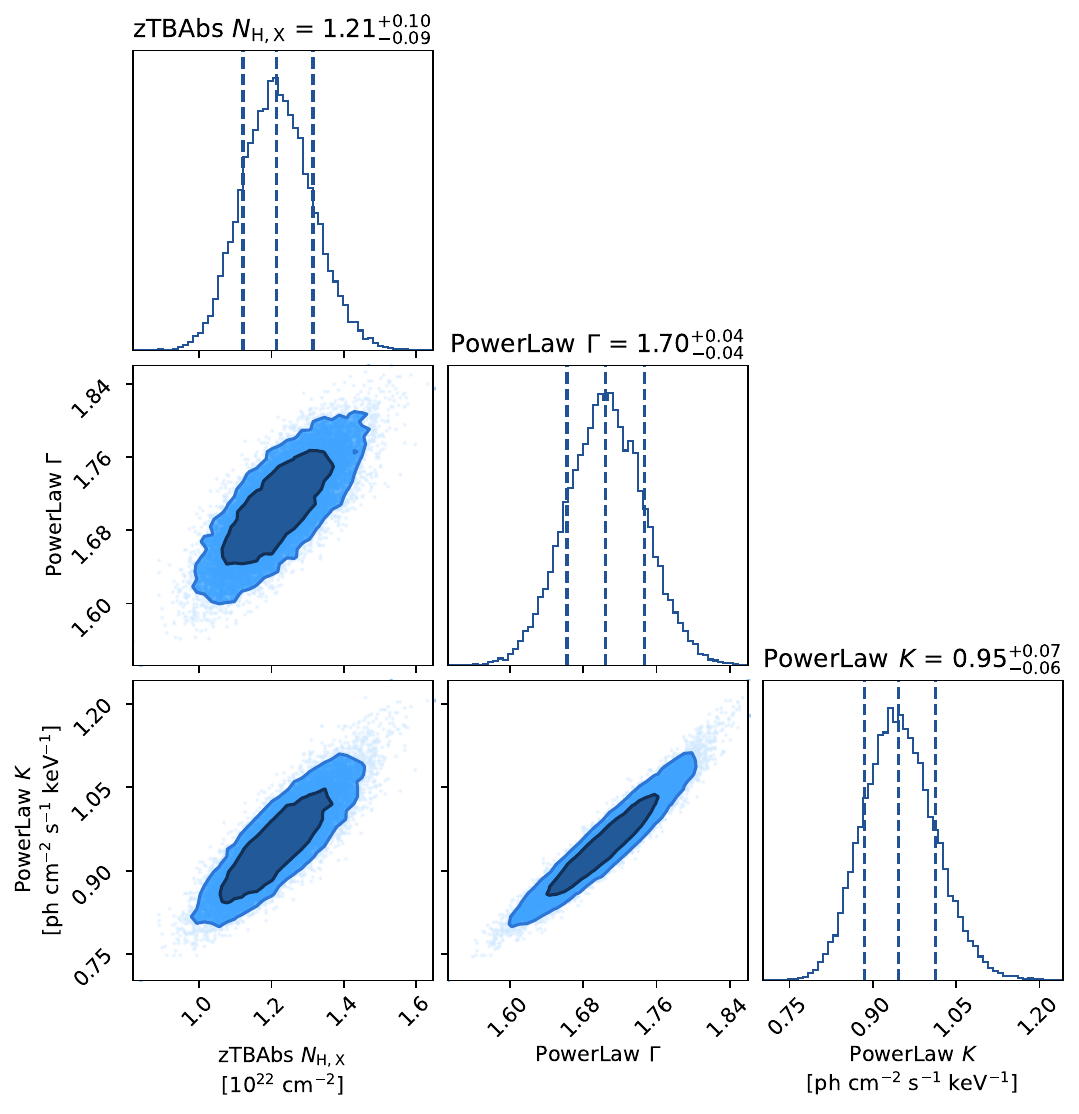}
        \caption{Posterior distributions of model parameters.}
    \end{subfigure}
    \vspace{1em}
    \begin{subfigure}[t]{0.49\textwidth}
        \centering
        \includegraphics[width=\linewidth]{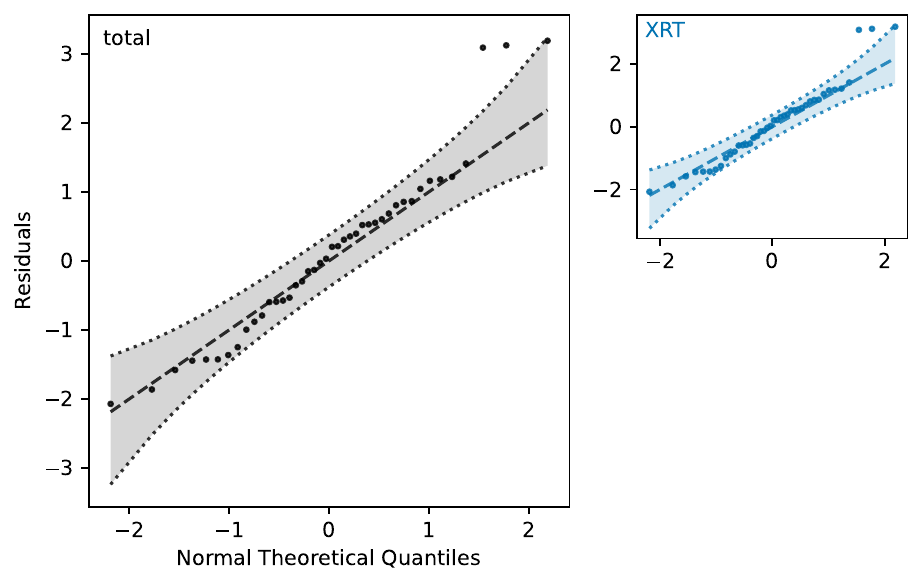}
        \caption{Q-Q plots of residuals.}
    \end{subfigure}
    \hfill
    \begin{subfigure}[t]{0.49\textwidth}
        \centering
        \includegraphics[width=\linewidth]{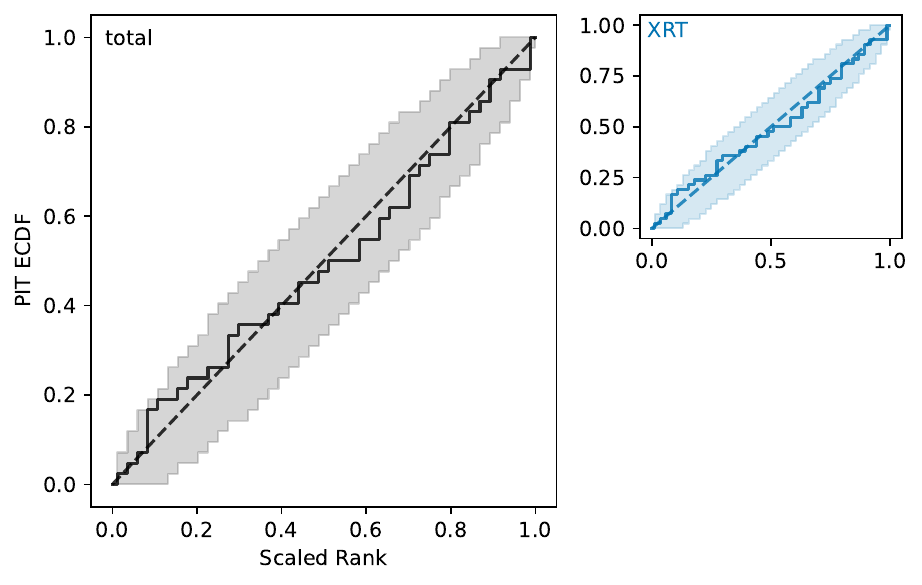}
        \caption{PIT-ECDF diagnostic.}
    \end{subfigure}
    \caption{XRT spectrum extracted from 808-940 s post-trigger, fitted with an absorbed PL model and shown with posterior diagnostics. This figure follows the format of Figure~\ref{fig:PowerLawSpec1}.}
    \label{fig:PowerLawSpec12}
\end{figure*}

\begin{figure*}[ht!]
    \centering
    \begin{subfigure}[t]{0.49\textwidth}
        \centering
        \includegraphics[width=\linewidth]{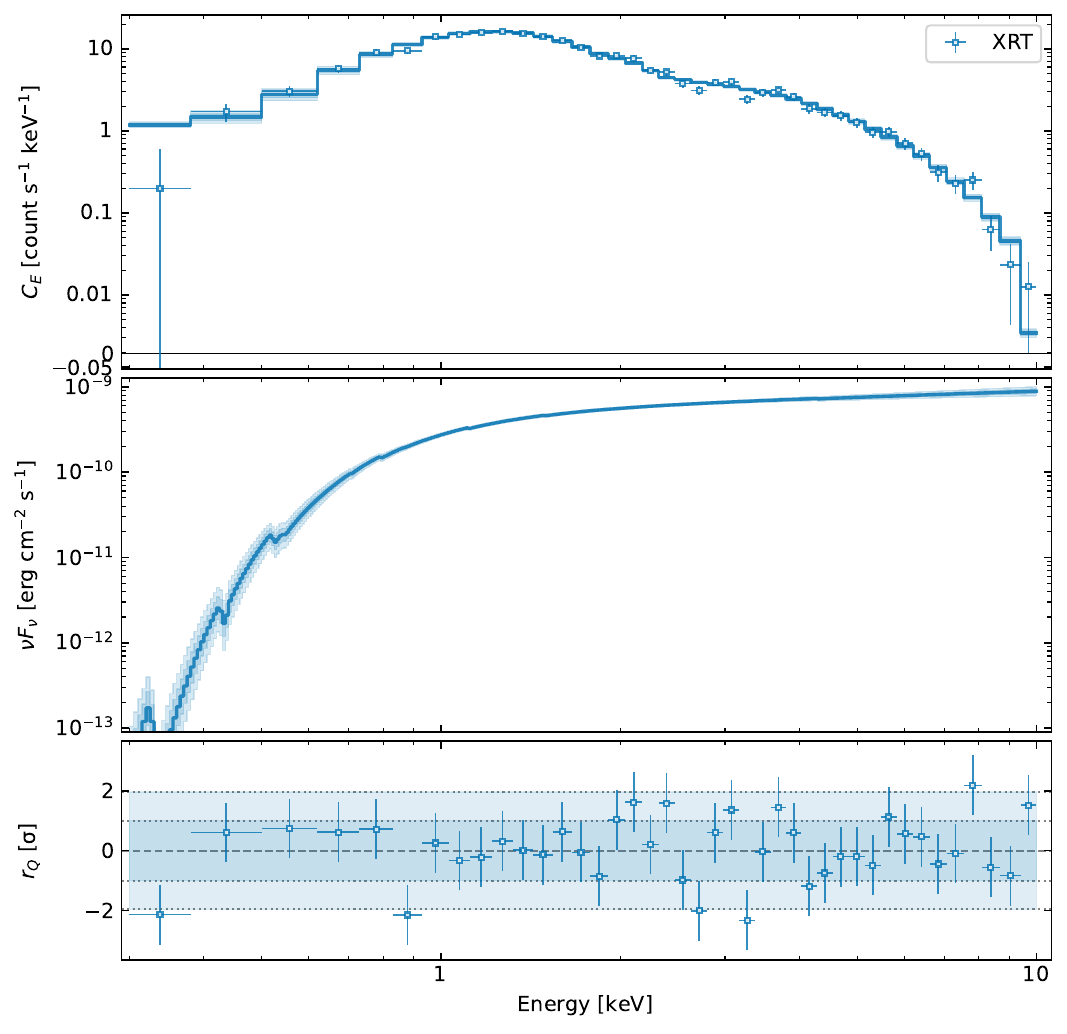}
        \caption{XRT spectrum, PL model, and residuals.}
    \end{subfigure}
    \hfill
    \begin{subfigure}[t]{0.49\textwidth}
        \centering
        \includegraphics[width=\linewidth]{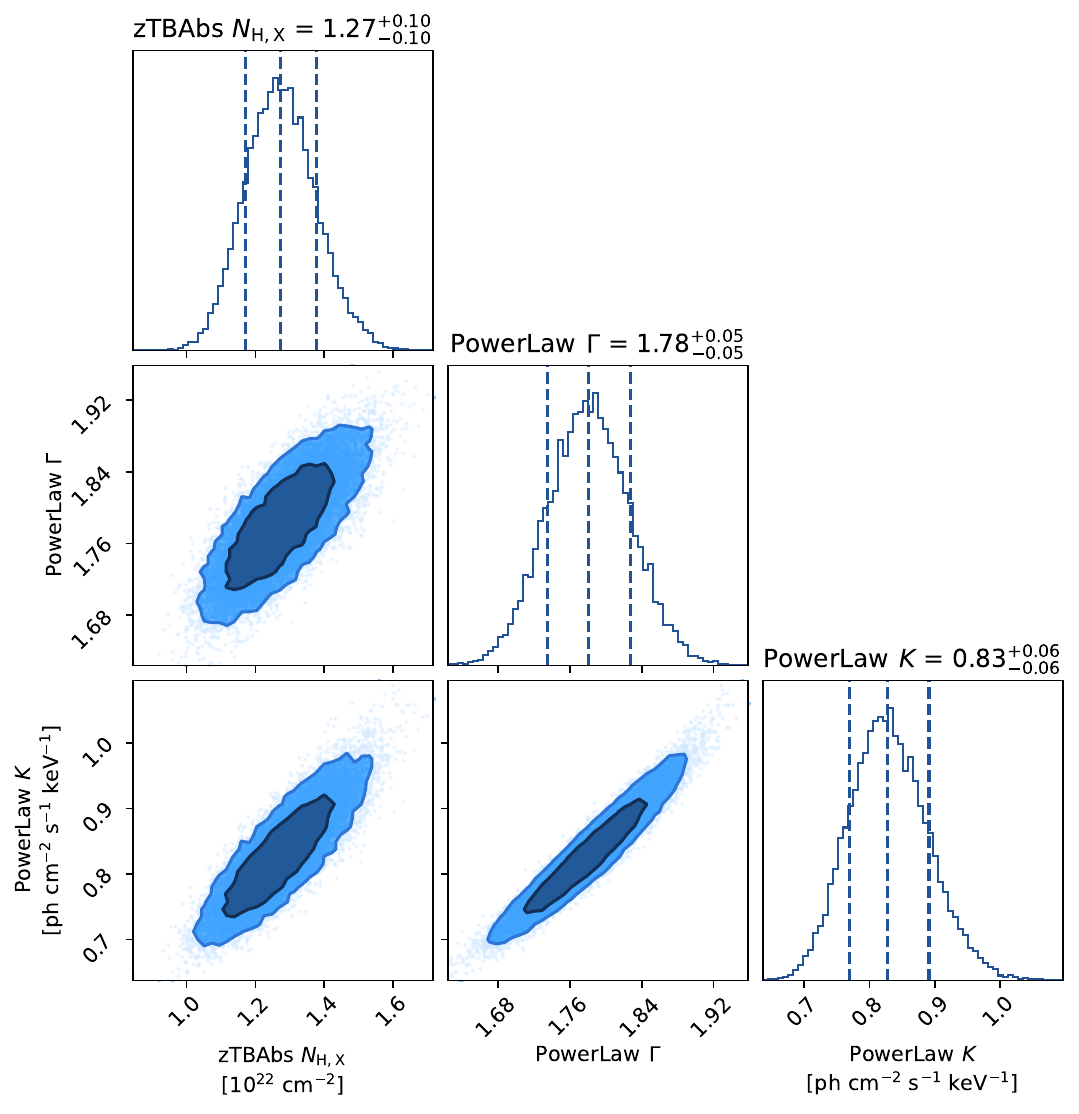}
        \caption{Posterior distributions of model parameters.}
    \end{subfigure}
    \vspace{1em}
    \begin{subfigure}[t]{0.49\textwidth}
        \centering
        \includegraphics[width=\linewidth]{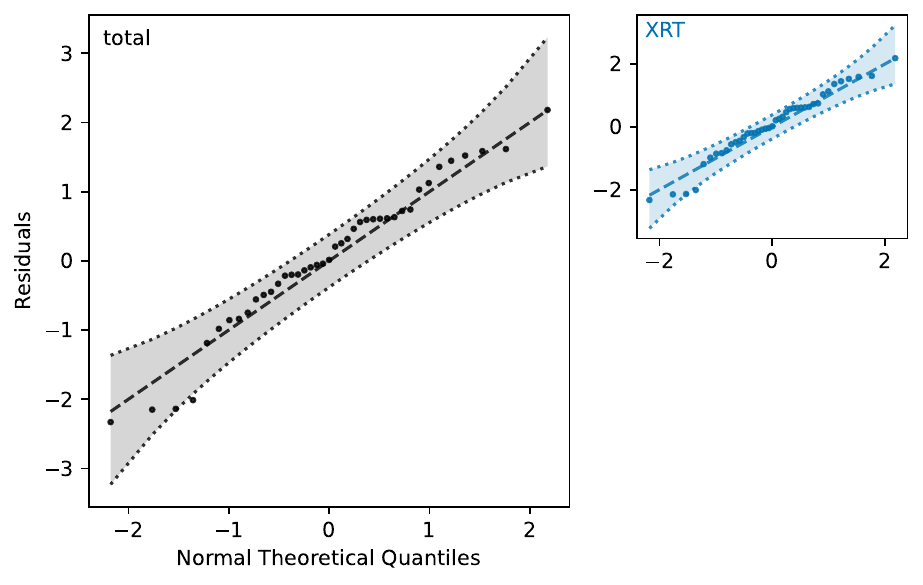}
        \caption{Q-Q plots of residuals.}
    \end{subfigure}
    \hfill
    \begin{subfigure}[t]{0.49\textwidth}
        \centering
        \includegraphics[width=\linewidth]{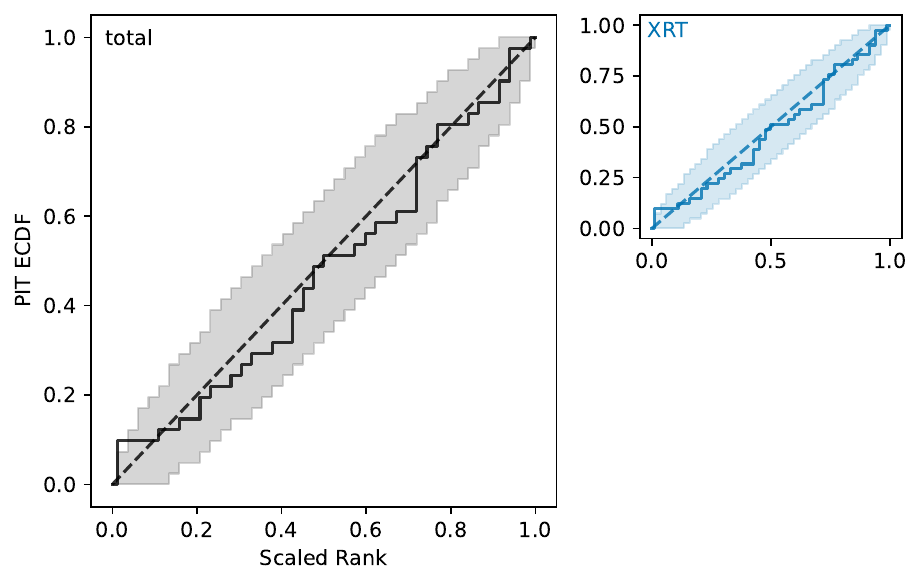}
        \caption{PIT-ECDF diagnostic.}
    \end{subfigure}
    \caption{XRT spectrum extracted from 940-1074 s post-trigger, fitted with an absorbed PL model and shown with posterior diagnostics. This figure follows the format of Figure~\ref{fig:PowerLawSpec1}.}
    \label{fig:PowerLawSpec13}
\end{figure*}

\begin{figure*}[ht!]
    \centering
    \begin{subfigure}[t]{0.49\textwidth}
        \centering
        \includegraphics[width=\linewidth]{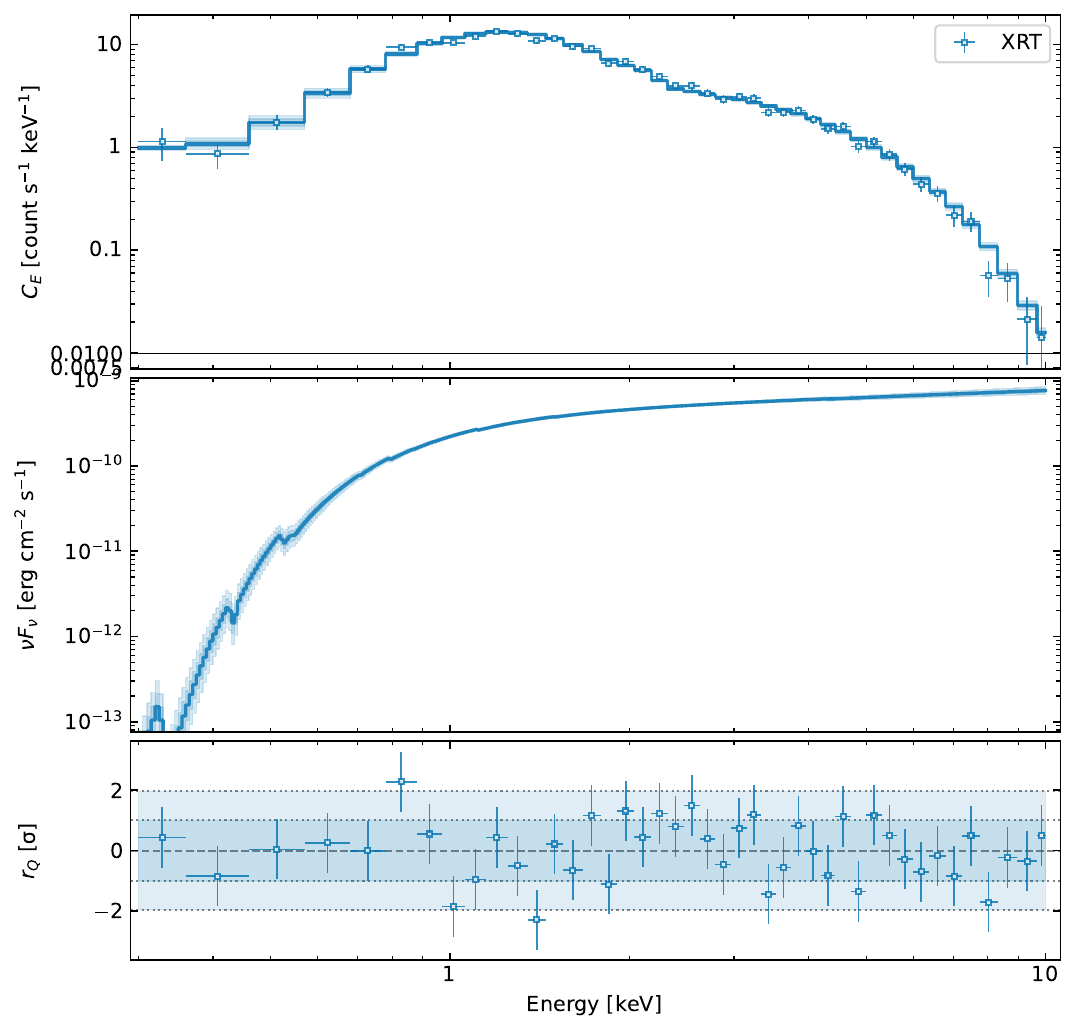}
        \caption{XRT spectrum, PL model, and residuals.}
    \end{subfigure}
    \hfill
    \begin{subfigure}[t]{0.49\textwidth}
        \centering
        \includegraphics[width=\linewidth]{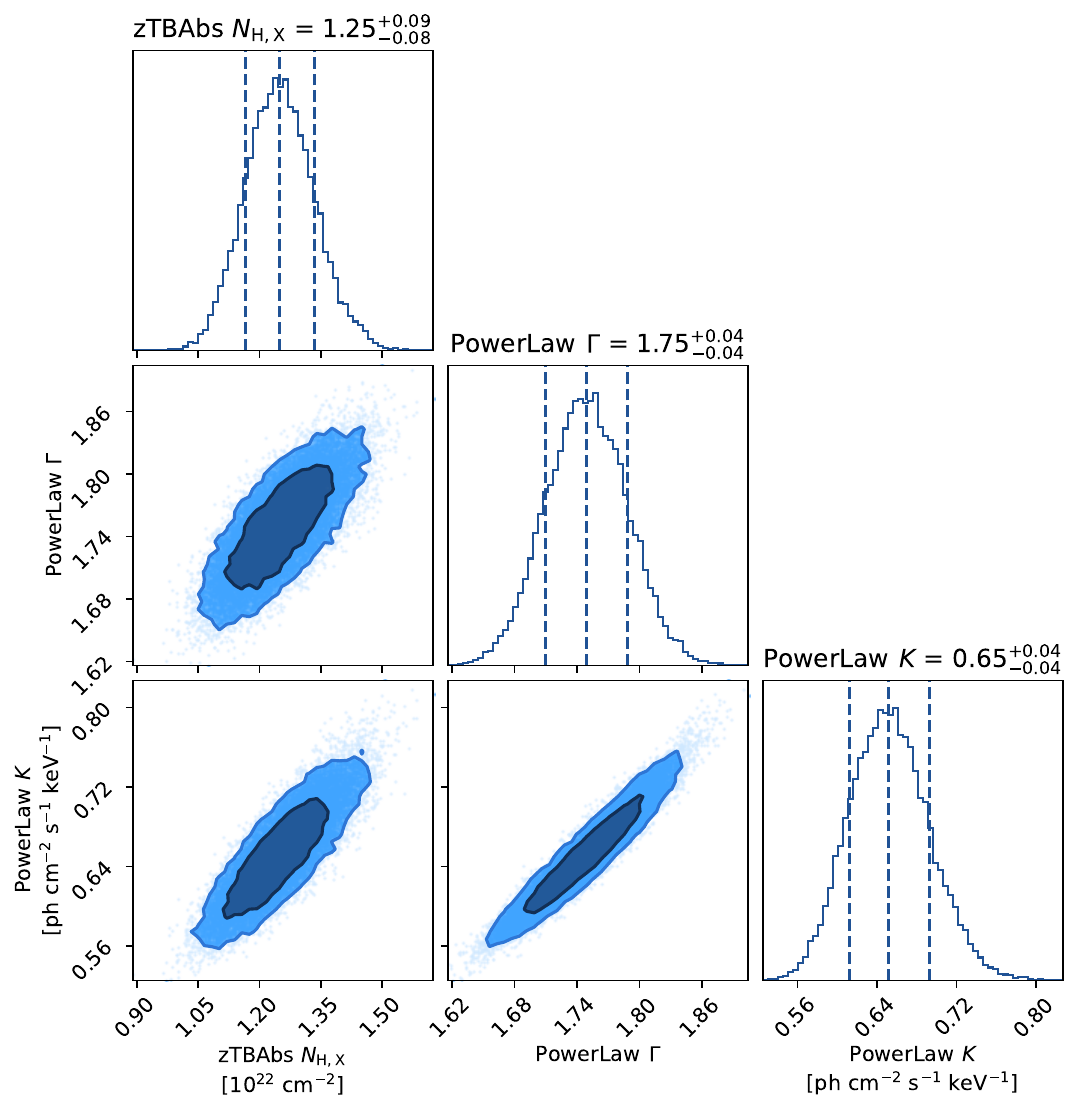}
        \caption{Posterior distributions of model parameters.}
    \end{subfigure}
    \vspace{1em}
    \begin{subfigure}[t]{0.49\textwidth}
        \centering
        \includegraphics[width=\linewidth]{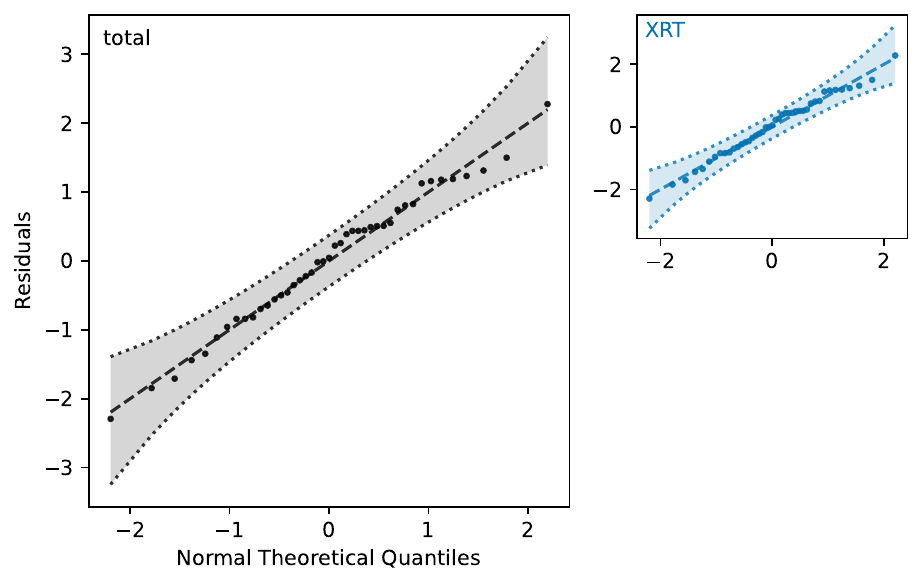}
        \caption{Q-Q plots of residuals.}
    \end{subfigure}
    \hfill
    \begin{subfigure}[t]{0.49\textwidth}
        \centering
        \includegraphics[width=\linewidth]{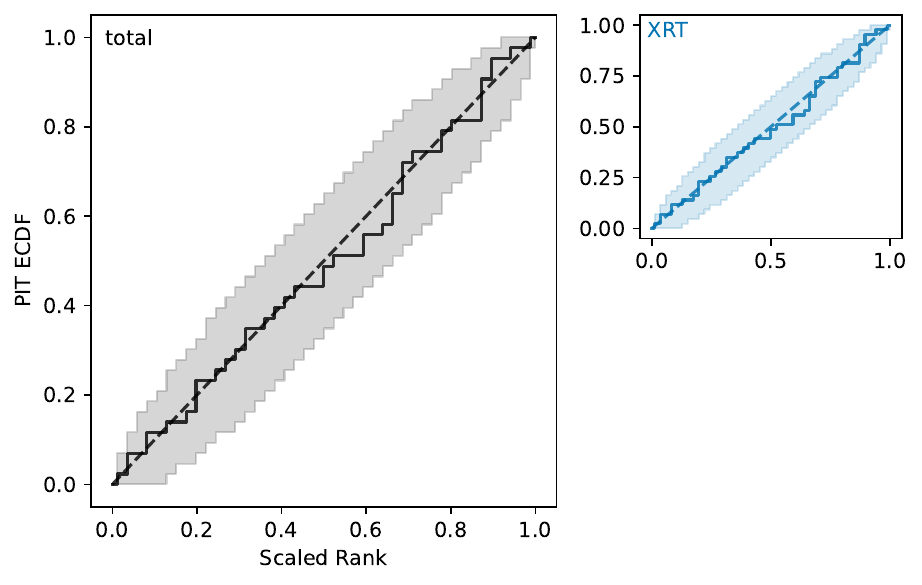}
        \caption{PIT-ECDF diagnostic.}
    \end{subfigure}
    \caption{XRT spectrum extracted from 1074-1295 s post-trigger, fitted with an absorbed PL model and shown with posterior diagnostics. This figure follows the format of Figure~\ref{fig:PowerLawSpec1}.}
    \label{fig:PowerLawSpec14}
\end{figure*}

\begin{figure*}[ht!]
    \centering
    \begin{subfigure}[t]{0.49\textwidth}
        \centering
        \includegraphics[width=\linewidth]{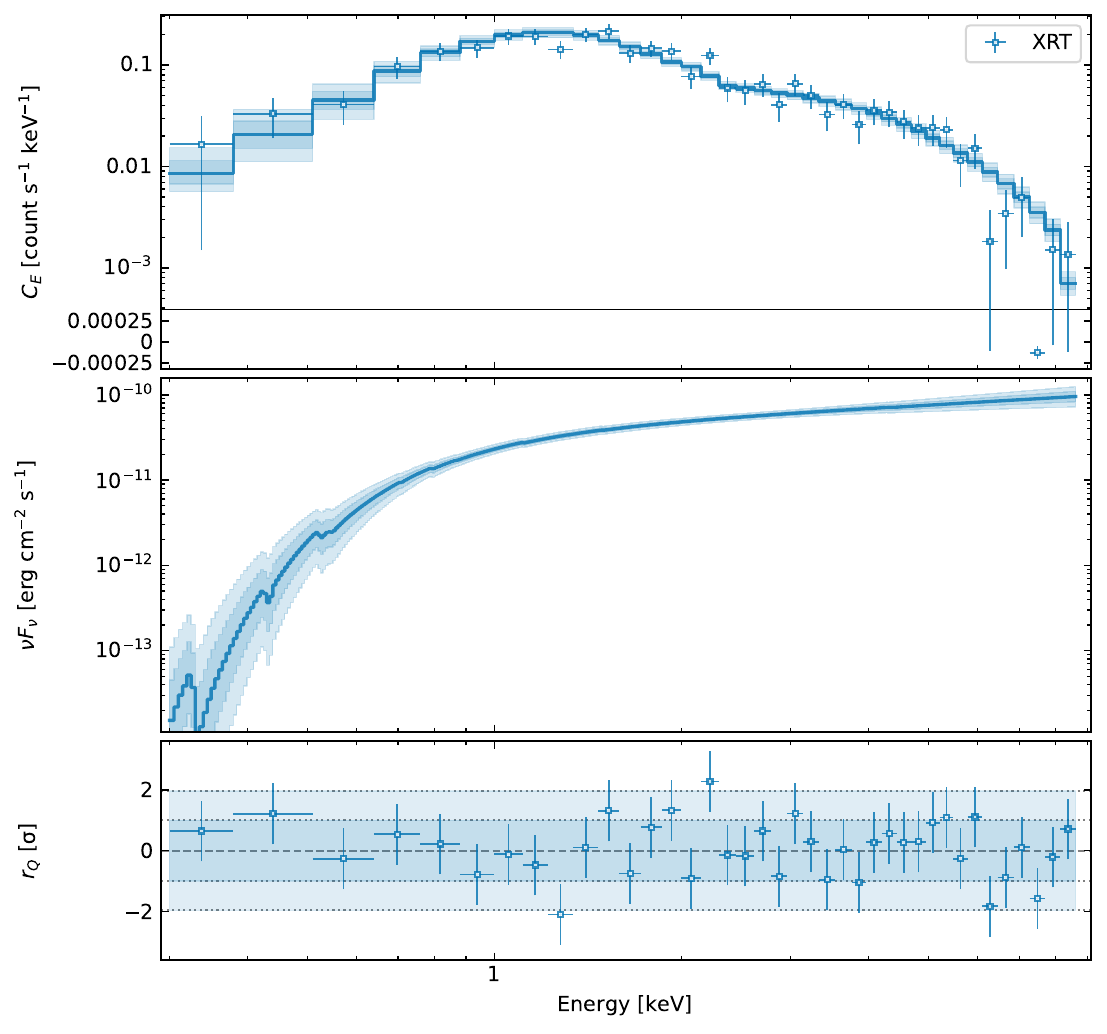}
        \caption{XRT spectrum, PL model, and residuals.}
    \end{subfigure}
    \hfill
    \begin{subfigure}[t]{0.49\textwidth}
        \centering
        \includegraphics[width=\linewidth]{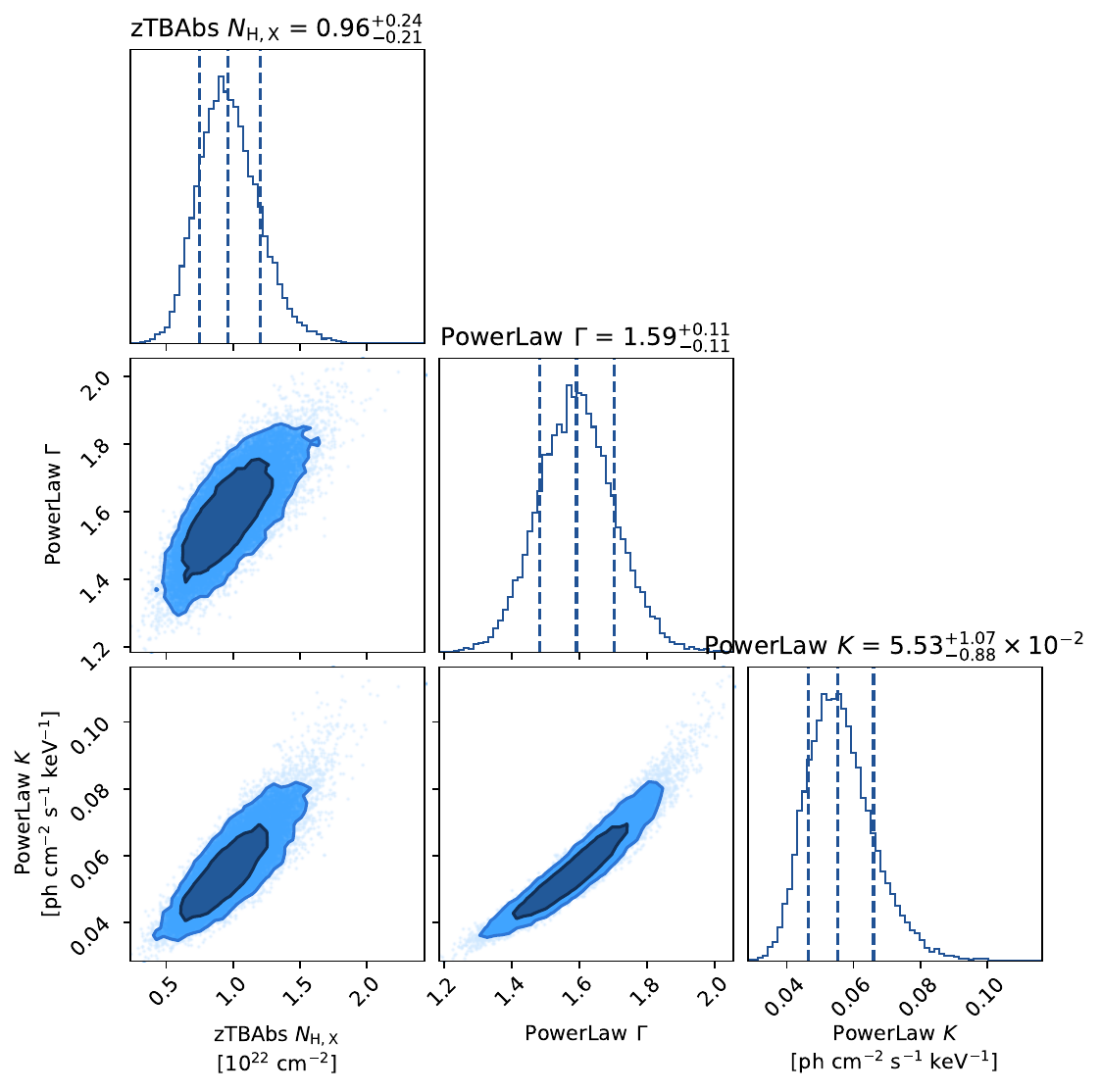}
        \caption{Posterior distributions of model parameters.}
    \end{subfigure}
    \vspace{1em}
    \begin{subfigure}[t]{0.49\textwidth}
        \centering
        \includegraphics[width=\linewidth]{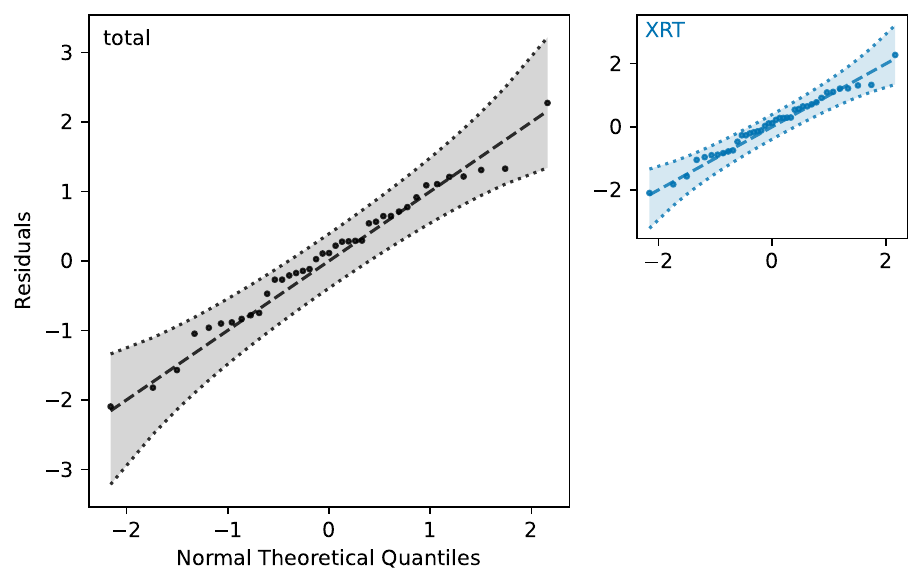}
        \caption{Q-Q plots of residuals.}
    \end{subfigure}
    \hfill
    \begin{subfigure}[t]{0.49\textwidth}
        \centering
        \includegraphics[width=\linewidth]{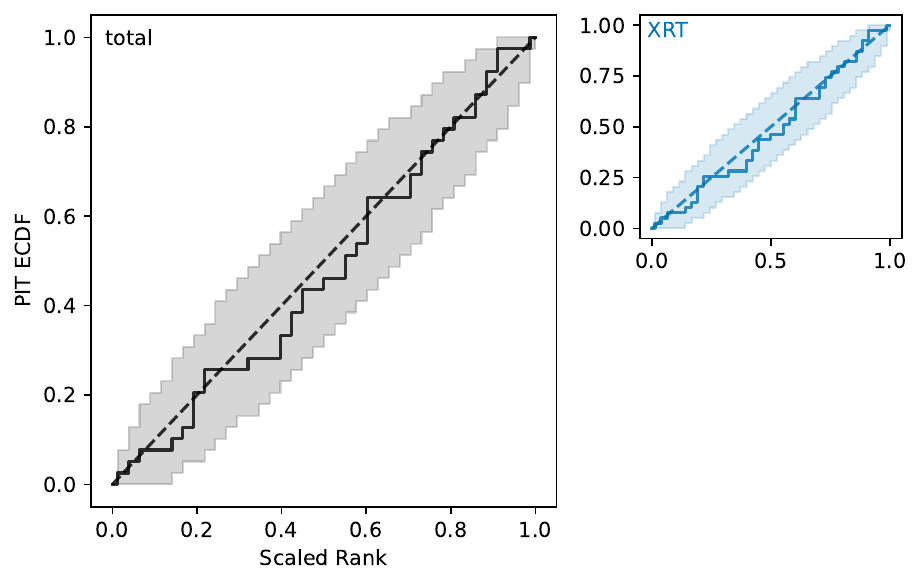}
        \caption{PIT-ECDF diagnostic.}
    \end{subfigure}
    \caption{XRT spectrum extracted from 4797-8430 s post-trigger, fitted with an absorbed PL model and shown with posterior diagnostics. This figure follows the format of Figure~\ref{fig:PowerLawSpec1}.}
    \label{fig:PowerLawSpec15}
\end{figure*}

\begin{figure*}[ht!]
    \centering
    \begin{subfigure}[t]{0.49\textwidth}
        \centering
        \includegraphics[width=\linewidth]{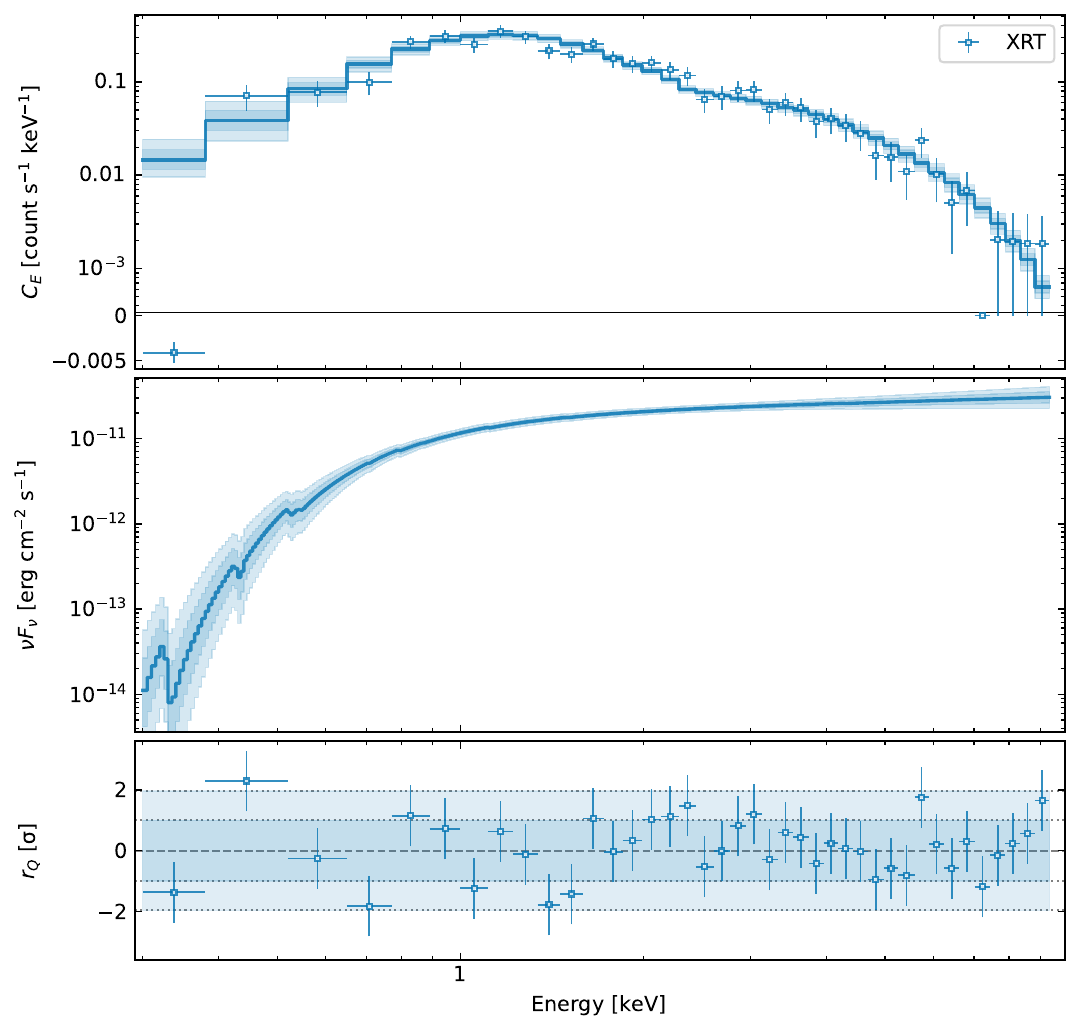}
        \caption{XRT spectrum, PL model, and residuals.}
    \end{subfigure}
    \hfill
    \begin{subfigure}[t]{0.49\textwidth}
        \centering
        \includegraphics[width=\linewidth]{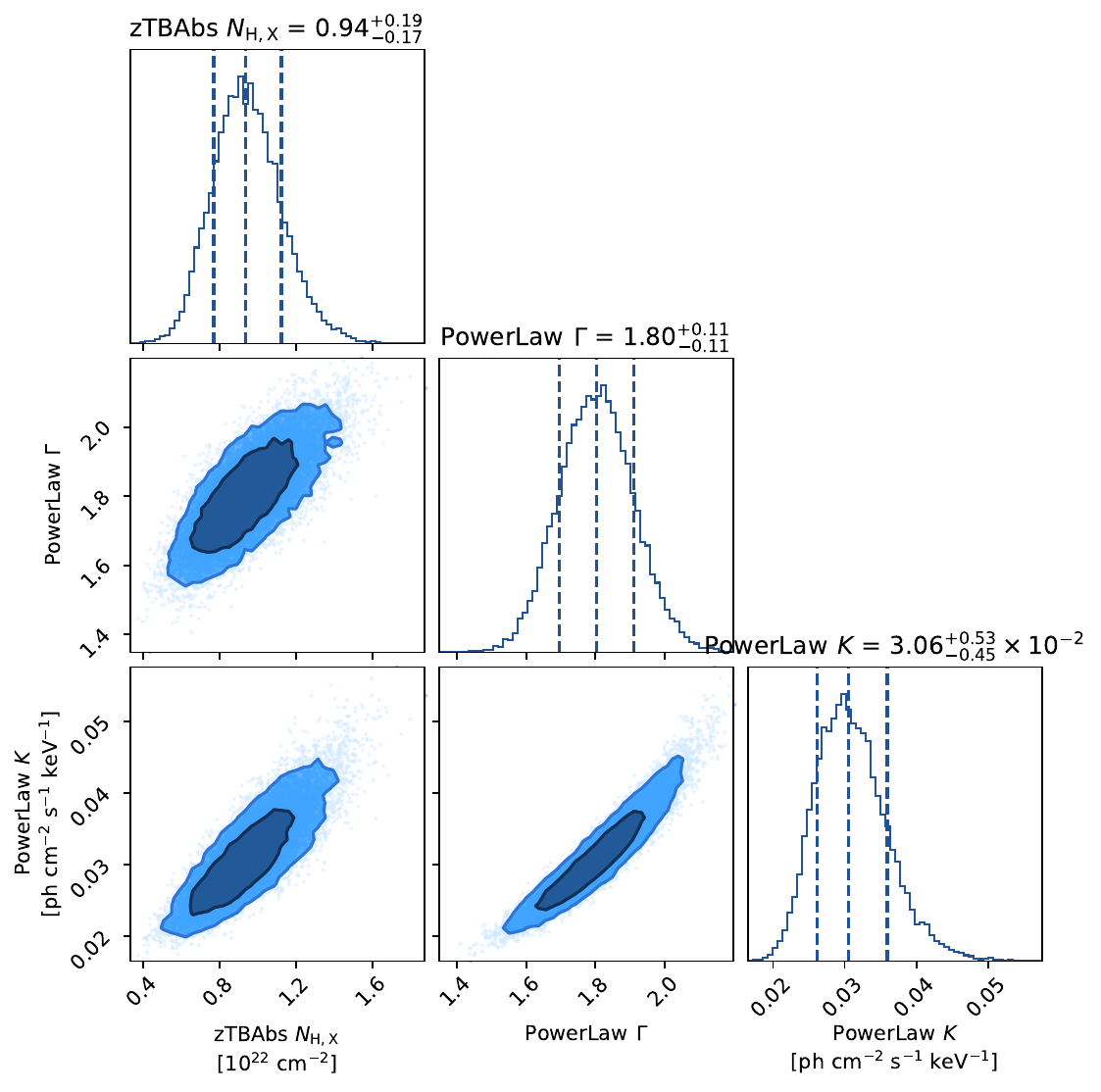}
        \caption{Posterior distributions of model parameters.}
    \end{subfigure}
    \vspace{1em}
    \begin{subfigure}[t]{0.49\textwidth}
        \centering
        \includegraphics[width=\linewidth]{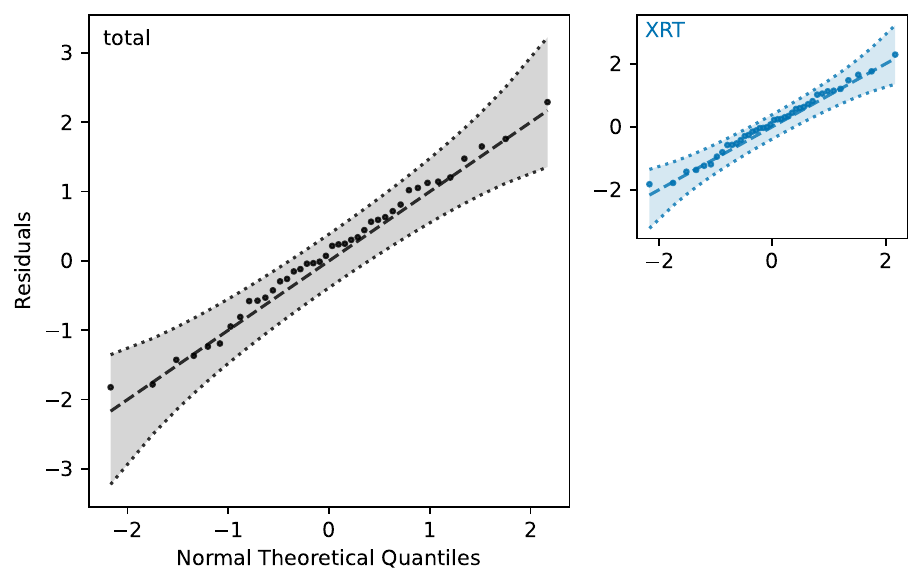}
        \caption{Q-Q plots of residuals.}
    \end{subfigure}
    \hfill
    \begin{subfigure}[t]{0.49\textwidth}
        \centering
        \includegraphics[width=\linewidth]{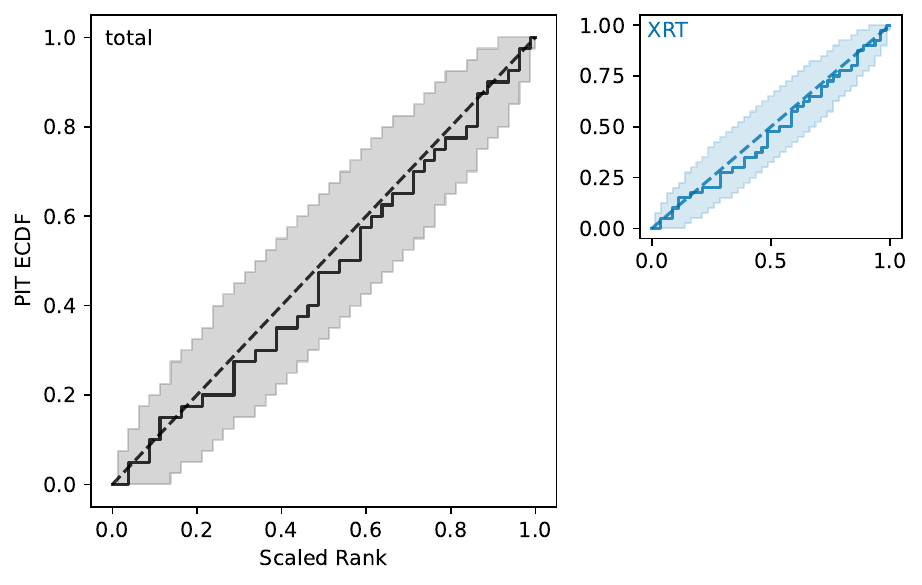}
        \caption{PIT-ECDF diagnostic.}
    \end{subfigure}
    \caption{XRT spectrum extracted from 8430-14129 s post-trigger, fitted with an absorbed PL model and shown with posterior diagnostics. This figure follows the format of Figure~\ref{fig:PowerLawSpec1}.}
    \label{fig:PowerLawSpec16}
\end{figure*}

\begin{figure*}[ht!]
    \centering
    \begin{subfigure}[t]{0.49\textwidth}
        \centering
        \includegraphics[width=\linewidth]{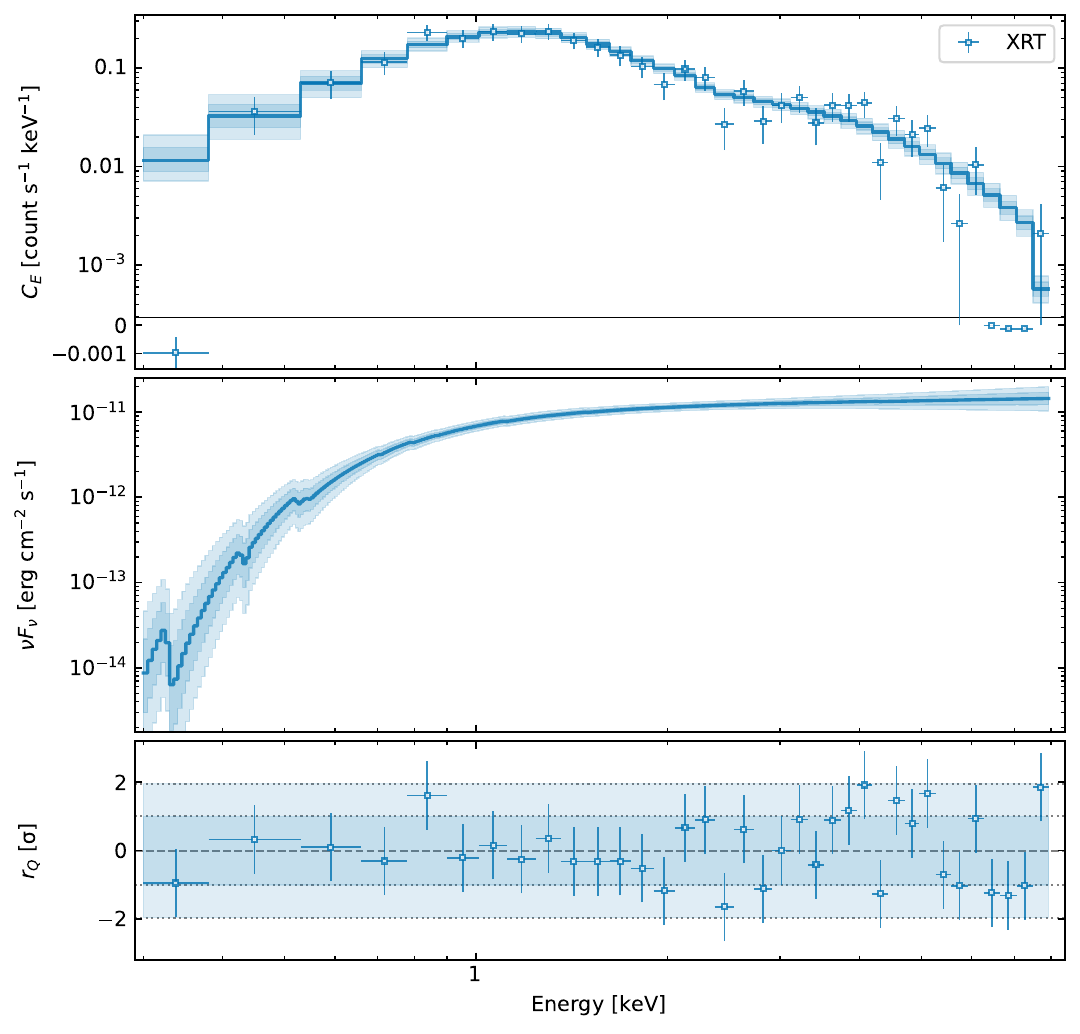}
        \caption{XRT spectrum, PL model, and residuals.}
    \end{subfigure}
    \hfill
    \begin{subfigure}[t]{0.49\textwidth}
        \centering
        \includegraphics[width=\linewidth]{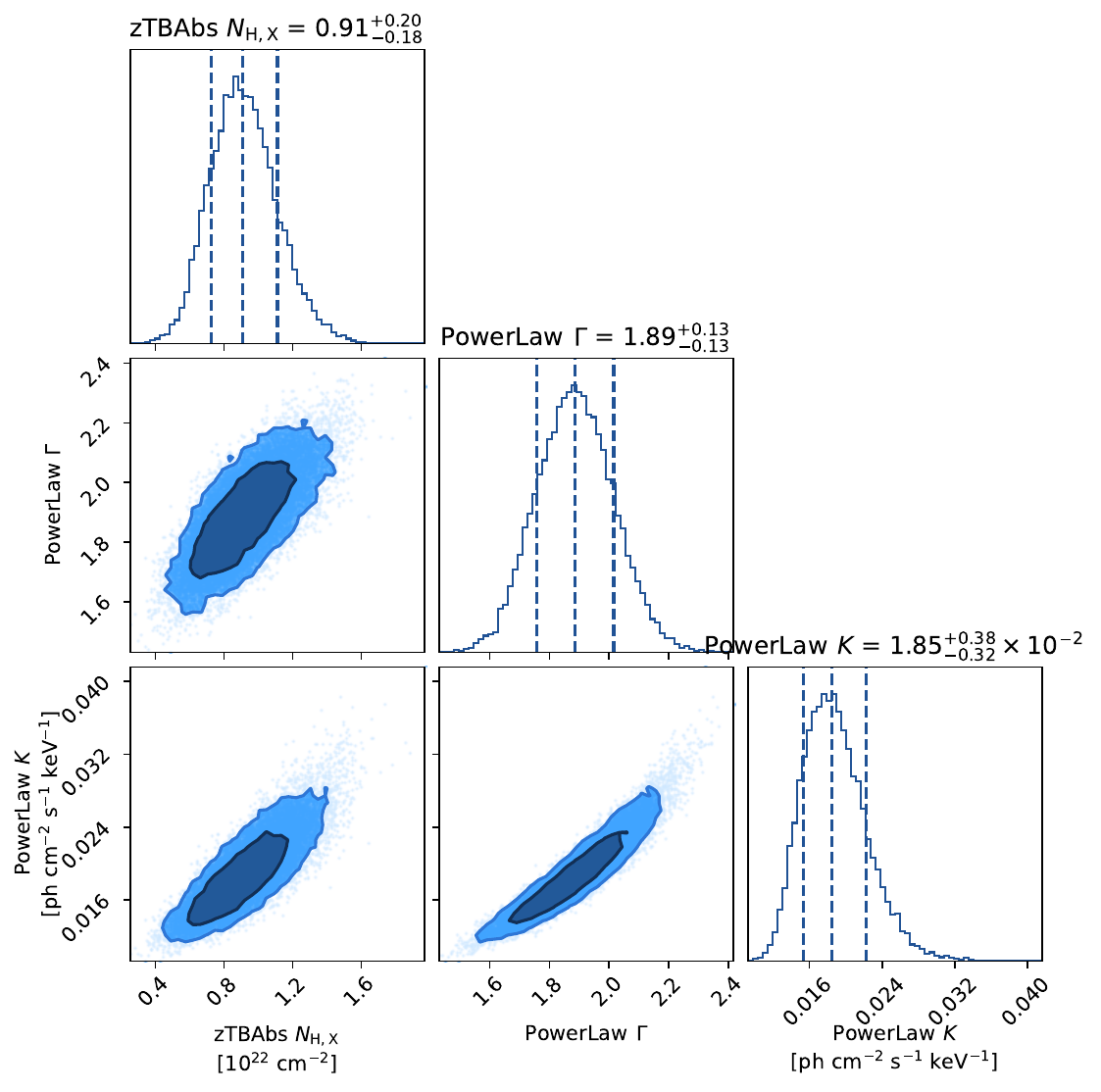}
        \caption{Posterior distributions of model parameters.}
    \end{subfigure}
    \vspace{1em}
    \begin{subfigure}[t]{0.49\textwidth}
        \centering
        \includegraphics[width=\linewidth]{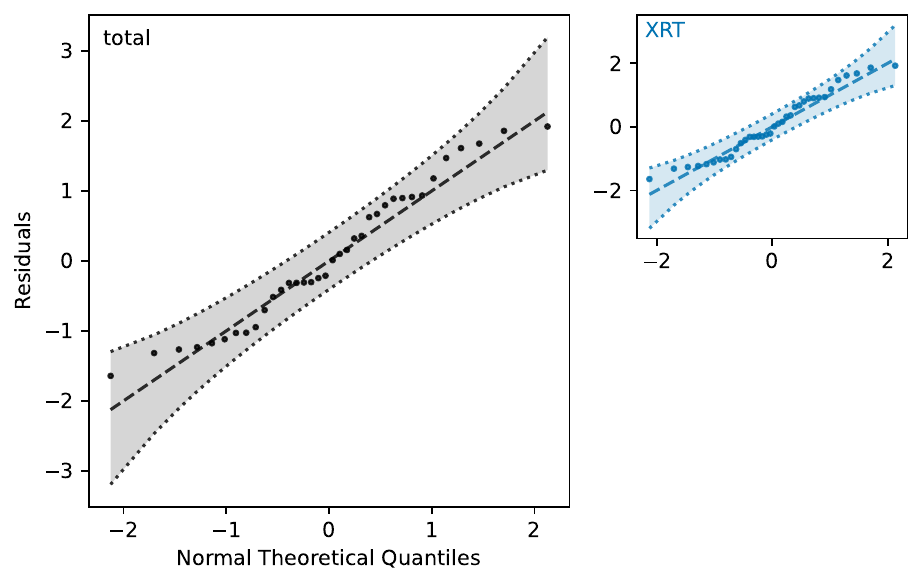}
        \caption{Q-Q plots of residuals.}
    \end{subfigure}
    \hfill
    \begin{subfigure}[t]{0.49\textwidth}
        \centering
        \includegraphics[width=\linewidth]{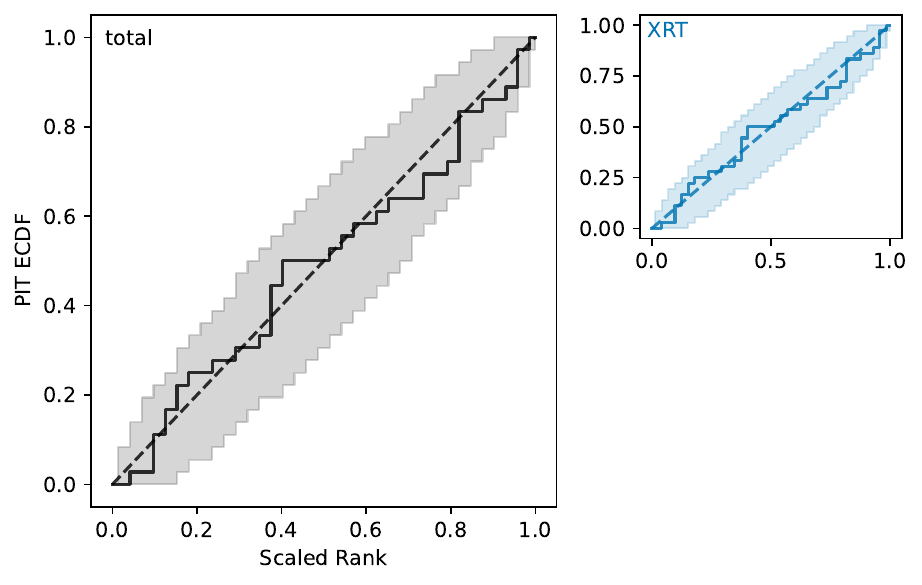}
        \caption{PIT-ECDF diagnostic.}
    \end{subfigure}
    \caption{XRT spectrum extracted from 14129-26067 s post-trigger, fitted with an absorbed PL model and shown with posterior diagnostics. This figure follows the format of Figure~\ref{fig:PowerLawSpec1}.}
    \label{fig:PowerLawSpec17}
\end{figure*}

\begin{figure*}[ht!]
    \centering
    \begin{subfigure}[t]{0.49\textwidth}
        \centering
        \includegraphics[width=\linewidth]{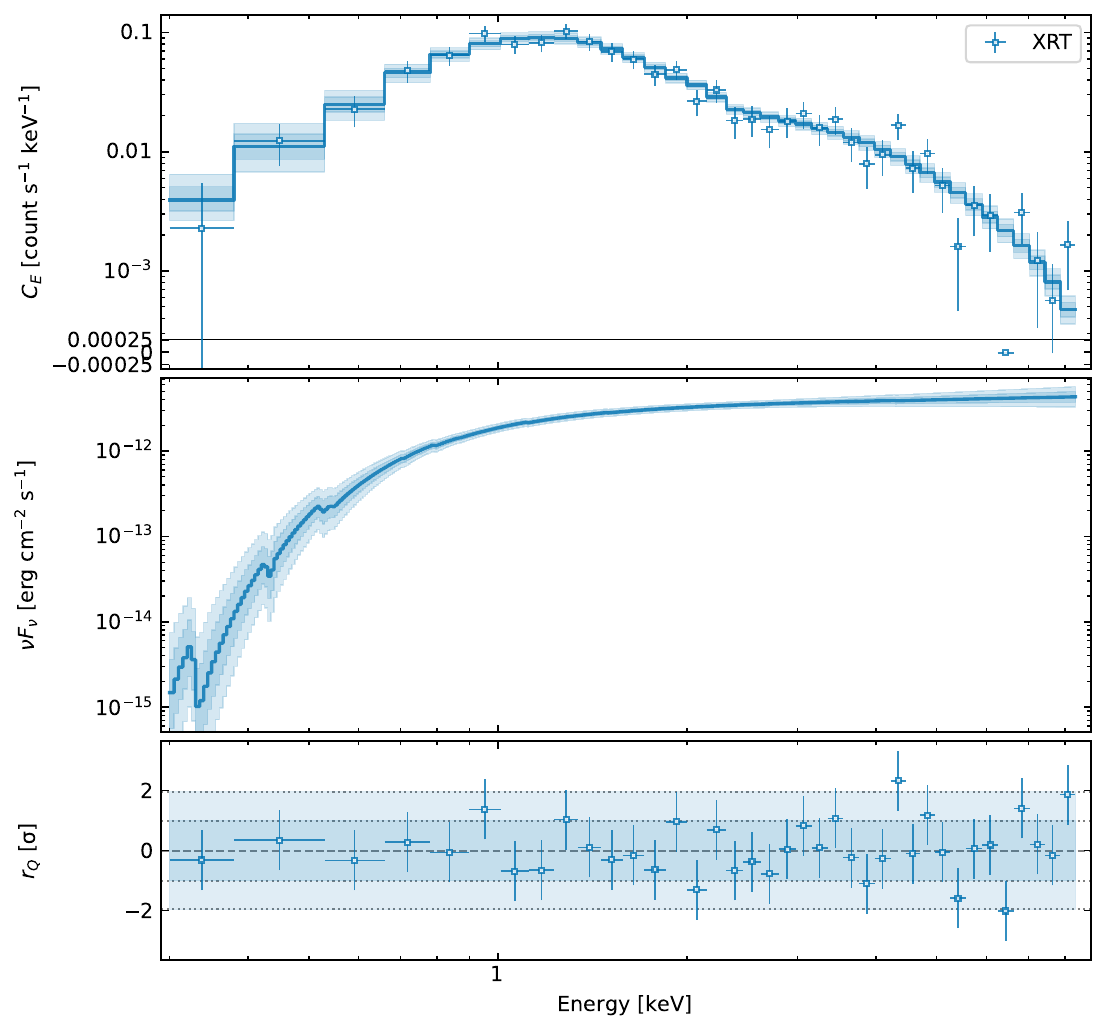}
        \caption{XRT spectrum, PL model, and residuals.}
    \end{subfigure}
    \hfill
    \begin{subfigure}[t]{0.49\textwidth}
        \centering
        \includegraphics[width=\linewidth]{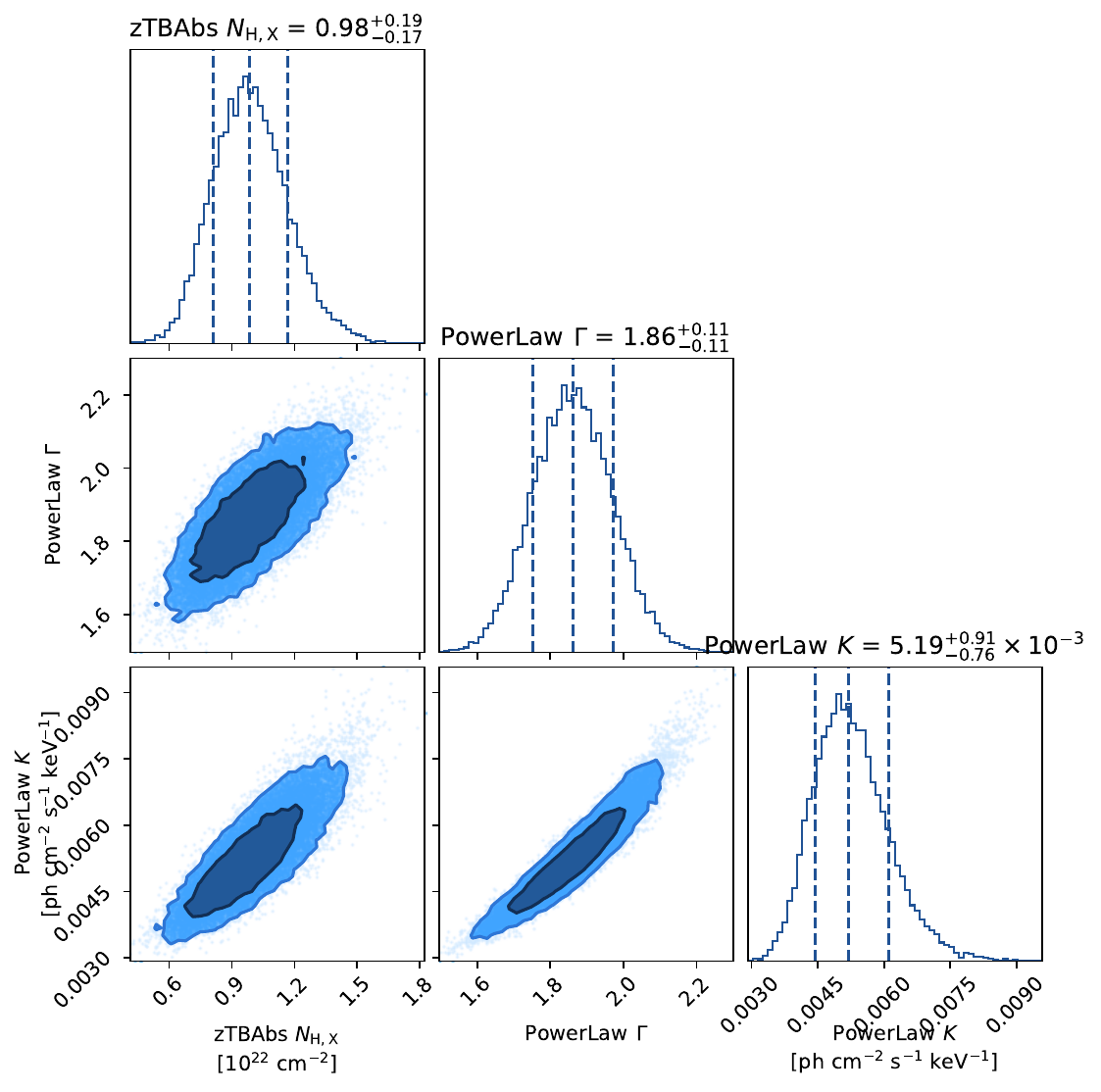}
        \caption{Posterior distributions of model parameters.}
    \end{subfigure}
    \vspace{1em}
    \begin{subfigure}[t]{0.49\textwidth}
        \centering
        \includegraphics[width=\linewidth]{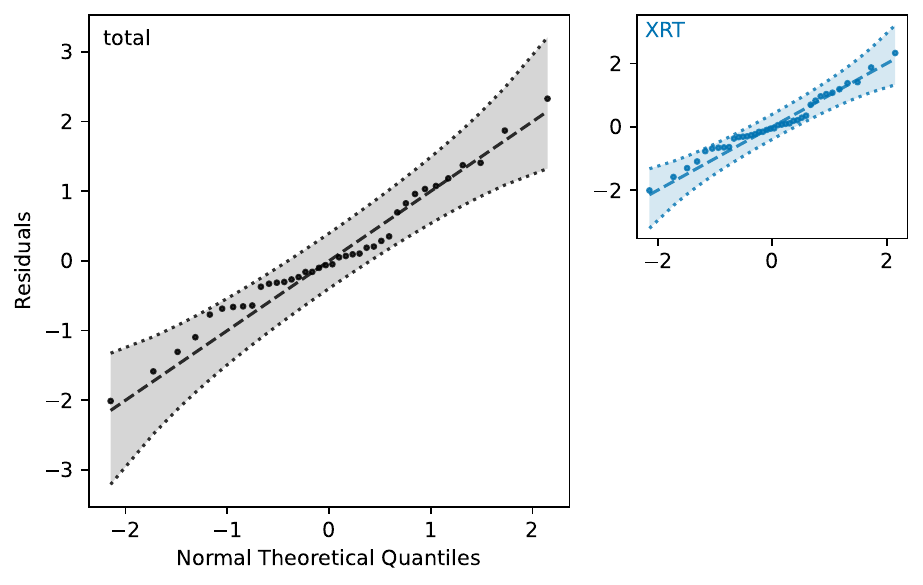}
        \caption{Q-Q plots of residuals.}
    \end{subfigure}
    \hfill
    \begin{subfigure}[t]{0.49\textwidth}
        \centering
        \includegraphics[width=\linewidth]{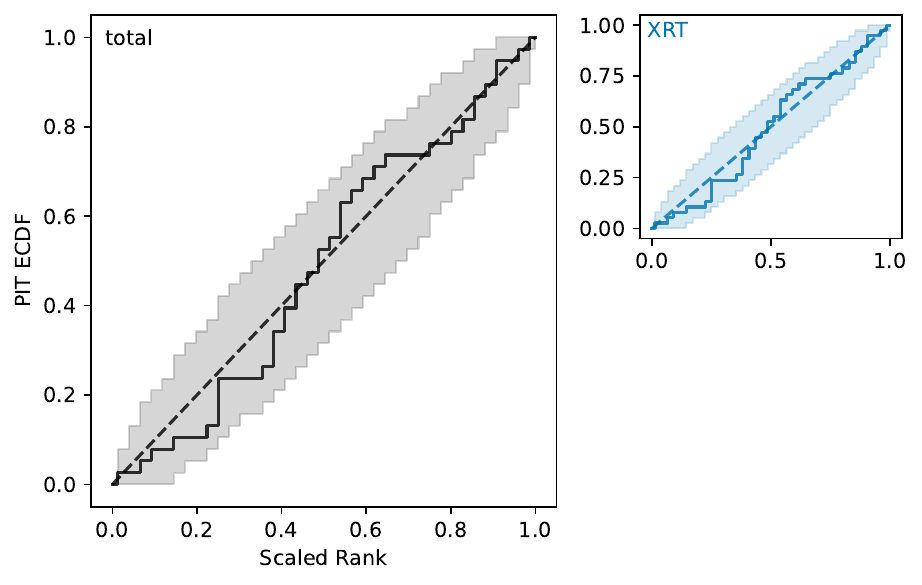}
        \caption{PIT-ECDF diagnostic.}
    \end{subfigure}
    \caption{XRT spectrum extracted from 26067-61951 s post-trigger, fitted with an absorbed PL model and shown with posterior diagnostics. This figure follows the format of Figure~\ref{fig:PowerLawSpec1}.}
    \label{fig:PowerLawSpec18}
\end{figure*}

\begin{figure*}[ht!]
    \centering
    \begin{subfigure}[t]{0.49\textwidth}
        \centering
        \includegraphics[width=\linewidth]{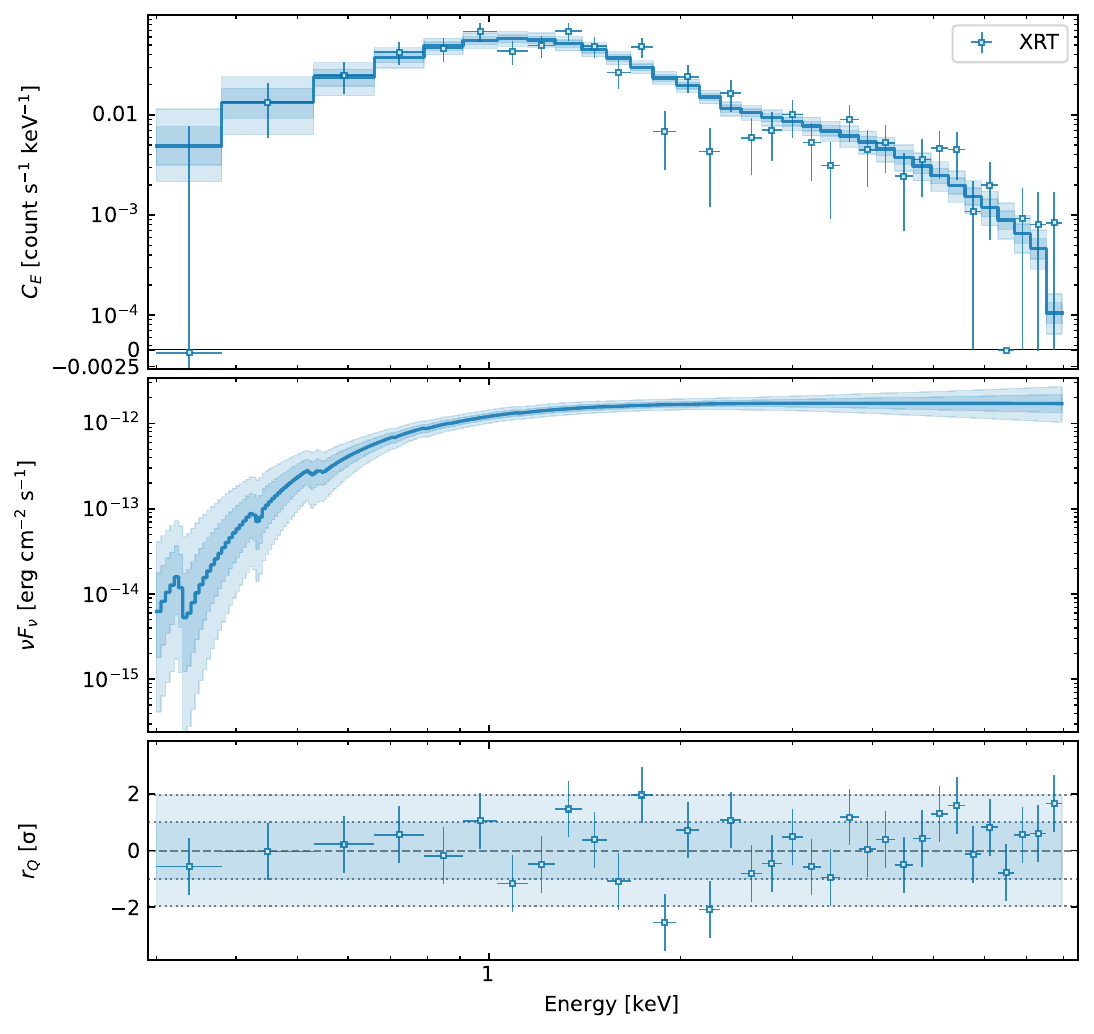}
        \caption{XRT spectrum, PL model, and residuals.}
    \end{subfigure}
    \hfill
    \begin{subfigure}[t]{0.49\textwidth}
        \centering
        \includegraphics[width=\linewidth]{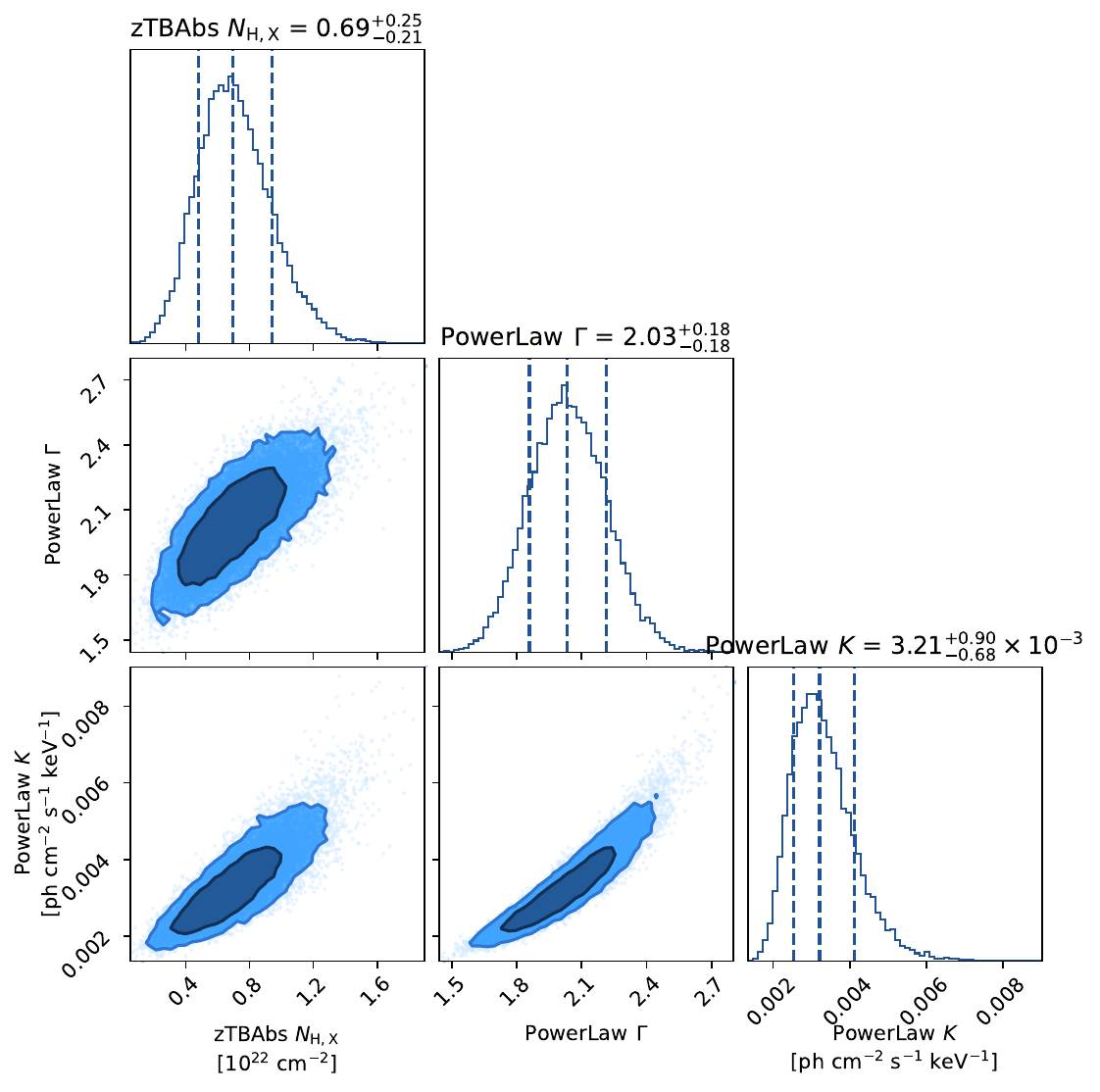}
        \caption{Posterior distributions of model parameters.}
    \end{subfigure}
    \vspace{1em}
    \begin{subfigure}[t]{0.49\textwidth}
        \centering
        \includegraphics[width=\linewidth]{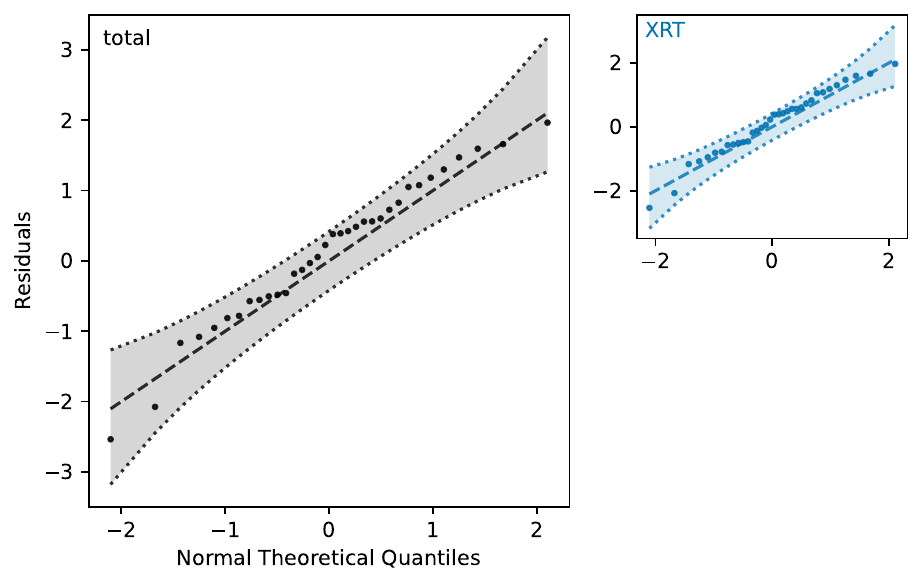}
        \caption{Q-Q plots of residuals.}
    \end{subfigure}
    \hfill
    \begin{subfigure}[t]{0.49\textwidth}
        \centering
        \includegraphics[width=\linewidth]{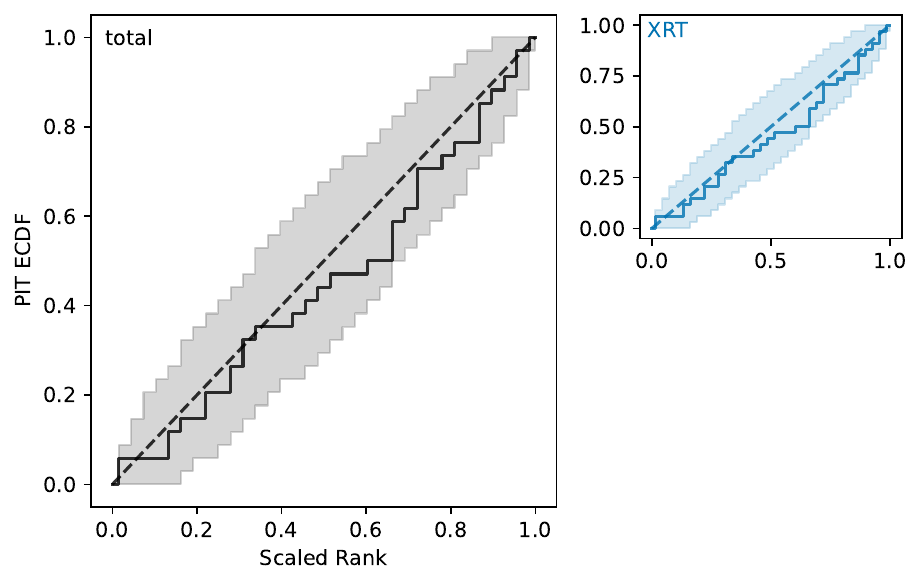}
        \caption{PIT-ECDF diagnostic.}
    \end{subfigure}
    \caption{XRT spectrum extracted from 61951-110958 s post-trigger, fitted with an absorbed PL model and shown with posterior diagnostics. This figure follows the format of Figure~\ref{fig:PowerLawSpec1}.}
    \label{fig:PowerLawSpec19}
\end{figure*}

\begin{figure*}[ht!]
    \centering
    \begin{subfigure}[t]{0.49\textwidth}
        \centering
        \includegraphics[width=\linewidth]{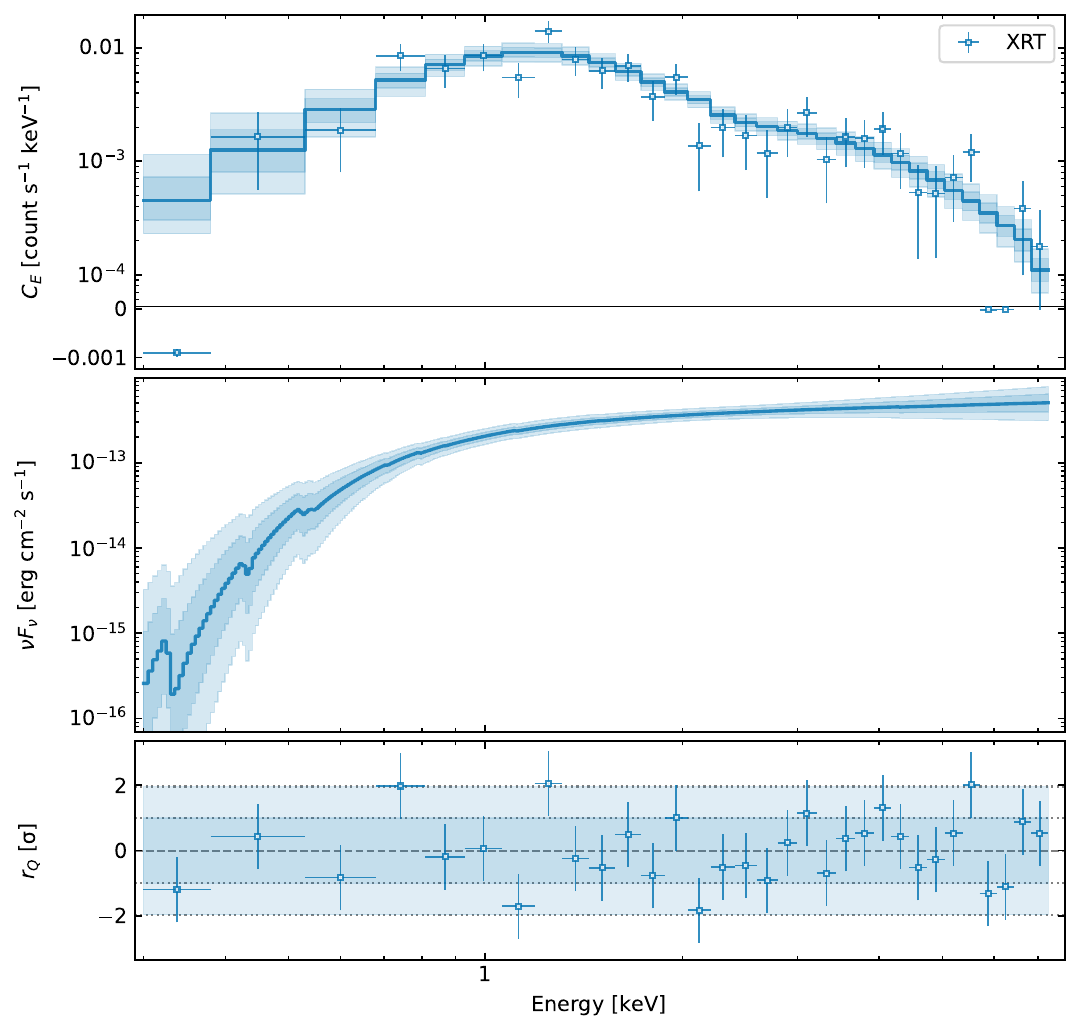}
        \caption{XRT spectrum, PL model, and residuals.}
    \end{subfigure}
    \hfill
    \begin{subfigure}[t]{0.49\textwidth}
        \centering
        \includegraphics[width=\linewidth]{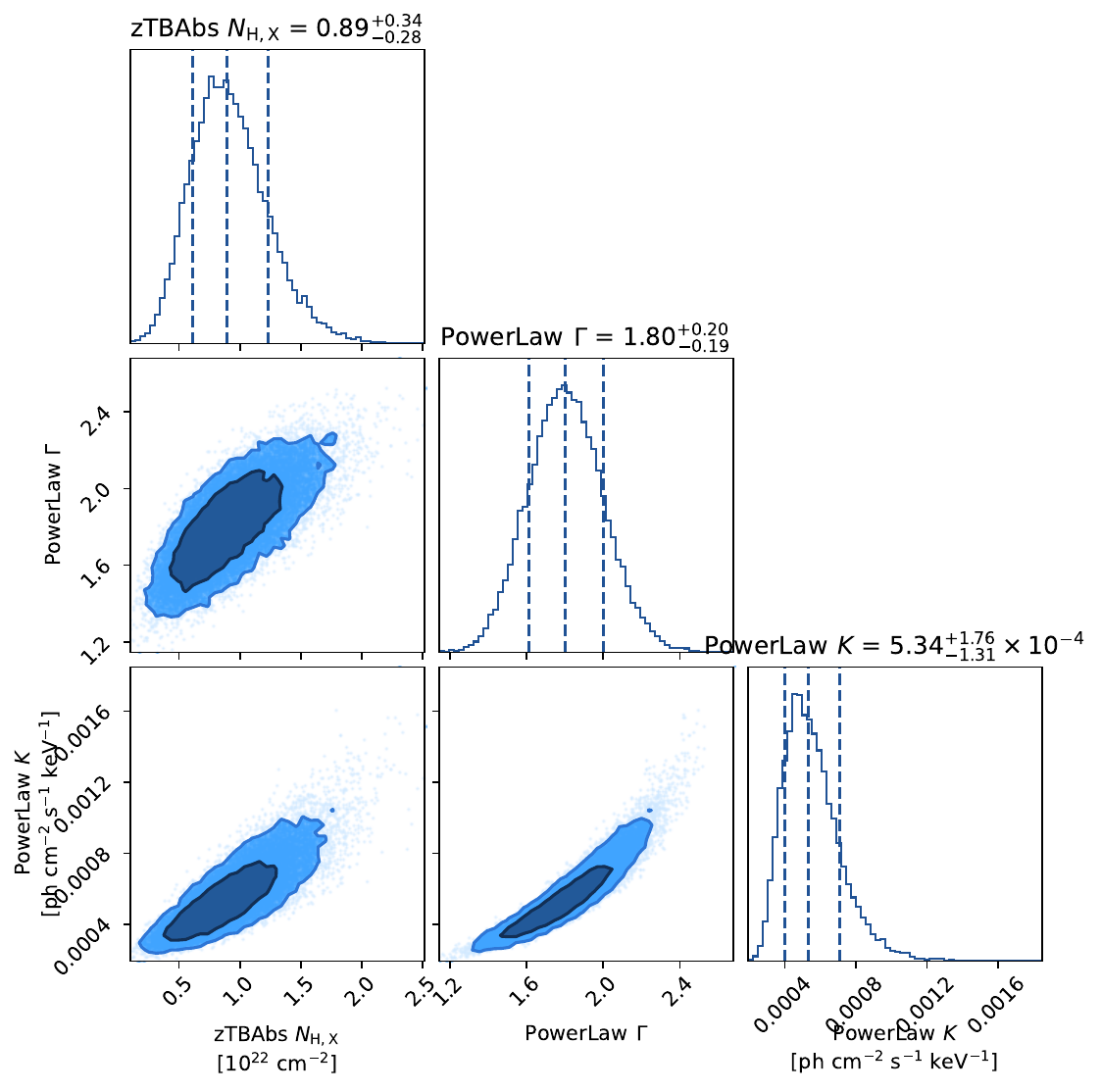}
        \caption{Posterior distributions of model parameters.}
    \end{subfigure}
    \vspace{1em}
    \begin{subfigure}[t]{0.49\textwidth}
        \centering
        \includegraphics[width=\linewidth]{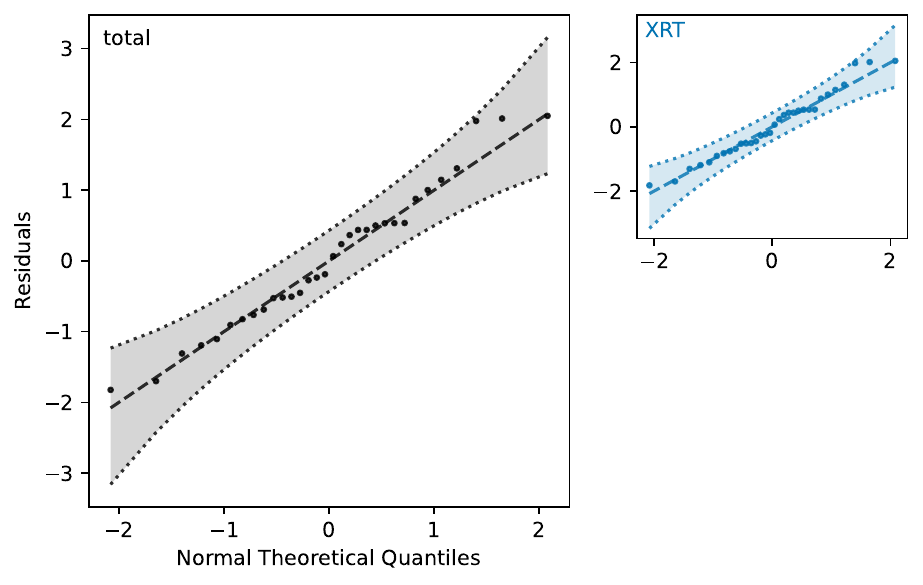}
        \caption{Q-Q plots of residuals.}
    \end{subfigure}
    \hfill
    \begin{subfigure}[t]{0.49\textwidth}
        \centering
        \includegraphics[width=\linewidth]{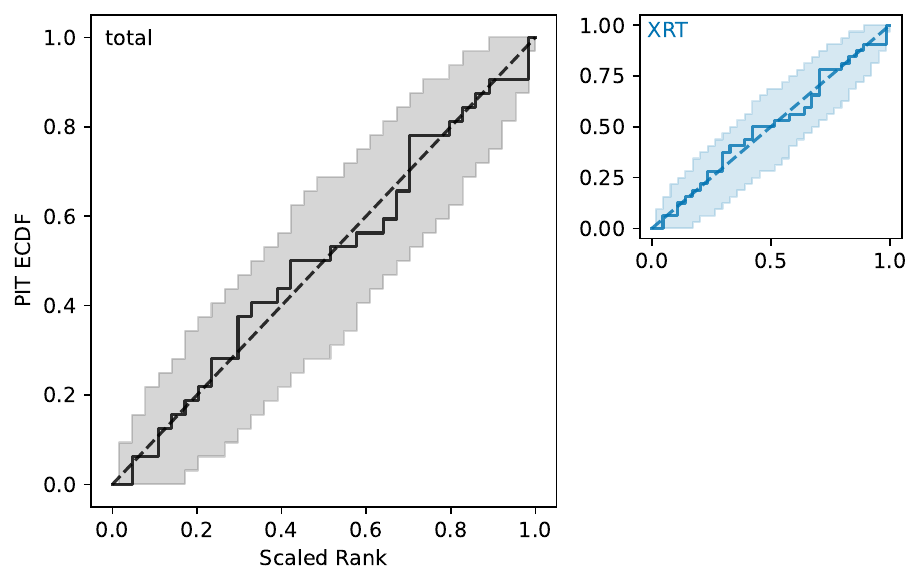}
        \caption{PIT-ECDF diagnostic.}
    \end{subfigure}
    \caption{XRT spectrum extracted from 110958-1637153 s post-trigger, fitted with an absorbed PL model and shown with posterior diagnostics. This figure follows the format of Figure~\ref{fig:PowerLawSpec1}.}
    \label{fig:PowerLawSpec20}
\end{figure*}

\begin{figure*}[ht!]
    \centering
    \begin{subfigure}[t]{0.49\textwidth}
        \centering
        \includegraphics[width=\linewidth]{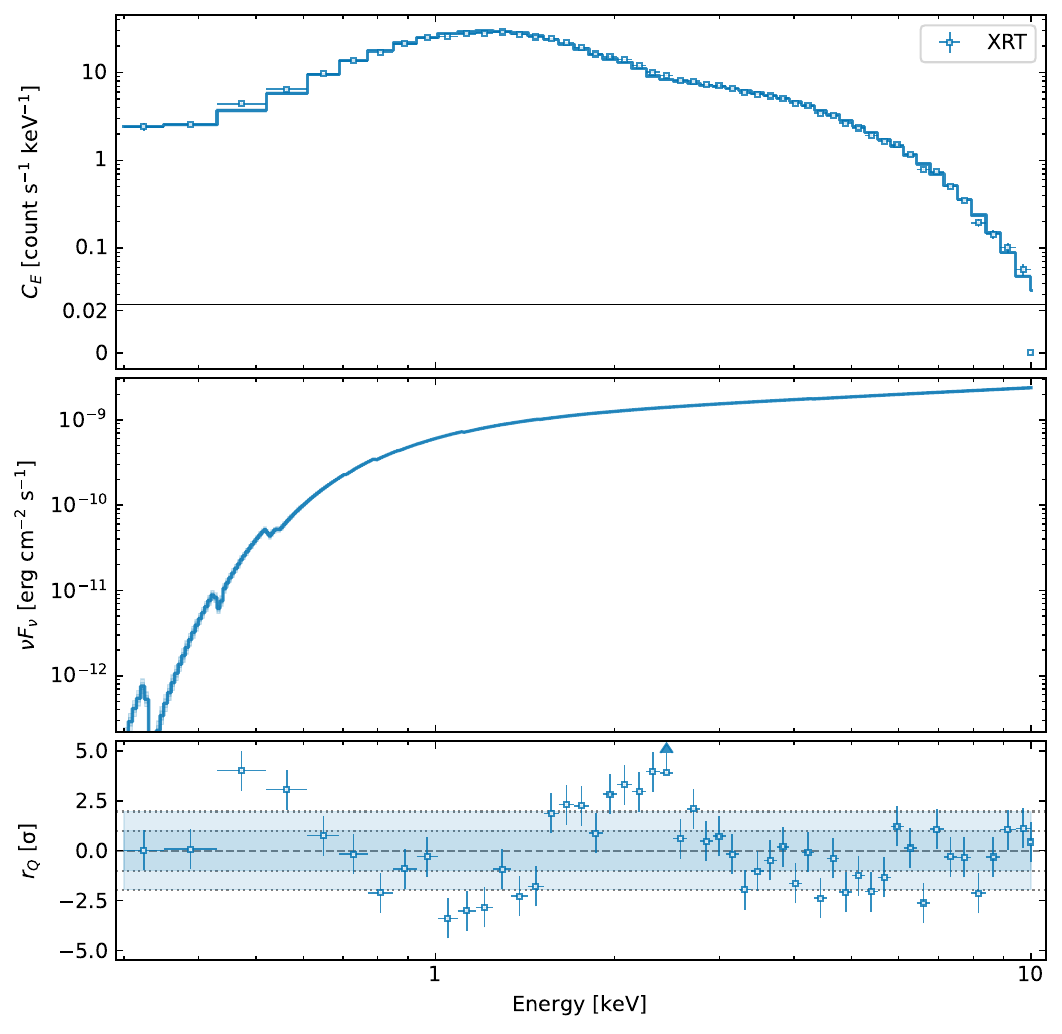}
        \caption{XRT spectrum, PL model, and residuals.}
    \end{subfigure}
    \hfill
    \begin{subfigure}[t]{0.49\textwidth}
        \centering
        \includegraphics[width=\linewidth]{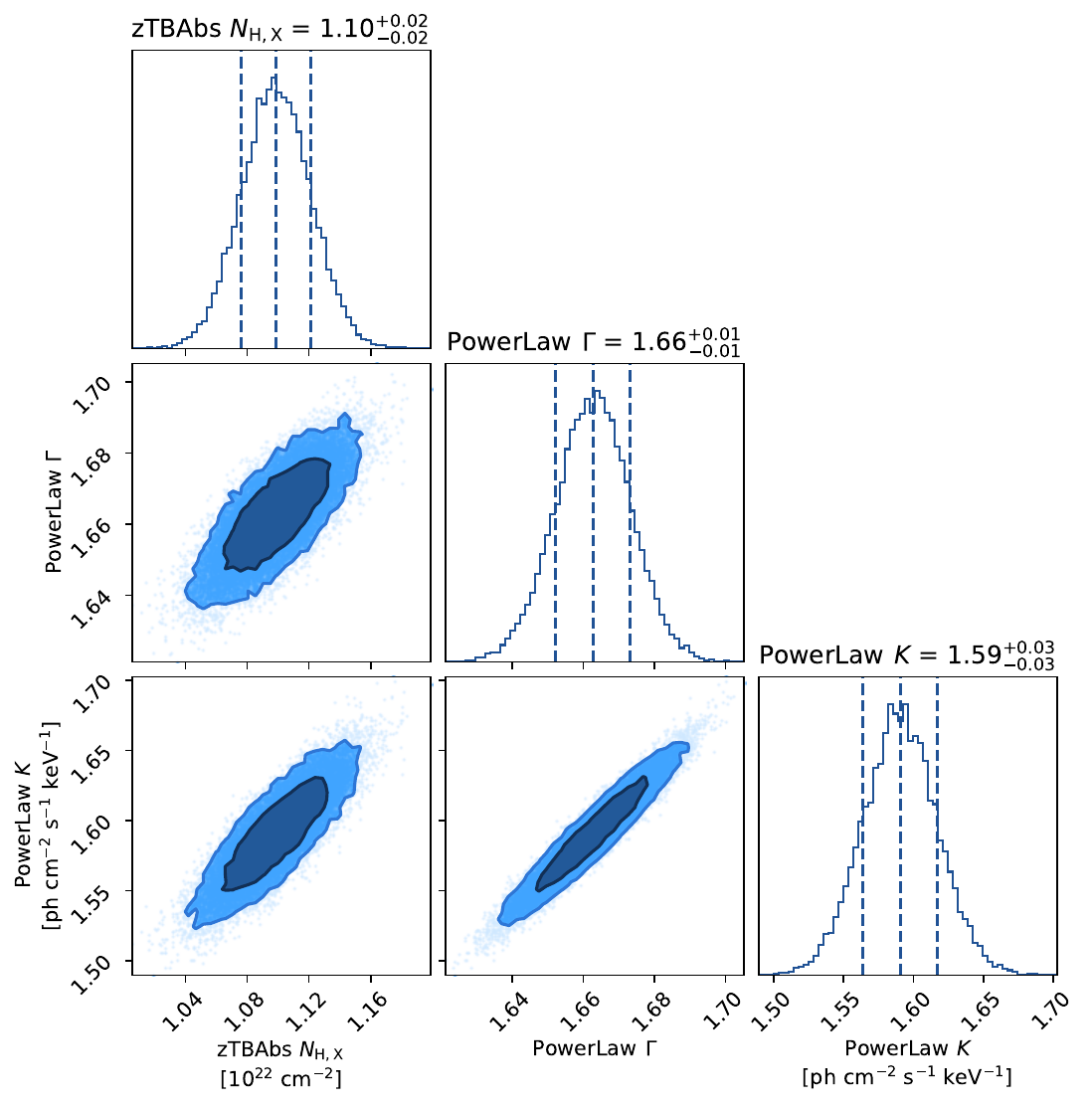}
        \caption{Posterior distributions of model parameters.}
    \end{subfigure}
    \vspace{1em}
    \begin{subfigure}[t]{0.49\textwidth}
        \centering
        \includegraphics[width=\linewidth]{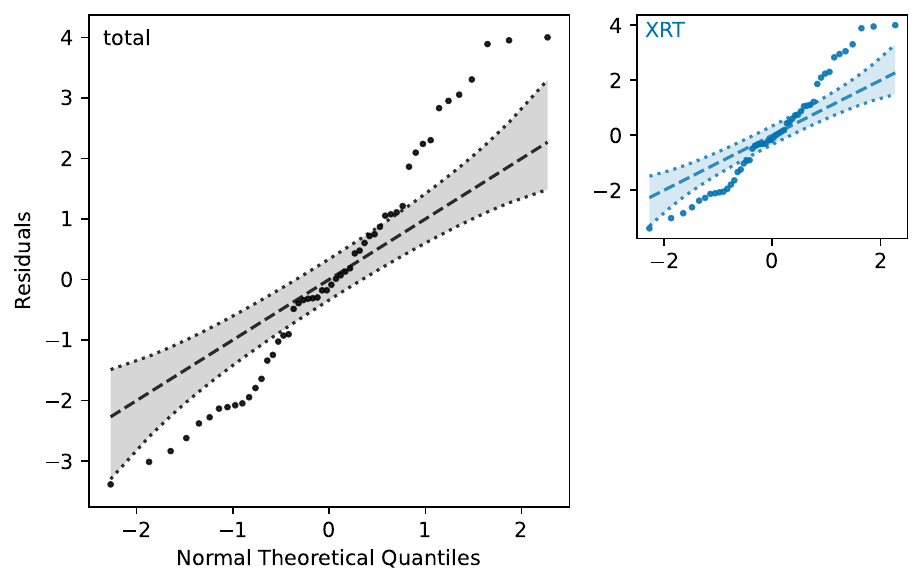}
        \caption{Q-Q plots of residuals.}
    \end{subfigure}
    \hfill
    \begin{subfigure}[t]{0.49\textwidth}
        \centering
        \includegraphics[width=\linewidth]{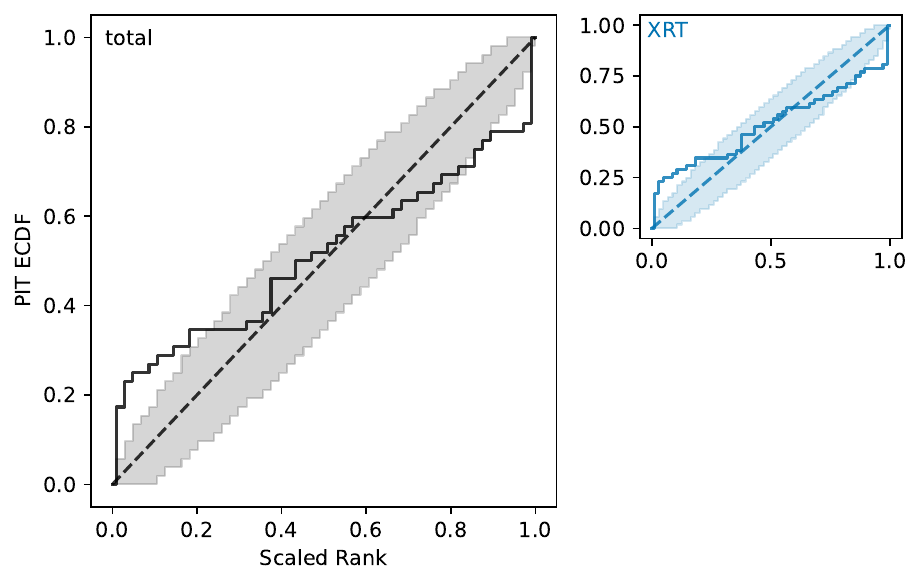}
        \caption{PIT-ECDF diagnostic.}
    \end{subfigure}
    \caption{XRT spectrum extracted from 86-1297 s post-trigger, fitted with an absorbed PL model and shown with posterior diagnostics. This figure follows the format of Figure~\ref{fig:PowerLawSpec1}.}
    \label{fig:PowerLawWT}
\end{figure*}

\begin{figure*}[ht!]
    \centering
    \begin{subfigure}[t]{0.49\textwidth}
        \centering
        \includegraphics[width=\linewidth]{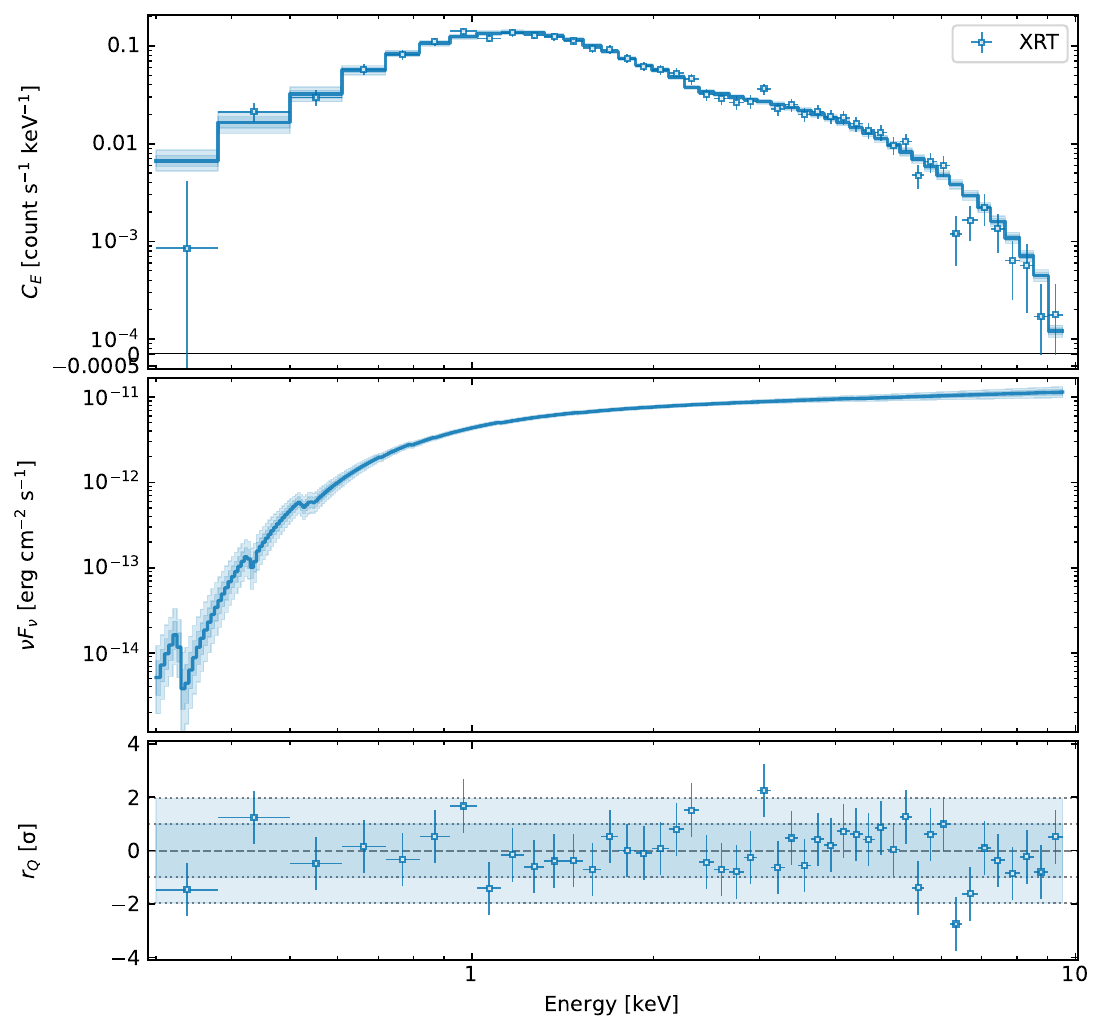}
        \caption{XRT spectrum, PL model, and residuals.}
    \end{subfigure}
    \hfill
    \begin{subfigure}[t]{0.49\textwidth}
        \centering
        \includegraphics[width=\linewidth]{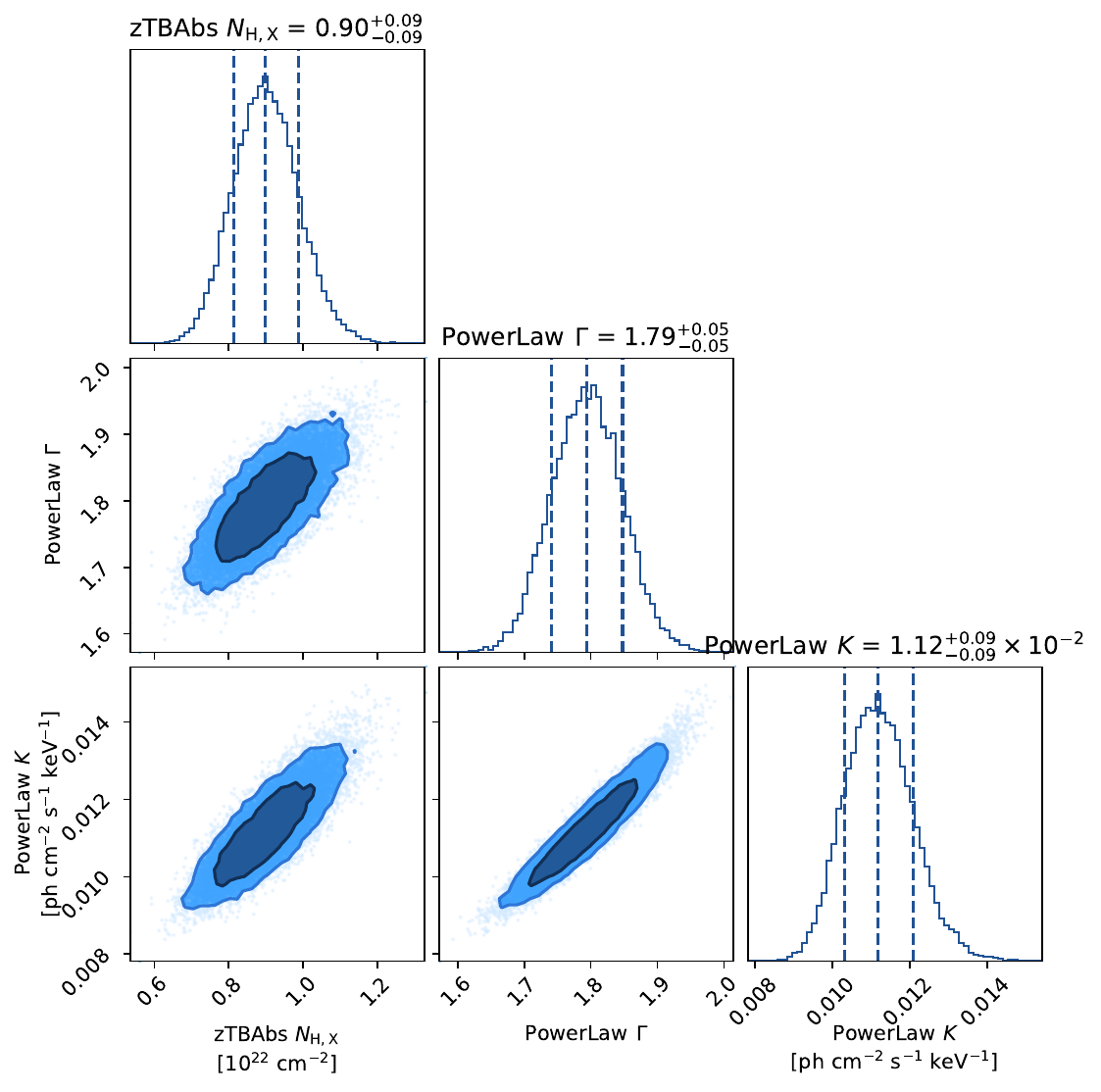}
        \caption{Posterior distributions of model parameters.}
    \end{subfigure}
    \vspace{1em}
    \begin{subfigure}[t]{0.49\textwidth}
        \centering
        \includegraphics[width=\linewidth]{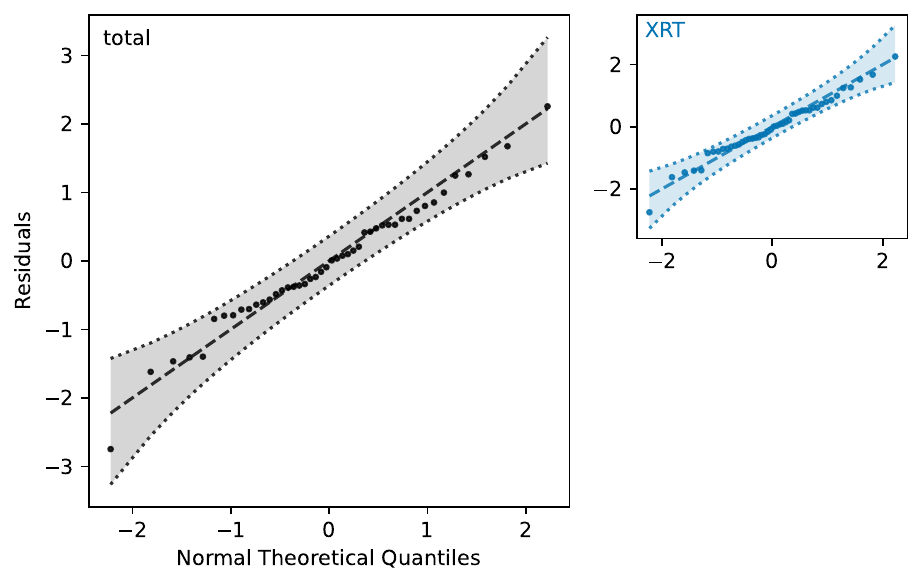}
        \caption{Q-Q plots of residuals.}
    \end{subfigure}
    \hfill
    \begin{subfigure}[t]{0.49\textwidth}
        \centering
        \includegraphics[width=\linewidth]{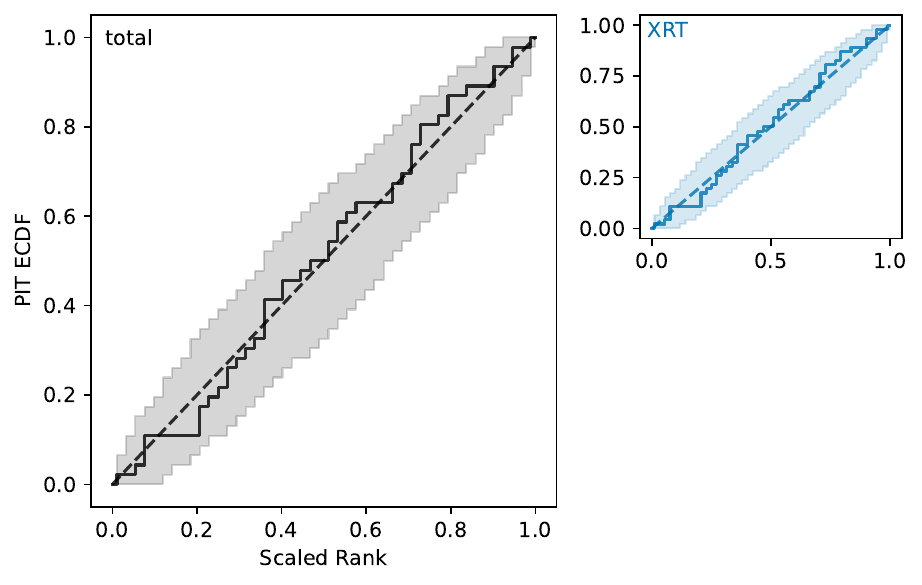}
        \caption{PIT-ECDF diagnostic.}
    \end{subfigure}
    \caption{XRT spectrum extracted from 4773-89889 s post-trigger, fitted with an absorbed PL model and shown with posterior diagnostics. This figure follows the format of Figure~\ref{fig:PowerLawSpec1}.}
    \label{fig:PowerLawPC}
\end{figure*}

\begin{figure*}[ht!]
    \centering
    \begin{subfigure}[t]{0.49\textwidth}
        \centering
        \includegraphics[width=\linewidth]{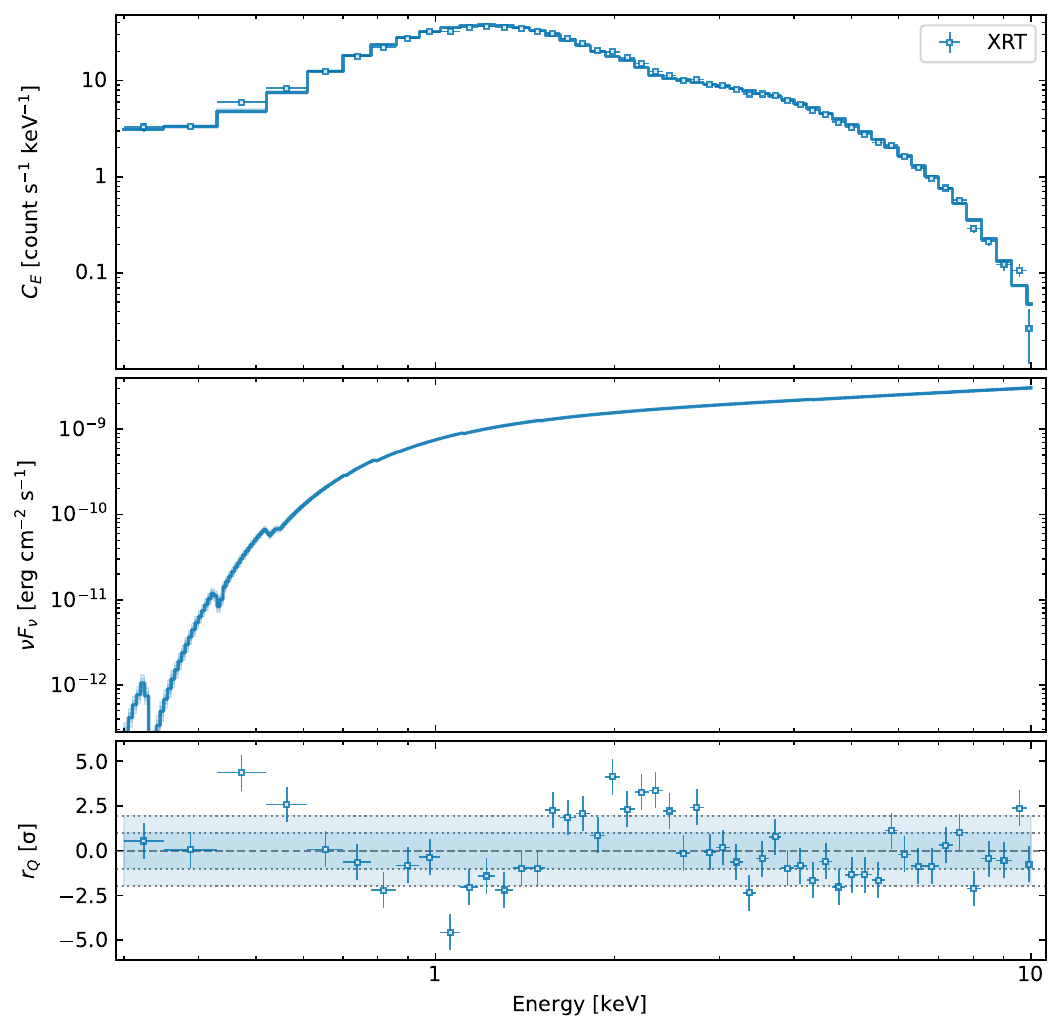}
        \caption{XRT spectrum, PL model, and residuals.}
    \end{subfigure}
    \hfill
    \begin{subfigure}[t]{0.49\textwidth}
        \centering
        \includegraphics[width=\linewidth]{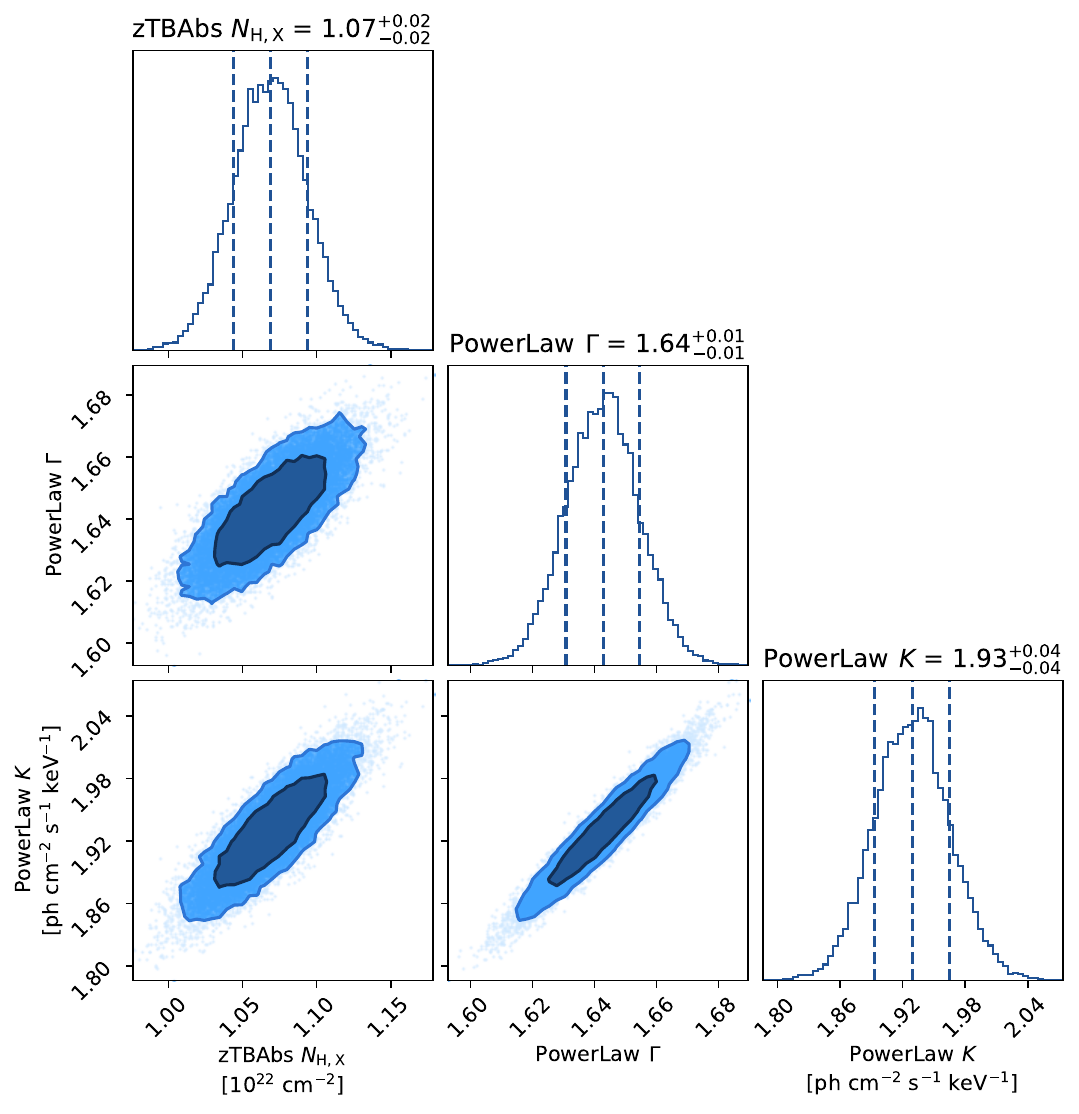}
        \caption{Posterior distributions of model parameters.}
    \end{subfigure}
    \vspace{1em}
    \begin{subfigure}[t]{0.49\textwidth}
        \centering
        \includegraphics[width=\linewidth]{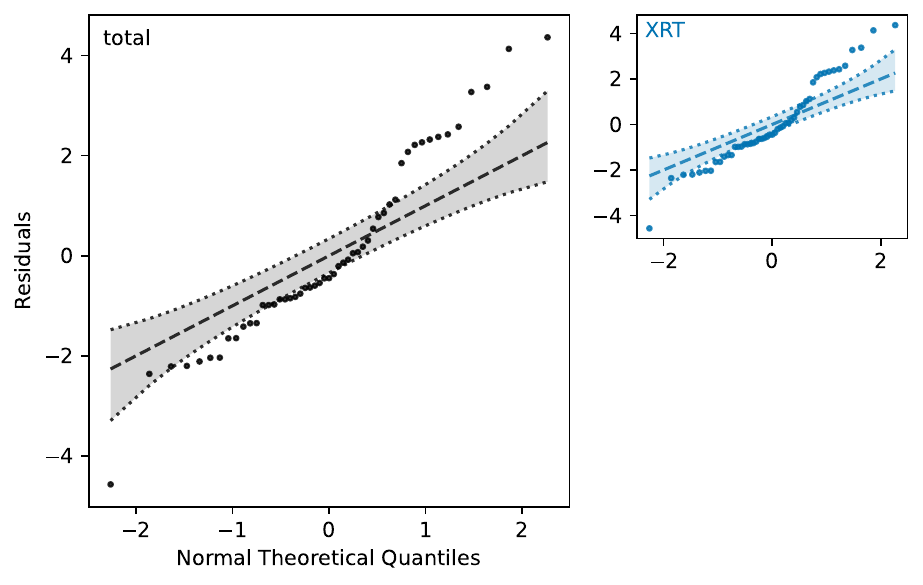}
        \caption{Q-Q plots of residuals.}
    \end{subfigure}
    \hfill
    \begin{subfigure}[t]{0.49\textwidth}
        \centering
        \includegraphics[width=\linewidth]{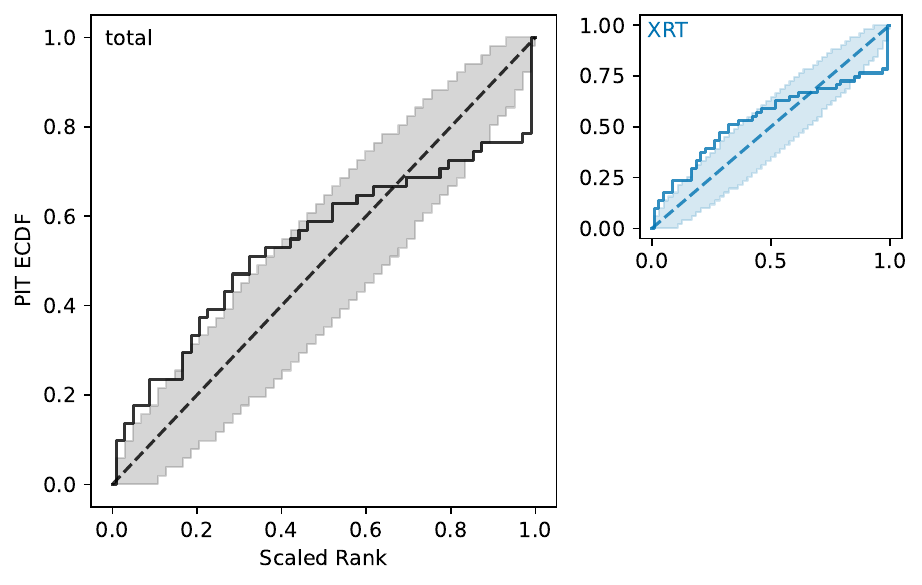}
        \caption{PIT-ECDF diagnostic.}
    \end{subfigure}
    \caption{XRT spectrum extracted from 87-832 s post-trigger, fitted with an absorbed PL model and shown with posterior diagnostics. This figure follows the format of Figure~\ref{fig:PowerLawSpec1}.}
    \label{fig:PowerLawXRT}
\end{figure*}

\begin{figure*}[ht!]
    \centering
    \begin{subfigure}[t]{0.49\textwidth}
        \centering
        \includegraphics[width=\linewidth]{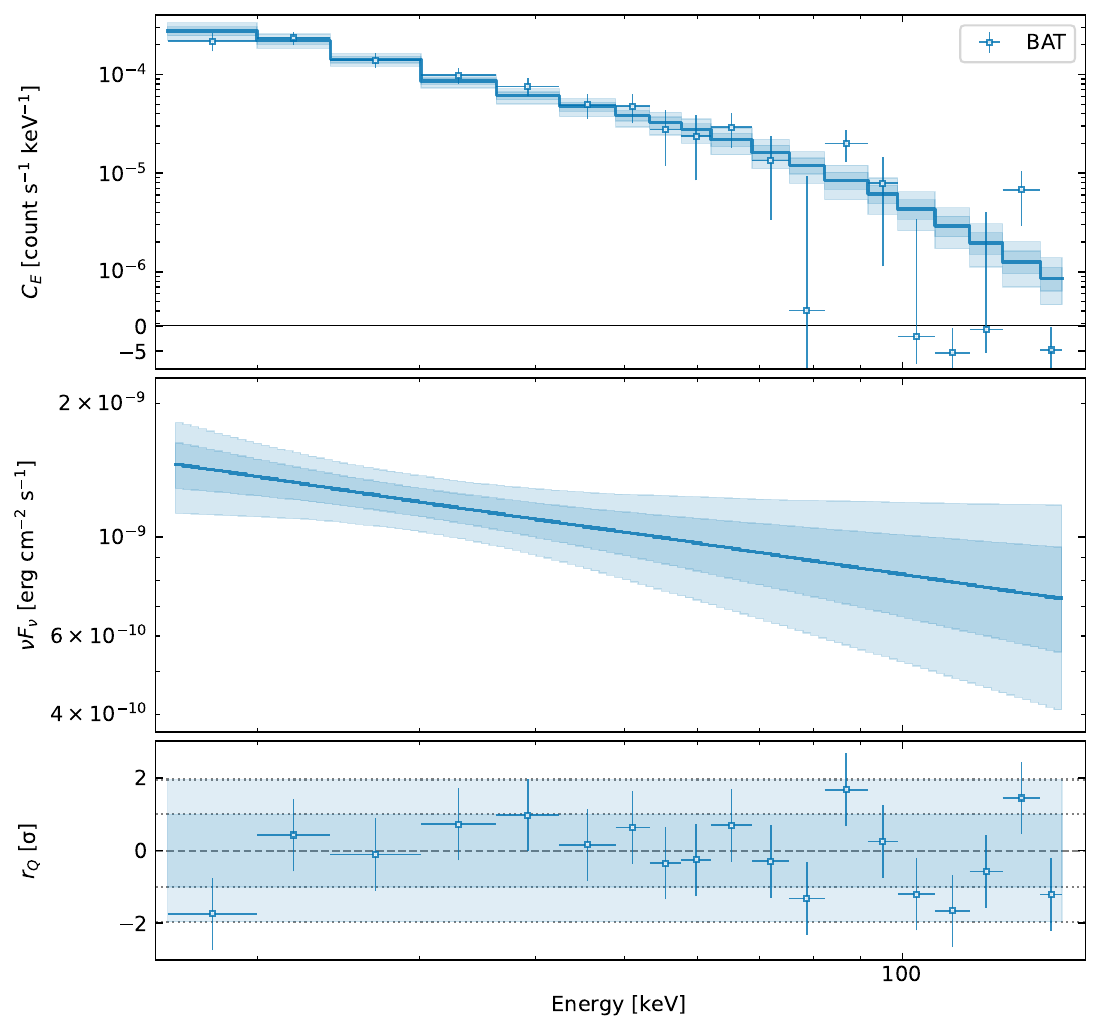}
        \caption{BAT spectrum, PL model, and residuals.}
    \end{subfigure}
    \hfill
    \begin{subfigure}[t]{0.49\textwidth}
        \centering
        \includegraphics[width=\linewidth]{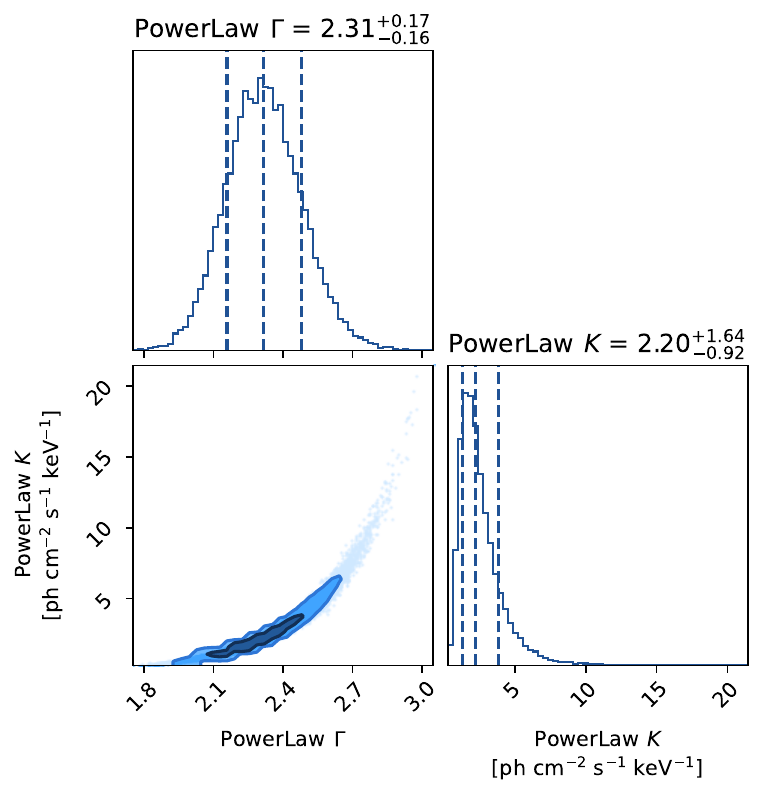}
        \caption{Posterior distributions of model parameters.}
    \end{subfigure}
    \vspace{1em}
    \begin{subfigure}[t]{0.49\textwidth}
        \centering
        \includegraphics[width=\linewidth]{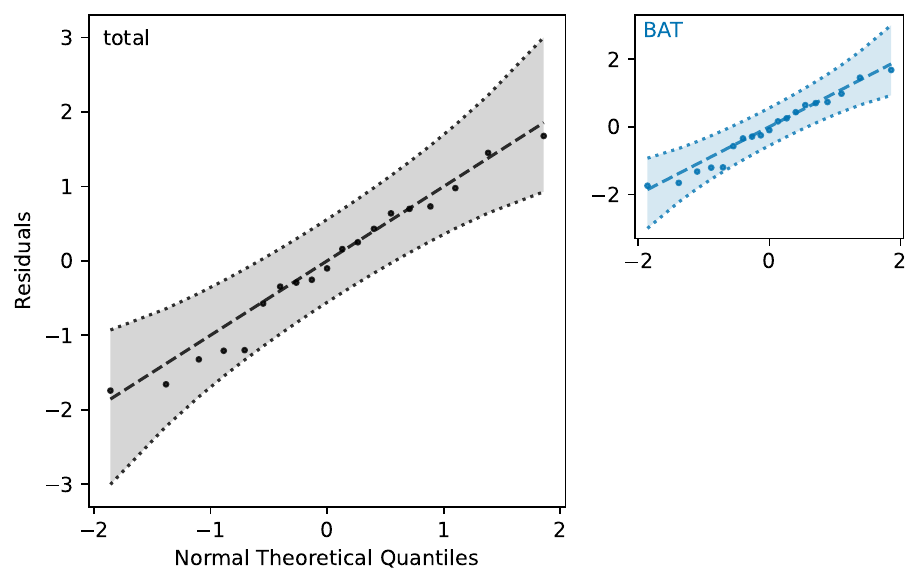}
        \caption{Q-Q plots of residuals.}
    \end{subfigure}
    \hfill
    \begin{subfigure}[t]{0.49\textwidth}
        \centering
        \includegraphics[width=\linewidth]{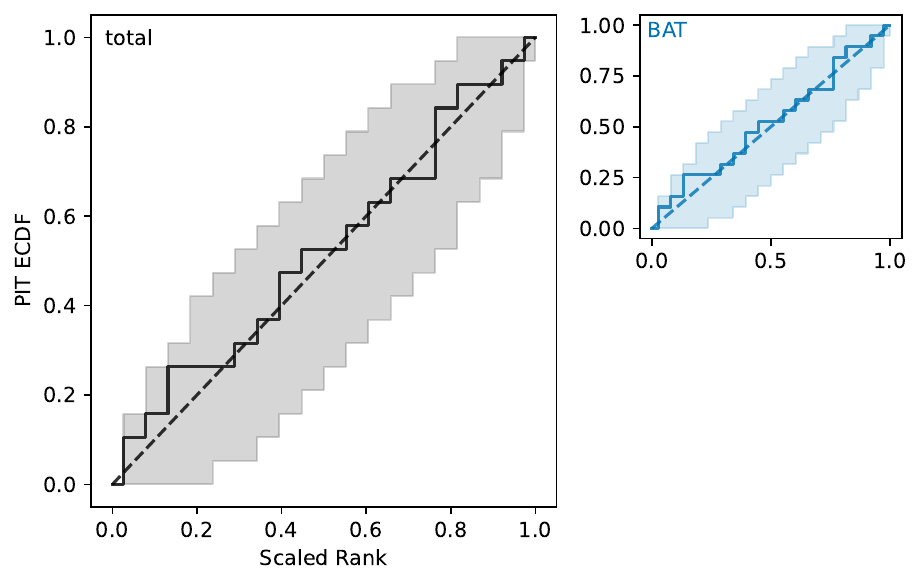}
        \caption{PIT-ECDF diagnostic.}
    \end{subfigure}
    \caption{BAT spectrum extracted from 87-832 s post-trigger, fitted with an unabsorbed PL model and shown with posterior diagnostics. This figure follows the format of Figure~\ref{fig:PowerLawSpec1}.}
    \label{fig:PowerLawBAT}
\end{figure*}

\begin{figure*}[ht!]
    \centering
    \begin{subfigure}[t]{0.49\textwidth}
        \centering
        \includegraphics[width=\linewidth]{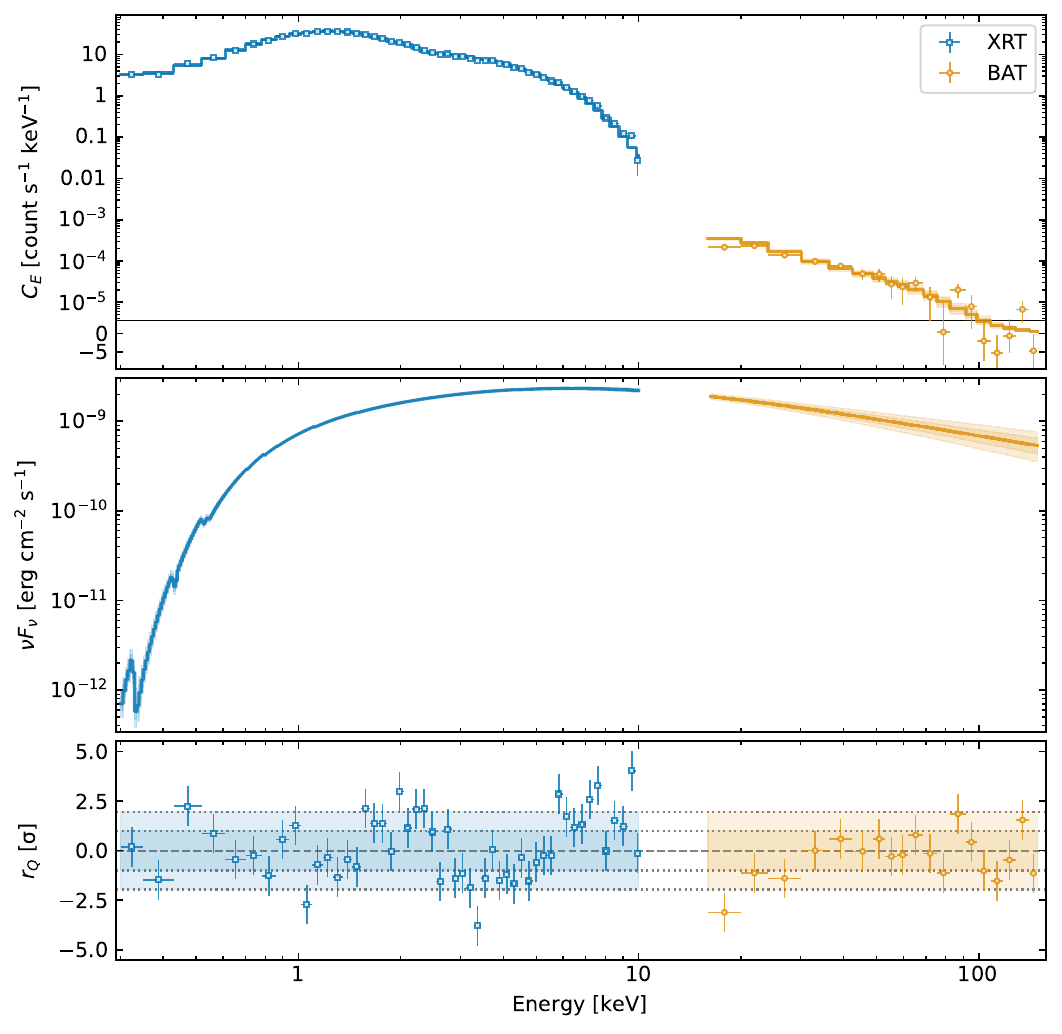}
        \caption{Joint spectrum from XRT and BAT, SBPL model, and residuals.}
    \end{subfigure}
    \hfill
    \begin{subfigure}[t]{0.49\textwidth}
        \centering
        \includegraphics[width=\linewidth]{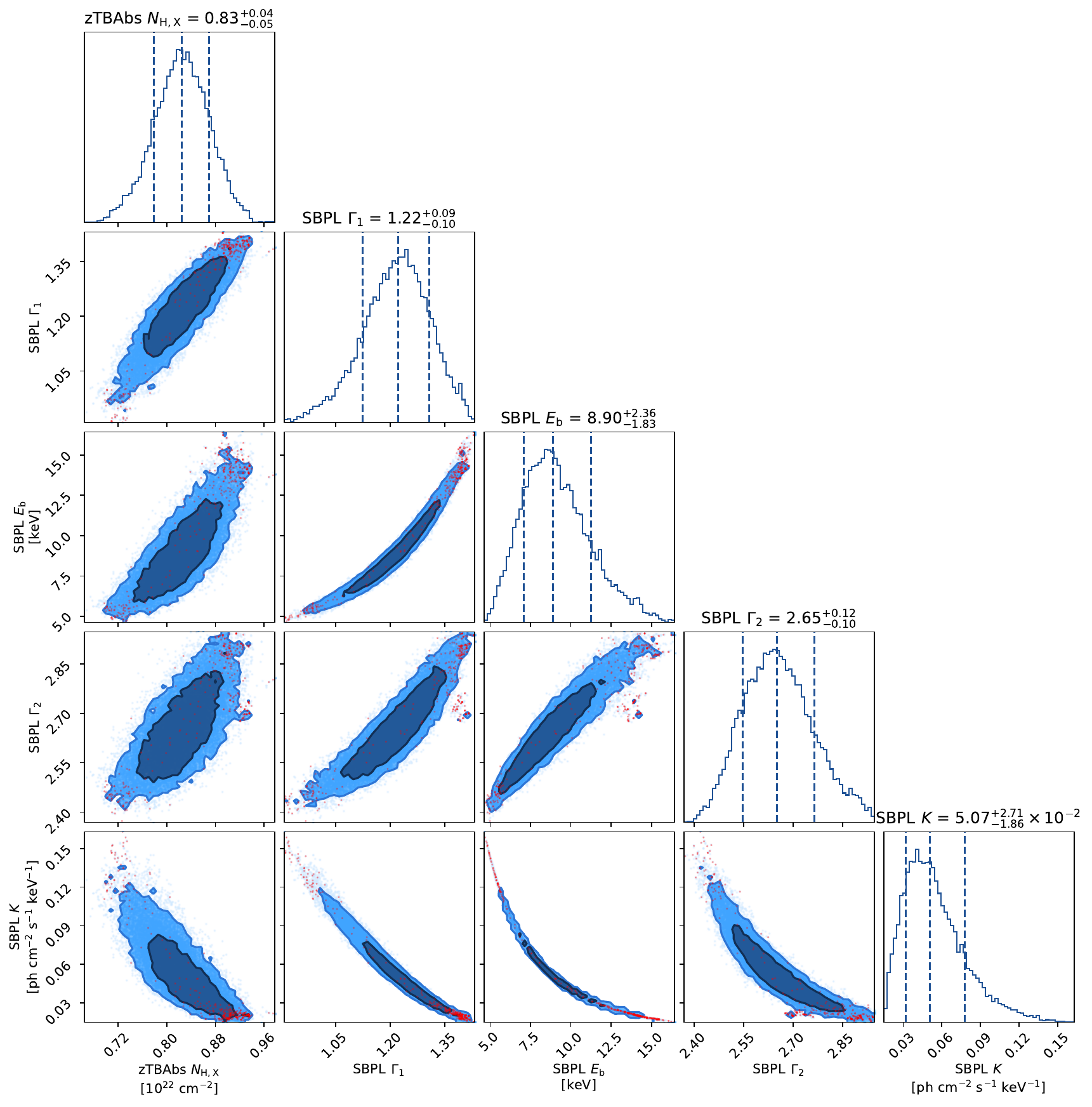}
        \caption{Posterior distributions of model parameters.}
    \end{subfigure}
    \vspace{1em}
    \begin{subfigure}[t]{0.49\textwidth}
        \centering
        \includegraphics[width=\linewidth]{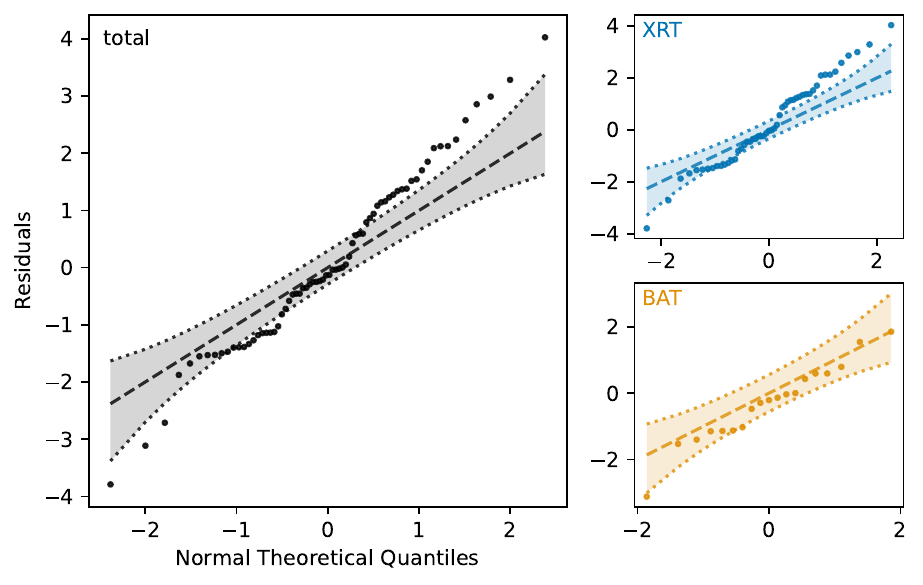}
        \caption{Q-Q plots of residuals.}
    \end{subfigure}
    \hfill
    \begin{subfigure}[t]{0.49\textwidth}
        \centering
        \includegraphics[width=\linewidth]{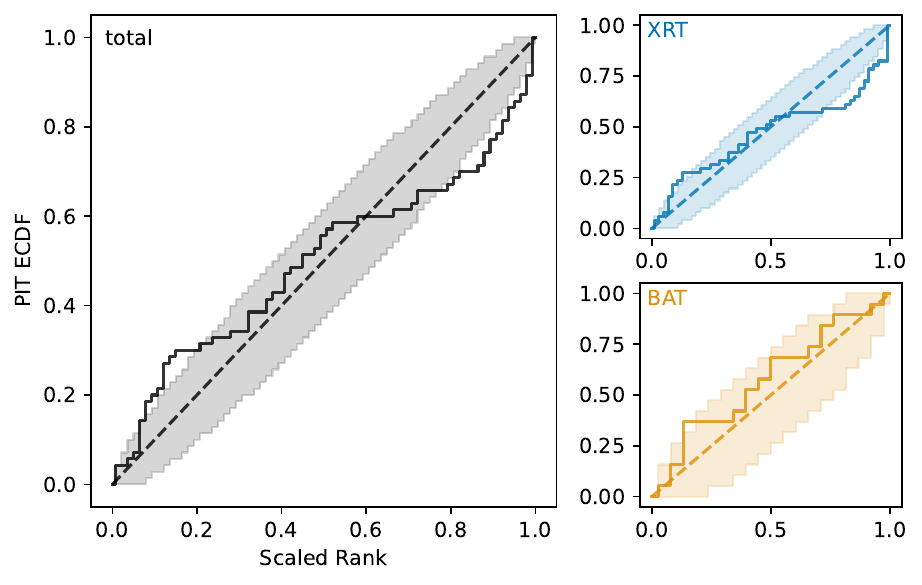}
        \caption{PIT-ECDF diagnostic.}
    \end{subfigure}
    \caption{BAT and XRT spectra extracted from 87-832 s post-trigger, fitted with an absorbed SBPL model and shown with posterior diagnostics. This figure follows the format of Figure~\ref{fig:PowerLawSpec1}.}
    \label{fig:SmoothlyBrokenPLJoint}
\end{figure*}

\begin{figure*}[ht!]
    \centering
    \begin{subfigure}[t]{0.49\textwidth}
        \centering
        \includegraphics[width=\linewidth]{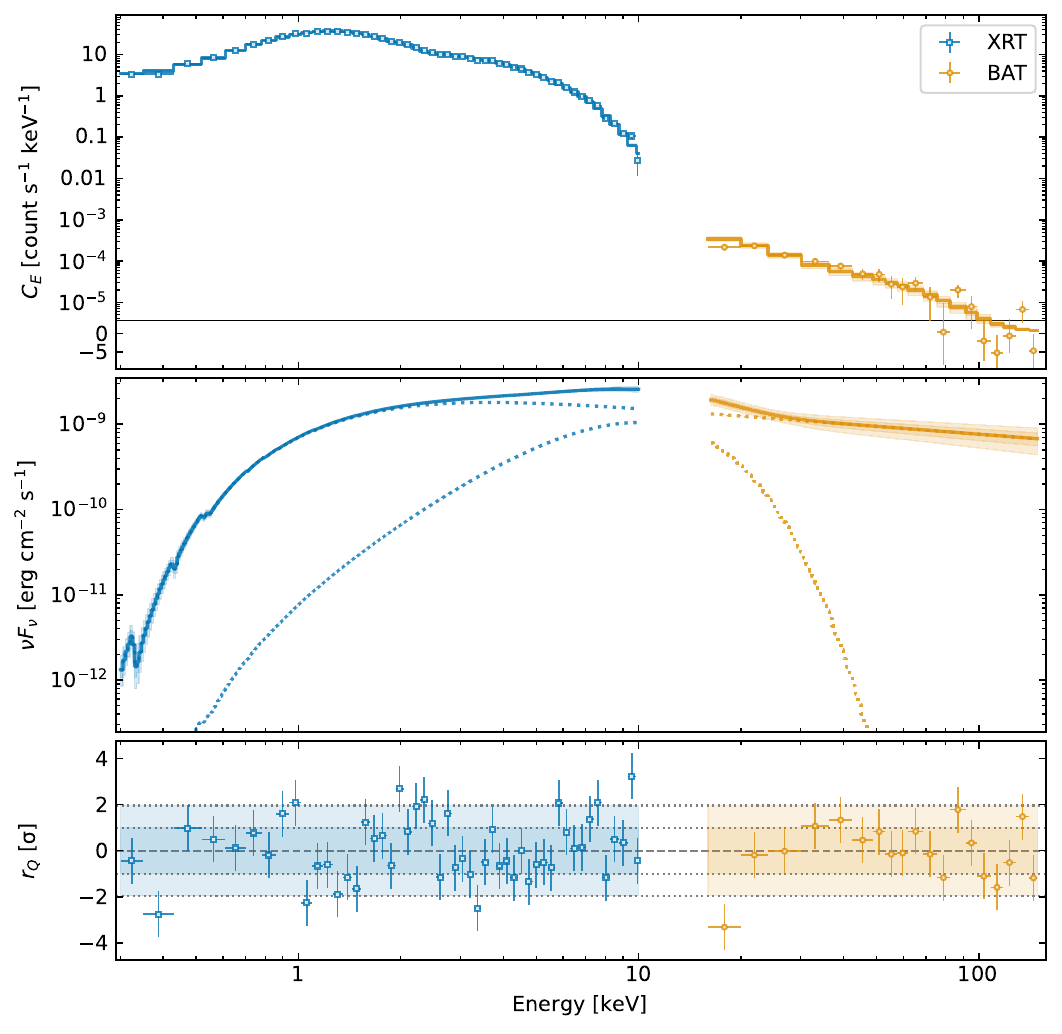}
        \caption{Joint spectrum from XRT and BAT, SBPL+BB model, and residuals.}
    \end{subfigure}
    \hfill
    \begin{subfigure}[t]{0.49\textwidth}
        \centering
        \includegraphics[width=\linewidth]{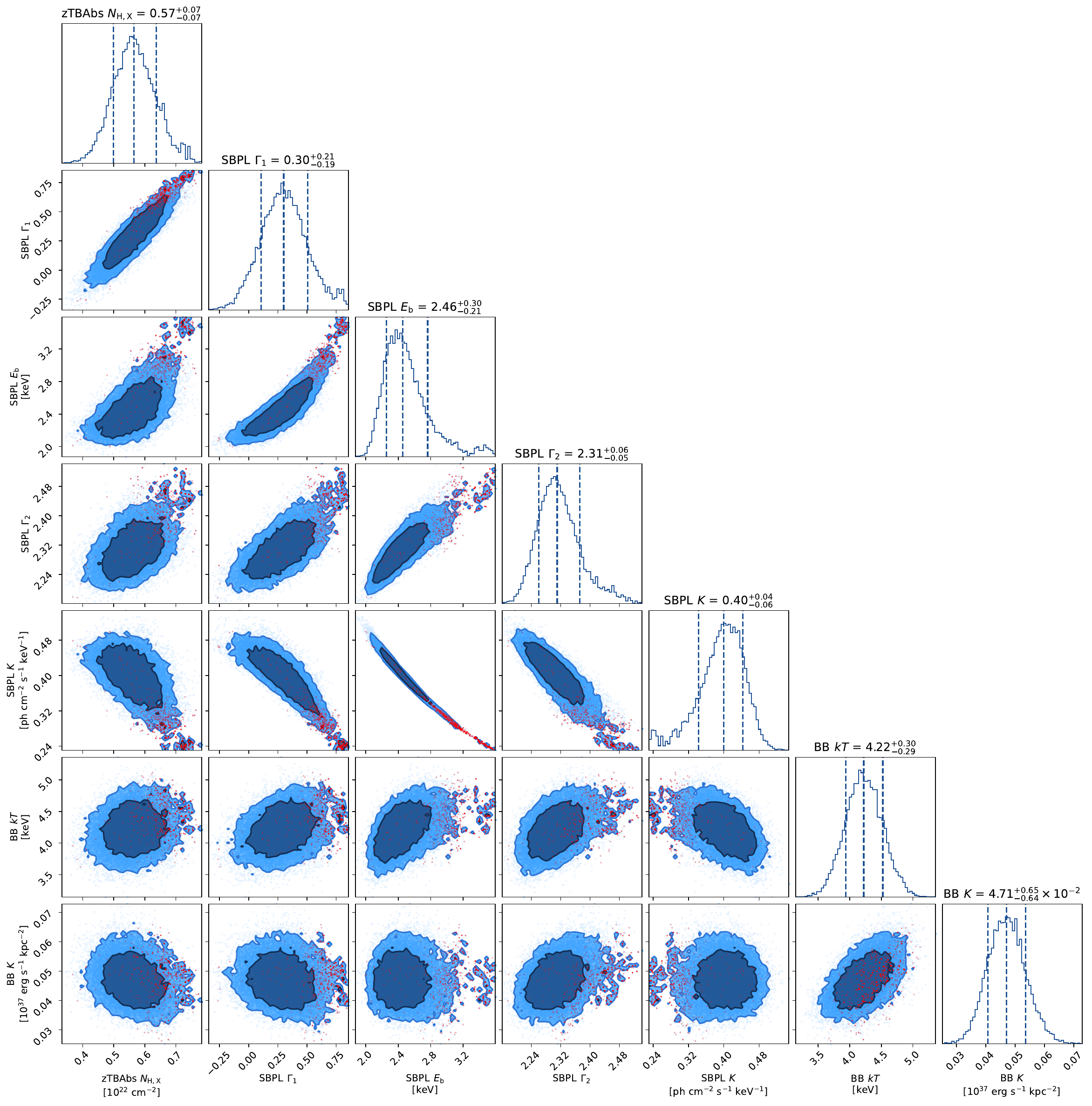}
        \caption{Marginal posterior distributions of model parameters.}
    \end{subfigure}
    \vspace{1em}
    \begin{subfigure}[t]{0.49\textwidth}
        \centering
        \includegraphics[width=\linewidth]{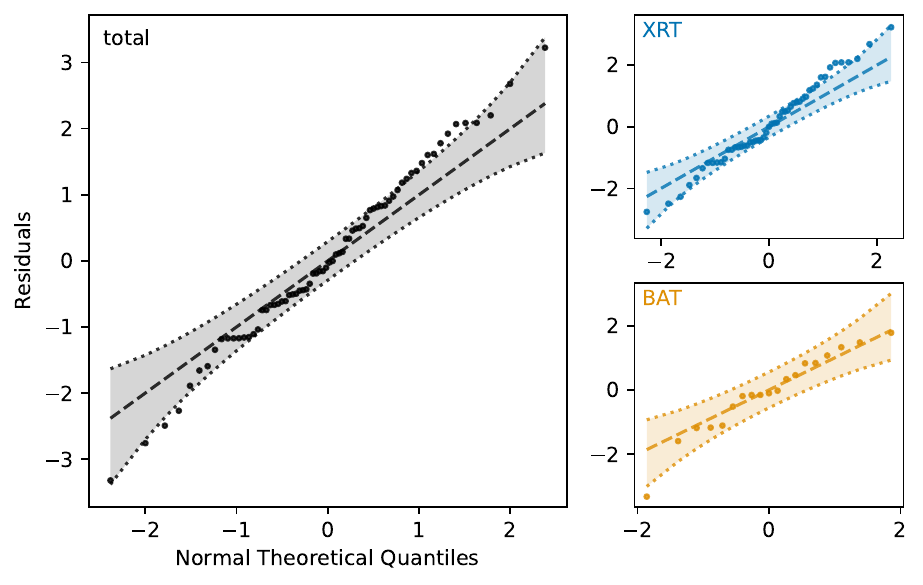}
        \caption{Q-Q plots of residuals.}
    \end{subfigure}
    \hfill
    \begin{subfigure}[t]{0.49\textwidth}
        \centering
        \includegraphics[width=\linewidth]{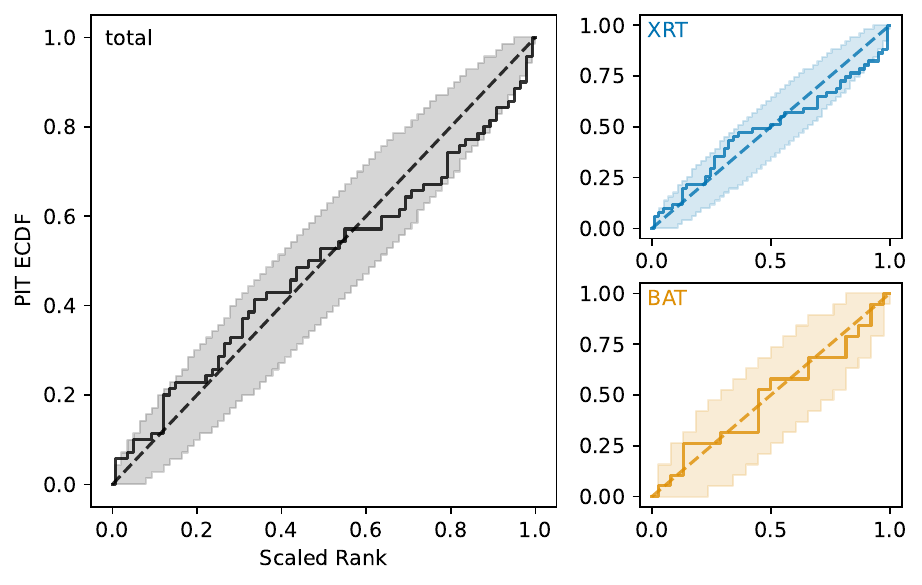}
        \caption{PIT-ECDF diagnostic.}
    \end{subfigure}
    \caption{BAT and XRT spectra extracted from 87-832 s post-trigger, fitted with an absorbed SBPL+BB model and shown with posterior diagnostics. This figure follows the format of Figure~\ref{fig:PowerLawSpec1}.}
    \label{fig:SmoothlyBrokenPLBlackbodyJoint}
\end{figure*}

\begin{figure}[htbp]
  \centering
  \includegraphics[width=\linewidth]{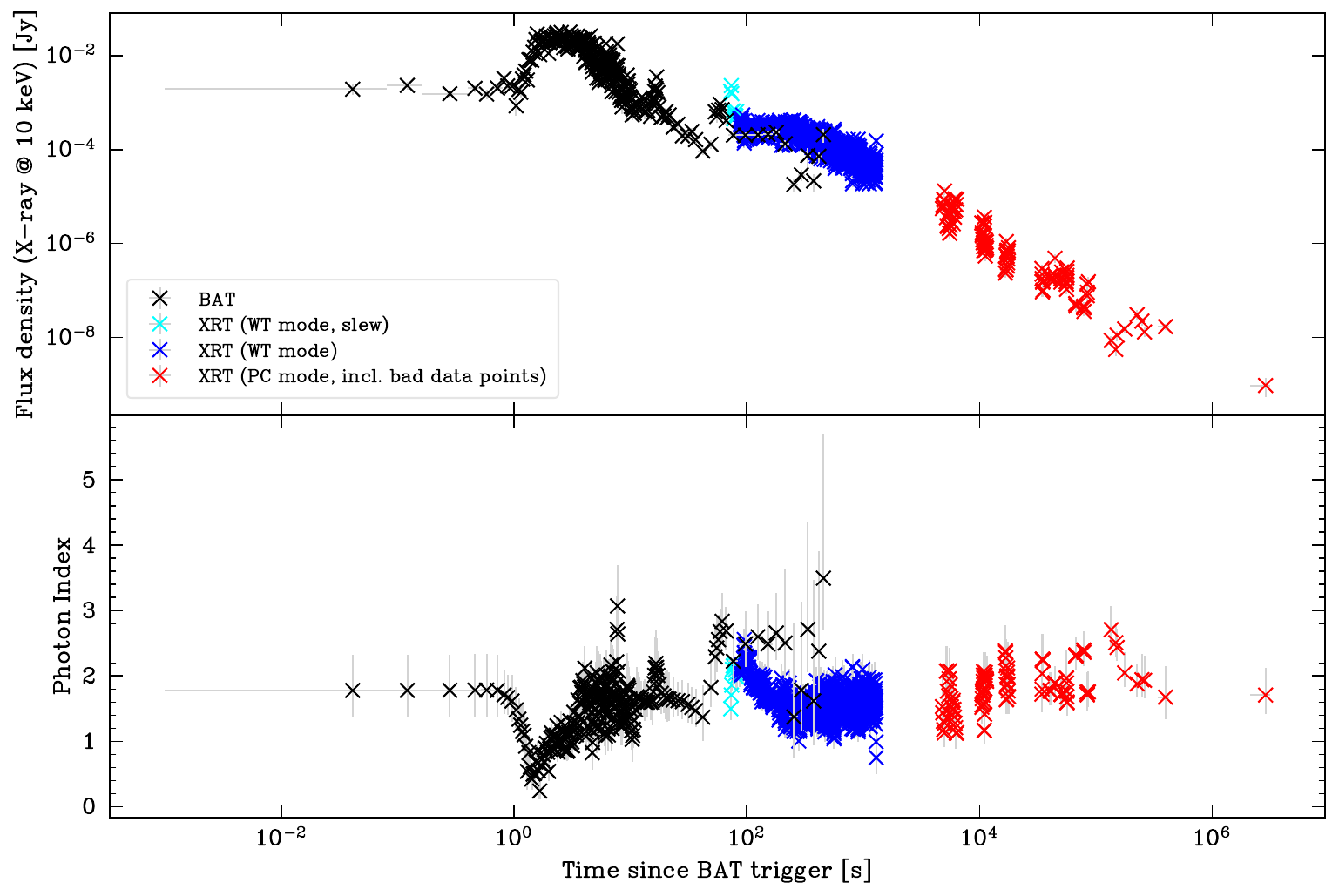}
  \caption{
    \textit{Upper panel:} Flux density light curve at 10\,keV, obtained from Swift/BAT (black crosses) and Swift/XRT observations. The XRT data are shown separately for the WT mode during spacecraft slew (cyan crosses), the standard WT mode (blue crosses), and the PC mode (red crosses, including bad data points).
    \textit{Lower panel:} The temporal evolution of the photon index. The measurements are color-coded as in the upper panel. This figure is taken from the burst analyser page of GRB 240825A (\url{https://www.swift.ac.uk/burst_analyser/01250617/}).
    }
  \label{fig:BAT_XRT_lc}
\end{figure}

The X-ray spectral fitting procedures are described in Section~\ref{sec:swift_data}. The corresponding results are presented in Tables~\ref{tab:time-sliced} and~\ref{tab:XRT+BAT}, following the same order as in the main text. The corresponding spectral plots (Figures~\ref{fig:PowerLawSpec1}-\ref{fig:SmoothlyBrokenPLBlackbodyJoint}) and the 10 keV flux density light curve (Figure~\ref{fig:BAT_XRT_lc}) are available at doi:\dataset[10.5281/zenodo.17059771]{\doi{10.5281/zenodo.17059771}} (The link is available, and all plots are presented below in the manuscript for ease of reference). The Leave-One-Out Information Criterion (LOOIC), the Widely Applicable Information Criterion \citep[WAIC;][]{2010arXiv1004.2316W,2015arXiv150704544V}, the Quantile-Quantile (Q-Q) plot of residuals, the Probability Integral Transformation-Empirical Cumulative Distribution Function (PIT-ECDF), and the Pareto $k$ diagnostic are used for posterior diagnostics.

% https://homepage.cs.uiowa.edu/~luke/classes/STAT4580/slides/qqpp.html
% \begin{rotatetable}
\startlongtable
\begin{deluxetable*}{ccc|cc|c|cc|cc|cccc|c}
\tablecaption{Time slices used for time-resolved XRT spectral analysis and its results.}
\tabletypesize{\scriptsize}%\tiny %\scriptsize
\setlength{\tabcolsep}{0.02in}
\tablewidth{0pt}
\tablehead{
\colhead{ID} &
\colhead{Time Slice} & 
\colhead{$T_\mathrm{mid}$} & 
\multicolumn{2}{c}{Parameters} &
\colhead{stat/(d.o.f.)} &
\multicolumn{4}{c}{Information Criterion} & 
\multicolumn{4}{c}{Pareto $k$ diagnostic} &
\colhead{Figure} \\
\colhead{} & 
\colhead{} & 
\colhead{} & 
\colhead{$N_\mathrm{H,X}$} & 
\colhead{$\Gamma$} & 
\colhead{} &
\multicolumn{2}{c}{LOOIC} &
\multicolumn{2}{c}{WAIC} &
\colhead{(-Inf,0.5]} & 
\colhead{(0.5,0.7]} & 
\colhead{(0.7,1]} &
\colhead{(1,Inf)} &
\colhead{} \\
\colhead{} & 
\colhead{[s]} & 
\colhead{[s]} & 
\colhead{[$10^{22}\mathrm{\,cm^{-2}}$]} & 
\colhead{} & 
\colhead{} &
\colhead{Deviance} & 
\colhead{p} & 
\colhead{Deviance} & 
\colhead{p} & 
\colhead{good} & 
\colhead{ok} & 
\colhead{bad} &
\colhead{very bad} &
\colhead{}
}
\startdata
1 & 87-97 & 91 & $1.60_{-0.17}^{+0.18}$ & $2.53_{-0.10}^{+0.11}$ & 33.47/30 & $36.37 \pm 7.31$ & 2.10 & $36.33 \pm 7.31$ & 2.07 & 100.0\% & 0 & 0 & 0 & \ref{fig:PowerLawSpec1} \\
2 & 97-109 & 102 & $1.15_{-0.11}^{+0.12}$ & $2.28_{-0.08}^{+0.08}$ & 41.09/35 & $45.36 \pm 10.50$ & 3.20 & $45.18 \pm 10.47$ & 3.11 & 97.4\% & 2.6\% & 0 & 0 & \ref{fig:PowerLawSpec2} \\
3 & 109-149 & 127 & $0.70_{-0.06}^{+0.06}$ & $1.71_{-0.04}^{+0.04}$ & 43.00/39 & $46.86 \pm 8.41$ & 2.94 & $46.78 \pm 8.40$ & 2.90 & 100.0\% & 0 & 0 & 0 & \ref{fig:PowerLawSpec3} \\
4 & 149-203 & 174 & $0.82_{-0.08}^{+0.08}$ & $1.62_{-0.04}^{+0.04}$ & 51.04/38 & $57.36 \pm 11.53$ & 5.05 & $57.19 \pm 11.48$ & 4.96 & 97.6\% & 2.4\% & 0 & 0 & \ref{fig:PowerLawSpec4} \\
5 & 203-265 & 232 & $1.05_{-0.07}^{+0.07}$ & $1.57_{-0.03}^{+0.03}$ & 91.35/41 & $99.06 \pm 16.77$ & 6.41 & $98.85 \pm 16.73$ & 6.30 & 97.7\% & 2.3\% & 0 & 0 & \ref{fig:PowerLawSpec5} \\
6 & 265-340 & 301 & $1.23_{-0.07}^{+0.08}$ & $1.58_{-0.03}^{+0.03}$ & 69.36/41 & $75.75 \pm 17.24$ & 5.15 & $75.61 \pm 17.20$ & 5.08 & 97.7\% & 2.3\% & 0 & 0 & \ref{fig:PowerLawSpec6} \\
7 & 340-421 & 379 & $1.15_{-0.08}^{+0.08}$ & $1.62_{-0.03}^{+0.03}$ & 93.48/41 & $106.28 \pm 26.95$ & 10.48 & $106.14 \pm 26.88$ & 10.41 & 97.7\% & 0 & 2.3\% & 0 & \ref{fig:PowerLawSpec7} \\
8 & 421-482 & 451 & $1.28_{-0.10}^{+0.11}$ & $1.66_{-0.04}^{+0.04}$ & 42.55/39 & $46.43 \pm 8.89$ & 2.99 & $46.35 \pm 8.88$ & 2.95 & 100.0\% & 0 & 0 & 0 & \ref{fig:PowerLawSpec8} \\
9 & 482-569 & 525 & $1.25_{-0.09}^{+0.09}$ & $1.65_{-0.04}^{+0.04}$ & 47.37/40 & $52.06 \pm 8.89$ & 3.62 & $51.94 \pm 8.87$ & 3.56 & 100.0\% & 0 & 0 & 0 & \ref{fig:PowerLawSpec9} \\
10 & 569-709 & 637 & $1.12_{-0.07}^{+0.07}$ & $1.64_{-0.03}^{+0.03}$ & 40.90/40 & $45.04 \pm 7.65$ & 3.13 & $44.91 \pm 7.61$ & 3.06 & 100.0\% & 0 & 0 & 0 & \ref{fig:PowerLawSpec10} \\
11 & 709-808 & 758 & $1.18_{-0.09}^{+0.10}$ & $1.68_{-0.04}^{+0.04}$ & 47.68/38 & $52.26 \pm 8.90$ & 3.56 & $52.14 \pm 8.88$ & 3.50 & 100.0\% & 0 & 0 & 0 & \ref{fig:PowerLawSpec11} \\
12 & 808-940 & 873 & $1.21_{-0.09}^{+0.10}$ & $1.70_{-0.04}^{+0.04}$ & 61.44/39 & $	68.05 \pm 15.82$ & 5.24 & $67.81 \pm 15.72$ & 5.12 & 100.0\% & 0 & 0 & 0 & \ref{fig:PowerLawSpec12} \\
13 & 940-1074 & 1005 & $1.27_{-0.10}^{+0.10}$ & $1.78_{-0.05}^{+0.05}$ & 45.18/38 & $	48.41 \pm 9.47$ & 2.41 & $48.36 \pm 9.46$ & 2.38 & 100.0\% & 0 & 0 & 0 & \ref{fig:PowerLawSpec13} \\
14 & 1074-1295 & 1182 & $1.25_{-0.08}^{+0.09}$ & $1.75_{-0.04}^{+0.04}$ & 41.35/40 & $	45.22 \pm 8.40$ & 2.96 & $45.14 \pm 8.37$ & 2.92 & 100.0\% & 0 & 0 & 0 & \ref{fig:PowerLawSpec14} \\
15 & 4797-8430 & 5568 & $0.96_{-0.21}^{+0.24}$ & $1.59_{-0.11}^{+0.11}$ & 36.16/36 & $39.46 \pm 8.38$ & 2.40 & $39.39 \pm 8.37$ & 2.36 & 100.0\% & 0 & 0 & 0 & \ref{fig:PowerLawSpec15} \\
16 & 8430-14129 & 10986 & $0.94_{-0.17}^{+0.19}$ & $1.80_{-0.11}^{+0.11}$ & 37.85/37 & $43.01 \pm 8.26$ & 3.80 & $42.74 \pm 8.14$ & 3.67 & 97.5\% & 2.5\% & 0 & 0 & \ref{fig:PowerLawSpec16} \\
17 & 14129-26067 & 17255 & $0.91_{-0.18}^{+0.20}$ & $1.89_{-0.13}^{+0.13}$ & 42.24/33 & $45.26 \pm 6.92$ & 2.17 & $45.21 \pm 6.92$ & 2.15 & 100.0\% & 0 & 0 & 0 & \ref{fig:PowerLawSpec17} \\
18 & 26067-61951 & 45473 & $0.98_{-0.17}^{+0.19}$ & $1.86_{-0.11}^{+0.11}$ & 34.54/35 & $37.30 \pm 8.89$ & 1.94 & $37.26 \pm 8.89$ & 1.92 & 100.0\% & 0 & 0 & 0 & \ref{fig:PowerLawSpec18} \\
19 & 61951-110958 & 77128 & $0.69_{-0.21}^{+0.25}$ & $2.03_{-0.18}^{+0.18}$ & 38.60/31 & $41.87 \pm 8.81$ & 2.39 & $41.81 \pm 8.81$ & 2.35 & 97.1\% & 2.9\% & 0 & 0 & \ref{fig:PowerLawSpec19} \\
20 & 110958-1637153 & 246172 & $0.89_{-0.28}^{+0.34}$ & $1.80_{-0.19}^{+0.20}$ & 36.40/29 & $40.66 \pm 7.68$ & 3.14 & $40.46 \pm 7.62$ & 3.04 & 96.9\% & 3.1\% & 0 & 0 & \ref{fig:PowerLawSpec20} \\
\tableline
21 & 86-1297 & 515 & $1.10_{-0.02}^{+0.02}$ & $1.66_{-0.01}^{+0.01}$ & 183.12/49 & $197.86 \pm 32.85$ & 12.72 & $197.39 \pm 32.75$ & 12.49 & 96.2\% & 0 & 3.8\% & 0 & \ref{fig:PowerLawWT} \\ % WT
22 & 4773-89889 & 26164 & $0.90_{-0.09}^{+0.09}$ & $1.79_{-0.05}^{+0.05}$ & 41.40/43 & $44.49 \pm 9.92$ & 2.25 & $44.42 \pm 9.91$ & 2.22 & 100.0\% & 0 & 0 & 0 & \ref{fig:PowerLawPC} \\ % PC
% \tableline
% phase1 & 87-102 & 94 & $1.35_{-0.11}^{+0.11}$ & $2.43_{-0.07}^{+0.07}$ & 37.52/35 & $ 41.02 \pm 7.70$ & 2.60 & $40.93 \pm 7.69$ & 2.56 & 100.0\% & 0 & 0 & 0 \\
% phase2 & 102-157 & 128 & $0.78_{-0.05}^{+0.05}$ & $1.76_{-0.03}^{+0.03}$ & 54.36/40 & $58.47 \pm 11.16$ & 3.23 & $58.41 \pm 11.15$ & 3.20 & 100.0\% & 0 & 0 & 0 \\
% phase3 & 157-407 & 282 & $1.09_{-0.04}^{+0.04}$ & $1.59_{-0.02}^{+0.02}$ & 147.90/45 & $161.89 \pm 30.95$ & 11.81 & $161.29 \pm 30.68$ & 11.51 & 97.9\% & 0 & 2.1\% & 0 \\
% phase4 & 407-1288 & 756 & $1.20_{-0.03}^{+0.03}$ & $1.68_{-0.01}^{+0.01}$ & 98.65/47 & $107.42 \pm 20.59$ & 7.30 & $107.14 \pm 20.52$ & 7.16 & 98.0\% & 2.0\% & 0 & 0 \\
% phase5 & 1288-1297 & 1292 & $2.31_{-0.99}^{+1.41}$ & $1.54_{-0.28}^{+0.34}$ & 41.94/24 & $49.50 \pm 10.49$ & 5.87 & $49.23 \pm 10.47$ & 5.74 & 96.3\% & 3.7\% & 0 & 0 \\
% phase6 & 4773-3129684 & 54328 & $0.97_{-0.09}^{+0.09}$ & $1.82_{-0.05}^{+0.05}$ & 51.48/43 & $55.03 \pm 12.01$ & 2.67 & $54.94 \pm 12.01$ & 2.63 & 100.0\% & 0 & 0 & 0 \\
\enddata
\tablecomments{The uncertainties for the above parameters are reported at the $1\sigma$ credible interval.}
\tablecomments{The time slices numbered 21 and 22 are the time-averaged spectra in the WT and PC modes, respectively.}
% \tablecomments{The spectra labeled as phases 1-6 are available on the XRT Live Catalog page for GRB~240825A (\url{https://www.swift.ac.uk/xrt_live_cat/01250617/}). The corresponding spectral analysis results are also listed there.}
\label{tab:time-sliced}
\end{deluxetable*}
% \end{rotatetable}

\begin{sidewaystable}
\startlongtable
\begin{deluxetable*}{ccc|ccccc|c|cc|cc|cccc|c}
\tablecaption{Time slices used for BAT or/and XRT spectral analysis and its results.}
\tabletypesize{\scriptsize}%\tiny %\scriptsize
\setlength{\tabcolsep}{0.02in}
\tablewidth{0pt}
\tablehead{
\colhead{Spectrum} &
\colhead{Time Slice} & 
\colhead{$T_\mathrm{mid}$} & 
\multicolumn{5}{c}{Parameters} &
\colhead{stat/(d.o.f.)} &
\multicolumn{4}{c}{Information Criterion} & 
\multicolumn{4}{c}{Pareto $k$ diagnostic} &
\colhead{Figure} \\
\colhead{} & 
\colhead{} & 
\colhead{} & 
\colhead{$N_\mathrm{H,X}$} & 
\colhead{$\Gamma_1$} & 
\colhead{$E_\mathrm{b}$} &
\colhead{$\Gamma_2$} & 
\colhead{$kT$} & 
\colhead{} &
\multicolumn{2}{c}{LOOIC} &
\multicolumn{2}{c}{WAIC} &
\colhead{(-Inf,0.5]} & 
\colhead{(0.5,0.7]} & 
\colhead{(0.7,1]} &
\colhead{(1,Inf)} &
\colhead{} \\
\colhead{} & 
\colhead{[s]} & 
\colhead{[s]} & 
\colhead{[$10^{22}\mathrm{\,cm^{-2}}$]} & 
\colhead{} & 
\colhead{[keV]} &
\colhead{} & 
\colhead{[keV]} &
\colhead{} &
\colhead{Deviance} & 
\colhead{p} & 
\colhead{Deviance} & 
\colhead{p} & 
\colhead{good} & 
\colhead{ok} & 
\colhead{bad} &
\colhead{very bad} &
\colhead{}
}
\startdata
XRT & 87-832 & 393 & $1.07_{-0.02}^{+0.02}$ & $1.64_{-0.01}^{+0.01}$ & --- & --- & --- & 159.94/48 & $174.27 \pm 33.22$ & 12.15 & $173.63 \pm 32.97$ & 11.84 & 96.1\% & 2.0\% & 2.0\% & 0 & \ref{fig:PowerLawXRT} \\
BAT & 87-832 & 393 & --- & --- & --- & $2.31_{-0.16}^{+0.17}$ & --- & 	17.63/17 & $20.68 \pm 4.56$ & 1.80 & $20.22 \pm 4.33$ & 1.88 & 94.7\% & 0 & 5.3\% & 0 & \ref{fig:PowerLawBAT} \\
XRT+BAT & 87-832 & 393 & $0.83_{-0.05}^{+0.04}$ & $1.22_{-0.10}^{+0.09}$ & $8.90_{-1.83}^{+2.36}$ & $2.65_{-0.10}^{+0.12}$ & --- & 153.70/65 & $166.53 \pm 25.77$ & 10.95 & $166.07 \pm 25.72$ & 10.73 & 97.1\% & 2.9\% & 0 & 0 & \ref{fig:SmoothlyBrokenPLJoint} \\
XRT+BAT & 87-832 & 393 & $0.57_{-0.07}^{+0.07}$ & $0.30_{-0.19}^{+0.21}$ & $2.46_{-0.21}^{+0.31}$ & $2.31_{-0.05}^{+0.06}$ & $4.22_{-0.29}^{+0.30}$ & 112.51/63 & $126.93 \pm 19.50$ & 11.62 & $125.85 \pm 19.11$ & 11.08 & 97.1\% & 1.4\% & 1.4\% & 0 & \ref{fig:SmoothlyBrokenPLBlackbodyJoint} \\
\enddata
\vspace{2em}
\tablecomments{The uncertainties for the above parameters are reported at the $1\sigma$ credible interval.}
\tablecomments{For the XRT data alone, an absorbed power-law model was used for the spectral analysis, with the intrinsic column density and photon index denoted as $N_\mathrm{H,X}$ and $\Gamma_1$, respectively (see the first row of the table above). For the BAT data alone, an unabsorbed power-law model was used for the spectral analysis, with the photon index denoted as $\Gamma_2$ (see the second row of the table above).}
\tablecomments{The joint XRT and BAT spectrum was first fitted with an absorbed SBPL model (see the third row of the table above). We then applied an absorbed SBPL+BB model to fit the same spectrum (see the fourth row of the table above).}
\label{tab:XRT+BAT}
\end{deluxetable*}
\end{sidewaystable}
\clearpage

\clearpage

\begin{center}
    Online Materials
\end{center}

The detailed procedures of the spectral fitting are described in Section~\ref{sec:swift_data} of the main text. The results of the spectral fitting are presented in Tables~\ref{tab:time-sliced} and~\ref{tab:XRT+BAT} of the main text, and are listed here in the same order as presented in the article.

% The Quantile-Quantile (Q-Q) plot of residuals, the Probability Integral Transformation-Empirical Cumulative Distribution Function (PIT-ECDF) are used for posterior diagnostics.

% The Leave-One-Out Information Criterion (LOOIC), the Widely Applicable Information Criterion (WAIC) \citep[][]{2010arXiv1004.2316W,2015arXiv150704544V}, the Quantile-Quantile (Q-Q) plot of residuals, the Probability Integral Transformation-Empirical Cumulative Distribution Function (PIT-ECDF), and the Pareto $k$ diagnostic are used for posterior diagnostics.

\clearpage

The early spectral curvature was identified in the X-ray emission of GRB 240825A. Therefore, the 10 keV flux density light curve, constructed from both BAT and XRT data, was extracted from the burst analyser page of GRB 240825A (see Figure~\ref{fig:BAT_XRT_lc}). The 10\,keV flux density light curve allows a direct comparison between BAT and XRT data, while minimizing the effect of extrapolation.

\clearpage
\section{Photometric Data on the Optical Afterglow of GRB 240825A}\label{sec:opt_data}
The GMG data were collected using two filters, including the standard Johnson-Cousins $R$ and the Sloan $r$ bands. All raw images underwent customized pre-processing tailored to the specifications of the telescope, including steps such as bias correction, flat-fielding, and cosmic-ray removal. The initial astrometric solution of each frame was derived using the \texttt{Astrometry.net} service\footnote{\url{https://nova.astrometry.net}} \citep{2010AJ....139.1782L}. Three frames were acquired on the fourth night. However, only two were resampled and co-added using the \texttt{SWarp} software \citep{2010ascl.soft10068B} to enhance the SNR of the images, as one frame was of poor quality. For the first frame obtained on the first night, with a midtime of 1.91 hours post-trigger, we performed transient detection, refined astrometry, and point-spread function (PSF) photometry at the transient position \citep{2024GCN.37280....1L} using \texttt{STDPipe} \citep{2021ascl.soft12006K}. For the frames obtained on the subsequent three nights, PSF photometry was conducted at the transient position. 
To perform photometric calibration, field stars from two reference catalogs were chosen according to the filter system used in the observations. For frames obtained with Johnson-Cousins filters, the Gaia DR3 Syntphot catalog \citep{2023A&A...674A..33G} was employed, which generates synthetic photometry based on low-resolution stellar spectra (XP spectra) observed by the Gaia mission. For frames captured with Sloan filters, the Pan-STARRS DR1 catalog \citep{2016arXiv161205560C} was utilized.

The optical photometric data of GRB~240825A, combined from GMG and GCN Circulars, are compiled in Table~\ref{tab:gcn-data}.

\startlongtable
\begin{deluxetable*}{ccccc}
\tablecaption{Follow-up observation data of GRB 240825A, combined from GMG and GCN Circulars.}
\tabletypesize{\scriptsize}
% \setlength{\tabcolsep}{0.02in}
% \tablewidth{0pt}
\tablehead{
\colhead{$T_\mathrm{mid}$} & 
\colhead{Filter} & 
\colhead{Magnitude} & 
\colhead{Facility} & 
\colhead{Reference}\\ 
\colhead{[$\mathrm{s}$]} & 
\colhead{} & 
\colhead{[mag]} & 
\colhead{ } & 
\colhead{}\\
\colhead{(1)} & 
\colhead{(2)} & 
\colhead{(3)} & 
\colhead{(4)} & 
\colhead{(5)}
}
\startdata
% $167 \pm 75$ & $White$ & $15.57 \pm 0.14$ & Swift-UVOT & \citet{2024GCN.37274....1G} \\
$168 \pm 73.50$ & $White$ & $15.58 \pm 0.02$ & Swift-UVOT & \citet{2024GCN.37296....1K} \\
% $187 \pm 30$ & $r$ & $14.46 \pm 0.01$ & Mephisto & \citet{2024GCN.37278....1Z} \\
% $247 \pm 60$ & $z$ & $13.61 \pm 0.01$ & Mephisto & \citet{2024GCN.37278....1Z} \\
$360 \pm 7$ & $I$ & $14.96 \pm 0.04$ & Skynet & \citet{2024GCN.37276....1D} \\
$428 \pm 123$ & $U$ & $16.93 \pm 0.05$ & Swift-UVOT & \citet{2024GCN.37296....1K} \\
$571 \pm 10$ & $B$ & $17.54 \pm 0.15$ & Swift-UVOT & \citet{2024GCN.37296....1K} \\
$713 \pm 29$ & $UVW1$ & $18.55 \pm 0.30$ & Swift-UVOT & \citet{2024GCN.37296....1K} \\
$817 \pm 10$ & $V$ & $17.02 \pm 0.20$ & Swift-UVOT & \citet{2024GCN.37296....1K} \\
$849 \pm 30$ & $G$ & $17$ & Nanshan/HMT & \citet{2024GCN.37275....1J} \\
% $970 \pm 30$ & $g$ & $17.49 \pm 0.03$ & Mephisto & \citet{2024GCN.37278....1Z} \\
% $1030 \pm 60$ & $i$ & $16.42 \pm 0.01$ & Mephisto & \citet{2024GCN.37278....1Z} \\
% $1090 \pm 90$ & $u$ & $19.15 \pm 0.09$ & Mephisto & \citet{2024GCN.37278....1Z} \\
$4590 \pm 630$ & $Ic$ & $17.30 \pm 0.15$ & AKO & \citet{2024GCN.37299....1O} \\
$5725 \pm 1170$ & $Ic$ & $17.60 \pm 0.16$ & AKO & \citet{2024GCN.37277....1O} \\
$6030 \pm 630$ & $Ic$ & $17.90 \pm 0.18$ & AKO & \citet{2024GCN.37299....1O} \\
% $6876 \pm 300$ & $R$ & $18.97 \pm 0.04$ & GMG & \citet{2024GCN.37280....1L} \\
$6880\pm300$ & $R$ & $18.93 \pm 0.04$ & GMG & This work \\
$7470 \pm 630$ & $Ic$ & $18.10 \pm 0.13$ & AKO & \citet{2024GCN.37299....1O} \\
$9270 \pm 630$ & $Ic$ & $18.20 \pm 0.13$ & AKO & \citet{2024GCN.37299....1O} \\
$10710 \pm 630$ & $Ic$ & $18.40 \pm 0.17$ & AKO & \citet{2024GCN.37299....1O} \\
$17640 \pm 180$ & $r$ & $20.10 \pm 0.10$ & LCO & \citet{2024GCN.37287....1I} \\
$17760 \pm 300$ & $z$ & $19.30 \pm 0.10$ & LCO & \citet{2024GCN.37287....1I} \\
$19188 \pm 1000$ & $Rc$ & $19.81 \pm 0.10$ & Montarrenti & \citet{2024GCN.37400....1L} \\
$20800 \pm 1000$ & $Rc$ & $19.81 \pm 0.10$ & Montarrenti & \citet{2024GCN.37291....1L} \\
$21312 \pm 1000$ & $Rc$ & $19.97 \pm 0.11$ & Montarrenti & \citet{2024GCN.37400....1L} \\
$23976 \pm 1000$ & $Rc$ & $20.26 \pm 0.12$ & Montarrenti & \citet{2024GCN.37400....1L} \\
$24660 \pm 450$ & $r^\prime$ & $20.63 \pm 0.03$ & MISTRAL/T193 OHP & \citet{2024GCN.37300....1L} \\
$26640$ & $r$ & $20$ & TNG & \citet{2024GCN.37310....1M} \\
$26748 \pm 1000$ & $Rc$ & $20.49 \pm 0.16$ & Montarrenti & \citet{2024GCN.37400....1L} \\
$29100 \pm 300$ & $J$ & $18.90 \pm 0.20$ & PRIME & \citet{2024GCN.37303....1G} \\
$29100 \pm 300$ & $H$ & $18.50 \pm 0.20$ & PRIME & \citet{2024GCN.37303....1G} \\
$29448 \pm 1000$ & $Rc$ & $20.78 \pm 0.22$ & Montarrenti & \citet{2024GCN.37400....1L} \\
$32220 \pm 1000$ & $Rc$ & $20.71 \pm 0.19$ & Montarrenti & \citet{2024GCN.37400....1L} \\
$32760$ & $r$ & $20.50 \pm 0.30$ & REM & \citet{2024GCN.37295....1B} \\
$33004$ & $g$ & $20.68 \pm 0.23$ & MeerLICHT & \citet{2024GCN.37372....1D} \\
$33004$ & $i$ & $20.24 \pm 0.36$ & MeerLICHT & \citet{2024GCN.37372....1D} \\
$33120$ & $J$ & $17.70 \pm 0.30$ & REM & \citet{2024GCN.37295....1B} \\
$40350 \pm 30$ & $r$ & $20.80$ & VLT & \citet{2024GCN.37293....1M} \\
$41328$ & Luminance & $21 \pm 0.20$ & iTelescope & \citet{2024GCN.37335....1G} \\
$50544$ & $R$ & $20.80 \pm 0.20$ & KAIT & \citet{2024GCN.37304....1Z} \\
% $84960 \pm 900$ & $r$ & $21.46 \pm 0.09$ & GMG & \citet{2024GCN.37306....1W} \\
$84966 \pm 900$ & $r$ & $21.33 \pm 0.09$ & GMG & This work \\
$95040$ & VT\_B & $22.39 \pm 0.20$ & SVOM-VT & \citet{2024GCN.37338....1S} \\
$95040$ & VT\_R & $21.12 \pm 0.10$ & SVOM-VT & \citet{2024GCN.37338....1S} \\
$167325 \pm 900$ & $r$ & $21.26 \pm 0.11$ & GMG &This work \\
$189630 \pm 2100$ & $R$ & $21.88 \pm 0.07$ & SAO RAS & \citet{2024GCN.37313....1M} \\
$257650 \pm 1800$ & $r$ & $21.71 \pm 0.07$ & GMG & This work \\
$274674$ & $R$ & $22.15 \pm 0.07$ & SAO RAS & \citet{2024GCN.37336....1M} \\
$1520640$ & $r'$ & $22.80 \pm 0.10$ & LBT & \citet{2024GCN.37638....1M}
\enddata
\tablecomments{In column (1), the $T_\mathrm{mid}$ are the delay between midtime of photometric observations and Swift-BAT trigger time $T_{0}$ in seconds. In column (3), the magnitudes are not corrected for the Galactic foreground and host-galaxy dust extinction.}
\label{tab:gcn-data}
\end{deluxetable*}

\section{Temporal Decay Indices of Light Curves}\label{sec:Temporal_Decay_Indices}

Since the optical afterglow of GRB 240825A was primarily measured in the $r$-band, we modeled the $r$-band light curve with a multi-segment smoothly broken power-law function:
\begin{equation}\label{eq:temporal_decay}
F_{\nu} = 10^{c}\cdot t^{-\alpha_1}\cdot\prod_{i=1}^{n-1}\left[1+\left(\frac{t}{t_{\mathrm{b}i}}\right)^{1/\delta_i}\right]^{-\delta_i(\alpha_{i+1}-\alpha_i)},
\end{equation} where the normalization is represented by the constant $c$. The temporal evolution is described by $n$ power-law segments. Two temporal decay indices before and after the break time $t_{\mathrm{b}i}$ are indicated by $\alpha_{i}$ and $\alpha_{i+1}$, respectively. The smoothness parameter of the corresponding break is denoted by $\delta_i$, with all $\delta_i$ values fixed at 0.1. The parameter space was explored via Bayesian inference. 
The best-fitting model of the $r$-band light curve is a three-segment smoothly broken power-law model, with two break times at 
$t_{\mathrm{b}1} = 472.40_{-131.64}^{+358.17}\mathrm{\,s}$ and  
$t_{\mathrm{b}2} = 7530.16_{-2643.43}^{+2712.80}\mathrm{\,s}$, respectively. 
The three corresponding temporal decay indices are 
$\alpha_1 = 1.66_{-0.19}^{+0.30}$, 
$\alpha_2 = 1.00_{-0.08}^{+0.05}$ and 
$\alpha_3 = 0.71_{-0.04}^{+0.04}$, respectively. The corresponding normalization constant is $c = 1.51_{-0.47}^{+0.70}$.

We also fitted the X-ray flux density light curves at 1.732\,keV and 10\,keV, respectively, using Equation~\ref{eq:temporal_decay}. The 1.732\,keV flux density light curve can be fitted by a six-segment smoothly broken power-law model, with five break times at 
$t_{\mathrm{b}1} = 97.90^{+5.74}_{-4.82}\mathrm{\,s}$, 
$t_{\mathrm{b}2} = 225.86^{+17.89}_{-18.57}\mathrm{\,s}$, 
$t_{\mathrm{b}3} = 384.15^{+24.79}_{-19.75}\mathrm{\,s}$, 
$t_{\mathrm{b}4} = 1888.95^{+353.66}_{-339.80}\mathrm{\,s}$ and 
$t_{\mathrm{b}5} = 3696.31^{+799.62}_{-685.50}\mathrm{\,s}$, respectively. 
The six corresponding temporal decay indices are 
$\alpha_1 = 3.51^{+0.40}_{-0.35}$, 
$\alpha_2 = 1.27^{+0.08}_{-0.08}$, 
$\alpha_3 = 0.65^{+0.08}_{-0.09}$, 
$\alpha_4 = 1.10^{+0.01}_{-0.01}$, 
$\alpha_5 = 2.01^{+0.28}_{-0.26}$ and 
$\alpha_6 = 1.27^{+0.01}_{-0.01}$, respectively. The corresponding normalization constant is $c = 4.07_{-0.67}^{+0.72}$.
  
The 10\,keV flux density light curve can be parameterized by a three-segment smoothly broken power-law model, with two break times at 
$t_{\mathrm{b}1} = 272.60^{+12.04}_{-11.23}\mathrm{\,s}$ and 
$t_{\mathrm{b}2} = 1325.52^{+424.61}_{-306.89}\mathrm{\,s}$, respectively. 
The three corresponding temporal decay indices are 
$\alpha_1 = -0.03 \pm 0.07$, 
$\alpha_2 = 1.27 \pm 0.03$ and 
$\alpha_3 = 1.55^{+0.03}_{-0.02}$, respectively. The corresponding normalization constant is $c = -3.66_{-0.17}^{+0.14}$.

\bibliography{main}{}

@string{june = {June}}

@article{1995A&AS..109..125V,
 adsurl = {https://ui.adsabs.harvard.edu/abs/1995A&AS..109..125V},
 author = {{Verner}, D.~A. and {Yakovlev}, D.~G.},
 journal = {\aaps},
 month = {January},
 pages = {125-133},
 title = {{Analytic FITS for partial photoionization cross sections.}},
 volume = {109},
 year = {1995}
}

@article{1996ApJ...465..487V,
 adsurl = {https://ui.adsabs.harvard.edu/abs/1996ApJ...465..487V},
 archiveprefix = {arXiv},
 author = {{Verner}, D.~A. and {Ferland}, G.~J. and {Korista}, K.~T. and {Yakovlev}, D.~G.},
 doi = {10.1086/177435},
 eprint = {astro-ph/9601009},
 journal = {\apj},
 month = {July},
 pages = {487},
 primaryclass = {astro-ph},
 title = {{Atomic Data for Astrophysics. II. New Analytic Fits for Photoionization Cross Sections of Atoms and Ions}},
 volume = {465},
 year = {1996}
}

@article{1998ApJ...494L..45P,
 adsurl = {https://ui.adsabs.harvard.edu/abs/1998ApJ...494L..45P},
 archiveprefix = {arXiv},
 author = {{Paczy{\'n}ski}, Bohdan},
 doi = {10.1086/311148},
 eprint = {astro-ph/9710086},
 journal = {\apjl},
 month = {February},
 number = {1},
 pages = {L45-L48},
 primaryclass = {astro-ph},
 title = {{Are Gamma-Ray Bursts in Star-Forming Regions?}},
 volume = {494},
 year = {1998}
}

@article{1998ApJ...497L..17S,
 adsurl = {https://ui.adsabs.harvard.edu/abs/1998ApJ...497L..17S},
 archiveprefix = {arXiv},
 author = {{Sari}, Re'em and {Piran}, Tsvi and {Narayan}, Ramesh},
 doi = {10.1086/311269},
 eprint = {astro-ph/9712005},
 journal = {\apjl},
 month = {April},
 number = {1},
 pages = {L17-L20},
 primaryclass = {astro-ph},
 title = {{Spectra and Light Curves of Gamma-Ray Burst Afterglows}},
 volume = {497},
 year = {1998}
}

@article{1998ApJ...504..405S,
 adsurl = {https://ui.adsabs.harvard.edu/abs/1998ApJ...504..405S},
 archiveprefix = {arXiv},
 author = {{Scargle}, Jeffrey D.},
 doi = {10.1086/306064},
 eprint = {astro-ph/9711233},
 journal = {\apj},
 month = {September},
 number = {1},
 pages = {405-418},
 primaryclass = {astro-ph},
 title = {{Studies in Astronomical Time Series Analysis. V. Bayesian Blocks, a New Method to Analyze Structure in Photon Counting Data}},
 volume = {504},
 year = {1998}
}

@article{2000ApJ...542..914W,
 adsurl = {https://ui.adsabs.harvard.edu/abs/2000ApJ...542..914W},
 archiveprefix = {arXiv},
 author = {{Wilms}, J. and {Allen}, A. and {McCray}, R.},
 doi = {10.1086/317016},
 eprint = {astro-ph/0008425},
 journal = {\apj},
 month = {October},
 number = {2},
 pages = {914-924},
 primaryclass = {astro-ph},
 title = {{On the Absorption of X-Rays in the Interstellar Medium}},
 volume = {542},
 year = {2000}
}

@article{2002ApJ...565..174R,
 adsurl = {https://ui.adsabs.harvard.edu/abs/2002ApJ...565..174R},
 archiveprefix = {arXiv},
 author = {{Reichart}, Daniel E. and {Price}, Paul A.},
 doi = {10.1086/324156},
 eprint = {astro-ph/0107547},
 journal = {\apj},
 month = {January},
 number = {1},
 pages = {174-181},
 primaryclass = {astro-ph},
 title = {{Evidence for a Molecular Cloud Origin of Gamma-Ray Bursts: Implications for the Nature of Star Formation in the Universe}},
 volume = {565},
 year = {2002}
}

@article{2002ApJ...569..780D,
 adsurl = {https://ui.adsabs.harvard.edu/abs/2002ApJ...569..780D},
 archiveprefix = {arXiv},
 author = {{Draine}, B.~T. and {Hao}, Lei},
 doi = {10.1086/339394},
 eprint = {astro-ph/0108243},
 journal = {\apj},
 month = {April},
 number = {2},
 pages = {780-791},
 primaryclass = {astro-ph},
 title = {{Gamma-Ray Burst in a Molecular Cloud: Destruction of Dust and H$_{2}$ and the Emergent Spectrum}},
 volume = {569},
 year = {2002}
}

@article{2003ApJ...591..288H,
 adsurl = {https://ui.adsabs.harvard.edu/abs/2003ApJ...591..288H},
 archiveprefix = {arXiv},
 author = {{Heger}, A. and {Fryer}, C.~L. and {Woosley}, S.~E. and {Langer}, N. and {Hartmann}, D.~H.},
 doi = {10.1086/375341},
 eprint = {astro-ph/0212469},
 journal = {\apj},
 month = {July},
 number = {1},
 pages = {288-300},
 primaryclass = {astro-ph},
 title = {{How Massive Single Stars End Their Life}},
 volume = {591},
 year = {2003}
}

@article{2004A&A...415..171V,
 adsurl = {https://ui.adsabs.harvard.edu/abs/2004A&A...415..171V},
 archiveprefix = {arXiv},
 author = {{Vergani}, S.~D. and {Molinari}, E. and {Zerbi}, F.~M. and {Chincarini}, G.},
 doi = {10.1051/0004-6361:20034612},
 eprint = {astro-ph/0310184},
 journal = {\aap},
 month = {February},
 pages = {171-177},
 primaryclass = {astro-ph},
 title = {{Dust and dark gamma-ray bursts: Mutual implications}},
 volume = {415},
 year = {2004}
}

@article{2004ApJ...611.1005G,
 adsurl = {https://ui.adsabs.harvard.edu/abs/2004ApJ...611.1005G},
 archiveprefix = {arXiv},
 author = {{Gehrels}, N. and {Chincarini}, G. and {Giommi}, P. and {Mason}, K.~O. and {Nousek}, J.~A. and {Wells}, A.~A. and {White}, N.~E. and {Barthelmy}, S.~D. and {Burrows}, D.~N. and {Cominsky}, L.~R. and {Hurley}, K.~C. and {Marshall}, F.~E. and {M{\'e}sz{\'a}ros}, P. and {Roming}, P.~W.~A. and {Angelini}, L. and {Barbier}, L.~M. and {Belloni}, T. and {Campana}, S. and {Caraveo}, P.~A. and {Chester}, M.~M. and {Citterio}, O. and {Cline}, T.~L. and {Cropper}, M.~S. and {Cummings}, J.~R. and {Dean}, A.~J. and {Feigelson}, E.~D. and {Fenimore}, E.~E. and {Frail}, D.~A. and {Fruchter}, A.~S. and {Garmire}, G.~P. and {Gendreau}, K. and {Ghisellini}, G. and {Greiner}, J. and {Hill}, J.~E. and {Hunsberger}, S.~D. and {Krimm}, H.~A. and {Kulkarni}, S.~R. and {Kumar}, P. and {Lebrun}, F. and {Lloyd-Ronning}, N.~M. and {Markwardt}, C.~B. and {Mattson}, B.~J. and {Mushotzky}, R.~F. and {Norris}, J.~P. and {Osborne}, J. and {Paczynski}, B. and {Palmer}, D.~M. and {Park}, H. -S. and {Parsons}, A.~M. and {Paul}, J. and {Rees}, M.~J. and {Reynolds}, C.~S. and {Rhoads}, J.~E. and {Sasseen}, T.~P. and {Schaefer}, B.~E. and {Short}, A.~T. and {Smale}, A.~P. and {Smith}, I.~A. and {Stella}, L. and {Tagliaferri}, G. and {Takahashi}, T. and {Tashiro}, M. and {Townsley}, L.~K. and {Tueller}, J. and {Turner}, M.~J.~L. and {Vietri}, M. and {Voges}, W. and {Ward}, M.~J. and {Willingale}, R. and {Zerbi}, F.~M. and {Zhang}, W.~W.},
 doi = {10.1086/422091},
 eprint = {astro-ph/0405233},
 journal = {\apj},
 month = {August},
 number = {2},
 pages = {1005-1020},
 primaryclass = {astro-ph},
 title = {{The Swift Gamma-Ray Burst Mission}},
 volume = {611},
 year = {2004}
}

@article{2004ApJ...617L..21J,
 adsurl = {https://ui.adsabs.harvard.edu/abs/2004ApJ...617L..21J},
 archiveprefix = {arXiv},
 author = {{Jakobsson}, P. and {Hjorth}, J. and {Fynbo}, J.~P.~U. and {Watson}, D. and {Pedersen}, K. and {Bj{\"o}rnsson}, G. and {Gorosabel}, J.},
 doi = {10.1086/427089},
 eprint = {astro-ph/0411036},
 journal = {\apjl},
 month = {December},
 number = {1},
 pages = {L21-L24},
 primaryclass = {astro-ph},
 title = {{Swift Identification of Dark Gamma-Ray Bursts}},
 volume = {617},
 year = {2004}
}

@inproceedings{2004SPIE.5165..201B,
 adsurl = {https://ui.adsabs.harvard.edu/abs/2004SPIE.5165..201B},
 author = {{Burrows}, David N. and {Hill}, Joanne E. and {Nousek}, John A. and {Wells}, Alan A. and {Chincarini}, Guido and {Abbey}, Anthony F. and {Beardmore}, Andrew P. and {Bosworth}, J. and {Br{\"a}uninger}, Heinrich W. and {Burkert}, Wolfgang and {Campana}, Sergio and {Capalbi}, Milvia and {Chang}, W. and {Citterio}, Oberto and {Freyberg}, Michael J. and {Giommi}, Paolo and {Hartner}, Gisela D. and {Killough}, Ronnie and {Kittle}, B. and {Klar}, R. and {Mangels}, C. and {McMeekin}, M. and {Miles}, B.~J. and {Moretti}, Alberto and {Mori}, Koji and {Morris}, Dave C. and {Mukerjee}, Kallol and {Osborne}, Julian P. and {Short}, Alexander D.~T. and {Tagliaferri}, Gianpiero and {Tamburelli}, Francesca and {Watson}, D.~J. and {Willingale}, Richard and {Zugger}, Michael E.},
 booktitle = {X-Ray and Gamma-Ray Instrumentation for Astronomy XIII},
 doi = {10.1117/12.504868},
 editor = {{Flanagan}, Kathryn A. and {Siegmund}, Oswald H.~W.},
 month = {February},
 pages = {201-216},
 series = {Society of Photo-Optical Instrumentation Engineers (SPIE) Conference Series},
 title = {{The Swift X-Ray Telescope}},
 volume = {5165},
 year = {2004}
}

@article{2005SSRv..120...95R,
 adsurl = {https://ui.adsabs.harvard.edu/abs/2005SSRv..120...95R},
 archiveprefix = {arXiv},
 author = {{Roming}, Peter W.~A. and {Kennedy}, Thomas E. and {Mason}, Keith O. and {Nousek}, John A. and {Ahr}, Lindy and {Bingham}, Richard E. and {Broos}, Patrick S. and {Carter}, Mary J. and {Hancock}, Barry K. and {Huckle}, Howard E. and {Hunsberger}, S.~D. and {Kawakami}, Hajime and {Killough}, Ronnie and {Koch}, T. Scott and {McLelland}, Michael K. and {Smith}, Kelly and {Smith}, Philip J. and {Soto}, Juan Carlos and {Boyd}, Patricia T. and {Breeveld}, Alice A. and {Holland}, Stephen T. and {Ivanushkina}, Mariya and {Pryzby}, Michael S. and {Still}, Martin D. and {Stock}, Joseph},
 doi = {10.1007/s11214-005-5095-4},
 eprint = {astro-ph/0507413},
 journal = {\ssr},
 month = {October},
 number = {3-4},
 pages = {95-142},
 primaryclass = {astro-ph},
 title = {{The Swift Ultra-Violet/Optical Telescope}},
 volume = {120},
 year = {2005}
}

@article{2005SSRv..120..143B,
 adsurl = {https://ui.adsabs.harvard.edu/abs/2005SSRv..120..143B},
 archiveprefix = {arXiv},
 author = {{Barthelmy}, Scott D. and {Barbier}, Louis M. and {Cummings}, Jay R. and {Fenimore}, Ed E. and {Gehrels}, Neil and {Hullinger}, Derek and {Krimm}, Hans A. and {Markwardt}, Craig B. and {Palmer}, David M. and {Parsons}, Ann and {Sato}, Goro and {Suzuki}, Masaya and {Takahashi}, Tadayuki and {Tashiro}, Makota and {Tueller}, Jack},
 doi = {10.1007/s11214-005-5096-3},
 eprint = {astro-ph/0507410},
 journal = {\ssr},
 month = {October},
 number = {3-4},
 pages = {143-164},
 primaryclass = {astro-ph},
 title = {{The Burst Alert Telescope (BAT) on the SWIFT Midex Mission}},
 volume = {120},
 year = {2005}
}

@article{2007A&A...469..379E,
 adsurl = {https://ui.adsabs.harvard.edu/abs/2007A&A...469..379E},
 archiveprefix = {arXiv},
 author = {{Evans}, P.~A. and {Beardmore}, A.~P. and {Page}, K.~L. and {Tyler}, L.~G. and {Osborne}, J.~P. and {Goad}, M.~R. and {O'Brien}, P.~T. and {Vetere}, L. and {Racusin}, J. and {Morris}, D. and {Burrows}, D.~N. and {Capalbi}, M. and {Perri}, M. and {Gehrels}, N. and {Romano}, P.},
 doi = {10.1051/0004-6361:20077530},
 eprint = {0704.0128},
 journal = {\aap},
 month = {July},
 number = {1},
 pages = {379-385},
 primaryclass = {astro-ph},
 title = {{An online repository of Swift/XRT light curves of {\ensuremath{\gamma}}-ray bursts}},
 volume = {469},
 year = {2007}
}

@article{2007ApJ...654L..17C,
 adsurl = {https://ui.adsabs.harvard.edu/abs/2007ApJ...654L..17C},
 archiveprefix = {arXiv},
 author = {{Campana}, S. and {Lazzati}, D. and {Ripamonti}, E. and {Perna}, R. and {Covino}, S. and {Tagliaferri}, G. and {Moretti}, A. and {Romano}, P. and {Cusumano}, G. and {Chincarini}, G.},
 doi = {10.1086/510719},
 eprint = {astro-ph/0611305},
 journal = {\apjl},
 month = {January},
 number = {1},
 pages = {L17-L20},
 primaryclass = {astro-ph},
 title = {{A Metal-rich Molecular Cloud Surrounds GRB 050904 at Redshift 6.3}},
 volume = {654},
 year = {2007}
}

@article{2007ApJ...663..320F,
 adsurl = {https://ui.adsabs.harvard.edu/abs/2007ApJ...663..320F},
 archiveprefix = {arXiv},
 author = {{Fitzpatrick}, E.~L. and {Massa}, D.},
 doi = {10.1086/518158},
 eprint = {0705.0154},
 journal = {\apj},
 month = {July},
 number = {1},
 pages = {320-341},
 primaryclass = {astro-ph},
 title = {{An Analysis of the Shapes of Interstellar Extinction Curves. V. The IR-through-UV Curve Morphology}},
 volume = {663},
 year = {2007}
}

@article{2007ApJ...663..407B,
 adsurl = {https://ui.adsabs.harvard.edu/abs/2007ApJ...663..407B},
 archiveprefix = {arXiv},
 author = {{Butler}, Nathaniel R. and {Kocevski}, Daniel},
 doi = {10.1086/518023},
 eprint = {astro-ph/0612564},
 journal = {\apj},
 month = {July},
 number = {1},
 pages = {407-419},
 primaryclass = {astro-ph},
 title = {{X-Ray Hardness Evolution in GRB Afterglows and Flares: Late-Time GRB Activity without N$_{H}$ Variations}},
 volume = {663},
 year = {2007}
}

@article{2007CSE.....9...90H,
 adsurl = {https://ui.adsabs.harvard.edu/abs/2007CSE.....9...90H},
 author = {{Hunter}, John D.},
 doi = {10.1109/MCSE.2007.55},
 journal = {Computing in Science and Engineering},
 month = {May},
 number = {3},
 pages = {90-95},
 title = {{Matplotlib: A 2D Graphics Environment}},
 volume = {9},
 year = {2007}
}

@article{2008MNRAS.386..608L,
 adsurl = {https://ui.adsabs.harvard.edu/abs/2008MNRAS.386..608L},
 archiveprefix = {arXiv},
 author = {{Lapi}, A. and {Kawakatu}, N. and {Bosnjak}, Z. and {Celotti}, A. and {Bressan}, A. and {Granato}, G.~L. and {Danese}, L.},
 doi = {10.1111/j.1365-2966.2008.13076.x},
 eprint = {0802.0787},
 journal = {\mnras},
 month = {May},
 number = {2},
 pages = {608-618},
 primaryclass = {astro-ph},
 title = {{Long gamma-ray bursts and their host galaxies at high redshift}},
 volume = {386},
 year = {2008}
}

@article{2009ApJ...693.1484C,
 adsurl = {https://ui.adsabs.harvard.edu/abs/2009ApJ...693.1484C},
 archiveprefix = {arXiv},
 author = {{Cenko}, S.~B. and {Kelemen}, J. and {Harrison}, F.~A. and {Fox}, D.~B. and {Kulkarni}, S.~R. and {Kasliwal}, M.~M. and {Ofek}, E.~O. and {Rau}, A. and {Gal-Yam}, A. and {Frail}, D.~A. and {Moon}, D. -S.},
 doi = {10.1088/0004-637X/693/2/1484},
 eprint = {0808.3983},
 journal = {\apj},
 month = {March},
 number = {2},
 pages = {1484-1493},
 primaryclass = {astro-ph},
 title = {{Dark Bursts in the Swift Era: The Palomar 60 Inch-Swift Early Optical Afterglow Catalog}},
 volume = {693},
 year = {2009}
}

@article{2009ApJ...699.1087V,
 adsurl = {https://ui.adsabs.harvard.edu/abs/2009ApJ...699.1087V},
 archiveprefix = {arXiv},
 author = {{van der Horst}, A.~J. and {Kouveliotou}, C. and {Gehrels}, N. and {Rol}, E. and {Wijers}, R.~A.~M.~J. and {Cannizzo}, J.~K. and {Racusin}, J. and {Burrows}, D.~N.},
 doi = {10.1088/0004-637X/699/2/1087},
 eprint = {0905.0524},
 journal = {\apj},
 month = {July},
 number = {2},
 pages = {1087-1091},
 primaryclass = {astro-ph.HE},
 title = {{Optical Classification of Gamma-Ray Bursts in the Swift Era}},
 volume = {699},
 year = {2009}
}

@article{2009ApJS..185..526F,
 adsurl = {https://ui.adsabs.harvard.edu/abs/2009ApJS..185..526F},
 archiveprefix = {arXiv},
 author = {{Fynbo}, J.~P.~U. and {Jakobsson}, P. and {Prochaska}, J.~X. and {Malesani}, D. and {Ledoux}, C. and {de Ugarte Postigo}, A. and {Nardini}, M. and {Vreeswijk}, P.~M. and {Wiersema}, K. and {Hjorth}, J. and {Sollerman}, J. and {Chen}, H. -W. and {Th{\"o}ne}, C.~C. and {Bj{\"o}rnsson}, G. and {Bloom}, J.~S. and {Castro-Tirado}, A.~J. and {Christensen}, L. and {De Cia}, A. and {Fruchter}, A.~S. and {Gorosabel}, J. and {Graham}, J.~F. and {Jaunsen}, A.~O. and {Jensen}, B.~L. and {Kann}, D.~A. and {Kouveliotou}, C. and {Levan}, A.~J. and {Maund}, J. and {Masetti}, N. and {Milvang-Jensen}, B. and {Palazzi}, E. and {Perley}, D.~A. and {Pian}, E. and {Rol}, E. and {Schady}, P. and {Starling}, R.~L.~C. and {Tanvir}, N.~R. and {Watson}, D.~J. and {Xu}, D. and {Augusteijn}, T. and {Grundahl}, F. and {Telting}, J. and {Quirion}, P. -O.},
 doi = {10.1088/0067-0049/185/2/526},
 eprint = {0907.3449},
 journal = {\apjs},
 month = {December},
 number = {2},
 pages = {526-573},
 primaryclass = {astro-ph.CO},
 title = {{Low-resolution Spectroscopy of Gamma-ray Burst Optical Afterglows: Biases in the Swift Sample and Characterization of the Absorbers}},
 volume = {185},
 year = {2009}
}

@article{2009MNRAS.397.1177E,
 adsurl = {https://ui.adsabs.harvard.edu/abs/2009MNRAS.397.1177E},
 archiveprefix = {arXiv},
 author = {{Evans}, P.~A. and {Beardmore}, A.~P. and {Page}, K.~L. and {Osborne}, J.~P. and {O'Brien}, P.~T. and {Willingale}, R. and {Starling}, R.~L.~C. and {Burrows}, D.~N. and {Godet}, O. and {Vetere}, L. and {Racusin}, J. and {Goad}, M.~R. and {Wiersema}, K. and {Angelini}, L. and {Capalbi}, M. and {Chincarini}, G. and {Gehrels}, N. and {Kennea}, J.~A. and {Margutti}, R. and {Morris}, D.~C. and {Mountford}, C.~J. and {Pagani}, C. and {Perri}, M. and {Romano}, P. and {Tanvir}, N.},
 doi = {10.1111/j.1365-2966.2009.14913.x},
 eprint = {0812.3662},
 journal = {\mnras},
 month = {August},
 number = {3},
 pages = {1177-1201},
 primaryclass = {astro-ph},
 title = {{Methods and results of an automatic analysis of a complete sample of Swift-XRT observations of GRBs}},
 volume = {397},
 year = {2009}
}

@article{2009Natur.461.1254T,
 adsurl = {https://ui.adsabs.harvard.edu/abs/2009Natur.461.1254T},
 archiveprefix = {arXiv},
 author = {{Tanvir}, N.~R. and {Fox}, D.~B. and {Levan}, A.~J. and {Berger}, E. and {Wiersema}, K. and {Fynbo}, J.~P.~U. and {Cucchiara}, A. and {Kr{\"u}hler}, T. and {Gehrels}, N. and {Bloom}, J.~S. and {Greiner}, J. and {Evans}, P.~A. and {Rol}, E. and {Olivares}, F. and {Hjorth}, J. and {Jakobsson}, P. and {Farihi}, J. and {Willingale}, R. and {Starling}, R.~L.~C. and {Cenko}, S.~B. and {Perley}, D. and {Maund}, J.~R. and {Duke}, J. and {Wijers}, R.~A.~M.~J. and {Adamson}, A.~J. and {Allan}, A. and {Bremer}, M.~N. and {Burrows}, D.~N. and {Castro-Tirado}, A.~J. and {Cavanagh}, B. and {de Ugarte Postigo}, A. and {Dopita}, M.~A. and {Fatkhullin}, T.~A. and {Fruchter}, A.~S. and {Foley}, R.~J. and {Gorosabel}, J. and {Kennea}, J. and {Kerr}, T. and {Klose}, S. and {Krimm}, H.~A. and {Komarova}, V.~N. and {Kulkarni}, S.~R. and {Moskvitin}, A.~S. and {Mundell}, C.~G. and {Naylor}, T. and {Page}, K. and {Penprase}, B.~E. and {Perri}, M. and {Podsiadlowski}, P. and {Roth}, K. and {Rutledge}, R.~E. and {Sakamoto}, T. and {Schady}, P. and {Schmidt}, B.~P. and {Soderberg}, A.~M. and {Sollerman}, J. and {Stephens}, A.~W. and {Stratta}, G. and {Ukwatta}, T.~N. and {Watson}, D. and {Westra}, E. and {Wold}, T. and {Wolf}, C.},
 doi = {10.1038/nature08459},
 eprint = {0906.1577},
 journal = {\nat},
 month = {October},
 number = {7268},
 pages = {1254-1257},
 primaryclass = {astro-ph.CO},
 title = {{A {\ensuremath{\gamma}}-ray burst at a redshift of z\raisebox{-0.5ex}\textasciitilde8.2}},
 volume = {461},
 year = {2009}
}

@article{2009RAA.....9.1103Z,
 adsurl = {https://ui.adsabs.harvard.edu/abs/2009RAA.....9.1103Z},
 archiveprefix = {arXiv},
 author = {{Zheng}, Wei-Kang and {Deng}, Jin-Song and {Wang}, Jing},
 doi = {10.1088/1674-4527/9/10/003},
 eprint = {0906.2244},
 journal = {Research in Astronomy and Astrophysics},
 month = {October},
 number = {10},
 pages = {1103-1118},
 primaryclass = {astro-ph.HE},
 title = {{Statistical studies of optically dark gamma-ray bursts in the Swift era}},
 volume = {9},
 year = {2009}
}

@article{2010AJ....139.1782L,
 adsurl = {https://ui.adsabs.harvard.edu/abs/2010AJ....139.1782L},
 archiveprefix = {arXiv},
 author = {{Lang}, Dustin and {Hogg}, David W. and {Mierle}, Keir and {Blanton}, Michael and {Roweis}, Sam},
 doi = {10.1088/0004-6256/139/5/1782},
 eprint = {0910.2233},
 journal = {\aj},
 month = {May},
 number = {5},
 pages = {1782-1800},
 primaryclass = {astro-ph.IM},
 title = {{Astrometry.net: Blind Astrometric Calibration of Arbitrary Astronomical Images}},
 volume = {139},
 year = {2010}
}

@article{2010ApJ...717..140M,
 adsurl = {https://ui.adsabs.harvard.edu/abs/2010ApJ...717..140M},
 archiveprefix = {arXiv},
 author = {{Mao}, Jirong},
 doi = {10.1088/0004-637X/717/1/140},
 eprint = {1005.1876},
 journal = {\apj},
 month = {July},
 number = {1},
 pages = {140-146},
 primaryclass = {astro-ph.HE},
 title = {{A Theoretical Investigation of Gamma-ray Burst Host Galaxies}},
 volume = {717},
 year = {2010}
}

@article{2010ApJ...719..378H,
 adsurl = {https://ui.adsabs.harvard.edu/abs/2010ApJ...719..378H},
 archiveprefix = {arXiv},
 author = {{Hashimoto}, T. and {Ohta}, K. and {Aoki}, K. and {Tanaka}, I. and {Yabe}, K. and {Kawai}, N. and {Aoki}, W. and {Furusawa}, H. and {Hattori}, T. and {Iye}, M. and {Kawabata}, K.~S. and {Kobayashi}, N. and {Komiyama}, Y. and {Kosugi}, G. and {Minowa}, Y. and {Mizumoto}, Y. and {Niino}, Y. and {Nomoto}, K. and {Noumaru}, J. and {Ogasawara}, R. and {Pyo}, T. -S. and {Sakamoto}, T. and {Sekiguchi}, K. and {Shirasaki}, Y. and {Suzuki}, M. and {Tajitsu}, A. and {Takata}, T. and {Tamagawa}, T. and {Terada}, H. and {Totani}, T. and {Watanabe}, J. and {Yamada}, T. and {Yoshida}, A.},
 doi = {10.1088/0004-637X/719/1/378},
 eprint = {1003.3717},
 journal = {\apj},
 month = {August},
 number = {1},
 pages = {378-384},
 primaryclass = {astro-ph.GA},
 title = {{``Dark'' GRB 080325 in a Dusty Massive Galaxy at z \raisebox{-0.5ex}\textasciitilde 2}},
 volume = {719},
 year = {2010}
}

@article{2010ApJ...720.1513K,
 adsurl = {https://ui.adsabs.harvard.edu/abs/2010ApJ...720.1513K},
 archiveprefix = {arXiv},
 author = {{Kann}, D.~A. and {Klose}, S. and {Zhang}, B. and {Malesani}, D. and {Nakar}, E. and {Pozanenko}, A. and {Wilson}, A.~C. and {Butler}, N.~R. and {Jakobsson}, P. and {Schulze}, S. and {Andreev}, M. and {Antonelli}, L.~A. and {Bikmaev}, I.~F. and {Biryukov}, V. and {B{\"o}ttcher}, M. and {Burenin}, R.~A. and {Castro Cer{\'o}n}, J.~M. and {Castro-Tirado}, A.~J. and {Chincarini}, G. and {Cobb}, B.~E. and {Covino}, S. and {D'Avanzo}, P. and {D'Elia}, V. and {Della Valle}, M. and {de Ugarte Postigo}, A. and {Efimov}, Yu. and {Ferrero}, P. and {Fugazza}, D. and {Fynbo}, J.~P.~U. and {G{\r{a}}lfalk}, M. and {Grundahl}, F. and {Gorosabel}, J. and {Gupta}, S. and {Guziy}, S. and {Hafizov}, B. and {Hjorth}, J. and {Holhjem}, K. and {Ibrahimov}, M. and {Im}, M. and {Israel}, G.~L. and {Je{\'l}inek}, M. and {Jensen}, B.~L. and {Karimov}, R. and {Khamitov}, I.~M. and {Kizilo{\v{g}}lu}, {\"U}. and {Klunko}, E. and {Kub{\'a}nek}, P. and {Kutyrev}, A.~S. and {Laursen}, P. and {Levan}, A.~J. and {Mannucci}, F. and {Martin}, C.~M. and {Mescheryakov}, A. and {Mirabal}, N. and {Norris}, J.~P. and {Ovaldsen}, J. -E. and {Paraficz}, D. and {Pavlenko}, E. and {Piranomonte}, S. and {Rossi}, A. and {Rumyantsev}, V. and {Salinas}, R. and {Sergeev}, A. and {Sharapov}, D. and {Sollerman}, J. and {Stecklum}, B. and {Stella}, L. and {Tagliaferri}, G. and {Tanvir}, N.~R. and {Telting}, J. and {Testa}, V. and {Updike}, A.~C. and {Volnova}, A. and {Watson}, D. and {Wiersema}, K. and {Xu}, D.},
 doi = {10.1088/0004-637X/720/2/1513},
 eprint = {0712.2186},
 journal = {\apj},
 month = {September},
 number = {2},
 pages = {1513-1558},
 primaryclass = {astro-ph},
 title = {{The Afterglows of Swift-era Gamma-ray Bursts. I. Comparing pre-Swift and Swift-era Long/Soft (Type II) GRB Optical Afterglows}},
 volume = {720},
 year = {2010}
}

@article{2010arXiv1004.2316W,
 adsurl = {https://ui.adsabs.harvard.edu/abs/2010arXiv1004.2316W},
 archiveprefix = {arXiv},
 author = {{Watanabe}, Sumio},
 doi = {10.48550/arXiv.1004.2316},
 eid = {arXiv:1004.2316},
 eprint = {1004.2316},
 journal = {arXiv e-prints},
 month = {April},
 pages = {arXiv:1004.2316},
 primaryclass = {cs.LG},
 title = {{Asymptotic Equivalence of Bayes Cross Validation and Widely Applicable Information Criterion in Singular Learning Theory}},
 year = {2010}
}

@software{2010ascl.soft10068B,
 adsurl = {https://ui.adsabs.harvard.edu/abs/2010ascl.soft10068B},
 author = {{Bertin}, Emmanuel},
 eid = {ascl:1010.068},
 howpublished = {Astrophysics Source Code Library, record ascl:1010.068},
 month = {October},
 title = {{SWarp: Resampling and Co-adding FITS Images Together}},
 year = {2010}
}

@article{2010MNRAS.402.2429C,
 adsurl = {https://ui.adsabs.harvard.edu/abs/2010MNRAS.402.2429C},
 archiveprefix = {arXiv},
 author = {{Campana}, S. and {Th{\"o}ne}, C.~C. and {de Ugarte Postigo}, A. and {Tagliaferri}, G. and {Moretti}, A. and {Covino}, S.},
 doi = {10.1111/j.1365-2966.2009.16006.x},
 eprint = {0911.1214},
 journal = {\mnras},
 month = {March},
 number = {4},
 pages = {2429-2435},
 primaryclass = {astro-ph.HE},
 title = {{The X-ray absorbing column densities of Swift gamma-ray bursts}},
 volume = {402},
 year = {2010}
}

@article{2011A&A...526A..30G,
 adsurl = {https://ui.adsabs.harvard.edu/abs/2011A&A...526A..30G},
 archiveprefix = {arXiv},
 author = {{Greiner}, J. and {Kr{\"u}hler}, T. and {Klose}, S. and {Afonso}, P. and {Clemens}, C. and {Filgas}, R. and {Hartmann}, D.~H. and {K{\"u}pc{\"u} Yolda{\c{s}}}, A. and {Nardini}, M. and {Olivares E.}, F. and {Rau}, A. and {Rossi}, A. and {Schady}, P. and {Updike}, A.},
 doi = {10.1051/0004-6361/201015458},
 eid = {A30},
 eprint = {1011.0618},
 journal = {\aap},
 month = {February},
 pages = {A30},
 primaryclass = {astro-ph.HE},
 title = {{The nature of ``dark'' gamma-ray bursts}},
 volume = {526},
 year = {2011}
}

@article{2011A&A...532A.143Z,
 adsurl = {https://ui.adsabs.harvard.edu/abs/2011A&A...532A.143Z},
 archiveprefix = {arXiv},
 author = {{Zafar}, T. and {Watson}, D. and {Fynbo}, J.~P.~U. and {Malesani}, D. and {Jakobsson}, P. and {de Ugarte Postigo}, A.},
 doi = {10.1051/0004-6361/201116663},
 eid = {A143},
 eprint = {1102.1469},
 journal = {\aap},
 month = {August},
 pages = {A143},
 primaryclass = {astro-ph.CO},
 title = {{The extinction curves of star-forming regions from z = 0.1 to 6.7 using GRB afterglow spectroscopy}},
 volume = {532},
 year = {2011}
}

@article{2011ApJ...734...96K,
 adsurl = {https://ui.adsabs.harvard.edu/abs/2011ApJ...734...96K},
 archiveprefix = {arXiv},
 author = {{Kann}, D.~A. and {Klose}, S. and {Zhang}, B. and {Covino}, S. and {Butler}, N.~R. and {Malesani}, D. and {Nakar}, E. and {Wilson}, A.~C. and {Antonelli}, L.~A. and {Chincarini}, G. and {Cobb}, B.~E. and {D'Avanzo}, P. and {D'Elia}, V. and {Della Valle}, M. and {Ferrero}, P. and {Fugazza}, D. and {Gorosabel}, J. and {Israel}, G.~L. and {Mannucci}, F. and {Piranomonte}, S. and {Schulze}, S. and {Stella}, L. and {Tagliaferri}, G. and {Wiersema}, K.},
 doi = {10.1088/0004-637X/734/2/96},
 eid = {96},
 eprint = {0804.1959},
 journal = {\apj},
 month = {June},
 number = {2},
 pages = {96},
 primaryclass = {astro-ph},
 title = {{The Afterglows of Swift-era Gamma-Ray Bursts. II. Type I GRB versus Type II GRB Optical Afterglows}},
 volume = {734},
 year = {2011}
}

@article{2011ApJ...736....7C,
 adsurl = {https://ui.adsabs.harvard.edu/abs/2011ApJ...736....7C},
 archiveprefix = {arXiv},
 author = {{Cucchiara}, A. and {Levan}, A.~J. and {Fox}, D.~B. and {Tanvir}, N.~R. and {Ukwatta}, T.~N. and {Berger}, E. and {Kr{\"u}hler}, T. and {K{\"u}pc{\"u} Yolda{\c{s}}}, A. and {Wu}, X.~F. and {Toma}, K. and {Greiner}, J. and {Olivares}, F.~E. and {Rowlinson}, A. and {Amati}, L. and {Sakamoto}, T. and {Roth}, K. and {Stephens}, A. and {Fritz}, Alexander and {Fynbo}, J.~P.~U. and {Hjorth}, J. and {Malesani}, D. and {Jakobsson}, P. and {Wiersema}, K. and {O'Brien}, P.~T. and {Soderberg}, A.~M. and {Foley}, R.~J. and {Fruchter}, A.~S. and {Rhoads}, J. and {Rutledge}, R.~E. and {Schmidt}, B.~P. and {Dopita}, M.~A. and {Podsiadlowski}, P. and {Willingale}, R. and {Wolf}, C. and {Kulkarni}, S.~R. and {D'Avanzo}, P.},
 doi = {10.1088/0004-637X/736/1/7},
 eid = {7},
 eprint = {1105.4915},
 journal = {\apj},
 month = {July},
 number = {1},
 pages = {7},
 primaryclass = {astro-ph.CO},
 title = {{A Photometric Redshift of z \raisebox{-0.5ex}\textasciitilde 9.4 for GRB 090429B}},
 volume = {736},
 year = {2011}
}

@article{2011ApJ...737..103S,
 adsurl = {https://ui.adsabs.harvard.edu/abs/2011ApJ...737..103S},
 archiveprefix = {arXiv},
 author = {{Schlafly}, Edward F. and {Finkbeiner}, Douglas P.},
 doi = {10.1088/0004-637X/737/2/103},
 eid = {103},
 eprint = {1012.4804},
 journal = {\apj},
 month = {August},
 number = {2},
 pages = {103},
 primaryclass = {astro-ph.GA},
 title = {{Measuring Reddening with Sloan Digital Sky Survey Stellar Spectra and Recalibrating SFD}},
 volume = {737},
 year = {2011}
}

@article{2011MNRAS.416.2078P,
 adsurl = {https://ui.adsabs.harvard.edu/abs/2011MNRAS.416.2078P},
 author = {{Page}, K.~L. and {Starling}, R.~L.~C. and {Fitzpatrick}, G. and {Pandey}, S.~B. and {Osborne}, J.~P. and {Schady}, P. and {McBreen}, S. and {Campana}, S. and {Ukwatta}, T.~N. and {Pagani}, C. and {Beardmore}, A.~P. and {Evans}, P.~A.},
 doi = {10.1111/j.1365-2966.2011.19183.x},
 journal = {\mnras},
 month = {September},
 number = {3},
 pages = {2078-2089},
 title = {{GRB 090618: detection of thermal X-ray emission from a bright gamma-ray burst}},
 volume = {416},
 year = {2011}
}

@article{2012ApJ...753...82Z,
 adsurl = {https://ui.adsabs.harvard.edu/abs/2012ApJ...753...82Z},
 archiveprefix = {arXiv},
 author = {{Zafar}, Tayyaba and {Watson}, Darach and {El{\'\i}asd{\'o}ttir}, {\'A}rd{\'\i}s and {Fynbo}, Johan P.~U. and {Kr{\"u}hler}, Thomas and {Schady}, Patricia and {Leloudas}, Giorgos and {Jakobsson}, P{\'a}ll and {Th{\"o}ne}, Christina C. and {Perley}, Daniel A. and {Morgan}, Adam N. and {Bloom}, Joshua and {Greiner}, Jochen},
 doi = {10.1088/0004-637X/753/1/82},
 eid = {82},
 eprint = {1205.0387},
 journal = {\apj},
 month = {July},
 number = {1},
 pages = {82},
 primaryclass = {astro-ph.CO},
 title = {{The Properties of the 2175 {\r{A}} Extinction Feature Discovered in GRB Afterglows}},
 volume = {753},
 year = {2012}
}

@article{2012MNRAS.421.1265M,
 adsurl = {https://ui.adsabs.harvard.edu/abs/2012MNRAS.421.1265M},
 archiveprefix = {arXiv},
 author = {{Melandri}, A. and {Sbarufatti}, B. and {D'Avanzo}, P. and {Salvaterra}, R. and {Campana}, S. and {Covino}, S. and {Vergani}, S.~D. and {Nava}, L. and {Ghisellini}, G. and {Ghirlanda}, G. and {Fugazza}, D. and {Mangano}, V. and {Capalbi}, M. and {Tagliaferri}, G.},
 doi = {10.1111/j.1365-2966.2011.20398.x},
 eprint = {1112.4480},
 journal = {\mnras},
 month = {April},
 number = {2},
 pages = {1265-1272},
 primaryclass = {astro-ph.HE},
 title = {{The dark bursts population in a complete sample of bright Swift long gamma-ray bursts}},
 volume = {421},
 year = {2012}
}

@article{2012MNRAS.421.1697C,
 adsurl = {https://ui.adsabs.harvard.edu/abs/2012MNRAS.421.1697C},
 archiveprefix = {arXiv},
 author = {{Campana}, S. and {Salvaterra}, R. and {Melandri}, A. and {Vergani}, S.~D. and {Covino}, S. and {D'Avanzo}, P. and {Fugazza}, D. and {Ghisellini}, G. and {Sbarufatti}, B. and {Tagliaferri}, G.},
 doi = {10.1111/j.1365-2966.2012.20428.x},
 eprint = {1112.5111},
 journal = {\mnras},
 month = {April},
 number = {2},
 pages = {1697-1702},
 primaryclass = {astro-ph.HE},
 title = {{The X-ray absorbing column density of a complete sample of bright Swift gamma-ray bursts}},
 volume = {421},
 year = {2012}
}

@article{2012MNRAS.427.2950S,
 adsurl = {https://ui.adsabs.harvard.edu/abs/2012MNRAS.427.2950S},
 archiveprefix = {arXiv},
 author = {{Starling}, R.~L.~C. and {Page}, K.~L. and {Pe'Er}, A. and {Beardmore}, A.~P. and {Osborne}, J.~P.},
 doi = {10.1111/j.1365-2966.2012.22116.x},
 eprint = {1207.1444},
 journal = {\mnras},
 month = {December},
 number = {4},
 pages = {2950-2964},
 primaryclass = {astro-ph.HE},
 title = {{A search for thermal X-ray signatures in gamma-ray bursts - I. Swift bursts with optical supernovae}},
 volume = {427},
 year = {2012}
}

@article{2012MNRAS.427.2965S,
 adsurl = {https://ui.adsabs.harvard.edu/abs/2012MNRAS.427.2965S},
 archiveprefix = {arXiv},
 author = {{Sparre}, Martin and {Starling}, Rhaana L.~C.},
 doi = {10.1111/j.1365-2966.2012.21858.x},
 eprint = {1207.1447},
 journal = {\mnras},
 month = {December},
 number = {4},
 pages = {2965-2974},
 primaryclass = {astro-ph.HE},
 title = {{A search for thermal X-ray signatures in gamma-ray bursts - II. The Swift sample}},
 volume = {427},
 year = {2012}
}

@article{2013A&A...558A..33A,
 adsurl = {https://ui.adsabs.harvard.edu/abs/2013A&A...558A..33A},
 archiveprefix = {arXiv},
 author = {{Astropy Collaboration} and {Robitaille}, Thomas P. and {Tollerud}, Erik J. and {Greenfield}, Perry and {Droettboom}, Michael and {Bray}, Erik and {Aldcroft}, Tom and {Davis}, Matt and {Ginsburg}, Adam and {Price-Whelan}, Adrian M. and {Kerzendorf}, Wolfgang E. and {Conley}, Alexander and {Crighton}, Neil and {Barbary}, Kyle and {Muna}, Demitri and {Ferguson}, Henry and {Grollier}, Fr{\'e}d{\'e}ric and {Parikh}, Madhura M. and {Nair}, Prasanth H. and {Unther}, Hans M. and {Deil}, Christoph and {Woillez}, Julien and {Conseil}, Simon and {Kramer}, Roban and {Turner}, James E.~H. and {Singer}, Leo and {Fox}, Ryan and {Weaver}, Benjamin A. and {Zabalza}, Victor and {Edwards}, Zachary I. and {Azalee Bostroem}, K. and {Burke}, D.~J. and {Casey}, Andrew R. and {Crawford}, Steven M. and {Dencheva}, Nadia and {Ely}, Justin and {Jenness}, Tim and {Labrie}, Kathleen and {Lim}, Pey Lian and {Pierfederici}, Francesco and {Pontzen}, Andrew and {Ptak}, Andy and {Refsdal}, Brian and {Servillat}, Mathieu and {Streicher}, Ole},
 doi = {10.1051/0004-6361/201322068},
 eid = {A33},
 eprint = {1307.6212},
 journal = {\aap},
 month = {October},
 pages = {A33},
 primaryclass = {astro-ph.IM},
 title = {{Astropy: A community Python package for astronomy}},
 volume = {558},
 year = {2013}
}

@article{2013ApJ...764..167S,
 adsurl = {https://ui.adsabs.harvard.edu/abs/2013ApJ...764..167S},
 archiveprefix = {arXiv},
 author = {{Scargle}, Jeffrey D. and {Norris}, Jay P. and {Jackson}, Brad and {Chiang}, James},
 doi = {10.1088/0004-637X/764/2/167},
 eid = {167},
 eprint = {1207.5578},
 journal = {\apj},
 month = {February},
 number = {2},
 pages = {167},
 primaryclass = {astro-ph.IM},
 title = {{Studies in Astronomical Time Series Analysis. VI. Bayesian Block Representations}},
 volume = {764},
 year = {2013}
}

@article{2013ApJ...774..132W,
 adsurl = {https://ui.adsabs.harvard.edu/abs/2013ApJ...774..132W},
 archiveprefix = {arXiv},
 author = {{Wang}, Xiang-Gao and {Liang}, En-Wei and {Li}, Liang and {Lu}, Rui-Jing and {Wei}, Jian-Yan and {Zhang}, Bing},
 doi = {10.1088/0004-637X/774/2/132},
 eid = {132},
 eprint = {1307.5939},
 journal = {\apj},
 month = {September},
 number = {2},
 pages = {132},
 primaryclass = {astro-ph.HE},
 title = {{A Comprehensive Study of Gamma-Ray Burst Optical Emission. III. Brightness Distributions and Luminosity Functions of Optical Afterglows}},
 volume = {774},
 year = {2013}
}

@article{2013ApJ...779...66S,
 adsurl = {https://ui.adsabs.harvard.edu/abs/2013ApJ...779...66S},
 archiveprefix = {arXiv},
 author = {{Stratta}, G. and {Gendre}, B. and {Atteia}, J.~L. and {Bo{\"e}r}, M. and {Coward}, D.~M. and {De Pasquale}, M. and {Howell}, E. and {Klotz}, A. and {Oates}, S. and {Piro}, L.},
 doi = {10.1088/0004-637X/779/1/66},
 eid = {66},
 eprint = {1306.1699},
 journal = {\apj},
 month = {December},
 number = {1},
 pages = {66},
 primaryclass = {astro-ph.HE},
 title = {{The Ultra-long GRB 111209A. II. Prompt to Afterglow and Afterglow Properties}},
 volume = {779},
 year = {2013}
}

@article{2013MNRAS.431..394W,
 adsurl = {https://ui.adsabs.harvard.edu/abs/2013MNRAS.431..394W},
 archiveprefix = {arXiv},
 author = {{Willingale}, R. and {Starling}, R.~L.~C. and {Beardmore}, A.~P. and {Tanvir}, N.~R. and {O'Brien}, P.~T.},
 doi = {10.1093/mnras/stt175},
 eprint = {1303.0843},
 journal = {\mnras},
 month = {May},
 number = {1},
 pages = {394-404},
 primaryclass = {astro-ph.HE},
 title = {{Calibration of X-ray absorption in our Galaxy}},
 volume = {431},
 year = {2013}
}

@article{2014A&A...571A..16P,
 adsurl = {https://ui.adsabs.harvard.edu/abs/2014A&A...571A..16P},
 archiveprefix = {arXiv},
 author = {{Planck Collaboration} and {Ade}, P.~A.~R. and {Aghanim}, N. and {Armitage-Caplan}, C. and {Arnaud}, M. and {Ashdown}, M. and {Atrio-Barandela}, F. and {Aumont}, J. and {Baccigalupi}, C. and {Banday}, A.~J. and et al.},
 doi = {10.1051/0004-6361/201321591},
 eid = {A16},
 eprint = {1303.5076},
 journal = {\aap},
 month = {November},
 pages = {A16},
 primaryclass = {astro-ph.CO},
 title = {{Planck 2013 results. XVI. Cosmological parameters}},
 volume = {571},
 year = {2014}
}

@article{2014MNRAS.440.1810M,
 adsurl = {https://ui.adsabs.harvard.edu/abs/2014MNRAS.440.1810M},
 archiveprefix = {arXiv},
 author = {{Morgan}, Adam N. and {Perley}, Daniel A. and {Cenko}, S. Bradley and {Bloom}, Joshua S. and {Cucchiara}, Antonino and {Richards}, Joseph W. and {Filippenko}, Alexei V. and {Haislip}, Joshua B. and {LaCluyze}, Aaron and {Corsi}, Alessandra and {Melandri}, Andrea and {Cobb}, Bethany E. and {Gomboc}, Andreja and {Horesh}, Assaf and {James}, Berian and {Li}, Weidong and {Mundell}, Carole G. and {Reichart}, Daniel E. and {Steele}, Iain},
 doi = {10.1093/mnras/stu344},
 eprint = {1305.1928},
 journal = {\mnras},
 month = {May},
 number = {2},
 pages = {1810-1823},
 primaryclass = {astro-ph.HE},
 title = {{Evidence for dust destruction from the early-time colour change of GRB 120119A}},
 volume = {440},
 year = {2014}
}

@article{2015A&A...579A..74J,
 adsurl = {https://ui.adsabs.harvard.edu/abs/2015A&A...579A..74J},
 archiveprefix = {arXiv},
 author = {{Japelj}, J. and {Covino}, S. and {Gomboc}, A. and {Vergani}, S.~D. and {Goldoni}, P. and {Selsing}, J. and {Cano}, Z. and {D'Elia}, V. and {Flores}, H. and {Fynbo}, J.~P.~U. and {Hammer}, F. and {Hjorth}, J. and {Jakobsson}, P. and {Kaper}, L. and {Kopa{\v{c}}}, D. and {Kr{\"u}hler}, T. and {Melandri}, A. and {Piranomonte}, S. and {S{\'a}nchez-Ram{\'\i}rez}, R. and {Tagliaferri}, G. and {Tanvir}, N.~R. and {de Ugarte Postigo}, A. and {Watson}, D. and {Wijers}, R.~A.~M.~J.},
 doi = {10.1051/0004-6361/201525665},
 eid = {A74},
 eprint = {1503.03623},
 journal = {\aap},
 month = {July},
 pages = {A74},
 primaryclass = {astro-ph.HE},
 title = {{Spectrophotometric analysis of gamma-ray burst afterglow extinction curves with X-Shooter}},
 volume = {579},
 year = {2015}
}

@article{2015ApJ...806..250H,
 adsurl = {https://ui.adsabs.harvard.edu/abs/2015ApJ...806..250H},
 archiveprefix = {arXiv},
 author = {{Hashimoto}, Tetsuya and {Perley}, Daniel A. and {Ohta}, Kouji and {Aoki}, Kentaro and {Tanaka}, Ichi and {Niino}, Yuu and {Yabe}, Kiyoto and {Kawai}, Nobuyuki},
 doi = {10.1088/0004-637X/806/2/250},
 eid = {250},
 eprint = {1411.3357},
 journal = {\apj},
 month = {June},
 number = {2},
 pages = {250},
 primaryclass = {astro-ph.GA},
 title = {{The Star Formation Rate and Metallicity of the Host Galaxy of the Dark GRB 080325 at z=1.78}},
 volume = {806},
 year = {2015}
}

@article{2015arXiv150704544V,
 adsurl = {https://ui.adsabs.harvard.edu/abs/2015arXiv150704544V},
 archiveprefix = {arXiv},
 author = {{Vehtari}, Aki and {Gelman}, Andrew and {Gabry}, Jonah},
 doi = {10.48550/arXiv.1507.04544},
 eid = {arXiv:1507.04544},
 eprint = {1507.04544},
 journal = {arXiv e-prints},
 month = {July},
 pages = {arXiv:1507.04544},
 primaryclass = {stat.CO},
 title = {{Practical Bayesian model evaluation using leave-one-out cross-validation and WAIC}},
 year = {2015}
}

@article{2015MNRAS.449.2919L,
 adsurl = {https://ui.adsabs.harvard.edu/abs/2015MNRAS.449.2919L},
 archiveprefix = {arXiv},
 author = {{Littlejohns}, O.~M. and {Butler}, N.~R. and {Cucchiara}, A. and {Watson}, A.~M. and {Fox}, O.~D. and {Lee}, W.~H. and {Kutyrev}, A.~S. and {Richer}, M.~G. and {Klein}, C.~R. and {Prochaska}, J.~X. and {Bloom}, J.~S. and {Troja}, E. and {Ramirez-Ruiz}, E. and {de Diego}, J.~A. and {Georgiev}, L. and {Gonz{\'a}lez}, J. and {Rom{\'a}n-Z{\'u}{\~n}iga}, C.~G. and {Gehrels}, N. and {Moseley}, H.},
 doi = {10.1093/mnras/stv479},
 eprint = {1412.6530},
 journal = {\mnras},
 month = {May},
 number = {3},
 pages = {2919-2936},
 primaryclass = {astro-ph.HE},
 title = {{A detailed study of the optical attenuation of gamma-ray bursts in the Swift era}},
 volume = {449},
 year = {2015}
}

@article{2015PhR...561....1K,
 adsurl = {https://ui.adsabs.harvard.edu/abs/2015PhR...561....1K},
 archiveprefix = {arXiv},
 author = {{Kumar}, Pawan and {Zhang}, Bing},
 doi = {10.1016/j.physrep.2014.09.008},
 eprint = {1410.0679},
 journal = {\physrep},
 month = {February},
 pages = {1-109},
 primaryclass = {astro-ph.HE},
 title = {{The physics of gamma-ray bursts \& relativistic jets}},
 volume = {561},
 year = {2015}
}

@article{2016ApJ...831..111M,
 adsurl = {https://ui.adsabs.harvard.edu/abs/2016ApJ...831..111M},
 archiveprefix = {arXiv},
 author = {{Mu}, Hui-Jun and {Lin}, Da-Bin and {Xi}, Shao-Qiang and {Lin}, Ting-Ting and {Wang}, Yuan-Zhu and {Liang}, Yun-Feng and {L{\"u}}, Lian-Zhong and {Zhang}, Jin and {Liang}, En-Wei},
 doi = {10.3847/0004-637X/831/1/111},
 eid = {111},
 eprint = {1608.05028},
 journal = {\apj},
 month = {November},
 number = {1},
 pages = {111},
 primaryclass = {astro-ph.HE},
 title = {{The History of GRB Outflows: Ejection Lorentz Factor and Radiation Location of X-Ray Flares}},
 volume = {831},
 year = {2016}
}

@article{2016arXiv161205560C,
 adsurl = {https://ui.adsabs.harvard.edu/abs/2016arXiv161205560C},
 archiveprefix = {arXiv},
 author = {{Chambers}, K.~C. and {Magnier}, E.~A. and {Metcalfe}, N. and {Flewelling}, H.~A. and {Huber}, M.~E. and {Waters}, C.~Z. and {Denneau}, L. and {Draper}, P.~W. and {Farrow}, D. and {Finkbeiner}, D.~P. and {Holmberg}, C. and {Koppenhoefer}, J. and {Price}, P.~A. and {Rest}, A. and {Saglia}, R.~P. and {Schlafly}, E.~F. and {Smartt}, S.~J. and {Sweeney}, W. and {Wainscoat}, R.~J. and {Burgett}, W.~S. and {Chastel}, S. and {Grav}, T. and {Heasley}, J.~N. and {Hodapp}, K.~W. and {Jedicke}, R. and {Kaiser}, N. and {Kudritzki}, R. -P. and {Luppino}, G.~A. and {Lupton}, R.~H. and {Monet}, D.~G. and {Morgan}, J.~S. and {Onaka}, P.~M. and {Shiao}, B. and {Stubbs}, C.~W. and {Tonry}, J.~L. and {White}, R. and {Ba{\~n}ados}, E. and {Bell}, E.~F. and {Bender}, R. and {Bernard}, E.~J. and {Boegner}, M. and {Boffi}, F. and {Botticella}, M.~T. and {Calamida}, A. and {Casertano}, S. and {Chen}, W. -P. and {Chen}, X. and {Cole}, S. and {Deacon}, N. and {Frenk}, C. and {Fitzsimmons}, A. and {Gezari}, S. and {Gibbs}, V. and {Goessl}, C. and {Goggia}, T. and {Gourgue}, R. and {Goldman}, B. and {Grant}, P. and {Grebel}, E.~K. and {Hambly}, N.~C. and {Hasinger}, G. and {Heavens}, A.~F. and {Heckman}, T.~M. and {Henderson}, R. and {Henning}, T. and {Holman}, M. and {Hopp}, U. and {Ip}, W. -H. and {Isani}, S. and {Jackson}, M. and {Keyes}, C.~D. and {Koekemoer}, A.~M. and {Kotak}, R. and {Le}, D. and {Liska}, D. and {Long}, K.~S. and {Lucey}, J.~R. and {Liu}, M. and {Martin}, N.~F. and {Masci}, G. and {McLean}, B. and {Mindel}, E. and {Misra}, P. and {Morganson}, E. and {Murphy}, D.~N.~A. and {Obaika}, A. and {Narayan}, G. and {Nieto-Santisteban}, M.~A. and {Norberg}, P. and {Peacock}, J.~A. and {Pier}, E.~A. and {Postman}, M. and {Primak}, N. and {Rae}, C. and {Rai}, A. and {Riess}, A. and {Riffeser}, A. and {Rix}, H.~W. and {R{\"o}ser}, S. and {Russel}, R. and {Rutz}, L. and {Schilbach}, E. and {Schultz}, A.~S.~B. and {Scolnic}, D. and {Strolger}, L. and {Szalay}, A. and {Seitz}, S. and {Small}, E. and {Smith}, K.~W. and {Soderblom}, D.~R. and {Taylor}, P. and {Thomson}, R. and {Taylor}, A.~N. and {Thakar}, A.~R. and {Thiel}, J. and {Thilker}, D. and {Unger}, D. and {Urata}, Y. and {Valenti}, J. and {Wagner}, J. and {Walder}, T. and {Walter}, F. and {Watters}, S.~P. and {Werner}, S. and {Wood-Vasey}, W.~M. and {Wyse}, R.},
 doi = {10.48550/arXiv.1612.05560},
 eid = {arXiv:1612.05560},
 eprint = {1612.05560},
 journal = {arXiv e-prints},
 month = {December},
 pages = {arXiv:1612.05560},
 primaryclass = {astro-ph.IM},
 title = {{The Pan-STARRS1 Surveys}},
 year = {2016}
}

@article{2017A&A...601A..83H,
 adsurl = {https://ui.adsabs.harvard.edu/abs/2017A&A...601A..83H},
 archiveprefix = {arXiv},
 author = {{Heintz}, K.~E. and {Fynbo}, J.~P.~U. and {Jakobsson}, P. and {Kr{\"u}hler}, T. and {Christensen}, L. and {Watson}, D. and {Ledoux}, C. and {Noterdaeme}, P. and {Perley}, D.~A. and {Rhodin}, H. and {Selsing}, J. and {Schulze}, S. and {Tanvir}, N.~R. and {M{\o}ller}, P. and {Goldoni}, P. and {Xu}, D. and {Milvang-Jensen}, B.},
 doi = {10.1051/0004-6361/201730702},
 eid = {A83},
 eprint = {1703.07109},
 journal = {\aap},
 month = {May},
 pages = {A83},
 primaryclass = {astro-ph.GA},
 title = {{Steep extinction towards GRB 140506A reconciled from host galaxy observations: Evidence that steep reddening laws are local}},
 volume = {601},
 year = {2017}
}

@article{2017ApJ...835...37L,
 adsurl = {https://ui.adsabs.harvard.edu/abs/2017ApJ...835...37L},
 archiveprefix = {arXiv},
 author = {{Lapi}, A. and {Mancuso}, C. and {Celotti}, A. and {Danese}, L.},
 doi = {10.3847/1538-4357/835/1/37},
 eid = {37},
 eprint = {1612.01304},
 journal = {\apj},
 month = {January},
 number = {1},
 pages = {37},
 primaryclass = {astro-ph.GA},
 title = {{Galaxy Evolution at High Redshift: Obscured Star Formation, GRB Rates, Cosmic Reionization, and Missing Satellites}},
 volume = {835},
 year = {2017}
}

@article{2017ApJ...850..161T,
 adsurl = {https://ui.adsabs.harvard.edu/abs/2017ApJ...850..161T},
 archiveprefix = {arXiv},
 author = {{Tsvetkova}, A. and {Frederiks}, D. and {Golenetskii}, S. and {Lysenko}, A. and {Oleynik}, P. and {Pal'shin}, V. and {Svinkin}, D. and {Ulanov}, M. and {Cline}, T. and {Hurley}, K. and {Aptekar}, R.},
 doi = {10.3847/1538-4357/aa96af},
 eid = {161},
 eprint = {1710.08746},
 journal = {\apj},
 month = {December},
 number = {2},
 pages = {161},
 primaryclass = {astro-ph.HE},
 title = {{The Konus-Wind Catalog of Gamma-Ray Bursts with Known Redshifts. I. Bursts Detected in the Triggered Mode}},
 volume = {850},
 year = {2017}
}

@article{2018A&A...617A.141C,
 adsurl = {https://ui.adsabs.harvard.edu/abs/2018A&A...617A.141C},
 archiveprefix = {arXiv},
 author = {{Corre}, D. and {Buat}, V. and {Basa}, S. and {Boissier}, S. and {Japelj}, J. and {Palmerio}, J. and {Salvaterra}, R. and {Vergani}, S.~D. and {Zafar}, T.},
 doi = {10.1051/0004-6361/201832926},
 eid = {A141},
 eprint = {1807.00635},
 journal = {\aap},
 month = {September},
 pages = {A141},
 primaryclass = {astro-ph.GA},
 title = {{Investigation of dust attenuation and star formation activity in galaxies hosting GRBs}},
 volume = {617},
 year = {2018}
}

@article{2018AJ....156..123A,
 adsurl = {https://ui.adsabs.harvard.edu/abs/2018AJ....156..123A},
 archiveprefix = {arXiv},
 author = {{Astropy Collaboration} and {Price-Whelan}, A.~M. and {Sip{\H{o}}cz}, B.~M. and {G{\"u}nther}, H.~M. and {Lim}, P.~L. and {Crawford}, S.~M. and {Conseil}, S. and {Shupe}, D.~L. and {Craig}, M.~W. and {Dencheva}, N. and {Ginsburg}, A. and {VanderPlas}, J.~T. and {Bradley}, L.~D. and {P{\'e}rez-Su{\'a}rez}, D. and {de Val-Borro}, M. and {Aldcroft}, T.~L. and {Cruz}, K.~L. and {Robitaille}, T.~P. and {Tollerud}, E.~J. and {Ardelean}, C. and {Babej}, T. and {Bach}, Y.~P. and {Bachetti}, M. and {Bakanov}, A.~V. and {Bamford}, S.~P. and {Barentsen}, G. and {Barmby}, P. and {Baumbach}, A. and {Berry}, K.~L. and {Biscani}, F. and {Boquien}, M. and {Bostroem}, K.~A. and {Bouma}, L.~G. and {Brammer}, G.~B. and {Bray}, E.~M. and {Breytenbach}, H. and {Buddelmeijer}, H. and {Burke}, D.~J. and {Calderone}, G. and {Cano Rodr{\'\i}guez}, J.~L. and {Cara}, M. and {Cardoso}, J.~V.~M. and {Cheedella}, S. and {Copin}, Y. and {Corrales}, L. and {Crichton}, D. and {D'Avella}, D. and {Deil}, C. and {Depagne}, {\'E}. and {Dietrich}, J.~P. and {Donath}, A. and {Droettboom}, M. and {Earl}, N. and {Erben}, T. and {Fabbro}, S. and {Ferreira}, L.~A. and {Finethy}, T. and {Fox}, R.~T. and {Garrison}, L.~H. and {Gibbons}, S.~L.~J. and {Goldstein}, D.~A. and {Gommers}, R. and {Greco}, J.~P. and {Greenfield}, P. and {Groener}, A.~M. and {Grollier}, F. and {Hagen}, A. and {Hirst}, P. and {Homeier}, D. and {Horton}, A.~J. and {Hosseinzadeh}, G. and {Hu}, L. and {Hunkeler}, J.~S. and {Ivezi{\'c}}, {\v{Z}}. and {Jain}, A. and {Jenness}, T. and {Kanarek}, G. and {Kendrew}, S. and {Kern}, N.~S. and {Kerzendorf}, W.~E. and {Khvalko}, A. and {King}, J. and {Kirkby}, D. and {Kulkarni}, A.~M. and {Kumar}, A. and {Lee}, A. and {Lenz}, D. and {Littlefair}, S.~P. and {Ma}, Z. and {Macleod}, D.~M. and {Mastropietro}, M. and {McCully}, C. and {Montagnac}, S. and {Morris}, B.~M. and {Mueller}, M. and {Mumford}, S.~J. and {Muna}, D. and {Murphy}, N.~A. and {Nelson}, S. and {Nguyen}, G.~H. and {Ninan}, J.~P. and {N{\"o}the}, M. and {Ogaz}, S. and {Oh}, S. and {Parejko}, J.~K. and {Parley}, N. and {Pascual}, S. and {Patil}, R. and {Patil}, A.~A. and {Plunkett}, A.~L. and {Prochaska}, J.~X. and {Rastogi}, T. and {Reddy Janga}, V. and {Sabater}, J. and {Sakurikar}, P. and {Seifert}, M. and {Sherbert}, L.~E. and {Sherwood-Taylor}, H. and {Shih}, A.~Y. and {Sick}, J. and {Silbiger}, M.~T. and {Singanamalla}, S. and {Singer}, L.~P. and {Sladen}, P.~H. and {Sooley}, K.~A. and {Sornarajah}, S. and {Streicher}, O. and {Teuben}, P. and {Thomas}, S.~W. and {Tremblay}, G.~R. and {Turner}, J.~E.~H. and {Terr{\'o}n}, V. and {van Kerkwijk}, M.~H. and {de la Vega}, A. and {Watkins}, L.~L. and {Weaver}, B.~A. and {Whitmore}, J.~B. and {Woillez}, J. and {Zabalza}, V. and {Astropy Contributors}},
 doi = {10.3847/1538-3881/aabc4f},
 eid = {123},
 eprint = {1801.02634},
 journal = {\aj},
 month = {September},
 number = {3},
 pages = {123},
 primaryclass = {astro-ph.IM},
 title = {{The Astropy Project: Building an Open-science Project and Status of the v2.0 Core Package}},
 volume = {156},
 year = {2018}
}

@article{2018ApJ...860L..21Z,
 adsurl = {https://ui.adsabs.harvard.edu/abs/2018ApJ...860L..21Z},
 archiveprefix = {arXiv},
 author = {{Zafar}, T. and {Heintz}, K.~E. and {Fynbo}, J.~P.~U. and {Malesani}, D. and {Bolmer}, J. and {Ledoux}, C. and {Arabsalmani}, M. and {Kaper}, L. and {Campana}, S. and {Starling}, R.~L.~C. and {Selsing}, J. and {Kann}, D.~A. and {de Ugarte Postigo}, A. and {Schweyer}, T. and {Christensen}, L. and {M{\o}ller}, P. and {Japelj}, J. and {Perley}, D. and {Tanvir}, N.~R. and {D'Avanzo}, P. and {Hartmann}, D.~H. and {Hjorth}, J. and {Covino}, S. and {Sbarufatti}, B. and {Jakobsson}, P. and {Izzo}, L. and {Salvaterra}, R. and {D'Elia}, V. and {Xu}, D.},
 doi = {10.3847/2041-8213/aaca3f},
 eid = {L21},
 eprint = {1806.00293},
 journal = {\apjl},
 month = {June},
 number = {2},
 pages = {L21},
 primaryclass = {astro-ph.GA},
 title = {{The 2175 {\r{A}} Extinction Feature in the Optical Afterglow Spectrum of GRB 180325A at z = 2.25}},
 volume = {860},
 year = {2018}
}

@article{2018ApJ...863...95H,
 adsurl = {https://ui.adsabs.harvard.edu/abs/2018ApJ...863...95H},
 archiveprefix = {arXiv},
 author = {{Hashimoto}, Tetsuya and {Chaudhary}, Ravi and {Ohta}, Kouji and {Goto}, Tomotsugu and {Hammer}, Francois and {Kong}, Albert K.~H. and {Nomoto}, Ken'ichi and {Mao}, Jirong},
 doi = {10.3847/1538-4357/aad2d1},
 eid = {95},
 eprint = {1808.08969},
 journal = {\apj},
 month = {August},
 number = {1},
 pages = {95},
 primaryclass = {astro-ph.GA},
 title = {{Why Are Some Gamma-Ray Bursts Hosted by Oxygen-rich Galaxies?}},
 volume = {863},
 year = {2018}
}

@article{2018MNRAS.479.1542Z,
 adsurl = {https://ui.adsabs.harvard.edu/abs/2018MNRAS.479.1542Z},
 archiveprefix = {arXiv},
 author = {{Zafar}, T. and {Watson}, D. and {M{\o}ller}, P. and {Selsing}, J. and {Fynbo}, J.~P.~U. and {Schady}, P. and {Wiersema}, K. and {Levan}, A.~J. and {Heintz}, K.~E. and {de Ugarte Postigo}, A. and {D'Elia}, V. and {Jakobsson}, P. and {Bolmer}, J. and {Japelj}, J. and {Covino}, S. and {Gomboc}, A. and {Cano}, Z.},
 doi = {10.1093/mnras/sty1380},
 eprint = {1805.07016},
 journal = {\mnras},
 month = {September},
 number = {2},
 pages = {1542-1554},
 primaryclass = {astro-ph.GA},
 title = {{VLT/X-shooter GRBs: Individual extinction curves of star-forming regions}},
 volume = {479},
 year = {2018}
}

@article{2018MNRAS.480..108Z,
 adsurl = {https://ui.adsabs.harvard.edu/abs/2018MNRAS.480..108Z},
 archiveprefix = {arXiv},
 author = {{Zafar}, T. and {M{\o}ller}, P. and {Watson}, D. and {Lattanzio}, J. and {Hopkins}, A.~M. and {Karakas}, A. and {Fynbo}, J.~P.~U. and {Tanvir}, N.~R. and {Selsing}, J. and {Jakobsson}, P. and {Heintz}, K.~E. and {Kann}, D.~A. and {Groves}, B. and {Kulkarni}, V. and {Covino}, S. and {D'Elia}, V. and {Japelj}, J. and {Corre}, D. and {Vergani}, S.},
 doi = {10.1093/mnras/sty1876},
 eprint = {1807.03597},
 journal = {\mnras},
 month = {October},
 number = {1},
 pages = {108-118},
 primaryclass = {astro-ph.GA},
 title = {{X-shooting GRBs at high redshift: probing dust production history*}},
 volume = {480},
 year = {2018}
}

@article{2019MNRAS.486.2063H,
 adsurl = {https://ui.adsabs.harvard.edu/abs/2019MNRAS.486.2063H},
 archiveprefix = {arXiv},
 author = {{Heintz}, K.~E. and {Zafar}, T. and {De Cia}, A. and {Vergani}, S.~D. and {Jakobsson}, P. and {Fynbo}, J.~P.~U. and {Watson}, D. and {Japelj}, J. and {M{\o}ller}, P. and {Covino}, S. and {Kaper}, L. and {Andersen}, A.~C.},
 doi = {10.1093/mnras/stz1012},
 eprint = {1904.04301},
 journal = {\mnras},
 month = {June},
 number = {2},
 pages = {2063-2074},
 primaryclass = {astro-ph.GA},
 title = {{On the dust properties of high-redshift molecular clouds and the connection to the 2175 {\r{A}} extinction bump}},
 volume = {486},
 year = {2019}
}

@article{2019RAA....19..149W,
 adsurl = {https://ui.adsabs.harvard.edu/abs/2019RAA....19..149W},
 archiveprefix = {arXiv},
 author = {{Wang}, Chuan-Jun and {Bai}, Jin-Ming and {Fan}, Yu-Feng and {Mao}, Ji-Rong and {Chang}, Liang and {Xin}, Yu-Xin and {Zhang}, Ju-Jia and {Lun}, Bao-Li and {Wang}, Jian-Guo and {Zhang}, Xi-Liang and {Ying}, Mei and {Lu}, Kai-Xing and {Wang}, Xiao-Li and {Ji}, Kai-Fan and {Xiong}, Ding-Rong and {Yu}, Xiao-Guang and {Ding}, Xu and {Ye}, Kai and {Xing}, Li-Feng and {Yi}, Wei-Min and {Xu}, Liang and {Zheng}, Xiang-Ming and {Feng}, Yuan-Jie and {He}, Shou-Sheng and {Wang}, Xue-Li and {Liu}, Zhong and {Chen}, Dong and {Xu}, Jun and {Qin}, Song-Nian and {Zhang}, Rui-Long and {Tan}, Hui-Song and {Li}, Zhi and {Lou}, Ke and {Li}, Jian and {Liu}, Wei-Wei},
 doi = {10.1088/1674-4527/19/10/149},
 eid = {149},
 eprint = {1905.05915},
 journal = {Research in Astronomy and Astrophysics},
 month = {October},
 number = {10},
 pages = {149},
 primaryclass = {astro-ph.IM},
 title = {{Lijiang 2.4-meter Telescope and its instruments}},
 volume = {19},
 year = {2019}
}

@article{2020ApJ...895...16H,
 adsurl = {https://ui.adsabs.harvard.edu/abs/2020ApJ...895...16H},
 archiveprefix = {arXiv},
 author = {{Hoang}, Thiem and {Giang}, Nguyen Chau and {Tram}, Le Ngoc},
 doi = {10.3847/1538-4357/ab8ae1},
 eid = {16},
 eprint = {1912.03803},
 journal = {\apj},
 month = {May},
 number = {1},
 pages = {16},
 primaryclass = {astro-ph.GA},
 title = {{Gamma-Ray Burst Afterglows: Time-varying Extinction, Polarization, and Colors due to Rotational Disruption of Dust Grains}},
 volume = {895},
 year = {2020}
}

@article{2020Natur.585..357H,
 adsurl = {https://ui.adsabs.harvard.edu/abs/2020Natur.585..357H},
 archiveprefix = {arXiv},
 author = {{Harris}, Charles R. and {Millman}, K. Jarrod and {van der Walt}, St{\'e}fan J. and {Gommers}, Ralf and {Virtanen}, Pauli and {Cournapeau}, David and {Wieser}, Eric and {Taylor}, Julian and {Berg}, Sebastian and {Smith}, Nathaniel J. and {Kern}, Robert and {Picus}, Matti and {Hoyer}, Stephan and {van Kerkwijk}, Marten H. and {Brett}, Matthew and {Haldane}, Allan and {del R{\'\i}o}, Jaime Fern{\'a}ndez and {Wiebe}, Mark and {Peterson}, Pearu and {G{\'e}rard-Marchant}, Pierre and {Sheppard}, Kevin and {Reddy}, Tyler and {Weckesser}, Warren and {Abbasi}, Hameer and {Gohlke}, Christoph and {Oliphant}, Travis E.},
 doi = {10.1038/s41586-020-2649-2},
 eprint = {2006.10256},
 journal = {\nat},
 month = {September},
 number = {7825},
 pages = {357-362},
 primaryclass = {cs.MS},
 title = {{Array programming with NumPy}},
 volume = {585},
 year = {2020}
}

@article{2020RAA....20..149X,
 adsurl = {https://ui.adsabs.harvard.edu/abs/2020RAA....20..149X},
 archiveprefix = {arXiv},
 author = {{Xin}, Yu-Xin and {Bai}, Jin-Ming and {Lun}, Bao-Li and {Fan}, Yu-Feng and {Wang}, Chuan-Jun and {Liu}, Xiao-Wei and {Yu}, Xiao-Guang and {Ye}, Kai and {Song}, Teng-Fei and {Chang}, Liang and {He}, Shou-Sheng and {Mao}, Ji-Rong and {Xu}, Liang and {Xiong}, Ding-Rong and {Zhang}, Xi-Liang and {Wang}, Jian-Guo and {Ding}, Xu and {Feng}, Hai-Cheng and {Liu}, Xiang-Kun and {Huang}, Yang and {Chen}, Bing-Qiu},
 doi = {10.1088/1674-4527/20/9/149},
 eid = {149},
 eprint = {2004.12128},
 journal = {Research in Astronomy and Astrophysics},
 month = {September},
 number = {9},
 pages = {149},
 primaryclass = {astro-ph.IM},
 title = {{Astronomical Site Monitoring System at Lijiang Observatory}},
 volume = {20},
 year = {2020}
}

@inproceedings{2020SPIE11445E..7MY,
 adsurl = {https://ui.adsabs.harvard.edu/abs/2020SPIE11445E..7MY},
 author = {{Yuan}, Xiangyan and {Li}, Zhengyang and {Liu}, Xiaowei and {Niu}, Dongsheng and {Lu}, Qishui and {Jiang}, Fanghua and {Wang}, Yuefei and {Li}, Xiaoyan and {Liang}, YongJun and {Wang}, Hai and {Zhang}, Chao and {Wang}, Jinfeng and {Li}, Bo and {Tian}, Jie and {Lu}, Haiping and {Chen}, Bingqiu and {Huang}, Yang and {Liu}, Xiangkun and {Yao}, Zhengqiu and {Cui}, Xiangqun and {Li}, Guoping},
 booktitle = {Ground-based and Airborne Telescopes VIII},
 doi = {10.1117/12.2562334},
 editor = {{Marshall}, Heather K. and {Spyromilio}, Jason and {Usuda}, Tomonori},
 eid = {114457M},
 month = {December},
 pages = {114457M},
 series = {Society of Photo-Optical Instrumentation Engineers (SPIE) Conference Series},
 title = {{Development of the Multi-channel Photometric Survey telescope}},
 volume = {11445},
 year = {2020}
}

@article{2021A&A...649A.135C,
 adsurl = {https://ui.adsabs.harvard.edu/abs/2021A&A...649A.135C},
 archiveprefix = {arXiv},
 author = {{Campana}, Sergio and {Lazzati}, Davide and {Perna}, Rosalba and {Grazia Bernardini}, Maria and {Nava}, Lara},
 doi = {10.1051/0004-6361/202140439},
 eid = {A135},
 eprint = {2103.05735},
 journal = {\aap},
 month = {May},
 pages = {A135},
 primaryclass = {astro-ph.HE},
 title = {{The variable absorption in the X-ray spectrum of GRB 190114C}},
 volume = {649},
 year = {2021}
}

@software{2021ascl.soft12006K,
 adsurl = {https://ui.adsabs.harvard.edu/abs/2021ascl.soft12006K},
 author = {{Karpov}, Sergey},
 eid = {ascl:2112.006},
 howpublished = {Astrophysics Source Code Library, record ascl:2112.006},
 month = {December},
 title = {{STDPipe: Simple Transient Detection Pipeline}},
 year = {2021}
}

@article{2021NatCo..12.4040R,
 adsurl = {https://ui.adsabs.harvard.edu/abs/2021NatCo..12.4040R},
 archiveprefix = {arXiv},
 author = {{Ronchini}, Samuele and {Oganesyan}, Gor and {Branchesi}, Marica and {Ascenzi}, Stefano and {Bernardini}, Maria Grazia and {Brighenti}, Francesco and {Dall'Osso}, Simone and {D'Avanzo}, Paolo and {Ghirlanda}, Giancarlo and {Ghisellini}, Gabriele and {Ravasio}, Maria Edvige and {Salafia}, Om Sharan},
 doi = {10.1038/s41467-021-24246-x},
 eid = {4040},
 eprint = {2009.03913},
 journal = {Nature Communications},
 month = {January},
 pages = {4040},
 primaryclass = {astro-ph.HE},
 title = {{Spectral index-flux relation for investigating the origins of steep decay in {\ensuremath{\gamma}}-ray bursts}},
 volume = {12},
 year = {2021}
}

@article{2022ApJ...935..167A,
 adsurl = {https://ui.adsabs.harvard.edu/abs/2022ApJ...935..167A},
 archiveprefix = {arXiv},
 author = {{Astropy Collaboration} and {Price-Whelan}, Adrian M. and {Lim}, Pey Lian and {Earl}, Nicholas and {Starkman}, Nathaniel and {Bradley}, Larry and {Shupe}, David L. and {Patil}, Aarya A. and {Corrales}, Lia and {Brasseur}, C.~E. and {N{\"o}the}, Maximilian and {Donath}, Axel and {Tollerud}, Erik and {Morris}, Brett M. and {Ginsburg}, Adam and {Vaher}, Eero and {Weaver}, Benjamin A. and {Tocknell}, James and {Jamieson}, William and {van Kerkwijk}, Marten H. and {Robitaille}, Thomas P. and {Merry}, Bruce and {Bachetti}, Matteo and {G{\"u}nther}, H. Moritz and {Aldcroft}, Thomas L. and {Alvarado-Montes}, Jaime A. and {Archibald}, Anne M. and {B{\'o}di}, Attila and {Bapat}, Shreyas and {Barentsen}, Geert and {Baz{\'a}n}, Juanjo and {Biswas}, Manish and {Boquien}, M{\'e}d{\'e}ric and {Burke}, D.~J. and {Cara}, Daria and {Cara}, Mihai and {Conroy}, Kyle E. and {Conseil}, Simon and {Craig}, Matthew W. and {Cross}, Robert M. and {Cruz}, Kelle L. and {D'Eugenio}, Francesco and {Dencheva}, Nadia and {Devillepoix}, Hadrien A.~R. and {Dietrich}, J{\"o}rg P. and {Eigenbrot}, Arthur Davis and {Erben}, Thomas and {Ferreira}, Leonardo and {Foreman-Mackey}, Daniel and {Fox}, Ryan and {Freij}, Nabil and {Garg}, Suyog and {Geda}, Robel and {Glattly}, Lauren and {Gondhalekar}, Yash and {Gordon}, Karl D. and {Grant}, David and {Greenfield}, Perry and {Groener}, Austen M. and {Guest}, Steve and {Gurovich}, Sebastian and {Handberg}, Rasmus and {Hart}, Akeem and {Hatfield-Dodds}, Zac and {Homeier}, Derek and {Hosseinzadeh}, Griffin and {Jenness}, Tim and {Jones}, Craig K. and {Joseph}, Prajwel and {Kalmbach}, J. Bryce and {Karamehmetoglu}, Emir and {Ka{\l}uszy{\'n}ski}, Miko{\l}aj and {Kelley}, Michael S.~P. and {Kern}, Nicholas and {Kerzendorf}, Wolfgang E. and {Koch}, Eric W. and {Kulumani}, Shankar and {Lee}, Antony and {Ly}, Chun and {Ma}, Zhiyuan and {MacBride}, Conor and {Maljaars}, Jakob M. and {Muna}, Demitri and {Murphy}, N.~A. and {Norman}, Henrik and {O'Steen}, Richard and {Oman}, Kyle A. and {Pacifici}, Camilla and {Pascual}, Sergio and {Pascual-Granado}, J. and {Patil}, Rohit R. and {Perren}, Gabriel I. and {Pickering}, Timothy E. and {Rastogi}, Tanuj and {Roulston}, Benjamin R. and {Ryan}, Daniel F. and {Rykoff}, Eli S. and {Sabater}, Jose and {Sakurikar}, Parikshit and {Salgado}, Jes{\'u}s and {Sanghi}, Aniket and {Saunders}, Nicholas and {Savchenko}, Volodymyr and {Schwardt}, Ludwig and {Seifert-Eckert}, Michael and {Shih}, Albert Y. and {Jain}, Anany Shrey and {Shukla}, Gyanendra and {Sick}, Jonathan and {Simpson}, Chris and {Singanamalla}, Sudheesh and {Singer}, Leo P. and {Singhal}, Jaladh and {Sinha}, Manodeep and {Sip{\H{o}}cz}, Brigitta M. and {Spitler}, Lee R. and {Stansby}, David and {Streicher}, Ole and {{\v{S}}umak}, Jani and {Swinbank}, John D. and {Taranu}, Dan S. and {Tewary}, Nikita and {Tremblay}, Grant R. and {de Val-Borro}, Miguel and {Van Kooten}, Samuel J. and {Vasovi{\'c}}, Zlatan and {Verma}, Shresth and {de Miranda Cardoso}, Jos{\'e} Vin{\'\i}cius and {Williams}, Peter K.~G. and {Wilson}, Tom J. and {Winkel}, Benjamin and {Wood-Vasey}, W.~M. and {Xue}, Rui and {Yoachim}, Peter and {Zhang}, Chen and {Zonca}, Andrea and {Astropy Project Contributors}},
 doi = {10.3847/1538-4357/ac7c74},
 eid = {167},
 eprint = {2206.14220},
 journal = {\apj},
 month = {August},
 number = {2},
 pages = {167},
 primaryclass = {astro-ph.IM},
 title = {{The Astropy Project: Sustaining and Growing a Community-oriented Open-source Project and the Latest Major Release (v5.0) of the Core Package}},
 volume = {935},
 year = {2022}
}

@software{2022ascl.soft08005G,
 adsurl = {https://ui.adsabs.harvard.edu/abs/2022ascl.soft08005G},
 author = {{Gobat}, Caden},
 eid = {ascl:2208.005},
 howpublished = {Astrophysics Source Code Library, record ascl:2208.005},
 month = {August},
 title = {{Asymmetric Uncertainty: Handling nonstandard numerical uncertainties}},
 year = {2022}
}

@article{2022JApA...43...11G,
 adsurl = {https://ui.adsabs.harvard.edu/abs/2022JApA...43...11G},
 archiveprefix = {arXiv},
 author = {{Gupta}, Rahul and {Kumar}, Amit and {Pandey}, Shashi Bhushan and {Castro-Tirado}, A.~J. and {Ghosh}, Ankur and {Dimple} and {Hu}, Y. -D. and {Fern{\'a}ndez-Garc{\'\i}a}, E. and {Caballero-Garc{\'\i}a}, M.~D. and {Castro-Tirado}, M. {\'A}. and {Hedrosa}, R.~P. and {Hermelo}, I. and {Vico}, I. and {Misra}, Kuntal and {Kumar}, Brajesh and {Aryan}, Amar and {Tiwari}, Sugriva Nath},
 doi = {10.1007/s12036-021-09794-4},
 eid = {11},
 eprint = {2111.11795},
 journal = {Journal of Astrophysics and Astronomy},
 month = {June},
 number = {1},
 pages = {11},
 primaryclass = {astro-ph.HE},
 title = {{Revealing nature of GRB 210205A, ZTF21aaeyldq (AT2021any) and follow-up observations with the 4K<inline-formula id=``IEq1''><mml:math><mml:mo>{\texttimes}</mml:mo></mml:math></inline-formula>4K CCD imager + 3.6m DOT}},
 volume = {43},
 year = {2022}
}

@article{2023A&A...674A..33G,
 adsurl = {https://ui.adsabs.harvard.edu/abs/2023A&A...674A..33G},
 archiveprefix = {arXiv},
 author = {{Gaia Collaboration} and {Montegriffo}, P. and {Bellazzini}, M. and {De Angeli}, F. and {Andrae}, R. and {Barstow}, M.~A. and {Bossini}, D. and {Bragaglia}, A. and {Burgess}, P.~W. and {Cacciari}, C. and et al.},
 doi = {10.1051/0004-6361/202243709},
 eid = {A33},
 eprint = {2206.06215},
 journal = {\aap},
 month = {June},
 pages = {A33},
 primaryclass = {astro-ph.SR},
 title = {{Gaia Data Release 3. The Galaxy in your preferred colours: Synthetic photometry from Gaia low-resolution spectra}},
 volume = {674},
 year = {2023}
}

@article{2023ApJ...944...73V,
 adsurl = {https://ui.adsabs.harvard.edu/abs/2023ApJ...944...73V},
 archiveprefix = {arXiv},
 author = {{Valan}, Vlasta and {Larsson}, Josefin and {Ahlgren}, Bj{\"o}rn},
 doi = {10.3847/1538-4357/acafe4},
 eid = {73},
 eprint = {2301.09292},
 journal = {\apj},
 month = {February},
 number = {1},
 pages = {73},
 primaryclass = {astro-ph.HE},
 title = {{Investigating Time Variability of X-Ray Absorption in Swift GRBs}},
 volume = {944},
 year = {2023}
}

@article{2023MNRAS.520.6104K,
 adsurl = {https://ui.adsabs.harvard.edu/abs/2023MNRAS.520.6104K},
 archiveprefix = {arXiv},
 author = {{Komesh}, Toktarkhan and {Grossan}, Bruce and {Maksut}, Zhanat and {Abdikamalov}, Ernazar and {Krugov}, Maxim and {Smoot}, George F.},
 doi = {10.1093/mnras/stad538},
 eprint = {2211.03029},
 journal = {\mnras},
 month = {April},
 number = {4},
 pages = {6104-6110},
 primaryclass = {astro-ph.HE},
 title = {{Evolution of the afterglow optical spectral shape of GRB 201015A in the first hour: evidence for dust destruction}},
 volume = {520},
 year = {2023}
}

@article{2023MNRAS.523..775G,
 adsurl = {https://ui.adsabs.harvard.edu/abs/2023MNRAS.523..775G},
 archiveprefix = {arXiv},
 author = {{Gobat}, Caden and {van der Horst}, Alexander J. and {Fitzpatrick}, David},
 doi = {10.1093/mnras/stad1189},
 eprint = {2304.09122},
 journal = {\mnras},
 month = {July},
 number = {1},
 pages = {775-784},
 primaryclass = {astro-ph.HE},
 title = {{Optical darkness in short-duration {\ensuremath{\gamma}}-ray bursts}},
 volume = {523},
 year = {2023}
}

@software{2023zndo...8126529L,
 adsurl = {https://ui.adsabs.harvard.edu/abs/2023zndo...8126529L},
 author = {{Li}, Jiaxuan},
 doi = {10.5281/zenodo.8126529},
 eid = {10.5281/zenodo.8126529},
 month = {July},
 publisher = {Zenodo},
 title = {{AstroJacobLi/smplotlib: v0.0.9}},
 version = {v0.0.9},
 year = {2023}
}

@software{2023zndo..10107975T,
 adsurl = {https://ui.adsabs.harvard.edu/abs/2023zndo..10107975T},
 author = {{The pandas development Team}},
 doi = {10.5281/zenodo.10107975},
 eid = {10.5281/zenodo.10107975},
 month = {November},
 publisher = {Zenodo},
 title = {{pandas-dev/pandas: Pandas}},
 version = {v2.1.3},
 year = {2023}
}

@article{2024A&A...690A.373R,
 adsurl = {https://ui.adsabs.harvard.edu/abs/2024A&A...690A.373R},
 archiveprefix = {arXiv},
 author = {{Rakotondrainibe}, N.~A. and {Buat}, V. and {Turpin}, D. and {Dornic}, D. and {Le Floc'h}, E. and {Vergani}, S.~D. and {Basa}, S.},
 doi = {10.1051/0004-6361/202450601},
 eid = {A373},
 eprint = {2409.03570},
 journal = {\aap},
 month = {October},
 pages = {A373},
 primaryclass = {astro-ph.HE},
 title = {{A simple model of dust extinction in gamma-ray burst host galaxies}},
 volume = {690},
 year = {2024}
}

@article{2024AJ....168..223L,
 adsurl = {https://ui.adsabs.harvard.edu/abs/2024AJ....168..223L},
 archiveprefix = {arXiv},
 author = {{Li}, Rui-Zhi and {Chen}, Bing-Qiu and {Li}, Guang-Xing and {Wang}, Bo-Ting and {Ren}, Hao-Ming and {Guo}, Qi-Ning},
 doi = {10.3847/1538-3881/ad77a3},
 eid = {223},
 eprint = {2409.02057},
 journal = {\aj},
 month = {November},
 number = {5},
 pages = {223},
 primaryclass = {astro-ph.GA},
 title = {{The Correlation Between Dust and Gas Contents in Molecular Clouds}},
 volume = {168},
 year = {2024}
}

@article{2024GCN.37273....1F,
 adsurl = {https://ui.adsabs.harvard.edu/abs/2024GCN.37273....1F},
 author = {{Fermi GBM Team}},
 journal = {GRB Coordinates Network},
 month = {August},
 pages = {1},
 title = {{GRB 240825A: Fermi GBM Final Real-time Localization}},
 volume = {37273},
 year = {2024}
}

@article{2024GCN.37274....1G,
 adsurl = {https://ui.adsabs.harvard.edu/abs/2024GCN.37274....1G},
 author = {{Gupta}, R. and {Brivio}, R. and {Dichiara}, S. and {Ferro}, M. and {Kennea}, J.~A. and {Page}, K.~L. and {Palmer}, D.~M. and {Sbarrato}, T. and {Neil Gehrels Swift Observatory Team}},
 journal = {GRB Coordinates Network},
 month = {August},
 pages = {1},
 title = {{GRB 240825A: Swift detection of a burst with a bright optical counterpart}},
 volume = {37274},
 year = {2024}
}

@article{2024GCN.37275....1J,
 adsurl = {https://ui.adsabs.harvard.edu/abs/2024GCN.37275....1J},
 author = {{Jiang}, S.~Q. and {Liu}, X. and {An}, J. and {Zhu}, Z.~P. and {Fu}, S.~Y. and {Xu}, D. and {Gao}, X. and {Liu}, J.~Z.},
 journal = {GRB Coordinates Network},
 month = {August},
 pages = {1},
 title = {{GRB 240825A: Nanshan/HMT optical afterglow observations}},
 volume = {37275},
 year = {2024}
}

@article{2024GCN.37276....1D,
 adsurl = {https://ui.adsabs.harvard.edu/abs/2024GCN.37276....1D},
 author = {{Dutton}, Dylan and {Dubay}, Megan and {Schlekat}, Donovan and {Fu}, Ruide and {Reichart}, Daniel and {Haislip}, Joshua and {Kouprianov}, Vladimir and {Kennewall}, John and {Veever}, Arie and {Janzen}, Daryl},
 journal = {GRB Coordinates Network},
 month = {August},
 pages = {1},
 title = {{GRB 240825A: Skynet Optical Afterglow observations}},
 volume = {37276},
 year = {2024}
}

@article{2024GCN.37277....1O,
 adsurl = {https://ui.adsabs.harvard.edu/abs/2024GCN.37277....1O},
 author = {{Odeh}, Mohammad and {Guessoum}, Nidhal},
 journal = {GRB Coordinates Network},
 month = {August},
 pages = {1},
 title = {{GRB 240825A: AKO Optical Afterglow Detection}},
 volume = {37277},
 year = {2024}
}

@article{2024GCN.37278....1Z,
 adsurl = {https://ui.adsabs.harvard.edu/abs/2024GCN.37278....1Z},
 author = {{Zhang}, Jinghua and {Du}, Guowang and {Guo}, Helong and {Kumar}, Brajesh and {Wang}, Tao and {Qin}, Zhenfei and {Jin}, Yicheng and {Zou}, Xingzhu and {Pan}, Yu and {Chen}, Xinlei and {Fang}, Yuan and {Han}, Xuhui and {Zhang}, Pinpin and {Xin}, Liping and {Wu}, Chao and {Yang}, Yuanpei and {Er}, Xinzhong and {Liu}, Xiangkun and {Liu}, Xiaowei and {Mephisto Team}},
 journal = {GRB Coordinates Network},
 month = {August},
 pages = {1},
 title = {{GRB 240825A: 1.6m Mephisto multi-band detection}},
 volume = {37278},
 year = {2024}
}

@article{2024GCN.37279....1L,
 adsurl = {https://ui.adsabs.harvard.edu/abs/2024GCN.37279....1L},
 author = {{Lipunov}, V. and {Kornilov}, V. and {Gorbovskoy}, E. and {Zhirkov}, K. and {Tyurina}, N. and {Balanutsa}, P. and {Kuznetsov}, A. and {Senik}, V. and {Vlasenko}, D. and {Antipov}, G. and {Zimnukhov}, D. and {Minkina}, E. and {Chasovnikov}, A. and {Topolev}, V. and {Kuvshinov}, D. and {Cheryasov}, D. and {Kechin}, Ya. and {Tselik}, Yu. and {Sosnovskij}, A. and {Podesta}, R. and {Lopez}, C. and {Podesta}, F. and {Francile}, C. and {Rebolo}, R. and {Serra}, M. and {Buckley}, D. and {Gress}, O.~A. and {Budnev}, N.~M. and {Ershova}, O. and {Carrasco}, L. and {Valdes}, J.~R. and {Chavushyan}, V. and {Alvarez}, Patino V.~M. and {Martinez}, J. and {Corella}, A.~R. and {Rodriguez}, L.~H. and {Tlatov}, A. and {Dormidontov}, D. and {Yurkov}, V. and {Gabovich}, A.},
 journal = {GRB Coordinates Network},
 month = {August},
 pages = {1},
 title = {{Swift GRB 240825A: Global MASTER-Net observations report}},
 volume = {37279},
 year = {2024}
}

@article{2024GCN.37280....1L,
 adsurl = {https://ui.adsabs.harvard.edu/abs/2024GCN.37280....1L},
 author = {{Li}, R. -Z. and {Wang}, B. -T. and {Song}, F. -F. and {Mao}, J. and {Xin}, Y. -X. and {Bai}, J. -M.},
 journal = {GRB Coordinates Network},
 month = {August},
 pages = {1},
 title = {{GRB 240825A: GMG Optical Observation}},
 volume = {37280},
 year = {2024}
}

@article{2024GCN.37283....1L,
 adsurl = {https://ui.adsabs.harvard.edu/abs/2024GCN.37283....1L},
 author = {{Lipunov}, V. and {Kornilov}, V. and {Gorbovskoy}, E. and {Zhirkov}, K. and {Tyurina}, N. and {Balanutsa}, P. and {Kuznetsov}, A. and {Senik}, V. and {Vlasenko}, D. and {Antipov}, G. and {Zimnukhov}, D. and {Minkina}, E. and {Chasovnikov}, A. and {Topolev}, V. and {Kuvshinov}, D. and {Cheryasov}, D. and {Kechin}, Ya. and {Tselik}, Yu. and {Sosnovskij}, A. and {Podesta}, R. and {Lopez}, C. and {Podesta}, F. and {Francile}, C. and {Rebolo}, R. and {Serra}, M. and {Buckley}, D. and {Gress}, O.~A. and {Budnev}, N.~M. and {Ershova}, O. and {Carrasco}, L. and {Valdes}, J.~R. and {Chavushyan}, V. and {Alvarez}, Patino V.~M. and {Martinez}, J. and {Corella}, A.~R. and {Rodriguez}, L.~H. and {Tlatov}, A. and {Dormidontov}, D. and {Yurkov}, V. and {Gabovich}, A.},
 journal = {GRB Coordinates Network},
 month = {August},
 pages = {1},
 title = {{Fermi GRB 240825A: Global MASTER-Net observations report}},
 volume = {37283},
 year = {2024}
}

@article{2024GCN.37287....1I,
 adsurl = {https://ui.adsabs.harvard.edu/abs/2024GCN.37287....1I},
 author = {{Izzo}, L. and {Malesani}, D.~B.},
 journal = {GRB Coordinates Network},
 month = {August},
 pages = {1},
 title = {{GRB 240825A: LCO optical observations}},
 volume = {37287},
 year = {2024}
}

@article{2024GCN.37289....1L,
 adsurl = {https://ui.adsabs.harvard.edu/abs/2024GCN.37289....1L},
 author = {{Lipunov}, V. and {Buckley}, D. and {Zhirkov}, K. and {Gorbunov}, I. and {Balanutsa}, P. and {Antipov}, G. and {Kuznetsov}, A. and {Tiurina}, N. and {Gorbovskoy}, E. and {Vlasenko}, D. and {Kechin}, Ya. and {Tselik}, Yu. and {Senik}, V. . and {Gress}, O. and {Budnev}, N. ''ISU'' and {Sosnovskij}, A. and {Podesta}, C. Francile. F. and {Podesta}, R. and {Rebolo}, R. and {Serra}, M. ''The Instituto de Astrofisica de Canarias'' and {Tlatov}, A. and {Dormidontov}, D. and {Gabovich}, A. and {Yurkov}, V.},
 journal = {GRB Coordinates Network},
 month = {August},
 pages = {1},
 title = {{GRB 240825A: MASTER optical counterpart observation}},
 volume = {37289},
 year = {2024}
}

@article{2024GCN.37290....1E,
 adsurl = {https://ui.adsabs.harvard.edu/abs/2024GCN.37290....1E},
 author = {{Evans}, P.~A. and {Goad}, M.~R. and {Osborne}, J.~P. and {Beardmore}, A.~P. and {Swift-XRT Team.}},
 journal = {GRB Coordinates Network},
 month = {August},
 pages = {1},
 title = {{GRB 240825A: Enhanced Swift-XRT position}},
 volume = {37290},
 year = {2024}
}

@article{2024GCN.37291....1L,
 adsurl = {https://ui.adsabs.harvard.edu/abs/2024GCN.37291....1L},
 author = {{Leonini}, S. and {Conti}, M. and {Rosi}, P. and {Tinjaca Ramirez}, L.~M. and {Dainotti}, M.~G. and {Niino}, Y. and {Kalinowski}, K. and {De Simone}, B.},
 journal = {GRB Coordinates Network},
 month = {August},
 pages = {1},
 title = {{GRB 240825A: Montarrenti Observatory optical observations}},
 volume = {37291},
 year = {2024}
}

@article{2024GCN.37292....1S,
 adsurl = {https://ui.adsabs.harvard.edu/abs/2024GCN.37292....1S},
 author = {{SVOM/C-GFT Team} and {Wu}, Chao and {Kang}, Zhe and {Xin}, Liping and {Han}, Xuhui and {Zhang}, Pinpin and {Lu}, Xiaomeng and {Li}, Zhenwei and {Lv}, You and {Zhang}, Ruosong and {Xiao}, Yujiei and {SVOM JSWG} and {Wei}, Jian-Yan and {Cordier}, Bertrand and {Zhang}, Shuang-Nan and {Basa}, Stephane and {Att{\'e}ia}, Jean-Luc and {Claret}, Arnaud and {Dai}, Zi-Gao and {Daigne}, Frederic and {Deng}, Jin-Song and {Goldwurm}, Andrea and {G{\"o}tz}, Diego and {Han}, Xu-Hui and {Lachaud}, Cyril and {Liang}, En-Wei and {Qiu}, Yu-Lei and {Vergani}, Susanna and {Wang}, Jing and {Wu}, Chao and {Xin}, Li-Ping and {Zhang}, Bing},
 journal = {GRB Coordinates Network},
 month = {August},
 pages = {1},
 title = {{GRB 240825A: SVOM/C-GFT optical observations}},
 volume = {37292},
 year = {2024}
}

@article{2024GCN.37293....1M,
 adsurl = {https://ui.adsabs.harvard.edu/abs/2024GCN.37293....1M},
 author = {{Martin-Carrillo}, A. and {Schneider}, B. and {Pugliese}, G. and {Izzo}, L. and {Malesani}, D.~B. and {Saccardi}, A. and {Laskar}, T. and {Agui Fernandez}, J.~F. and {Vergani}, S.~D. and {Stargate Collaboration}},
 journal = {GRB Coordinates Network},
 month = {August},
 pages = {1},
 title = {{GRB 240825A: VLT/X-shooter redshift}},
 volume = {37293},
 year = {2024}
}

@article{2024GCN.37294....1G,
 adsurl = {https://ui.adsabs.harvard.edu/abs/2024GCN.37294....1G},
 author = {{Gropp}, J.~D. and {Osborne}, J.~P. and {Page}, K.~L. and {Beardmore}, A.~P. and {Perri}, M. and {D'Elia}, V. and {Sbarufatti}, B. and {Dichiara}, S. and {Kennea}, J.~A. and {Evans}, P.~A. and {Swift-XRT Team}},
 journal = {GRB Coordinates Network},
 month = {August},
 pages = {1},
 title = {{GRB 240825A: Swift-XRT refined Analysis}},
 volume = {37294},
 year = {2024}
}

@article{2024GCN.37295....1B,
 adsurl = {https://ui.adsabs.harvard.edu/abs/2024GCN.37295....1B},
 author = {{Brivio}, R. and {Ferro}, M. and {D'Avanzo}, P. and {Covino}, S. and {Fugazza}, D. and {REM Team}},
 journal = {GRB Coordinates Network},
 month = {August},
 pages = {1},
 title = {{GRB 240825A: REM detection of the optical/NIR afterglow}},
 volume = {37295},
 year = {2024}
}

@article{2024GCN.37296....1K,
 adsurl = {https://ui.adsabs.harvard.edu/abs/2024GCN.37296....1K},
 author = {{Kuin}, N.~P.~M. and {Gupta}, R. and {Swift/UVOT Team}},
 journal = {GRB Coordinates Network},
 month = {August},
 pages = {1},
 title = {{GRB 240825A: Swift/UVOT Detection}},
 volume = {37296},
 year = {2024}
}

@article{2024GCN.37299....1O,
 adsurl = {https://ui.adsabs.harvard.edu/abs/2024GCN.37299....1O},
 author = {{Odeh}, Mohammad and {Guessoum}, Nidhal},
 journal = {GRB Coordinates Network},
 month = {August},
 pages = {1},
 title = {{GRB 240825A: AKO Optical Follow-Up Observations}},
 volume = {37299},
 year = {2024}
}

@article{2024GCN.37300....1L,
 adsurl = {https://ui.adsabs.harvard.edu/abs/2024GCN.37300....1L},
 author = {{Le Floc'h}, E. and {Adami}, C. and {Schneider}, B. and {Saccardi}, A. and {Basa}, S. and {Dennefeld}, M. and {Sch{\"u}ssler}, F. and {Mistral Grb Collaboration}},
 journal = {GRB Coordinates Network},
 month = {August},
 pages = {1},
 title = {{GRB 240825A : MISTRAL/T193 OHP optical follow-up of the afterglow}},
 volume = {37300},
 year = {2024}
}

@article{2024GCN.37302....1F,
 adsurl = {https://ui.adsabs.harvard.edu/abs/2024GCN.37302....1F},
 author = {{Frederiks}, D. and {Lysenko}, A. and {Ridnaia}, A. and {Svinkin}, D. and {Tsvetkova}, A. and {Ulanov}, M. and {Cline}, T. and {Konus-Wind Team}},
 journal = {GRB Coordinates Network},
 month = {August},
 pages = {1},
 title = {{Konus-Wind detection of GRB 240825A (bright/long)}},
 volume = {37302},
 year = {2024}
}

@article{2024GCN.37303....1G,
 adsurl = {https://ui.adsabs.harvard.edu/abs/2024GCN.37303....1G},
 author = {{Guiffreda}, O. and {Durbak}, J. and {Atri}, S. and {Kutyrev}, A.~S. and {Troja}, E. and {De}, K. and {Cenko}, S.~B.},
 journal = {GRB Coordinates Network},
 month = {August},
 pages = {1},
 title = {{GRB 240825A: PRIME near-infrared detection}},
 volume = {37303},
 year = {2024}
}

@article{2024GCN.37304....1Z,
 adsurl = {https://ui.adsabs.harvard.edu/abs/2024GCN.37304....1Z},
 author = {{Zheng}, WeiKang and {Filippenko}, Alexei V. and {KAIT GRB team}},
 journal = {GRB Coordinates Network},
 month = {August},
 pages = {1},
 title = {{GRB 240825A: KAIT optical observations}},
 volume = {37304},
 year = {2024}
}

@article{2024GCN.37306....1W,
 adsurl = {https://ui.adsabs.harvard.edu/abs/2024GCN.37306....1W},
 author = {{Wang}, B. -T. and {Song}, F. -F. and {Li}, R. -Z. and {Mao}, J. and {Xin}, Y. -X. and {Bai}, J. -M.},
 journal = {GRB Coordinates Network},
 month = {August},
 pages = {1},
 title = {{GRB 240825A: GMG Continued Optical Observation on the Second Night}},
 volume = {37306},
 year = {2024}
}

@article{2024GCN.37307....1M,
 adsurl = {https://ui.adsabs.harvard.edu/abs/2024GCN.37307....1M},
 author = {{Maksut}, Z. and {Grossan}, B. and {Komesh}, T. and {Abdullayev}, Z. and {Krugov}, M. and {Abdikamalov}, E.},
 journal = {GRB Coordinates Network},
 month = {August},
 pages = {1},
 title = {{GRB 240825A: NUTTelA-TAO / BSTI Optical Limits}},
 volume = {37307},
 year = {2024}
}

@article{2024GCN.37310....1M,
 adsurl = {https://ui.adsabs.harvard.edu/abs/2024GCN.37310....1M},
 author = {{Melandri}, A. and {D'Avanzo}, P. and {Ferro}, M. and {D'Elia}, V. and {Malesani}, D.~B. and {Rossi}, A. and {Harutyunyan}, A. and {Carosati}, D. and {CIBO Collaboration}},
 journal = {GRB Coordinates Network},
 month = {August},
 pages = {1},
 title = {{GRB 240825A: TNG optical photometry and spectroscopy}},
 volume = {37310},
 year = {2024}
}

@article{2024GCN.37313....1M,
 adsurl = {https://ui.adsabs.harvard.edu/abs/2024GCN.37313....1M},
 author = {{Moskvitin}, A.~S. and {Spiridonova}, O.~I. and {GRB follow-up Team.}},
 journal = {GRB Coordinates Network},
 month = {August},
 pages = {1},
 title = {{GRB 240825A: SAO RAS optical observations}},
 volume = {37313},
 year = {2024}
}

@article{2024GCN.37335....1G,
 adsurl = {https://ui.adsabs.harvard.edu/abs/2024GCN.37335....1G},
 author = {{Romanov}, Filipp Dmitrievich},
 journal = {GRB Coordinates Network},
 month = {August},
 pages = {1},
 title = {{GRB 240825A: iTelescope optical observations}},
 volume = {37335},
 year = {2024}
}

@article{2024GCN.37336....1M,
 adsurl = {https://ui.adsabs.harvard.edu/abs/2024GCN.37336....1M},
 author = {{Moskvitin}, A.~S. and {Spiridonova}, O.~I. and {GRB follow-up Team}},
 journal = {GRB Coordinates Network},
 month = {August},
 pages = {1},
 title = {{GRB 240825A: further SAO RAS optical observations}},
 volume = {37336},
 year = {2024}
}

@article{2024GCN.37338....1S,
 adsurl = {https://ui.adsabs.harvard.edu/abs/2024GCN.37338....1S},
 author = {{SVOM/VT Team} and {Qiu}, Y.~L. and {Xin}, L.~P. and {Li}, H.~L. and {Wu}, C. and {Han}, X.~H. and {Cai}, H.~B. and {Xu}, Y. and {Xiao}, Y.~J. and {Zhang}, P.~P. and {Lan}, L. and {Xie}, W.~J. and {Lu}, X.~M. and {Zhang}, R.~S. and {Xu}, D.~W. and {Li}, G.~W. and {Zhang}, J. and {Dan}, L.~J. and {Zou}, G.~Y. and {Wang}, C.~J. and {Du}, Y.~F. and {Huang}, C. and {Jesse}, P. and {SVOM JSWG} and {Wei}, Jian-Yan and {Cordier}, Bertrand and {Zhang}, Shuang-Nan and {Basa}, Stephane and {Att{\'e}ia}, Jean-Luc and {Claret}, Arnaud and {Dai}, Zi-Gao and {Daigne}, Frederic and {Deng}, Jin-Song and {Goldwurm}, Andrea and {G{\"o}tz}, Diego and {Han}, Xu-Hui and {Lachaud}, Cyril and {Liang}, En-Wei and {Qiu}, Yu-Lei and {Vergani}, Susanna and {Wang}, Jing and {Wu}, Chao and {Xin}, Li-Ping and {Zhang}, Bing and {SVOM Team}},
 journal = {GRB Coordinates Network},
 month = {August},
 pages = {1},
 title = {{GRB 240825A: SVOM/VT optical detection}},
 volume = {37338},
 year = {2024}
}

@article{2024GCN.37355....1M,
 adsurl = {https://ui.adsabs.harvard.edu/abs/2024GCN.37355....1M},
 author = {{Moss}, M.~J. and {Barthelmy}, S.~D. and {Gupta}, R. and {Krimm}, H.~A. and {Laha}, S. and {Lien}, A.~Y. and {Markwardt}, C.~B. and {Palmer}, D.~M. and {Parsotan}, T. and {Sadaula}, D. and {Sakamoto}, T.},
 journal = {GRB Coordinates Network},
 month = {August},
 pages = {1},
 title = {{GRB 240825A: Swift-BAT refined analysis}},
 volume = {37355},
 year = {2024}
}

@article{2024GCN.37361....1F,
 adsurl = {https://ui.adsabs.harvard.edu/abs/2024GCN.37361....1F},
 author = {{Freeburn}, J. and {Andreoni}, I.},
 journal = {GRB Coordinates Network},
 month = {August},
 pages = {1},
 title = {{GRB 240825A: SOAR observations}},
 volume = {37361},
 year = {2024}
}

@article{2024GCN.37367....1B,
 adsurl = {https://ui.adsabs.harvard.edu/abs/2024GCN.37367....1B},
 author = {{Belkin}, S. and {Reva}, I. and {Pozanenko}, A. and {Pankov}, N. and {GRB FuN}, the IKI},
 journal = {GRB Coordinates Network},
 month = {September},
 pages = {1},
 title = {{GRB 240825A: TSHAO Optical Upper Limit}},
 volume = {37367},
 year = {2024}
}

@article{2024GCN.37372....1D,
 adsurl = {https://ui.adsabs.harvard.edu/abs/2024GCN.37372....1D},
 author = {{de Wet}, S. and {Vreeswijk}, P.~M. and {Groot}, P.~J. and {Meerlicht Consortium}},
 journal = {GRB Coordinates Network},
 month = {September},
 pages = {1},
 title = {{GRB 240825A: MeerLICHT afterglow detection}},
 volume = {37372},
 year = {2024}
}

@article{2024GCN.37373....1S,
 adsurl = {https://ui.adsabs.harvard.edu/abs/2024GCN.37373....1S},
 author = {{SVOM/C-GFT Team} and {Chao Wu} and {Kang}, Zhe and {Xin}, Liping and {Han}, Xuhui and {Zhang}, Pinpin and {Lu}, Xiaomeng and {Li}, Zhenwei and {Lv}, You and {Zhang}, Ruosong and {Xiao}, Yujie and {SVOM JSWG} and {Wei}, Jian-Yan and {Cordier}, Bertrand and {Zhang}, Shuang-Nan and {Basa}, St{\'e}phane and {Att{\'e}ia}, Jean-Luc and {Claret}, Arnaud and {Dai}, Zi-Gao and {Daigne}, Fr{\'e}d{\'e}ric and {Deng}, Jin-Song and {Goldwurm}, Andrea and {G{\"o}tz}, Diego and {Han}, Xu-Hui and {Lachaud}, Cyril and {Liang}, En-Wei and {Qiu}, Yu-Lei and {Vergani}, Susanna and {Wang}, Jing and {Wu}, Chao and {Xin}, Li-Ping and {Zhang}, Bing and {SVOM team}},
 journal = {GRB Coordinates Network},
 month = {September},
 pages = {1},
 title = {{GRB 240825A: Magnitude Correction in SVOM/C-GFT Optical Observations}},
 volume = {37373},
 year = {2024}
}

@article{2024GCN.37400....1L,
 adsurl = {https://ui.adsabs.harvard.edu/abs/2024GCN.37400....1L},
 author = {{Leonini}, S. and {Conti}, M. and {Rosi}, P. and {Tinjaca Ramirez}, L.~M. and {Lorini}, A. and {Verna}, G. and {Bonnoli}, G.},
 journal = {GRB Coordinates Network},
 month = {September},
 pages = {1},
 title = {{GRB 240825A: Montarrenti Observatory optical follow-up observations}},
 volume = {37400},
 year = {2024}
}

@article{2024GCN.37454....1P,
 adsurl = {https://ui.adsabs.harvard.edu/abs/2024GCN.37454....1P},
 author = {{Paek}, Gregory S.~H. and {Im}, Myungshin and {Choi}, Hyeonho and {Chang}, Seo-Won and {Kim}, Ji Hoon and {7-Dimensional Telescope Collaboration}},
 journal = {GRB Coordinates Network},
 month = {September},
 pages = {1},
 title = {{GRB 240825A: 7DT Optical Upper Limit}},
 volume = {37454},
 year = {2024}
}

@article{2024GCN.37638....1M,
 adsurl = {https://ui.adsabs.harvard.edu/abs/2024GCN.37638....1M},
 author = {{Maiorano}, E. and {Rossi}, A. and {Palazzi}, E. and {Paris}, D. and {Malesani}, D.~B. and {De Pasquale}, M. and {CIBO Collaboration}},
 journal = {GRB Coordinates Network},
 month = {September},
 pages = {1},
 title = {{GRB 240825A: LBT optical observations}},
 volume = {37638},
 year = {2024}
}

@article{2025ApJ...979...38C,
 adsurl = {https://ui.adsabs.harvard.edu/abs/2025ApJ...979...38C},
 archiveprefix = {arXiv},
 author = {{Cheng}, Yehao and {Pan}, Yu and {Yang}, Yuan-Pei and {Zhang}, Jinghua and {Du}, Guowang and {Fang}, Yuan and {Kumar}, Brajesh and {Guo}, Helong and {Er}, Xinzhong and {Chen}, Xinlei and {Liu}, Chenxu and {Wang}, Tao and {Qin}, Zhenfei and {Jin}, Yicheng and {Zou}, Xingzhu and {Han}, Xuhui and {Zhang}, Pinpin and {Xin}, Liping and {Wu}, Chao and {Lian}, Jianhui and {Liu}, Xiangkun and {Liu}, Xiaowei},
 doi = {10.3847/1538-4357/ad9ea1},
 eid = {38},
 eprint = {2409.14716},
 journal = {\apj},
 month = {jan},
 number = {1},
 pages = {38},
 primaryclass = {astro-ph.HE},
 title = {{Simultaneous Multiband Photometry of the Early Optical Afterglow of GRB 240825A with Mephisto}},
 volume = {979},
 year = {2025}
}

@article{2025arXiv250702806W,
 adsurl = {https://ui.adsabs.harvard.edu/abs/2025RAA....25j5003W},
 archiveprefix = {arXiv},
 author = {{Wu}, Chao and {Wang}, Yun and {Li}, Hua-Li and {Xin}, Li-Ping and {Xu}, Dong and {Schneider}, Benjamin and {de Ugarte Postigo}, Antonio and {Lamb}, Gavin and {Reguitti}, Andrea and {Saccardi}, Andrea and {Gao}, Xing and {Li}, Xing-Ling and {Wang}, Qiu-Li and {Zhang}, Bing and {Wei}, Jian-Yan and {Zhang}, Shuang-Nan and {Daigne}, Fr{\'e}d{\'e}ric and {Atteia}, Jean-Luc and {Bernardini}, Maria-Grazia and {Cai}, Hong-bo and {Claret}, Arnaud and {Cordier}, Bertrand and {Deng}, Jin-Song and {Godet}, Olivier and {G{\"o}tz}, Diego and {Han}, Xu-Hui and {Kang}, Zhe and {Li}, Guang-Wei and {Li}, Zhen-Wei and {Liu}, Cheng-Zhi and {Lu}, Xiao-Meng and {Lv}, You and {Osborne}, Julian P. and {Palmerio}, Jesse T. and {Qiu}, Yu-Lei and {Schanne}, St{\'e}phane and {Turpin}, Damien and {Vergani}, Susanna Diana and {Wang}, Jing and {Xiao}, Yu-Jie and {Xie}, Wen-Jin and {Xu}, Yang and {Yao}, Zhu-Heng and {Zhang}, Pin-Pin and {Zhang}, Ruo-Son and {Zhu}, Cheng-Wei and {Brivio}, Riccardo and {Covino}, Stefano and {D'Avanzo}, Paolo and {Ferro}, Matteo and {Melandri}, Andrea and {Rossi}, Andrea and {Ag{\"u}{\'\i} Fern{\'a}ndez}, Jos{\'e} Feliciano and {Th{\"o}ne}, Christina C. and {Bai}, Chun-Hai and {Esamdin}, Ali and {Iskandar}, Abdusamatjan and {Yaqup}, Shahidin and {Zhang}, Yu and {Zhong}, Tu-Hong and {Fu}, Shao-Yu and {Jiang}, Shuai-Qing and {Liu}, Xing and {An}, Jie and {Zhu}, Zi-Pei and {Cao}, Jia-Xin and {Liang}, En-Wei and {Lin}, Da-Bin and {Wang}, Xiang-Gao and {Du}, Guo-Wang and {Er}, Xin-Zhong and {Fang}, Yuan and {Liu}, Xiao-Wei and {Adami}, Christophe and {Dennefeld}, Michel and {Le Floc'h}, Emeric and {Fynbo}, Johan Peter Uldall and {Jakobsson}, P{\'a}ll and {Malesani}, Daniele Bj{\o}rn and {Jin}, Zhi-Ping and {Ren}, Jia and {Wang}, Hao and {Wei}, Da-Ming and {Zhou}, Hao and {Campana}, Sergio and {Kobayashi}, Shiho and {De Pasquale}, Massimiliano},
 doi = {10.1088/1674-4527/adeaf0},
 eid = {105003},
 eprint = {2507.02806},
 journal = {Research in Astronomy and Astrophysics},
 month = {October},
 number = {10},
 pages = {105003},
 primaryclass = {astro-ph.HE},
 title = {{GRB 240825A: Early Reverse Shock and Its Physical Implications}},
 volume = {25},
 year = {2025}
}

@article{pymc2023,
 author = {Abril-Pla Oriol and Andreani Virgile and Carroll Colin and Dong Larry and Fonnesbeck Christopher J. and Kochurov Maxim and Kumar Ravin and Lao Jupeng and Luhmann Christian C. and Martin Osvaldo A. and Osthege Michael and Vieira Ricardo and Wiecki Thomas and Zinkov Robert},
 doi = {10.7717/peerj-cs.1516},
 journal = {{PeerJ} Computer Science},
 pages = {e1516},
 publisher = {{PeerJ}},
 title = {PyMC: A Modern and Comprehensive Probabilistic Programming Framework in Python},
 volume = {9},
 year = {2023}
}

@article{Karpov_2025,
 abstractnote = {
We present a simple web-based tool, STDWeb, for a quick-look photometry and transient detection in astronomical images. It tries to implement a self-consistent and mostly automatic data analysis workflow that would work on any image uploaded to it, allowing to perform basic interactive masking, object detection, astrometric calibration of the image, and building the photometric solution based on a selection of catalogues and supported filters, optionally including the colour term and positionally varying zero point. It also allows you to do image subtraction using either user-provided or automatically downloaded template images, and do a forced photometry for a specified target in either original or difference images, as well as transient detection with basic rejection of artefacts. The tool may be easily deployed allowing its integration into the infrastructure of robotic telescopes or data archives for an effortless analysis of their images.},
 author = {Karpov, Sergey},
 doi = {10.14311/AP.2025.65.0050},
 journal = {Acta Polytechnica},
 month = {Mar.},
 number = {1},
 pages = {50–64},
 place = {Prague, Czech Republic},
 title = {STDweb: simple transient detection pipeline for the web},
 url = {https://ojs.cvut.cz/ojs/index.php/ap/article/view/9969},
 volume = {65},
 year = {2025}
}

@software{ELISA,
  author       = {Xue, Wang-Chen and
                  Xie, Sheng-Lun and
                  Zheng, Chao and
                  Xiong, Shao-Lin and
                  Li, Xiao-Bo and
                  Chen, Yong and
                  Zhang, Peng and
                  Zhang, Yan-Qiu and
                  Liu, Jia-Cong and
                  Wang, Chen-Wei and
                  Tan, Wen-Jun and
                  Wang, Yue and
                  Zhang, Jin-Peng and
                  Yu, Zheng-Hang},
  title        = {ELISA: Efficient Library for Spectral Analysis in
                  High-Energy Astrophysics},
  month        = sep,
  year         = 2025,
  publisher    = {Zenodo},
  version      = {0.2.1},
  doi          = {10.5281/zenodo.11284456},
  url          = {https://doi.org/10.5281/zenodo.11284456},
}
\bibliographystyle{aasjournalv7}

%% This command is needed to show the entire author+affiliation list when
%% the collaboration and author truncation commands are used.  It has to
%% go at the end of the manuscript.
%\allauthors

%% Include this line if you are using the \added, \replaced, \deleted
%% commands to see a summary list of all changes at the end of the article.
%\listofchanges

\end{CJK*}
\end{document}